%% file: main.tex
\documentclass[11pt,openright,oneside,letterpaper,onecolumn]{report}  

\newcommand{\thesistitle}{Quantum Algorithms for Scientific Computing and Approximate Optimization}
\newcommand{\thesisauthor}{Stuart Andrew Hadfield}  
\newcommand{\thesisyear}{2018}

\input{header}    

\begin{document}
\pagestyle{empty}
\thesistitlepage
\thesiscopyrightpage

    \thispagestyle{empty}
    \begin{center}
    \textbf{\LARGE ABSTRACT} \\[1cm]
     \textbf{\LARGE \thesistitle} \\[1cm]
     \textbf{\LARGE \thesisauthor} \\[1cm]
    \end{center}
    \input{abstract}
    \cleardoublepage

\input{preamble1}

\input{ack.tex}     \cleardoublepage

\thispagestyle{plain}
\strut \vfill
\vskip -3pc
\centerline{
\Large 
This thesis is dedicated to my family, 
I couldn't have done it without you.
}
\vskip 3pc
\centerline{
\Large 
In memory of  J. F. Traub.}
\vfill \strut
\cleardoublepage


\input{epigraph.tex}

 \cleardoublepage

\input{preamble2}

\input{_ch_intro}

\input{_ch_sciComp}  
\input{_ch_ExcitedStates}

\input{_ch_HamSim}

\input{_ch_QAOA1}

\input{_ch_QAOA2}


\input{_ch_conclusions}

\cleardoublepage  
\addcontentsline{toc}{chapter}{Bibliography}
\bibliographystyle{acm} 
\bibliography{master.bib}

\begin{appendices}  
\input{_app_QC}
\input{_app_QAOtoolkit}

\input{_app_sciComp}

\input{_app_HamSim}
\input{_app_QAOA}
\end{appendices}

\end{document}

%% file: header.tex
\usepackage[titletoc]{appendix}

\usepackage{etoolbox}

\usepackage{graphics}
\usepackage{epsfig}
\usepackage{times}
\usepackage{amsmath}
\usepackage{float}
\usepackage{caption}
\usepackage{subfig}
\usepackage{comment}
\usepackage{amsfonts}
\usepackage{amssymb}
\usepackage{algorithm}
\usepackage{algorithmic}   
\usepackage{authblk}
\usepackage{amsthm}
\usepackage{graphicx, rotating}

\usepackage{tikz}

\usepackage{cite}

\usepackage{multicol}
\usepackage{hhline}

\input{Qcircuit}
\input{macros}

\usepackage{amsmath}
\usepackage{fancyhdr}
\usepackage{afterpage}

\usepackage[driverfallback=dvipdfm,pdftitle={\thesistitle},pdfauthor={\thesisauthor},pdfpagemode={UseOutlines},letterpaper,bookmarks,bookmarksopen=true,pdfstartview={FitH},bookmarksnumbered=true,]{hyperref}

\newcommand{\doublespace}{\renewcommand{\baselinestretch}{1.5} \small \normalsize}
\newcommand{\normalspace}{\doublespace}
\newcommand{\thesistitlepage}{
    \normalspace
    \thispagestyle{empty}
    \begin{center}
        \textbf{\LARGE \thesistitle} \\[1cm]
        \textbf{\LARGE \thesisauthor} \\[8cm]
        Submitted in partial fulfillment of the \\
        requirements for the degree \\
        of Doctor of Philosophy \\
        in the Graduate School of Arts and Sciences \\[4cm]
        \textbf{\Large COLUMBIA UNIVERSITY} \\[5mm]
        \thesisyear
    \end{center}
    \clearpage
}

\newcommand{\thesiscopyrightpage}{
    \thispagestyle{empty}
    \strut \vfill
    \begin{center}
      \copyright \thesisyear \\
      \thesisauthor \\
      All Rights Reserved
    \end{center}
    \cleardoublepage
}



%% file: macros.tex
\newcommand{\x}{  {\hat{x}_{s_1}}     }
\newcommand{\xh}{  {\hat{x}_{s_1}}     }
\newcommand{\p}{ {\mathcal{P} }}
\newcommand{\dd}{ {\delta }}

\definecolor{nblue}{rgb}{0.3,0.3,1.0}
\definecolor{ngreen}{rgb}{0.2,0.7,0.2}
\definecolor{nred}{rgb}{0.9,0.1,0}
\definecolor{nmagenta}{rgb}{255,0,255}
\definecolor{nviol}{rgb}{0.7,0.1,0.7}

\newcommand{\Braket}[2]{\left<#1|#2\right>}
\newcommand{\braket}[2]{\left<#1|#2\right>}

\newcommand{\ketbra}[2]{\ket{#1}\!\bra{#2}}

\newcommand{\naturals} {{\mathbb N}}
\newcommand{\nat} {{\mathbb N}}
 
\newcommand{\complex}{{\mathbb C}}
\newcommand{\reals}{{\mathbb R}}
\newcommand{\integers}{{\mathbb Z}}

\newtheorem{theorem}{Theorem}
\newtheorem{rem}{Remark}
\newtheorem{lem}{Lemma}
\newtheorem{prop}{Proposition}
\newtheorem{cor}{Corollary}

\newcommand\e{\varepsilon}

\newcommand{\norm}[1]{\| #1\|}


%% file: abstract.tex
Quantum computation appears to offer significant advantages over classical computation and
this has generated a tremendous interest in the field.
In this thesis we study the application of quantum computers to  computational problems in science and engineering, and to combinatorial optimization problems. We outline the results below.

Algorithms for scientific computing require modules, i.e., building blocks, implementing elementary numerical functions that have well-controlled numerical error, are uniformly scalable and reversible, and that can be implemented efficiently. We derive quantum algorithms and circuits for computing square roots, logarithms, and arbitrary fractional powers, and derive worst-case error and cost bounds.
We describe a 
modular approach to quantum algorithm design as a first step
towards numerical standards and mathematical libraries 
for quantum 
scientific 
computing. 

A fundamental but computationally hard problem in 
physics is to solve the time-independent Schr\"odinger equation. This is accomplished by computing the eigenvalues of the corresponding 
Hamiltonian operator.  The eigenvalues describe the different energy levels of a system.
The cost of classical deterministic algorithms computing these eigenvalues grows exponentially with the number of system 
degrees of freedom. The number of degrees of freedom is typically proportional to the number of particles in a physical system. 
We show 
an efficient quantum algorithm for approximating 
a constant number 
of low-order eigenvalues of a Hamiltonian using a perturbation approach. We apply this algorithm to a special case of the Schr\"odinger equation and show that 
our algorithm succeeds with high probability, 
and has cost 
that scales polynomially with the number of degrees of freedom and the reciprocal of the desired accuracy. 
This improves and extends earlier results on quantum algorithms for estimating the ground state energy.  

We consider the simulation of quantum mechanical systems on a quantum computer. 
We show a novel divide and conquer approach for Hamiltonian simulation. 
Using the Hamiltonian structure, we can obtain faster simulation algorithms. 
Considering a sum of Hamiltonians we split them into groups, simulate each group separately, and combine the partial results. Simulation is customized to take advantage of the properties of each group, and hence yield refined bounds to the overall simulation cost. We illustrate our results using the electronic structure problem of quantum chemistry, where we obtain significantly improved cost estimates under 
mild assumptions. 

We turn to 
combinatorial optimization problems. An important open question is whether quantum computers provide advantages for the approximation of classically hard combinatorial problems. A promising recently proposed approach of Farhi et al. is the 
Quantum Approximate Optimization Algorithm (QAOA). 
We study the application of QAOA to the Maximum Cut problem, and derive analytic performance bounds for the lowest circuit-depth realization, for both general and special classes of graphs. 
Along the way, we develop a general procedure for analyzing the performance of QAOA for other problems, 
and show an example demonstrating the difficulty of obtaining similar results for greater depth. 

We 
show a generalization of QAOA and 
its application to wider classes of 
combinatorial optimization 
problems, in particular, problems with feasibility constraints. We introduce the Quantum Alternating Operator Ansatz, 
which utilizes 
more general unitary operators than the original QAOA proposal.  
Our framework 
facilitates low-resource 
implementations for many applications which may be particularly suitable for early quantum computers. 
We specify design criteria, and 
develop a set of results and tools for mapping diverse problems to 
explicit quantum circuits. 
We 
derive 
constructions for several important prototypical problems including Maximum Independent Set, Graph Coloring, and the Traveling 
Salesman problem, and 
show appealing 
resource cost estimates for their implementations. 

%% file: preamble1.tex

\pagenumbering{roman}
\pagestyle{plain}

\setlength{\footskip}{0.5in}

\setcounter{tocdepth}{2}
\renewcommand{\contentsname}{Table of Contents}
\tableofcontents
\cleardoublepage

\addcontentsline{toc}{section}{List of Figures}
\listoffigures
\cleardoublepage

\addcontentsline{toc}{section}{List of Tables}
\listoftables 
\cleardoublepage

%% file: ack.tex
~\\[1in] 
\textbf{\LARGE Acknowledgments}\\
\noindent

\noindent I am immensely thankful to 
my wonderful academic advisors Professor Alfred V. Aho, Professor Joseph F. Traub, and Dr. Anargyros Papageorgiou for their support, 
mentorship, and patience. 
Together, they taught me the importance of seeking impactful research problems, and perhaps more importantly, perseverance.  
It has been a privilege and a pleasure to work with each of them. 

I am very grateful to Dr. Eleanor G. Rieffel and the other members of the NASA Quantum Artificial Intelligence Laboratory, where I spent a stimulating summer in 2016, 
for countless 
insightful interactions. 
Two chapters of this thesis resulted from research collaborations which began there. 

I would like to especially thank my dissertation committee members Prof. Aho, Dr. Papageorgiou, Dr. Rieffel, Professor Mihalis Yannakakis, and Professor Rocco Servedio for their service and valuable comments on this thesis. 

I am thankful to my many colleagues from the computer science department, and from the wider Columbia community, for countless stimulating discussions and interactions, both professionally and socially.  
Additionally, I wish to generally thank my collaborators and the many members of the broader quantum computing community who have 
positively affected 
my growth as a researcher along the way. 

Chapter \ref{ch:sciComp} of this thesis is a joint work with with Mihir Bhaskar, Anargyros Papageorgiou, and Iasonas Petras \cite{hadfield2016scientific}. The results of Chapters \ref{ch:ExcStates} and \ref{ch:HamSim} are collaborations with Anargyros Papageorgiou \cite{hadfield2015approximating,hadfield2017hamsim}. 
Chapters \ref{ch:QAOAperformance} and \ref{ch:QACOA} are based on joint works with Eleanor Rieffel, Zhihui Wang, Zhang Jiang, Bryan O'Gorman, Davide Venturelli, and Rupak Biswas \cite{wang2017quantum,hadfield2017qaoaPMES,hadfield2017quantum}.

Finally, most importantly, I thank my family and friends,  
especially Bahareh, for their unending patience and understanding during the 
often arduous PhD process. 
Without your constant love, support, and encouragement, I would not be where I am today.

%% file: epigraph.tex
~\\[0.5in]
\textbf{\LARGE Epigraph}\\

\vskip 3pc

\begin{quote}
\textit{The importance of accelerating approximating and computing mathematics by factors like 10,000 or more, lies not only in that one might thereby do in 10,000 times less time problems which one is now doing...but rather in that one will be able to handle problems which are considered completely unassailable at present.}
 
\textit{The projected device, or rather the species of which it is to be the first representative, is so radically new that many of its uses will become clear only after it has been put into operation, and after we have adjusted our mathematical habits and ways of thinking to its existence and possibilities. Furthermore, these uses which are not, or not easily, predictable now, are likely to be the most important ones...because they are farthest removed from what is now feasible.} 
\end{quote}
\hfill John von Neumann to Lewis L. Strauss, 1945 \cite{von2005john}

\vskip 9pc

\begin{quote}
 \textit{The \lq paradox\rq\ is only a conflict between reality and your feeling of what reality \lq ought to be.\rq\ }
 \end{quote}
\hfill Richard Feynman, \textit{The Feynman Lectures on Physics}, 1965 \cite{feynman1965lectures}

%% file: preamble2.tex
\pagestyle{headings}
\pagenumbering{arabic}

%
%
\setlength{\textheight}{8.5in}
\setlength{\footskip}{0in}

\fancypagestyle{plain} {%
\fancyhf{}
\fancyhead[LE,RO]{\thepage}
\fancyhead[RE,LO]{\itshape \leftmark}
\renewcommand{\headrulewidth}{0pt}
}
\pagestyle{plain}

%% file: _ch_intro.tex
\chapter{Introduction}
\label{ch:Intro}

The potential advantages of quantum over classical computers 
for solving hard problems has generated tremendous 
interest in quantum computation. 
Efficient quantum algorithms have been derived not only for 
discrete problems \cite{mosca2012quantum}, such as, famously, integer factorization \cite{Shor},  
but also for important computational problems in science and engineering, such as 
quantum simulation, eigenvalue estimation, 
integration, 
partial differential equations, and 
numerical linear algebra problems \cite{NC,georgescu2014quantum,Qcontinuous}. 
Thus quantum computers have 
immense potential impact on fields ranging from quantum chemistry to computer security to machine learning \cite{lanyon2010towards,broadbent2016quantum,biamonte2016quantum}. 

Indeed, as prototype quantum 
computers begin to emerge over the next few years \cite{IBM_Q,Sete16,Boixo16,Google2017nature}, quantum computing 
is finally posed to transition from a theoretical model to impactful practical computing devices. 
It is 
important to both characterize the power of such devices, 
and to find new and improved algorithms particularly applicable to both small- and medium-scale quantum hardware to take advantage of the emerging technologies. 
In this thesis, 
we take a number of steps toward these goals. 
We study five problems. The first three deal with quantum algorithms for computational problems in science and engineering. The remaining two deal with quantum algorithms for approximate optimization. 
In particular, we study quantum algorithms and circuits for scientific computing, the approximation of ground  and excited state energies of the Sch\"odinger equation, algorithms for Hamiltonian simulation, 
performance analysis of 
the quantum approximate optimization algorithm (QAOA), and a generalization of QAOA particularly suitable for constrained optimization problems and low-resource implementations. 
We briefly describe each of these problems below, providing motivation and summarizing the respective results. 
For the interest of the reader, a brief overview of quantum computation is included as Appendix~\ref{ch:QC}.

\section{Quantum Algorithms and Circuits for Scientific Computing}
The need for 
quantum algorithms for scientific computing, 
to be used as modules in other quantum algorithms, is apparent.
For example, a recent paper deals with the solution of linear systems on a quantum computer \cite{Harrow}.
The authors present an algorithm that requires the (approximate) calculation of the reciprocal of a number followed by the calculation of 
trigonometric functions 
needed in a controlled rotation 
on the way to the final result. 
However, the paper does not give any details about how these operations are to be 
implemented. From a complexity theory point of view this may not be a complication, but certainly
there is a lot of work that is left to be done before one is able to implement the linear systems algorithm
and eventually use it
on a quantum computer. 

In solving scientific and engineering problems, classical algorithms typically use floating point arithmetic and numerical
libraries of special functions. The IEEE Standard for Floating 
Point Arithmetic (IEEE 754-2008) \cite{IEEE754-2008} ensures that such calculations are performed 
with well-defined precision. 
A similar standard is needed 
for quantum computing. 
Many quantum algorithms use the quantum circuit model of computation, typically employing a 
fixed-precision representation of numbers.
Yet there is no standard specifying how arithmetic operations between numbers (of possibly disproportionate 
magnitudes) held in registers of finite length are to be performed, and how to deal with 
error. 
In designing algorithms for scientific computing the most challenging task is to control the error propagation. 
Since registers have finite length it is not reasonable to expect to calculate exactly and propagate all the results of intermediate calculations 
throughout all the stages of an algorithm. Intermediate approximations have to be made.
Most importantly, there are no existing libraries of quantum circuits with performance guarantees, implementing 
functions such as the square root of a number, 
the logarithm, 
or other similar elementary functions. 
Such quantum circuits should be uniformly scalable, and at the same time make efficient use of quantum resources 
to meet physical constraints of potential quantum computing devices of the foreseeable future.

It is worthwhile 
remarking on the direct applicability of classical algorithms to quantum computation. It is known that classical computation is subsumed by quantum computation, i.e., that for any classical algorithm, there exists a quantum algorithm which performs the same computation \cite{NC}. 
This follows from the fact that any classical algorithm (or circuit) can be implemented reversibly, in principle, but with the additional overhead of a possibly  large number of 
{\it ancilla} qubits that must be carried forward throughout the computation. 
For simple circuits consisting of the composition of basic logical operations, this overhead grows with the number of gates.
On the other hand, scientific computing algorithms are quite different. They typically  involve a large number of floating point arithmetic operations computed approximately according to rules that take into account the relative  magnitudes of the operands requiring mantissa shifting, normalization and rounding. Thus they are quite  expensive to implement reversibly because this would require many registers of large size to store all the intermediate results. Moreover, a mechanism is necessary for dealing with roundoff error and overflow which are not reversible operations. 
Hence, the direct simulation on a quantum computer of classical algorithms for scientific computing that have
been implemented in floating point arithmetic quickly becomes quite complicated and prohibitively expensive. 

With this motivation, 
in Chapter \ref{ch:sciComp} we show a modular approach 
forming a basis towards a standard for quantum scientific computing.  
We give efficient quantum algorithms and circuits for computing square roots, logarithms, and arbitrary fractional powers.  We derive worst-case error and cost bounds, providing performance guarantees with respect to the desired accuracy. 
We further illustrate the performance of our algorithms 
with tests comparing them to the respective floating point implementations found in widely used numerical software. 
Our results are important first steps towards mathematical libraries and numerical standards for quantum computing.

\section{Approximating Ground and Excited State Energies on a Quantum Computer}
Quantum mechanical systems are governed by the Schr\"odinger equation, 
where the 
evolution in time of a quantum system 
is determined by 
the \textit{Hamiltonian} operator. 
Hamiltonian eigenvalues describe the 
system energy levels. 
For example, 
computing the energy levels is one of the  
most important tasks in chemistry because they are required for predicting reaction rates and electronic structure; both of which, in particular, depend principally on the low-order energy levels. 
Computing the energy levels is a very hard problem in general.  
The best classical algorithms known 
have costs that grow exponentially in the number of system degrees of freedom~\cite{lanyon2010towards}.  
Therefore, efficient quantum algorithms would be an extremely powerful tool for new science and technology, having tremendous potential impact on the design of new medicines and advanced materials, and 
improving the efficiency of important chemical processes such as nitrogen fixation~\cite{reiher2017elucidating}. 

On the other hand, there are a number of recent results in discrete complexity theory suggesting that many eigenvalue problems 
are very hard even for quantum computers because they are QMA-complete 
\cite{KempeLocalHam,intBosons,schuchDFT,childs2013bose}. 
(Roughly speaking, QMA is the quantum analog of NP, i.e.,  the class of decision problems that can be efficiently \textit{verified} on a quantum computer.) 
However, discrete complexity theory deals with the worst case over large classes of Hamiltonians. It does not provide methods or necessary conditions determining when an  eigenvalue problem is hard. In fact, there is a dichotomy between theory and practice. As stated in  \cite{love2012back}, \lq\lq complexity theoretic proofs of the advantage of many widely used classical algorithms are few and far between.\rq\rq\
Therefore, it is important to develop new quantum algorithms and
to use them for solving  eigenvalue problems for which quantum computers 
can be shown to have a significant advantage over classical 
computers. 

In \cite{GS} the authors developed an algorithm and proved a strong exponential quantum speedup for approximating the ground state energy (i.e., the smallest eigenvaue) of the time-independent Schr\"odinger equation under certain assumptions. In \cite{Qspeedup} it is explained why this problem is different from the QMA-complete problems of discrete complexity theory. In \cite{Convex} an important assumption of \cite{GS} was relaxed and the results extended to the ground state energy approximation for 
the time-independent Schr\"odinger equation with a convex potential. 

An important advance would be to obtain analogous results for approximating excited state energies under weakened assumptions. The techniques used previously for the ground state energy do not extend to excited state energies. Similarly, in computational chemistry, for instance, Hohenberg-Kohn density functional theory (DFT) is strictly limited to ground states \cite{Oliv05,HK1}. There are other flavors of DFT that may provide approximations of excited state energies. However, in general, approximate methods in computational chemistry often succeed in predicting chemical properties yet their level of accuracy varies with the nature of the species and may fail in important instances; see \cite{aspuruGuzik2005yq,lanyon2010towards} and the references therein. Obtaining conditions allowing one to approximate excited state energies with a guaranteed accuracy and a reasonable
cost would provide a valuable insight into the complexity of these problems.

To this end, in Chapter \ref{ch:ExcStates}, under general conditions, and using a perturbation approach, we provide a quantum algorithm that produces estimates of a constant number 
of different low-order eigenvalues. The algorithm relies on a set of trial eigenvectors, whose construction depends on the particular Hamiltonian properties. We illustrate our results by considering a special case of the time-independent Schr\"odinger equation  with $d$ degrees of freedom. Our algorithm computes  estimates of a constant number 
of different low-order 
energy levels with error 
$\e$ and success probability at least~$3/4$, 
with cost polynomial\footnote{We say a quantity $c(n)$ is \textit{polynomial} in 
$n$ if as $n$ becomes large it grows at most as 
$c(n)=O(n^k)$ for some~$k\in\naturals$. We will often write $c = {\rm poly}(n)$ to denote such polynomial scaling.}
in~$\e^{-1}$ and~$d$. This extends earlier results on algorithms for estimating the ground state energy. The technique we present is sufficiently general to apply to 
problems beyond the application studied in Chapter~\ref{ch:ExcStates}. 

\section{Divide and Conquer Hamiltonian Simulation}
Simulating quantum mechanical systems using classical computers appears to be a very hard problem. 
The description of general quantum states grows exponentially with the system size and so does the computational cost of the best classical algorithms known for simulation. This difficulty led Feynman to propose simulation on a quantum computer, i.e., using one quantum system to simulate another. 
He conjectured that quantum computers might be able to carry out 
the simulation more efficiently than 
classical algorithms. 
Subsequently, a long line of research has shown efficient quantum algorithms for simulation for 
a variety of important applications. 
For an overview of quantum simulation see, e.g., 
\cite{Feynman,buluta2009quantum}. 
Early simulation results can be found in \cite{lloyd1996universal,Zalka1,Zalka2,aharonov2003adiabatic}. 
More recent developments can be found in  \cite{berry2007efficient,PZ12,berry2012black,berry2014exponential,berry2015hamiltonian}. 
Related applications to physics and chemistry can be found in 
\cite{AbramsLloydFermi,ortiz2001quantum,byrnes2006simulating,kassal2008polynomial,Kassal,whitfield2011simulation,jordan2014quantum,jordan2014quantum2,wecker2014gate,wecker2015solving,vivsvnak2015quantum}.

In the Hamiltonian simulation problem one is given a Hamiltonian $H$ acting on $q$ qubits, 
a time~$t\in\reals$, and an accuracy demand~$\e$, 
and the goal is to derive an algorithm 
that constructs an operator~$\widetilde{U}$ 
that approximates the unitary operator~$e^{-iHt}$ with error $\|\widetilde{U}-e^{-iHt}\| \leq \e$ measured in the spectral norm. 
The operator~$e^{-iHt}$ corresponds to quantum evolution under the Hamiltonian $H$ for time $t$. 
When the Hamiltonian is given explicitly, the size of the quantum circuit realizing the algorithm is its cost. 
In particular, the cost depends on the complexity parameters $q$, $t$ and $\e^{-1}$. 
On the other hand, when the Hamiltonian is given 
by an oracle, the number of queries (oracle calls) used by the algorithm plays a major role in  its cost, in addition to the number of qubits and the other necessary quantum operations. 
Different types of queries have been considered in the literature. 

Of particular interest are Hamiltonians $H$ that can be expressed as a sum 
$$H=\sum_{j=1}^m H_j,$$
%
where the $H_j$ are Hamiltonians which each 
can be simulated efficiently, 
i.e., we have an 
explicit quantum circuit for implementing each exponential (query) $e^{-iH_jt}$, $j=1,\dots,m$. 
The Born-Oppenheimer electronic Hamiltonian in the second-quantized form, which describes molecular systems, 
enjoys this property, see, e.g., \cite{whitfield2011simulation}. 
Suzuki-Trotter formulas \cite{Suzuki90,Suzuki91} are typically used in the simulation of these 
Hamiltonians. 
Using them Berry et al. \cite{berry2007efficient} showed a quantum algorithm for Hamiltonian simulation with cost (number of exponentials $e^{-iH_jt}$) polynomial in $m$, $t$, $ \|H\|$ and in~$\e^{-1}$. These results were subsequently improved in \cite{PZ12} where the authors observed that if we order the Hamiltonians so that $\| H_1\|\geq \| H_2\|\geq \cdots \| H_m\|$ and if
$\|H_2\| \rightarrow 0$ then 
a single exponential would suffice for the simulation of $H$, and hence showed  improved cost bounds polynomial in $m$, $t$, $\| H_1\|$, $\| H_2\|$ and in~$\e^{-1}$. The dependence of the cost on the norms of the individual Hamiltonians composing $H$ is very important since it can reduce the simulation cost significantly. 

For applications such as simulating the second-quantized electronic Hamiltonian where $m$ is typically very large, the simulation cost can be prohibitive \cite{wecker2014gate}. Moreover, if many Hamiltonians have small, or even negligible, norms compared to the largest, it may be possible to take advantage of this disparity to derive faster algorithms. Indeed, this situation is common in chemistry applications involving the second-quantized electronic Hamiltonian, and heuristics have been proposed to take advantage of this and reduce the simulation cost; see, e.g., \cite{wecker2014gate,poulin2014trotter,babbush2015chemical}. 

The goal, 
then, is to construct algorithms that improve known simulation cost estimates, particularly when $m$ is huge while relatively few Hamiltonians have large norms and many 
 have 
small norms. 
In Chapter 5, we accomplish this goal applying a divide and conquer approach. We partition the 
Hamiltonians into groups, simulate the sum of the Hamiltonians in each group separately, and then  combine the partial results. Simulation is customized to take advantage of the properties of each group, and hence yields refined bounds to the overall simulation cost that reflect these properties. 

We illustrate our results using the electronic structure problem of quantum chemistry. For the second-quantized electronic Hamiltonian describing a molecular system, the number of Hamiltonians in the sum is $m=\Theta(\mathcal{N}^4)$, where $\mathcal{N}$ is a parameter proportional to the number of particles.  
For many 
important problems in chemistry, simulating this Hamiltonian 
is well beyond the reach of the best classical algorithms. 
Standard quantum algorithms for simulation have polynomial cost that, roughly speaking, grows 
as 
$\mathcal{N}^8$ or $\mathcal{N}^9$. 
For moderately sized problems of interest where, say, $\mathcal{N}=100$, this cost dependence is already prohibitive and hence simulation is considered to be a cost bottleneck \cite{wecker2014gate}. 
Using our divide and conquer approach, 
we show 
under mild assumptions that the cost estimates for 
our algorithms scale with $\mathcal{N}$ in the range $\mathcal{N}^5$ to $\mathcal{N}^7$. 
Hence, our approach may reduce the simulation cost by several orders of magnitude, 
allowing the simulation of much larger chemical systems, 
especially on early quantum computers.

\section{Quantum Approximate Optimization} 
Combinatorial optimization problems are ubiquitous in science, engineering and operations research. 
Many important problems are not only NP-hard to solve, 
but are NP-hard even to approximate 
better than some factor. Two well-known examples are 
Maximum Satisfiability, where given a Boolean formula 
in conjunctive normal form 
we seek an assignment of 
the variables satisfying as many clauses as possible, and the Traveling 
Salesman problem, where 
given a list of cities and the distances between them 
we seek a tour visiting all cities with total length as small as possible.
Indeed,  
for many important applications 
we must settle for algorithms or heuristics producing approximate solutions, and this
has led to rich theories of approximation algorithms, 
approximation complexity,
 and the hardness of approximation  \cite{Vazirani,ShmoysBook,trevisan2004inapproximability,Ausiello2012complexity,AroraBarak}.

The difficulty in solving optimization problems has generated much excitement about the possibility of using quantum devices to approximately solve them. 
A particularly prominent 
metaheuristic is quantum annealing (or, more specifically, adiabatic quantum optimization)  \cite{nishimoriQA,farhi2000quantum}, which can be implemented on quantum devices that are much simpler than universal quantum computers and hence easier in principle to design and engineer. 
Currently, the commercially available family of D-WAVE machines provide quantum annealing with up to 2000 qubits, but with restricted connectivity and other limitations \cite{boixo2013quantum}. 
Google and the IARPA Quantum Enhanced Optimization program, for example, are working towards building the next generation of quantum annealing devices.  
However, 
these quantum annealers are heuristic solvers, since generally we do not have performance guarantees, and these machines must be characterized empirically; see, e.g., \cite{mcgeoch2014adiabatic}. 
These devices remain at an early stage, with 
hardware constraints severely limiting the classes of and sizes of problem to which they are applicable. 
It remains an important open problem whether or not such devices, 
or even
future improved versions, can provide advantages for real-world optimization problems \cite{boixo2013quantum,denchev2016computational}. 
Since the D-WAVE 
and other proposed quantum annealers are 
 not believed to be universal for quantum computation,\footnote{A closely related 
 computational model, adiabatic quantum computation \cite{aharonov2008adiabatic}, is universal.} 
the pertinent question is: 
do quantum computers based on the quantum circuit model of computation, which are universal and potentially more powerful than existing quantum annealers, offer significant advantages for approximate optimization? 

More concretely, we pose two general motivating questions. 
\begin{itemize}
\item Can quantum computers find good approximate solutions \textit{faster} than classical computers?
\item Do there exist important problems that can be approximated \textit{better} on a quantum computer than by any classical algorithm?
\end{itemize}

For some problems, we have tight classical results where the best algorithm known achieves the best possible 
approximation ratio\footnote{An algorithm achieves an $R$-approximation if it always produces a solution within a multiplicative factor of $R$ or better of the optimal solution, in time polynomial in the problem size; see e.g. \cite{Ausiello2012complexity}.}
 (under a standard assumption from computational complexity theory such as P$\neq$NP). 
For such cases, 
we do not believe
that quantum computers could 
achieve a much better approximation ratio, 
since this 
would imply that quantum computers could efficiently solve NP-hard problems; see e.g. 
 \cite{trevisan2004inapproximability,Ausiello2012complexity}. 
On the other hand, for some problems there exist 
substantial gaps between the ratio achieved by the best classical algorithms known and the sharpest hardness of approximation result; problems in this category are a promising class where quantum computers may offer an advantage. 
Indeed, an illuminating example is 
the problem Max-E3Lin2 where we seek to maximally satisfy a set of three-variable linear equations over  $\integers_2$. Remarkably, 
an efficient quantum algorithm producing a better approximation ratio than the best classical algorithm known was found \cite{Farhi2014b}, only to subsequently inspire an even 
better (by a logarithmic factor) classical algorithm~\cite{barak2015beating}. 
An important research direction, generally, is to find 
hard problems where quantum algorithms may provide 
advantages for approximation. 

Recently, Farhi et al.~\cite{Farhi2014}
proposed a new class of quantum algorithms, the Quantum Approximate
Optimization Algorithm (QAOA), to tackle
challenging approximate optimization problems on gate model
quantum computers. 
A handful of recent papers suggest 
QAOA circuits are powerful for computation 
\cite{Farhi2014b,Farhi2016,Shabani16,Jiang17}.
In QAOA, the algorithms require the determination of certain parameters, and their success 
relies on one being able to find a good set of such parameters. 
QAOA algorithms are characterized by their depth $p$, which we write QAOA$_p$. 
In particular, the algorithm consists of applications of a phase operator and a mixing operator, applied in alternation~$p$ times each. Generally, the performance of the algorithm improves with higher~$p$. 
The lowest depth version QAOA$_1$ has provable performance guarantees for certain problems \cite{Farhi2014,Farhi2014b,wang2017quantum}. 
Characterizing the performance of QAOA$_p$ circuits for $p>1$, which is 
the regime where we expect to see the most advantage for applications, 
remains the most important open problem
towards understanding whether these circuits can outperform classical algorithms. 

In Chapter \ref{ch:QAOAperformance}, we study the application of QAOA to the Maximum Cut problem, the original application considered in \cite{Farhi2014}. 
We derive 
novel analytic results towards characterizing the algorithm's performance and finding the optimal choice of 
parameters, for the $p=1$ case. 
In particular, we show bounds for the expected approximation ratio on both general and restricted classes of graphs. Our results extend earlier numerical results for special cases 
\cite{Farhi2014}. We apply our technique to a particular case of Maximum Cut for $p=2$, obtaining a complicated expression which demonstrates the difficulty of obtaining similar results for $p>1$. 
Along the way, we 
provide a procedure which can be used to derive similar results for other problems of interest. 
Our results significantly expand 
the known performance 
bounds for QAOA,  
and are important steps towards developing techniques to characterize its power in more general applications.

In Chapter \ref{ch:QACOA}, we show a generalization of QAOA, the \textit{Quantum Alternating Operator Ansatz}, enabling the exploration of QAOA approaches to a much broader selection of problems. 
In particular, we consider \textit{constrained} optimization problems, where we
are additionally given feasibility constraints and we seek the best solution 
within the feasible subset. 
An example 
application to such a problem was considered in \cite[Sec. VII]{Farhi2014}, but without 
giving the details. 
We carefully specify design criteria and requirements for general problems. 
Specifically, we 
allow for much more general mixing operators than those considered in \cite{Farhi2014}.
Importantly, we show constructions that restrict the evolution of the QAOA algorithm 
to the subspace of states corresponding to feasible solutions, which avoids altogether the difficulty and cost of dealing with infeasible states directly. 
(Without this property, many algorithm measurement outcomes could yield invalid solutions, which would have to be carefully accounted for in analyzing the success probability of the algorithm, or dealt with by some other means, often with considerable additional cost.) 
Our constructions facilitate low-resource implementations, which is particularly 
 advantageous 
 for early quantum computers. 
Moreover, we provide a toolkit of results for mapping Boolean and 
real functions to Hamiltonians, which 
may be useful for other quantum algorithms for approximation such as quantum annealing.  
We then realize our approach by deriving explicit constructions for a sequence of important prototypical problems, including Maximum Independent Set, several optimization problems related to Graph Coloring, and the Traveling Salesman problem.  
In each case we show that the numbers of qubits and basic quantum gates required are relatively low, e.g., scaling linearly or quadratically with the problem parameters, and hence  
these resource counts enable us to 
identify problems especially suitable for implementation on near-term quantum computing hardware.

%% file: _ch_sciComp.tex
\addtocounter{algorithm}{-1}   

\setlength{\belowcaptionskip}{4pt}
\captionsetup[table]{belowskip=4pt}
\setlength{\abovecaptionskip}{4pt}
\captionsetup[table]{aboveskip=4pt}

\chapter{Quantum Algorithms and Circuits for Scientific Computing} 
\label{ch:sciComp}

\section{Introduction}
Recent results \cite{Poisson,Harrow,TaShma} suggest that quantum computers may have a substantial advantage over classical computers for solving numerical  linear algebra problems and, more  generally, problems of computational science and engineering \cite{Qcontinuous}. Scientific computing applications require the evaluation of elementary functions like those found in mathematics libraries of programming languages, where the calculations are performed using floating point arithmetic.
Although in principle quantum computers can always directly simulate any classical algorithm, generally there is no guarantee that such simulations remain practical. 
Hence, it is important to develop efficient quantum algorithms and circuits implementing such functions, 
towards the goal of establishing a standard for numerical computation on quantum computers. 

In designing algorithms for scientific computing the most challenging task is to control the error propagation and to do so efficiently. In this sense, it is important to derive reusable modules with well-controlled error bounds.
In this chapter we continue this line of work of \cite{Poisson} 
by deriving quantum circuits which, given a number $w$ 
(represented as a fixed-precision binary number), 
compute the functions $w^{1/2^i}$ for $i=1,\dots,k$, $\ln(w)$ 
(and thereby the logarithm in different bases), and $w^f$ with $f\in [0,1)$. 
Our design is modular, combining a number of elementary quantum circuits to implement the functions, and 
for each circuit we provide cost and worst-case error estimates. 
We also illustrate the accuracy of our algorithms through examples comparing 
their error with that of widely used numerical software such as Matlab. In summary, our tests show 
that using a moderate amount of resources, our algorithms compute the values of the functions 
matching the corresponding values obtained using scientific computing software (using floating point arithmetic) with 12 to 16 decimal digits of accuracy. 
Our circuits complement those given in \cite{Poisson} for computing the reciprocal and basic trigonometric functions, and together are important first steps towards 
establishing libraries of quantum circuits for mathematical functions.  

We consider the quantum circuit model of computation where arithmetic operations are 
performed with fixed
precision.
We use a small number of elementary modules, or building blocks, to implement fundamental numerical functions.
Within each module the calculations are performed exactly. The results are logically truncated by selecting a desirable number of significant bits which are passed 
on as inputs to the next stage,
which is also implemented using an elementary module. We repeat this procedure until we obtain the final result. 
This way, it suffices to implement quantum mechanically a relatively small number of elementary modules, 
which can be done once, and then to combine 
them as necessary to obtain the quantum circuits implementing the different functions. 
The elementary modules carry out certain basic tasks such as shifting the bits of a number held in a 
quantum register, or counting bits, or computing 
expressions involving addition and/or multiplication of the inputs. The benefit of using only addition 
and multiplication is
that in fixed-precision arithmetic the format of the input specifies exactly the format of the output, i.e., 
the location of the decimal point in the result.
There exist numerous quantum circuits in the literature for addition and multiplication; 
see e.g.  \cite{SBN08,BASP96,ChuangCircuits,cuccaro2004new,draper2000addition,DKRS06,TK05,Takahashi,rieffel2011quantum}.

There are three advantages to this approach. The first is that one can derive error estimates 
by treating the elementary modules 
as black boxes and considering only the truncation error in the output of each. The second is that 
it is easy to obtain total resource estimates 
by adding the resources used by the individual modules. The third advantage is that the modular design 
allows one to modify or improve the implementation of the individual elementary modules in a transparent way. 
Such an approach was used in \cite{Poisson} that deals with a quantum algorithm and circuit design for solving the Poisson equation. 
The results of this chapter can also be found  in \cite{hadfield2016scientific}.

\section{Algorithms}  \label{sec:Algorithms}

We derive quantum algorithms and circuits computing
approximately $w^{1/2^i}$, $i=1,\dots,k$, $\ln(w)$ and $w^{f}$, $f\in [0,1)$, for a given input $w$.
We provide pseudocode%
\footnote{
We present our algorithms using standard high-level mathematical and programming language
expressions, traditionally known as \textit{Pidgin ALGOL} \cite{ahoAlgorithms}.}
 and show how the algorithms are obtained by combining  elementary quantum circuit modules. We provide error and cost estimates.

The input of the algorithms is a fixed-precision binary number. It is held in an $n$ qubit quantum register 
whose state is denoted $\ket w$ as shown in Figure \ref{fig:InputRegister}.
The $m$ left most qubits are used to represent the integer part of the number and the remaining $n-m$  qubits 
represent its fractional part. 
\vskip 1pc
\begin{figure}[H]
\centerline{
$\ket{w}=\underbrace{\ket{w^{(m-1)}} \otimes \ket{w^{(m-2)}} \otimes \cdots \otimes \ket{w^{(0)}}}_{{\rm integer\  part}} \otimes 
\underbrace{\ket{w^{(-1)}} \otimes \cdots \otimes \ket{w^{(m-n)}}}_{{\rm fractional\ part}},$
}
\caption{An $n$-qubit fixed-precision representation of a number $w\geq 0$ on a quantum register.}
\label{fig:InputRegister}
\end{figure}
Thus 
$\ket{w} =\ket{w^{(m-1)} w^{(m-2)}  \cdots w^{(0)} w^{(-1)} \cdots w^{(m-n)}}$,
where $w^{(j)}\in \{0,1\}$, $j = m-n, m-n+1, \ldots,0,\dots, m-1$ and $w = \sum_{j = m-n}^{m-1} w^{(j)}2^j$. 
Since fewer than $n$ bits may suffice for the representation of the input,   
a number of leftmost qubits in the register may be set to $\ket{0}$.
In general, 
a leading qubit may hold the sign of $w$, but since for the functions 
in this chapter $w$ is a nonnegative number we have
omitted the sign
qubit for simplicity.

Our algorithms use elementary modules that perform certain basic calculations. 
Some are used to shift the contents of registers, others are used as counters determining the position of the most significant bit of a 
number. 
An important elementary module computes expressions of the form $xy+z$ exactly in fixed-precision arithmetic.

\begin{figure}[H]
\centerline{
\Qcircuit @C = 1.2 em @R = 2 em{
\lstick{\ket{z}}    &     \multigate{2}{res=xy+z}     & \qw      & &    \ket{res} \\
\lstick{\ket{y}}    &      \ghost{res=ax+b}       &      \qw     &&     \ket{y}  \\
\lstick{\ket{x}}    &      \ghost{res=ax+b}       &      \qw     &  &   \ket{x}  \\
}
}
\vspace*{8pt}
\caption{Elementary module using fixed-precision arithmetic to implement exactly $res\leftarrow xy+z$ 
for $x$, $y$, and $z$. Note that register sizes, ancilla registers, and their values are not indicated.} 
\label{fig:basicModule}
\end{figure}
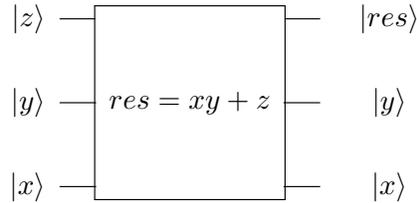

Following our convention concerning the fixed precision representation of numbers as we introduced it  in Fig. \ref{fig:InputRegister}, let  
$x$, $y$, and $z$ be represented using $n_1$-bits, of which $m_1$ bits  are used to represent the integer part. (It is not necessary to use the same number of bits to represent all
three numbers and this might be useful in cases where we know that their magnitudes are significantly different). 
The expression $xy+z$ can be computed exactly as long as we allocate $2n_1 +1$  bits to hold the result. 
In this case, the rightmost $2(n_1-m_1)$ bits hold the fractional part of the result. Such computations can be 
implemented reversibly.
There are numerous quantum circuit designs in the literature implementing addition and multiplication 
\cite{VBE96,BASP96,draper2000addition,cuccaro2004new,TK05,van2005fast,DKRS06,Takahashi,portugal2006reversible,SBN08,takahashi2008fast,rieffel2011quantum,saeedi2013synthesis,kepleyquantum}. 
Therefore,  we can use them to design a quantum circuit implementing $xy+z$. In fact, we can design a quantum 
circuit template for implementing such expressions and use it to derive the actual quantum circuit for any $n_1$, $m_1$ and
values of the $x$, $y$, $z$  represented with fixed precision.
We use Figure \ref{fig:basicModule} below to generically represent such a quantum circuit.
\vskip 1pc
Note that Fig. \ref{fig:basicModule} is an abstraction of an elementary module computing $res\leftarrow xy+z$. It is not meant to reveal or imply any of the implementation decisions including ancilla registers, saved values, and other details used for addition and multiplication.  
Any desired number $b$ of significant digits after the decimal point in the result $\ket{res}$ can be selected and passed 
on to the next stage of the computation. This corresponds to a truncation of the result to the desired accuracy.

We emphasize that although there are numerous ways of implementing 
basic arithmetic operations, 
there are trade-offs for the resulting quantum circuits in terms of the types of quantum gates used, the number of ancilla qubits required, and the resulting circuit size and depth.
 For example, Table~1 in \cite{Takahashi} summarizes some of the trade-offs for quantum circuits for addition. There we see that the circuit from \cite{draper2000addition} does not require Toffoli gates or ancilla qubits, but has size $O(n^2)$, while  the circuit in \cite{cuccaro2004new} uses one ancilla qubit, $O(n)$ Toffoli gates, and has size $O(n)$. It is reasonable to believe that low-level implementation details will have to be decided taking into account the target architecture because the unit cost of the different resources is unlikely to be equal.  
 Our algorithms, in addition to the applications of the elementary module of Fig. \ref{fig:basicModule}, use a constant number of extra arithmetic operations. 
Thus the cost of the algorithms can be expressed in a succinct and fairly accurate way by counting the number of arithmetic operations required, which is the approach taken in classical algorithms for scientific computing. The number of arithmetic operations leads to precise resource estimates once particular choices for the quantum circuits implementing these operations have been made. Hence, for the remainder of this chapter we will not be concerned with low-level implementation details or resource optimizations. 
 
We remark that particularly for the algorithms in this
chapter the integer parts of 
the inputs and outputs of the instances of the quantum circuit of  Fig. \ref{fig:basicModule} can be represented exactly using an equal number of qubits
to that used  for $\ket{w}$, i.e., $m$ qubits; we deal with the fractional parts separately. 
In \cite{Poisson}, such elementary modules were cascaded to derive the quantum algorithm, INV, computing the reciprocal function. 
The algorithms of this chapter compute approximations of functions which depend, to a certain 
extent, on INV, so we review its details in the next subsection. 

Without loss of generality and for brevity we only deal with the case $w>1$ in the functions computing the roots 
and the logarithm. Indeed, if $0<w<1$ one can suitably shift it to the left $\nu$ 
times to become 
$2^\nu w>1$. After obtaining the roots or the logarithm of the shifted number $2^\nu w$  
the final result for $w$ can be obtained in a straightforward way, either by shifting it to
the right in the case of the roots, or by subtracting $\nu \ln(2)$ in the case of the logarithm. 

In the following subsections we discuss each of our algorithms providing its details along with pseudocode and quantum circuits. We give 
theorems establishing the performance of each algorithm in terms of their accuracy in relation to the number of required qubits. To avoid distracting technical details, the proofs 
are deferred to Appendix \ref{app:SciComp}. 
Our results are summarized in Table~\ref{tab:SummaryOfResults1}.

\subsection{Reciprocal}
An algorithm computing the reciprocal of a number is shown in 
\cite{Poisson}. The algorithm is based on Newton iteration. Below we provide a slight modification of that algorithm. 

Recall that $w$ is represented with $n$ bits of which the first $m$ correspond to its
integer part.
Algorithm \ref{alg:inv} INV below approximates the reciprocal of a number $w\geq 1$, applying Newton iteration to the function $f(x) = \frac{1}{w} -x$.
This yields a sequence of numbers $x_i$, $i = 1,2,\ldots ,s$,  according to the iteration
\begin{equation}
\label{eq:NI.INV}
x_{i} =g_1(x_{i-1}) := -w\hat x_{i-1}^2 +2\hat x_{i-1}.
\end{equation}
Observe that the expression above can be computed using two applications of a quantum circuit of the type shown in Fig. \ref{fig:basicModule}. 
%
\input SummaryTable.tex   

\vskip 1pc
\begin{singlespace}
\begin{algorithm}
\caption{INV($w$, $n$, $m$, $b$)}  
\label{alg:inv}
\begin{algorithmic}[1]
\REQUIRE $w\geq 1$, held in an $n$ qubit register with $m$ qubits for its integer part.
\REQUIRE $b \in \nat$, $b \geq m$. We perform fixed-precision arithmetic and results are truncated to $b$ bits of accuracy after the decimal point. 
\IF{$w=1$} 
\RETURN 1
\ENDIF
\STATE $\hat{x}_0 \leftarrow 2^{-p}$, where $p\in\nat$ such that $2^p > w \geq 2^{p-1}$
\STATE $s \leftarrow \lceil \log_2 b \rceil$
\FOR{$i=1$ to $s$}
\STATE $x_i \leftarrow -w\hat{x}_{i-1}^2 + 2\hat{x}_{i-1} $
\STATE $\hat{x}_i \leftarrow x_i$ truncated to $b$ bits after the decimal point  
\ENDFOR
\RETURN $\hat{x}_{s}$
\end{algorithmic}
\end{algorithm}
\end{singlespace}

The initial approximation used is $\hat x_0 =2^{-p}$, with 
$2^p > w \geq 2^{p-1}$. The number of iterations~$s$ is specified in Algorithm \ref{alg:inv} INV. 
Note that $x_0<  1/w$ and the iteration converges to $1/w$ from below, i.e., $\hat x_i \le 1/w$.
Within each iterative step the arithmetic operations are performed in fixed precision and $x_i$ is computed exactly. We truncate $x_i$ to $b\ge n$ bits after the decimal point to obtain~$\hat x_i$ and pass it on as input to the next iterative step. Each iterative step is implemented using an elementary module of the form given in Fig. \ref{fig:basicModule} that requires only addition and multiplication. 
 The final approximation error is 
$$
| \hat{x}_s - \frac{1}{w}| \leq \frac{2+ \log_2 b}{2^b}.
$$
For the derivation of this error bound see Corollary \ref{cor0} in Appendix~\ref{app:SciComp}. 
We remark that although the iteration function (\ref{eq:NI.INV}) is well known in the literature \cite[Ex. 5-1]{traub1982iterative}, 
an important property of Algorithm \ref{alg:inv} INV is the fixed-precision implementation of Newton iteration for a specific initial approximation and a prescribed number of steps, so that the error bound of Corollary \ref{cor0} is satisfied. 

Turning to the cost, we iterate $O(\log_2 b)$ times  and as we mentioned each $x_i$ is computed exactly. Therefore, each iterative step requires $O(n+b)$ qubits and a  number of
quantum operations for implementing addition and  multiplication that is a low-degree 
polynomial in $n+b$.\footnote{For example, addition and multiplication can be performed with $O((n+b)^2)$ basic quantum gates using the grade-school algorithm.} 
The cost to obtain the initial approximation is 
relatively minor 
when compared to the overall cost of the multiplications and additions used in the algorithm.

\subsection{Square Root}
Computing approximately the square root $\sqrt{w}$, $w\ge 1$, can also be approached as a zero finding problem and one can 
apply Newton iteration to it.
However, the selection of the function whose zero is $\sqrt{w}$ has to be done carefully so that the resulting iterative steps
are easy to implement and analyze in terms of error and cost. Not all choices are equally good. 
For example, $f(x)= x^2-w$,
although well known in the literature \cite[Ex. 5-1]{traub1982iterative},
is not a particularly good choice. The resulting iteration is $x_{i+1} = x_i - (x_i^2 - w)/ (2x_i)$, $i=0,1,\dots$, which requires a division using an algorithm
such as  Algorithm~\ref{alg:inv}~INV at each iterative step. The division also requires circuits keeping 
track of the position of the decimal point in its result, because its location is not fixed but depends on 
the values $w$ and $x_i$. Since the result of a division may not be represented exactly using an a priori chosen 
fixed number of bits,
approximations are needed within each iterative step. This introduces error and overly complicates the analysis of the overall algorithm 
approximating $\sqrt{w}$ compared to an algorithm requiring only 
multiplication and addition in each iterative step. All these complications are avoided in our algorithm.
\input{fig-SQRToverall}

Each of the steps of the algorithm we present below can be implemented using only multiplication and addition in 
fixed-precision arithmetic as in Fig. \ref{fig:basicModule}.
This is accomplished by first approximating $1/w$ (applying iteration $g_1$ of equation (\ref{eq:NI.INV})) using steps essentially identical to those in Algorithm \ref{alg:inv} INV. 
Then we apply Newton iteration again to a suitably chosen function to approximate its zero and this yields the approximation of $\sqrt{w}$. 
In particular, in Algorithm \ref{alg:inv} INV we set $s=s_1=s_2$ and  first we approximate $1/w$ by $\hat x_s$ which has a fixed precision representation with $b$ bits after the decimal point
(steps 4--10 of Algorithm \ref{alg:inv} INV). 
Then applying the Newton method to $f(y) = \tfrac 1{y^2} - \tfrac 1{w}$ we obtain the
iteration 
\begin{equation}
\label{eq:NI.SQRT}
y_j = g_2(y_{j-1}) := \frac{1}{2} (3\hat y_{j-1} - \hat x_s \hat y^3_{j-1}),
\end{equation}
\noindent where $j = 1,2,\dots,s$. Each $y_i$ is computed exactly and then truncated  to $b$ bits after the decimal point to obtain $\hat y_i$, which is passed on as input to the next iterative step. See Algorithm \ref{alg:sqrt} SQRT for the values of the parameters and
other details. Steps 4 and 10 of the algorithm compute initial approximations for the two Newton iterations, the first computing 
the reciprocal and the second shown in (\ref{eq:NI.SQRT}).
They are implemented 
using the quantum circuits of  Figures \ref{fig-InitState1} and \ref{fig-InitState2}, respectively.
A block diagram of the overall circuit of the algorithm computing $\sqrt{w}$ is shown in Figure~\ref{fig-SQRToverall}.

\vskip 1pc
\begin{singlespace}
\begin{algorithm}
\caption{SQRT($w$, $n$, $m$, $b$)}
\label{alg:sqrt}
\begin{algorithmic}[1]
\REQUIRE $w\geq 1$, held in an $n$ qubit register with $m$ qubits for its integer part.
\REQUIRE $b\geq \max\{2m,4\}$. Results are truncated to $b$ bits of accuracy after the decimal point.
\IF{$w=1$} 
\RETURN 1
\ENDIF
\STATE $\hat{x}_0 \leftarrow 2^{-p}$, where $p\in\nat$ such that $2^p > w \geq 2^{p-1}$
\STATE $s \leftarrow \lceil \log_2 b \rceil$
\FOR{$i=1$ to $s$}
\STATE $x_i \leftarrow -w\hat{x}_{i-1}^2 + 2\hat{x}_{i-1} $
\STATE $\hat{x}_i \leftarrow x_i$ truncated to $b$ bits after the decimal point  
\ENDFOR
\STATE $\hat{y}_0 \leftarrow 2^{\lfloor (q-1)/2 \rfloor}$, where $q\in \nat$ such that $2^{1-q} > \hat{x}_{s}  \geq 2^{-q}$
\FOR{$j=1$ to $s$}
\STATE $y_j \leftarrow \frac12(3\hat{y}_{j-1} - \hat{x}_{s} \hat{y}^3_{j-1} )$
\STATE $\hat{y}_j \leftarrow y_j$ truncated to $b$ bits after the decimal point  
\ENDFOR
\RETURN $\hat{y}_{s}$
\end{algorithmic}
\end{algorithm}
\end{singlespace}

For $w\ge 1$ represented with $n$ bits of which the first $m$ give its integer part, Algorithm  \ref{alg:sqrt} SQRT 
computes $\sqrt{w}$ by $\hat y_s$ 
in fixed precision with $b\ge \max\{ 2m, 4\}$ bits after its decimal point and we have
$$
|\hat{y}_{s}-\sqrt{w}|  
\leq \left( \frac{3}{4} \right)^{b-2m}  \left( 2+ b + \log_2 b  \right).
$$
The proof can be found in Theorem \ref{SCIthm1} in Appendix~\ref{app:SciComp}. 

\vskip 2pc
\input{fig-InitState1} 

Turning to the cost of the quantum circuit in Fig. \ref{fig-SQRToverall} implementing Algorithm \ref{alg:sqrt} SQRT, we set the number of iterative steps of each of the two iterations to be
$s=s_1=s_2=O(\log_2 b)$. 
Observe that each of the iterations $g_1$  and $g_2$ can be computed using at most three applications of a quantum circuit of the type shown in Fig. \ref{fig:basicModule}. 
Therefore, each iterative step requires $O(n+b)$ qubits and a  number of
quantum operations for implementing addition and  multiplication that is a low-degree 
polynomial in $n+b$. 
Observe that the cost to obtain the initial approximations 
is again relatively minor compared to the overall cost of the additions and multiplications in the algorithm. 

\vskip 2pc
\input{fig-InitState2}

\subsection{$2^k$-Root}

Obtaining the roots, $w^{1/2^i}$, $i=1,\dots,k$, $k\in\naturals$, is straightforward. It is accomplished by calling Algorithm \ref{alg:sqrt} SQRT iteratively $k$ times, since
$w^{1/2^i}= \sqrt{ w^{1/2^{i-1}}}$, $i=1,\dots,k$. In particular, 
Algorithm \ref{alg:2krt} PowerOf2Roots calculates approximations of $w^{1/2^i}$, for $i=1,2, \ldots, k$. 
The circuit implementing this algorithm consists of $k$ repetitions of the circuit in Fig. \ref{fig-SQRToverall} approximating the square root.
The results are truncated to $b\ge \max\{ 2m, 4\}$ bits after the decimal point before passed on to the next stage.
The algorithm produces $k$ numbers $\hat z_i$, $i=1,\dots,k$. We have
$$|\hat{z}_{i}-w^{1/{2^i}}| \leq 2  \left( \frac{3}{4} \right)^{b-2m}  \left( 2+ b +  \log_2 b  \right),$$ 
$i=1,\dots,k$. 
The proof can be found in Theorem \ref{thm3} in Appendix~\ref{app:SciComp}. 
\vskip 1pc
\begin{singlespace}
\begin{algorithm}
\caption{PowerOf2Roots($w$, $k$, $n$, $m$, $b$)}
\label{alg:2krt}
\begin{algorithmic}[1]
\REQUIRE $w\geq 1$, held in an $n$ qubit register with $m$ qubits for its integer part.
\REQUIRE $k\geq 1$ an integer. The algorithm returns approximations of $w^\frac{1}{2^i}$, $i=1,\dots,k$.
\REQUIRE $b\geq \max\{2m,4\}$. Results are truncated to $b$ bits of accuracy after the decimal point.
\STATE $\hat z_1 \leftarrow$ SQRT($w$, $n$, $m$, $b$). Recall that SQRT returns a number with a fractional part $b$ bits long. The integer part of of $\hat z_1$ is represented by $m$ bits.
\FOR{$i=2$ to $k$}
\STATE $\hat z_i \leftarrow$ SQRT($\hat{z}_{i-1}$, $m+b$, $m$, $b$). Note that $\hat z_1$ and the $\hat z_i$ are held in registers of size $m+b$ bits of which the $b$ bits are for the fractional part.
\ENDFOR
\RETURN $\hat z_1$,$\hat z_2$,\dots,$\hat z_k$
\end{algorithmic}
\end{algorithm}
\end{singlespace}

Algorithm \ref{alg:2krt} PowerOf2Roots uses $k$ calls to Algorithm \ref{alg:sqrt} SQRT. Hence it requires $k\log b\cdot O(n+b)$ qubits and $k\log b\cdot p(n+b)$ quantum operations, where $p$ is a low-degree polynomial depending on the specific implementation of the circuit of Fig. \ref{fig:basicModule}.

\subsection{Logarithm}
To the best of our knowledge, the method presented in this section is entirely new. Let us first introduce the idea leading to the algorithm approximating $\ln(w)$, $w> 1$; the case $w=1$ is trivial. First we shift $w$ to the left, if necessary, to obtain the number $2^{-\nu} w \in [1,2)$.
It suffices to approximate $\ln (2^{-\nu} w)$ since 
$\ln( w) = \ln (2^{-\nu}  w) +\nu \ln 2$ and we can precompute $\ln 2$ up to any desirable number of bits. 

We use the following observation. When one takes the $2^\ell$-root of  a number that belongs to the interval  $(1,2)$  the fractional part $\delta$ of the result is 
roughly speaking proportional to $2^{-\ell}$, i.e, it is quite small for relatively  large $\ell$. 
Therefore, for $1+\dd = [2^{-\nu}w]^{1/2^\ell}$ we use the power series expansion for the logarithm to approximate $\ln(1+\dd)$ by $\dd- \dd^2/2$, with any desired accuracy 
since $\delta$ can be
made arbitrarily small by appropriately selecting $\ell$. 
Then the approximation of the logarithm follows from 
\begin{equation}
\label{eq:undolog}
\ln(w)  \approx \nu\ln 2 + 2^\ell(\dd- \frac {\dd^2}2).
\end{equation}

In particular, Algorithm \ref{alg:ln} LN approximates $\ln(w)$ for $w\ge 1$ represented by $n$ bits of which the first $m$ are used for its integer part. 
(The trivial case $w=1$ is dealt with first.)
In step 7, $p-1$ is the value of $\nu$, i.e., $\nu=  p-1$, where  $2^p> w\ge 2^{p-1}$, $p\in\naturals$. 
We compute 
$w_p=2^{1-p}w \in [1,2)$ using a right shift of $w$. We explain how to implement the shift operation below.
Then, $\hat t_p$, an approximation of $w_p^{1/(2\ell)}$, is calculated using Algorithm \ref{alg:2krt} PowerOf2Roots, 
where $\ell$ is a parameter that determines the error. 
Note that $\hat t_p$ can be written as $1+\dd$, for $0<\dd< 1$. We provide precise bounds for $\dd$ in Theorem \ref{thm3} in Appendix~\ref{app:SciComp}. 
Next, an approximation $\hat y_p$ of $\ln \hat t_p=\ln(1+\dd)$ is calculated, using the first two terms of the power series for the logarithm as we explained above. 
The whole procedure yields an estimate of $(\ln w_p)/2^\ell $. 
Finally the algorithm in steps 16 -- 20  uses equation (\ref{eq:undolog}), with $\nu=p-1$, to
derive $z_p+(p-1) r$ as an approximation to $\ln w$. 
All intermediate calculation results are truncated to $b$ bits after the decimal point. The value of $b$ is determined by the value of $\ell$ in step 1 of
the algorithm. Note that the precision of the algorithm grows with $\ell$, which is a user selected parameter that must satisfy $\ell\ge\lceil \log_2 8n\rceil$.

\begin{figure}[H]
\centerline{
\Qcircuit @C = 1.9 em @R = 1.6 em{
Digit & & & Initial\ State & & & & & & & & & & & Final\ State \\ 
2^{m-1} & & & \lstick{\ket{w^{(m-1)}}} & \ctrl{12} & \qw & \qw & \dots & & \qw & \qw & \qw & \qw &  \qw \\
2^{m-2} & & & \lstick{\ket{w^{(m-2)}}} & \qw & \ctrl{6} & \qw & \dots & & \qw & \qw & \qw & \qw & \qw\\
& & \vdots & & & & & \vdots & & & & & \vdots   \\
2^1 & & & \lstick{ \ket{w^{(1)}} } & \qw & \qw & \qw & \dots & & \ctrl{4} & \qw & \qw & \qw & \qw &  \raisebox{0 em}{$\ket{w}$} \\
2^{0} & & & \lstick{  \ket{w^{(0)}}    } & \qw & \qw & \qw & \dots & & \qw & \qw  & \qw & \qw & \qw \\
& & \vdots & & & & & \vdots & & & & & \vdots \\
2^{m-n} & & & \lstick{\ket{w^{(m-n)}}} & \qw & \qw & \qw & \dots & & \qw & \qw & \qw & \qw & \qw  \\
Ancilla & & & \lstick{\ket{0}} & \targ & \ctrlo{4} & \targ & \dots & & \ctrlo{2} & \targ  & \ctrlo{1} & \targ & \qw & \ket{0} \\
2^{0} & & & \lstick{\ket{0}} & \qw & \qw & \qw & \dots & & \qw & \qw  &  \targ & \ctrlo{-1} & \qw \\
2^{-1} & & & \lstick{\ket{0}} & \qw  & \qw & \qw & \dots & & \targ & \ctrl{-2} &  \qw & \qw & \qw \\
& & \vdots & & & & & \vdots & & & & & \vdots & &  &  \raisebox{0.6 em}{$\ket{ x}$}  \\
2^{-(m-2)} & & & \lstick{\ket{0}} & \qw & \targ & \ctrl{-4} & \dots & & \qw & \qw &  \qw & \qw & \qw \\
2^{-(m-1)} & & & \lstick{\ket{0}} & \targ & \qw & \qw & \dots & & \qw & \qw &  \qw & \qw & \qw 
\gategroup{2}{14}{8}{14}{1.5 em}{\}} \gategroup{10}{14}{14}{14}{1.2 em}{\}}
}
}
\vspace*{8pt}
\caption{For $w\ge 1 $ this quantum circuit computes $\ket{x}$ where $x$ is an $m$ bit number $x\in [2^{1-m},1]$. 
 For $w\ge 2$ we set $x=2^{1-p}$, where $p-1=\lfloor \log_2 w\rfloor \ge 1$.
For $1\le w <2$ we set $x=1$. 
Thus $m$ bits are needed for the representation of $x$,
with the first bit  $x^{(0)}$ denoting its integer part and all the remaining bits $x^{(-1)}, \dots,
x^{-(m-1)}$  denoting its fractional part.
This circuit is used in steps 6  -- 10 of Algorithm~3~LN to  derive $x=2^{1-p}$  so one can implement the shift of $w$ in terms of multiplication between $w$ and $x$,
i.e., $w_p=w\, x$. }
\label{fig-RShift}
\end{figure}
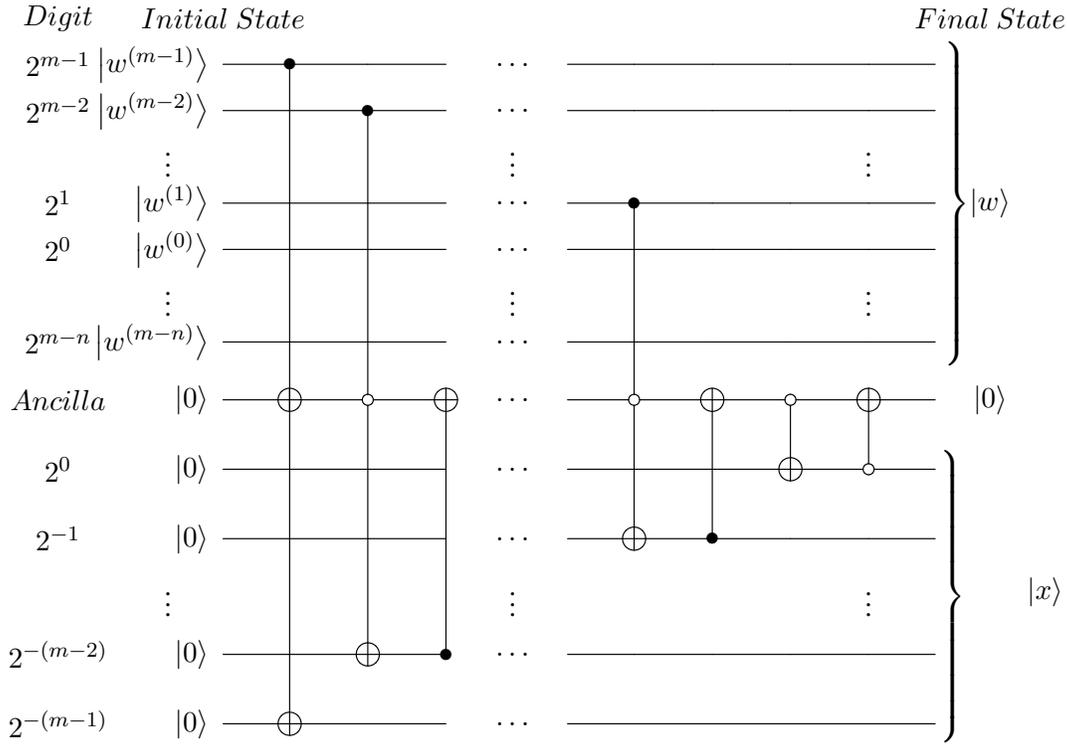

The block diagram of the overall circuit implementing Algorithm~\ref{alg:ln} LN can be found in Fig.~\ref{fig-LogOverall}. The first module is a right shift 
operation that calculates $w_p$. 
There are a number of ways to implement it. 
One is to implement the shift by a multiplication of $w$ by a suitable negative power of two. 
This is convenient since we 
have elementary quantum circuits for multiplication.  In Fig.~\ref{fig-RShift} we show a circuit computing  the necessary power of two, which we denote by $x$ in that figure. In particular, 
for $w\ge 2$ we set $x=2^{1-p}$, where $p-1=\lfloor \log_2 w\rfloor \ge 1$.
For $1\le w <2$ we set $x=1$. 
Thus $m$ bits are needed for the representation of $x$,
with the first bit  $x^{(0)}$ denoting its integer part and all the remaining bits $x^{(-1)}, \dots,
x^{-(m-1)}$  denoting its fractional part.
We implement the shift of $w$ in terms of multiplication between $w$ and $x$,
i.e., $w_p=w\, x$. Since $w$ and $x$ are held in registers of known size and we use fixed precision representation we know a priori the position of the decimal point in their product $w_p$. Moreover, since $w_p\in [1,2-2^{1-n}]$ ($w$ is an $n$ bit number) we have that $n$ bits (qubits), of which $1$ is used for its integer part, suffice to hold $w_p$ exactly.
\vskip 2pc
\input{fig-LogOverall}

The next module is the circuit for the PowerOf2Roots algorithm and calculates $\hat t_p$.
The third module calculates $\hat y_p$ and is comprised of modules that perform subtraction and multiplication in fixed precision. 
The fourth module of the circuit performs a series of left shift operations. 
Observe that $\ell$ is an accuracy parameter whose value is set once at the very beginning and does not change during the execution of the algorithm.
Thus the left shift $\ell$ times is, in general, much easier to implement than the right shift of $w$  at the beginning of the algorithm and we omit the details.

In its last step Algorithm \ref{alg:ln} LN computes the expression $\hat z:=z_p+(p-1)r$ which is the estimate of $\ln w$. The value of $p-1$ in this expression is obtained using
the quantum circuit of Fig. \ref{fig-compute_p}. The error of the algorithm satisfies
$$ |  \hat{z}  - \ln w | \leq   \left(\frac{3}{4}\right)^{5\ell/2} \left( m+ \frac{32}{9} + 2\left(\frac{32}{9} + \frac{n}{\ln 2} \right)^3 \right).$$
For the proof details see Theorem \ref{thm3} in Appendix~\ref{app:SciComp}.
%

\begin{singlespace}
\begin{algorithm}
\caption{LN($w$, $n$, $m$, $\ell$)}
\label{alg:ln}
\begin{algorithmic}[1]
\REQUIRE $w\geq 1$, held in an $n$ qubit register with $m$ qubits for its integer part.
\REQUIRE $\ell \geq \lceil \log_2 8n\rceil$ is a parameter upon which the error will depend and which we use to determine the number of bits $b$ after the decimal points in which arithmetic will be performed.
\STATE $b\leftarrow \max\{5\ell,25\}$. Results are truncated to $b$ bits of accuracy after the decimal point in the intermediate calculations.
\STATE $r$ $\leftarrow$ $\ln 2$ with $b$ bits of accuracy, i.e., $|r-\ln 2|\le 2^{-b}$. An approximation of $\ln 2$  can be precomputed with sufficiently many bits of accuracy and stored in a register, from which we take the first $b$ bits.
\IF{$w=1$} 
\RETURN 0
\ENDIF
\STATE Let $p\in \nat$ be such that $2^p > w \geq 2^{p-1}$
\IF {$p-1 =0$}
\STATE $w_p \leftarrow w$. In this case $w=w_p\in [1,2-2^{1-n}]$.
\ELSE 
\STATE $w_p \leftarrow w2^{1-p}$. Note that $w_p\in [1,2-2^{1-n}]$ for $w\ge 2$. The number of bits (qubits) used for $w_p$ is $n$ of which $1$ is for its integer part as explained in the text and the caption of  Fig.~\ref{fig-RShift}.
\STATE $x_p \leftarrow w_p - 1$. This is the fractional part of $w_p$. 
\ENDIF
\IF{$x_p=0$} 
\STATE $z_p \leftarrow 0$
\ELSE 
\STATE $\hat{t}_p \leftarrow $PowerOf2Roots($w_p$, $\ell$, $n$, $1$, $b$)$[\ell]$. The function $PowerOf2Roots$ returns a list of numbers and we take the last element, the $1/2^\ell$th root. Note that in this particular case  $1\leq \hat{t}_p < 2$. 
\STATE $\hat{y}_p \leftarrow (\hat{t}_p-1) - \frac12 (\hat{t}_p -1)^2$, computed to $b$ bits of accuracy after the decimal point. Note that $\hat{t}_p = 1 + \delta$ with $\delta \in (0,1)$, and we approximate $\ln(1+\delta)$ by $\delta-\frac12 \delta^2$, the first two terms of its power series expansion.
\STATE $z_p \leftarrow 2^{\ell}\hat{y}_p$. This corresponds to a logical right shift of the decimal point. 
\ENDIF
\RETURN $z_p+(p-1)r$ 
\end{algorithmic}
\end{algorithm}
\end{singlespace}

\vskip 1pc

Considering the cost of the quantum circuit in Fig. \ref{fig-LogOverall} implementing Algorithm \ref{alg:ln} LN, we have that the cost of computing the initial and the last shifts 
(first and fourth modules in Fig.~\ref{fig-LogOverall}),
as well as the cost of the arithmetic expression in the third module in Fig.~\ref{fig-LogOverall}, the cost of computing $p-1$ (see the circuits in Fig.~\ref{fig-RShift} and  Fig.~\ref{fig-compute_p} ) 
and the cost of the expression $z_p+(p-1)r$,
are each relatively minor compared to the other costs of the algorithm. 
The algorithm calls Algorithm~\ref{alg:2krt} PowerOf2Roots with input parameter $k:= \ell$, which requires 
 $O( \ell (n+b)\log b)$ qubits and $\ell \log b \cdot p(n+b)$ quantum operations, as explained in the previous subsection. Observe that the expression in Step 17 of Algorithm \ref{alg:ln} LN can be computed using a constant number of applications of a quantum circuit of the type shown in Fig. \ref{fig:basicModule}. Hence, the overall cost of Algorithm \ref{alg:ln} LN is $O( \ell (n+b)\log b)$ qubits and requires a number of quantum operations proportional to $\ell \log b \cdot p(n+b)$.

\vskip 2pc
\input{fig-compute_p}

%

\subsection{Fractional Power}
Another application of interest is the approximation of fractional powers of the form $w^f$, where $w\geq 1$ and
$f\in [0,1]$ a number whose binary form is $n_f$ bits long. The main idea is to calculate appropriate powers of 
the form $w^{1/2^i}$ according to the value of the bits of $f$ and multiply the results.
Hence initially, the algorithm PowersOf2Roots is used to derive a list of approximations $\hat w_i$ to the powers 
$w_i = w^{1/2^i}$, for $i = 1,2, \ldots ,n_f$.
The final result that approximates $w^{f}$
is $\Pi_{i\in \mathcal P} \hat w_i$, where $\mathcal P = \{1\leq i \leq n_f: f_i = 1\}$ and $f_i$ denotes the $i$th bit of the number $f$.
The process is described in detail in Algorithm \ref{alg:fracPower} Fractional Power.

\vskip 1pc
\begin{singlespace}
\begin{algorithm}
\caption{FractionalPower($w$, $f$, $n$, $m$, $n_f$, $\ell$)}
\label{alg:fracPower}
\begin{algorithmic}[1]
\REQUIRE $w\geq 1$, held in an $n$ qubit register with $m$ qubits for its integer part.
\REQUIRE $\ell \in \nat$ is a parameter upon which the error will depend. We use it to determine the number of bits $b$ after the decimal points in which arithmetic will be performed.
\REQUIRE $1\geq f \geq 0$ is a binary string corresponding to a fractional number given with $n_f$
bits of accuracy after the decimal point. 
The algorithm returns an approximation of $w^f$.
\STATE $b \leftarrow \max \{ n,n_f, \lceil 5(\ell + 2m +\ln n_f) \rceil,40\}$. Results are truncated to $b$ bits of accuracy after the decimal point.
\IF{$f=1$}
   \RETURN w
\ENDIF
\IF{$f=0$}
   \RETURN 1
\ENDIF
\STATE $\{ \hat{w}_i  \} \leftarrow $PowerOf2Roots($w$, $n_f$, $n$, $m$, $b$). The function returns a list of numbers  $ \hat{w}_i$  approximating $w_i =  w^{\frac{1}{2^i}}$, $i=1,\dots,n_f$.
\STATE $\hat{z} \leftarrow 1$
\FOR{$i=1$ to $n_f$}
   \IF{the $i$th bit of $f$ is $1$}
       \STATE $z \leftarrow \hat{z} \hat{w}_i$
       \STATE $\hat{z} \leftarrow z$ truncated to $b$ bits after the decimal point
   \ENDIF
\ENDFOR
\RETURN $\hat{z}$
\end{algorithmic}
\end{algorithm}
\end{singlespace}

For $w>1$ the value $\hat z$ returned by the algorithm satisfies
$$
|\hat{z} - w^{f}|  \leq \left( \frac{1}{2} \right)^{\ell  - 1  } , 
$$
where $w$ is represented by $n$ bits of which $m$ are used for its integer part, $n_f$ is number of bits in the representation of the exponent $f$, $\ell\in\naturals$ is a user selected parameter determining the accuracy of the result. The results of all intermediate steps are truncated to 
$b\ge\max\{ n,n_f,[5(\ell+2m+\ln n_f)],40\}$ bits before passing them  on to the next step. The proof can be found in Theorem \ref{thm4} in Appendix~\ref{app:SciComp}.

The algorithm can be extended to calculate $w^{p/q}$, $w>1$, where the exponent is a rational number $p/q \in [0,1]$. 
First $f$, an $n_f$ bit number, is calculated such that it approximates $p/q$ within $n_f$ bits of accuracy, 
namely $|f-\frac{p}{q}| \leq 2^{-n_f}$.
Then $f$ is used as the exponent in the parameters of Algorithm \ref{alg:fracPower} FractionalPower to 
get an approximation of $w^f$, which in turn is an approximation of~$w^{p/q}$.
The value $\hat z$ returned by the algorithm satisfies 
$$
|\hat{z} - w^{p/q}|  \leq     \left( \frac{1}{2} \right)^{\ell  - 1  }  + \frac{w\ln w}{2^{n_f}}. 
$$
The proof can be found in Corollary \ref{cor1} in Appendix~\ref{app:SciComp}.

\begin{rem}
For example, for $p/q=1/3$, one can use 
Algorithm \ref{alg:inv} INV to produce an approximation $f$ of $1/3$ and pass that to the algorithm.
In such a case, the approximation error $|w^{p/q} - w^f| \leq 2^{-n_f} w \ln w$, obtained using the mean value theorem for the function $w^x$,  appears as the last term of the equation above. For the details see Corollaries \ref{cor1} and  \ref{cor2} in Appendix~\ref{app:SciComp}. 
\end{rem}

When $w\in(0,1)$ we can shift it appropriately to obtain a number greater than one to which we can apply   Algorithm \ref{alg:fracPower} FractionalPower.
However, when approximating the fractional power, {\it undoing} the  initial shift to obtain an estimate of $w^f$ is a bit more involved than it is for the previously considered
functions. For this reason we provide the details in Algorithm~\ref{alg:fracPower2} FractionalPower2.
The algorithm first computes $k$ such that $2^k w \geq 1 > 2^{k-1}w$; see Fig. \ref{fig-ShiftInteger}. Using $k$ the algorithm shifts $w$ to the left to obtain $x:=2^k w$. The next step is
to approximate $x^f$. Observe that $x^f=2^{kf} w^f$. 
%
Therefore to undo the initial shift we have to divide by $2^{kf}= 2^c  2^{\{ c\} }$, where $c$ denotes the integer part of $kf$ and
$\{c\}$ denotes its fractional part. Dividing by $2^c$ is straightforward and is accomplished using shifts. 
Dividing by $2^{ \{ c\} }$ is accomplished by first approximating  $2^{ \{ c\} }$  and then
multiplying by its approximate reciprocal. See Algorithm \ref{alg:fracPower2} FractionalPower2 for the details. 

The value $\hat t$ the algorithm returns as an approximation of $w^f$, $w\in(0,1)$, satisfies
$$
|\hat{t} - w^{f}| \leq   \frac{1}{2^{\ell - 3 }},
$$
where $\ell$ is a user-defined parameter. 
The proof can be found in Theorem \ref{thm5} in Appendix~\ref{app:SciComp}.

Just like before we can approximate $w^p/q$, $w\in(0,1)$, and a rational exponent $p/q\in[0,1]$, by first approximating $p/q$ and then calling 
Algorithm \ref{alg:fracPower2}  FractionalPower2. The value $\hat t$ the algorithm returns satisfies
$$
|\hat{t} - w^{p/q}|  \leq     \left( \frac{1}{2} \right)^{\ell  - 2  }  + \frac{w\ln w}{2^{n_f}}.
$$
See Corollary \ref{cor2} in Appendix~\ref{app:SciComp}.

We now address the cost of our Algorithms computing $w^f$. First consider Algorithm \ref{alg:fracPower} FractionalPower, which calls Algorithm \ref{alg:2krt} PowerOf2Roots with input parameters $k:= n_f$ and~$b:=\max \{ n,n_f, \lceil 5(\ell + 2m +\ln n_f) \rceil,40\}$, which requires $O( n_f  (n+b)\log b)$ qubits and of order $n_f \log b \cdot p(n+b)$ quantum operations, as explained in the previous subsections. At most $n_f$ multiplications are then required, using a quantum circuit of the type shown in Fig. \ref{fig:basicModule}.
Therefore, Algorithm \ref{alg:fracPower} FractionalPower  requires a number of qubits and a number of quantum operations that is 
a low-degree polynomial in $n,n_f$, and $\ell$, respectively. 

\vskip 1pc
\begin{singlespace}
\begin{algorithm}
\caption{FractionalPower2($w$, $f$, $n$, $m$, $n_f$, $\ell$)}
\label{alg:fracPower2}
\begin{algorithmic}[1]
\REQUIRE $0 \leq w< 1$, held in an $n$ qubit register with $m$ qubits for its integer part. (For $w\geq 1$, use Algorithm \ref{alg:fracPower}.)
\REQUIRE $\ell \in \nat$ is a parameter upon which the error will depend. We use it to determine the number of bits $b$ after the decimal points in which arithmetic will be performed.
\REQUIRE 
$f\in [0,1]$ specified to $n_f$ 
bits. 
The algorithm returns an approximation of~$w^f$.
\STATE $b \leftarrow \max \{ n,n_f, \lceil 2\ell + 6m +2\ln n_f \rceil,40\}$. Results are truncated to $b$ bits of accuracy after the decimal point.
\IF{$w=0$}
   \RETURN 0
\ENDIF
\IF{$f=1$}
   \RETURN w
\ENDIF
\IF{$f=0$}
   \RETURN 1
\ENDIF
 \STATE $x \leftarrow 2^k w$, where $k$ is a positive integer such that  $2^k w \geq 1 > 2^{k-1} w$. This corresponds to a logical right shift of the decimal point. An example of the quantum circuit computing $k$ is given in Fig. \ref{fig-ShiftInteger}.
 \STATE $c \leftarrow kf$  
 \STATE $\hat{z} \leftarrow  $FractionalPower($x$, $f$, $n$, $m$, $n_f$, $\ell$). This computes an approximation of $z=x^f$.
 \STATE $\hat{y} \leftarrow  $FractionalPower($2$, ${\{c\}}$, $n$, $m$, $n_f$, $\ell$). This computes an approximation of $y=2^{\{c\}}$, where $\{c\} = c -  \lfloor c \rfloor$ (for $c\geq 0$) denotes the fractional part of $c$. Since we use fixed-precision arithmetic, the integer and fractional parts of numbers are readily available.
  \STATE $\hat{s} \leftarrow$ INV$(\hat{y}, n, 1,2\ell)$. This computes an approximation of $s=\hat{y}^{-1}$. 
\STATE $v \leftarrow 2^{- \lfloor c \rfloor} \hat{z}$. This corresponds to a logical left shift of the decimal point.
\STATE $t \leftarrow v \hat{s}$
\STATE $\hat{t} \leftarrow t$, truncated to $b$ bits after the decimal point.
\RETURN $\hat{t}$
\end{algorithmic}
\end{algorithm}
\end{singlespace}

Now consider Algorithm \ref{alg:fracPower2} FractionalPower2, which requires two calls to Algorithm \ref{alg:fracPower} FractionalPower, one call to Algorithm \ref{alg:inv} INV, and 
a constant number of
calls to  a quantum circuit of the type shown in Fig. \ref{fig:basicModule}. 
Hence, using the previously derived bounds for the costs of each of these modules, the cost of Algorithm \ref{alg:fracPower2} FractionalPower2 in terms of both the number of quantum operations and the number of qubits is  a low-degree polynomial in $n,n_f$, and $\ell$, respectively. 

\vskip 2pc
\input{fig-ShiftInteger}

\section{Numerical Results}  \label{sec:NumericalResults}

We present a number of tests comparing our algorithms to the respective ones implemented using floating point arithmetic in Matlab. To ensure that we compare against highly accurate values in Matlab we have used variable precision arithmetic (vpa) with 100 significant digits. In particular, we simulate the execution
of each of our algorithms and obtain the result. We obtain the corresponding result using a built-in function in Matlab. We compare the number of equal significant digits in the two results.
Our tests show  that using a moderate amount of resources, our algorithms compute values 
matching the corresponding ones obtained by commercial, widely used scientific computing software 
with 12 to 16 decimal digits. 
In our tests, the input $w$ was chosen randomly.

\subsection*{Algorithm: SQRT}
In Table \ref{tab:numSQRT}, 
all calculations for SQRT are performed with $b = 64$ bits (satisfying $b > 2m$ for all inputs). 
The first column in the table gives the different values of $w$ in our tests, the second and third columns give the computed value using Matlab and our algorithm respectively, and the last column gives the number of identical significant digits between the output of our algorithm and Matlab. 

\vskip 1pc
\begin{table}[H]
\begin{center}
\begin{tabular}{| c | c | c | c |}
	\hline
	$w$ & Matlab: $w^{1/2}$ & Our Algorithm: $w^{1/2}$ & \# of Identical Digits \\
	\hline
	$0.0198$ & $0.140712472794703$ & $0.140712472794703$ & $16$ \\
	\hline
	$48$ & $6.928203230275509$ &  $6.928203230275507$ & $15$  \\
	\hline
	$91338$ & $302.2217728754829$ & $302.2217728754835$ & $14$  \\
	\hline
	$171234050$ & $13085.64289593752$ &  $13085.64289596872$ & $12$ \\
	\hline
\end{tabular}
\end{center}
\caption{Numerical results for SQRT.}
\label{tab:numSQRT}
\end{table}

\begin{figure}[H]
\centering
\includegraphics[scale=0.6]{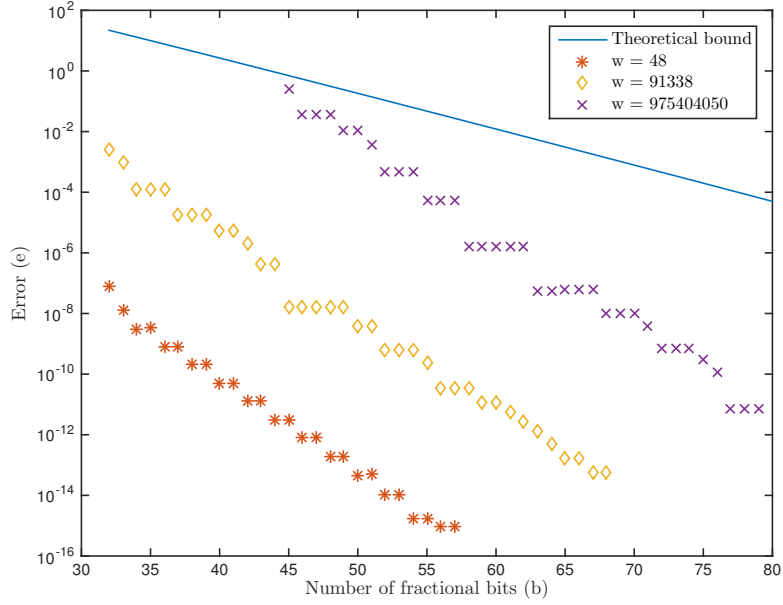}
\caption{Semilog plot showing the error 
of Algorithm~1 SQRT versus the number of precision bits~$b$. The solid blue line 
is a plot of the worst-case error of Theorem 1, for $n=2m=64$. 
The three data sets represent the absolute value of the difference between Matlab's floating point calculation of $\sqrt{w}$ 
and our algorithm's result for three different values of $w$.}
\label{fig-sqrtplot}
\end{figure}

In Fig. \ref{fig-sqrtplot} we give simulation results showing the error of our algorithm for different values of the parameter $b$. In particular, we show how the worst-case error of the algorithm depends on $b$, and then, using the results of Matlab as the correct values of $\sqrt w$, we show the actual error of our algorithm in relation to $b$ in a number of test cases.

\subsection*{Algorithm: PowerOf2Roots}
In Table \ref{tab:numPow2Roots},
we show only the result for the $2^{k}$th root (the highest root), where $k$ was chosen randomly such that $5\le k \le 10$.  Again, all our calculations are performed with $b = 64$ bits. \\

\begin{table}[h]
\begin{center}
\begin{tabular}{| c | c | c | c | c |}
	\hline
	$w$ & $k$ & Matlab: $w^{1/2^{k}}$ & Our Algorithm: $w^{1/2^{k}}$ & \# of Identical Digits \\
	\hline
	$0.3175$ & $6$ & $0.982233508377946$ & $0.982233508377946$ & $16$ \\
	\hline
	$28$ & $10$ & $1.003259406317532$ & $1.003259406317532$ & $16$ \\
	\hline
	$15762$ & $5$ & $1.352618595919273$ & $1.352618595919196$ & $13$ \\
	\hline
	$800280469$ & $8$ & $1.083373703681284$ & $1.083373703681403$ & $13$ \\
	\hline
\end{tabular}
\end{center}
\caption{Numerical results for PowerOf2Roots.}
\label{tab:numPow2Roots}
\end{table}

\subsection*{Algorithm: LN}
\begin{figure}[H]
\centering
\includegraphics[scale=0.6]{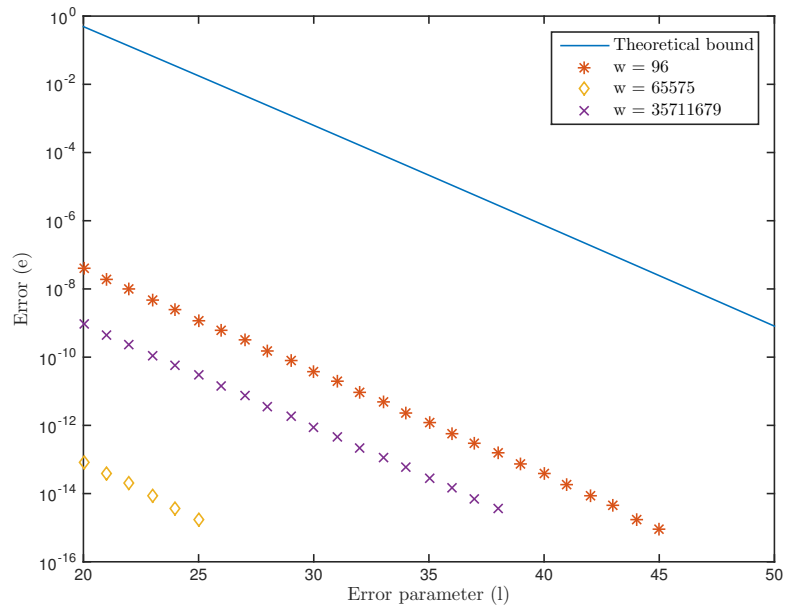}
\caption{Semilog plot showing the error of Algorithm 3 LN versus the 
parameter $\ell$ which controls the accuracy of the result. 
The solid blue line 
shows the worst-case error of Theorem 3, for $n=2m=64$, $b=\max\{ 5\ell,25\}$. 
The three data sets represent the absolute value of the difference between Matlab's floating point calculation of $\ln(w)$ and our algorithm's result for three different values of $w$.}
\label{fig-lnplot}
\end{figure}

In Fig. \ref{fig-lnplot}  we give simulation results showing the error of our algorithm for different values of the parameter $\ell$ that controls the desired accuracy. 
In particular, we show how the worst-case error of the algorithm depends on $\ell$, and then, using the results of Matlab as the correct values of $\ln(w)$, we show the actual error of our algorithm in relation to $\ell$ in a number of test cases.  

For the tests shown in Table~\ref{tab:numLN} below we have used $\ell=50$. 
%

%
\begin{table}[h]
\begin{center}
\begin{tabular}{| c | r | r | c |}
	\hline
	$w$ & Matlab: $\ln(w) \quad$& Our Algorithm: $\ln(w)$ & \# of Identical Digits \\
	\hline
	$96$ & $4.564348191467836$ & $4.564348191467836$ & $16$ \\
	\hline
	$65575$ & $11.090949804735075$ & $11.090949804735075$ & $17$ \\
	\hline
	$35711679$ & $17.390988336107455$ & $17.390988336107455$ & $17$ \\
	\hline
\end{tabular}
\end{center}
\caption{Numerical results for LN.}
\label{tab:numLN}
\end{table}

\subsection*{Algorithm: FractionalPower}
The table below shows tests for the approximation of $w^f$ for randomly generated $f\in (0,1)$. This result of this algorithm also depends on  a user-determined parameter $\ell$ controlling the accuracy.
We have used $\ell=50$ in the tests shown in Table~\ref{tab:numFracPow}. 

\begin{table}[h]
\begin{center}
\begin{tabular}{| c | c | r | r | c |}
	\hline
	$w$ & $f$ & Matlab: $w^f \qquad$ & Our Algorithm: $w^f$ & \# of Identical Digits \\
	\hline
	$0.7706$ & $0.1839$ & $0.953208384891996$ & $0.953208384891998$ & $15$ \\
	\hline
	$76$ & $0.7431$ & $24.982309269657478$ & $24.982309269657364$ & $14$ \\
	\hline
	$1826$ & $0.1091$ & $2.268975123215851$ & $2.268975123215838$ & $14$ \\
	\hline
	$631182688$ & $0.5136$ & $33094.79142555447$ & $33094.79142555433$ & $14$ \\
	\hline
\end{tabular}
\end{center}
\caption{Numerical results for FractionalPower.}
\label{tab:numFracPow}
\end{table}

\section{Discussion}
In designing algorithms for scientific computing the most challenging task is to control the error 
propagation and to do so efficiently. 
 In this sense, it is important to derive reusable modules with well-controlled error bounds. 
 We have given efficient, reversible, and scalable quantum algorithms for computing a variety of basic numerical functions and derived
worst-case error bounds.  
The designs are modular, combining a number of elementary quantum circuits to implement the functions, and the resulting algorithms have cost that is polynomial in the problem parameters.  

It is worthwhile to comment on the implementation of our algorithms. 
Since we are interested in quantum algorithms, the algorithms must be reversible. 
To the extent that the modules are based on classical computations, their reversibility is not a major issue since \cite{bennett1989time,levine1990note,portugal2006reversible} show how to simulate them reversibly. There are time/space trade-offs in 
an implementation that are of theoretical and practical importance. 
These considerations, as well as other constraints such as (fault-tolerant) gate sets or locality restrictions \cite{Takahashi}, should be optimized by the compiler, but we are not concerned with them here. 
For a discussion of the trade-offs, see, for example, \cite{parent2015reversible} and the references therein. 

We are at an early stage in the development of quantum computation, quantum programming languages and compilers. It is anticipated that implementation decisions will be made taking into account technological limitations concerning the different quantum computer architectures, but also it is expected that  things will change with time and some of the present constraints may no longer be as important.  
Nevertheless, simple but concrete implementation of the algorithms in this chapter are outlined in \cite{hadfield2016scientific}, though there are alternatives that one may opt for in practice.

Furthermore, although optimization of the individual quantum circuits realizing the algorithms for scientific 
computing described in this chapter is desirable, to a certain degree, it is not a panacea. Quantum 
algorithms may use one or more of our modules in different ways while performing their tasks. 
Therefore, global optimization of the resulting quantum circuit and resource management is much 
more important and will have to be performed by the compiler and optimizer. Quantum programming languages and compilers is an active area of research with many open questions.

In summary, our results are important first steps towards the goal of 
ultimately developing a comprehensive library of quantum circuits for scientific computing. 
The immediate next step is to further extend our results to a wider class of numerical functions. 
An important open question is whether or not, and for which numerical functions, quantum computers can exploit quantum mechanical effects to compute such functions in a novel way that is more efficient than classical algorithms. For instance, using the Fourier transform; see, e.g., \cite{wiebe2014quantum}.

Finally, we remark that our algorithms have applications beyond quantum computation. For instance, they can be used in 
signal processing and easily realized on an FPGA (Field Programmable Gate Array), providing superior performance with low cost. 
%
Moreover, there is resurgent 
interest in fixed-precision algorithms 
for low power/price applications such as mobile or embedded systems,
where hardware support for floating-point operations is often lacking~\cite {bocchieri2008fixed}. 

%% file: SummaryTable.tex

\begin{sidewaystable}[h!] 
\hskip-0.5cm 
\scalebox{1}{
\footnotesize
\begin{tabular}{ | l || l | l | l | l |}
\hline
Function             	    & Requirements                                  &Algorithm and Parameters                                  	 			& Idea 									& Error \\
\hline \hline
$1/w$               	    & $w\geq 1 $		 	   	& ${\rm INV}(w,n,m,b)$ 			 	 			& 1. Newton iteration. Calculate 						& $\leq \left( \frac{1}{2}\right)^{b} (2 + \log_2 b)$ \\
		      	    &						& $b \ge m$			 	  	 			& $x_i=- w\hat{x}^2_{i-1} +2\hat{x}_{i-1}$, $i=1,\ldots,s$ 				&  \\
		    	    &						& $s = \lceil \log_2 b \rceil	$ 	 	 			& 2. Return $\hat x_s$		& \\  
\hline
  $\sqrt w$          	    & $w\geq 1 $		 	   	&${\rm SQRT}(w,n,m,b)$		      	 			&  1. Call INV$(w,n,m,b)$	 						& $\leq \left(\frac{3}{4}\right)^{b-2m} (2+b+\log_2b)$ \\
		      	    &						&$b \geq \max\{2m,4\}$ 	 					&  2. Newton iteration. Calculate &  \\
		      	    &						&$s = \lceil \log_2 b \rceil	$	 	 			&  $y_j = \frac{1}{2} (3\hat y_{j-1} - \hat x_s \hat y^3_{j-1})$, $j = 1,\ldots,s$   & \\
		        	    &						&					 	 			& 3. Return $\hat y_s$	& \\
\hline
  $w^{1/2^i}$    	    & $w\geq 1 $		 	   	& ${\rm Powerof2Roots}(  w,k,  n,m,b)$    			&  1. $z_1 = \textrm{SQRT}(w,n,m,b)$ 				& $\leq \left(\frac{3}{4}\right)^{b-2m} 2(2+b+\log_2b)$ \\
$i = 1,\ldots,k$ 	    &						&  $b \geq \max\{2m,4\}$	 			&  2. Call SQRT() repeatedly, i.e., 						&  \\
		      	    &						&$s = \lceil \log_2 b \rceil	$ for	 	 			&   $z_i = \textrm{SQRT}(z_{i-1},m+b,m,b)$, $i = 1,\ldots,k$				&\\
		      	    &						& each call of SQRT()		 	 			& 3. Return $\{z_i\}$	& \\
\hline
  $\ln w$  	   	    & $w\geq 1 $		 	   	&${\rm LN}(	w,n,m,\ell)$	         		 			&  1. $w_p = w\cdot 2^{1-p}$ 						& $\leq \left( \frac{3}{4}\right)^{5\ell/2} \left(m + \frac{32}{9} +2 \left(\frac{32}{9} + \frac{n}{\ln 2}\right)^3 \right)$ \\
		 	    &						& $b = \max\{5\ell,25\}$			 			&  2. Call $\textrm{PowerOf2Roots}(w_p,\ell,n,1,b)$  								 	&  \\
		      	    &						&  $\ell \geq \lceil \log_2 8n \rceil$ 		  		&  and let $\hat t_p$ be the $\frac{1}{2^\ell}$th root of $w_p$ & \\
		      	    &						&$r \approx \ln 2$, with $b$ bits accuracy 			&   3. Approx. $\ln \hat t_p$ with $\hat y_p$, the first two terms			&\\
		      	    &						& $p = \lceil\log_2 w\rceil$			 			&  	of	its power series expansion. & \\
		      	    &						&				 					&  Return $z_p = 2^\ell \hat y_p + (p-1) r$  & \\
\hline
  $w^f$     		    & $w\geq 1 $		 	   	& ${\rm FractionalPower}(w,f,n,m,n_f,\ell)$		 	&  1. For $i = 1,\ldots,n_f$ calculate 	& $\leq \left( \frac{1}{2}\right)^{\ell -1}$ \\
			    & $f \in [0,1]$				& $b = \max\{n,n_f, \lceil 5(\ell, 2m,\ln n_f) \rceil\}$   	& $\hat w_i = \textrm{PowerOf2Roots}(w,n_f,n,m,b)$	& \\
   			    &$f$ is $n_f$ bits			& $\ell \in \nat$ determines the error						&   2. Return $\Pi_{i\in \mathcal P} \hat w_i$ 			 	&  \\
		      	    &						& 		&  	where $\mathcal P = \{1\leq i \leq n_f: f_i = 1\}$			& \\
\hline
$w^f$		    & $0 \leq w < 1$				&${\rm FractionalPower2}(w,f,n,m,n_f,\ell)$		 	& 1. Compute $w^\prime \geq 1$ 	by left shifting $w$			&       $\leq \left( \frac{1}{2}\right)^{\ell - 3}$\\
			    & $f \in [0,1]$				&$b = \max\{n,n_f, \lceil 2\ell + 6m + 2\ln n_f \rceil, 40\}$	& 2. Call $\textrm{FractionalPower}(w^\prime,f,n,m,n_f,\ell)$  		& \\
			    & $f$ is $n_f$ bits			& $\ell \in \nat$ determines the error			& 3. \textit{Undo} the initial shift of $w$ using right 	& \\
                &						&									& shifts, FractionalPower, and INV, and return 								& \\
\hline
\end{tabular}
}
\vspace*{13pt}
\caption{Summary of algorithms. All parameters are polynomial in $n$ and $b$ and so is the cost of all algorithms.}
\label{tab:SummaryOfResults1}
\end{sidewaystable}

\clearpage

%% file: fig-SQRToverall.tex
\begin{figure}[H]


$$\qquad\qquad \begin{array}{c}
\Qcircuit @C = 1 em @R = 1.9 em{
\lstick{\ket{\hat y_0}}  & \qw & \qw & \qw & \qw & \qw & \qw & \qw & \qw &\qw & \qw & \qw & \qw & \qw & \multigate{1}{g_2} & \qw & & \ket{\hat y_1} & & \dots & & \multigate{1}{g_2} & \qw & & \ket{\hat y_{s_2}}  & & & & 
\\
\lstick{\ket{\hat x_0}} & \multigate{1}{g_1} & \qw & & \ket{\hat x_1} & & \dots & & \multigate{1}{g_1} & \qw & & \ket{\hat x_{s_1}} & & & \ghost{g_2} & \qw & & \ket{\hat x_{s_1}} & & \dots & & \ghost{g_2} & \qw & & \ket{\hat x_{s_1}} & & & & 
\\
\lstick{\ket{w}} & \ghost{g_1} & \qw & & \ket{w} & & \dots & & \ghost{g_1} & \qw & & \ket{w}  & & & \qw & \qw & \qw & \qw & \qw& \qw &\qw & \qw & \qw & & \ket{w} \\
& & & & & \raisebox{-1.5 em}{$s_1\ \text{iterations}$} & & & & & & & & & & & & \raisebox{-1.1 em}{$s_2\ \text{iterations}$}
%
\gategroup{2}{2}{3}{9}{2.1 em}{_\}} \gategroup{2}{15}{3}{22}{2.1 em}{_\}}
}
\end{array}$$
\vspace*{2pt}
\caption{Block diagram of the overall circuit computing $\sqrt{w}$. Two stages of Newton's iteration using the functions $g_1$ and $g_2$ are applied $s_1$ and $s_2$ times respectively. The first stage outputs $\hat x_{s_1} \approx \frac{1}{w}$, which is then used by the second stage to compute $\hat y_{s_2}\approx \tfrac 1{\sqrt{\hat x_{s_1}}} \approx \sqrt{w}$. } 

\label{fig-SQRToverall}
\end{figure}
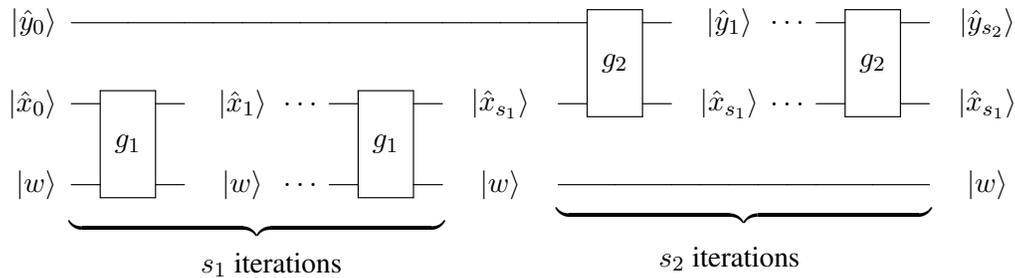

%% file: fig-InitState1.tex
\begin{figure}[H]
\centerline{
\Qcircuit @C = 2.0 em @R = 1.6 em{
Digit & & & Initial\ State & & & & & & & & & & & Final\ State \\ 
2^{m-1} & & & \lstick{\ket{w^{(m-1)}}} & \ctrl{13} & \qw & \qw & \dots & & \qw & \qw & \qw & \qw & \qw \\
2^{m-2} & & & \lstick{\ket{w^{(m-2)}}} & \qw & \ctrl{8} & \qw & \dots & & \qw & \qw & \qw & \qw & \qw \\
& & \vdots & & & & & \vdots & & & & & \vdots  \\
2^0 & & & \lstick{\ket{w^{(0)}}} & \qw & \qw & \qw & \dots & & \ctrl{6} & \qw & \qw & \qw & \qw & \raisebox{-4 em}{$\ket{w}$} \\
& & \vdots & & & & & \vdots & & & & & \vdots \\
2^{m-n} & & & \lstick{\ket{w^{(m-n)}}} & \qw & \qw & \qw & \dots & & \qw & \qw & \qw & \qw & \qw \\
2^{m-n-1} & & & \lstick{\ket{0}} & \qw & \qw & \qw & \dots & & \qw & \qw & \qw & \qw & \qw \\
& & \vdots & & & & & \vdots & & & & & \vdots \\
2^{-b} & & & \lstick{\ket{0}} & \qw & \qw & \qw & \dots & & \qw & \qw & \qw & \qw & \qw \\
Ancilla & & & \lstick{\ket{0}} & \targ & \ctrlo{3} & \targ & \dots & & \ctrlo{1} & \targ & \targ & \gate{X} & \qw & \ket{0} \\
2^{-1} & & & \lstick{\ket{0}} & \qw  & \qw & \qw & \dots & & \targ & \ctrl{-1} & \ctrlo{-1} & \qw & \qw \\
& & \vdots & & & & & \vdots & & & & & \vdots \\
2^{-(m-1)} & & & \lstick{\ket{0}} & \qw & \targ & \ctrl{-3} & \dots & & \qw & \qw & \ctrlo{-2} & \qw & \qw \\
2^{-m} & & & \lstick{\ket{0}} & \targ & \qw & \qw & \dots & & \qw & \qw & \ctrlo{-1} & \qw & \qw & \raisebox{3.5 em}{$\ket{\hat x_0}$} \\
& & \vdots & & & & & \vdots & & & & & \vdots \\
2^{-b} & & & \lstick{\ket{0}} & \qw & \qw & \qw & \dots & & \qw & \qw  & \qw & \qw & \qw
\gategroup{2}{14}{10}{14}{1.5 em}{\}} \gategroup{12}{14}{17}{14}{1.2 em}{\}}
}
}
\vspace*{8pt}
\caption{A quantum circuit computing the initial state $\ket{\hat x_0} = \ket{2^{-p}}$, for $\ket{w}$ given by $n$ bits of which $m$ are for its integer part, where $p\in \nat$ and $2^p > w \geq 2^{p-1}$. Here 
we have taken $b\ge n-m$. This circuit is used in step 4 of Algorithm 1 SQRT. 
Each horizontal line (\lq\lq wire\rq\rq) represents a qubit. 
This circuit consists of controlled-not (CNOT) and controlled-controlled-not (Toffoli) gates. Each crossed circle indicates a target qubit, controlled by the qubit(s) indicated by black dots. White dots indicate inverted control. 
See Appendix \ref{ch:QC} for a review of some important quantum gates. } 

\label{fig-InitState1}
\end{figure}
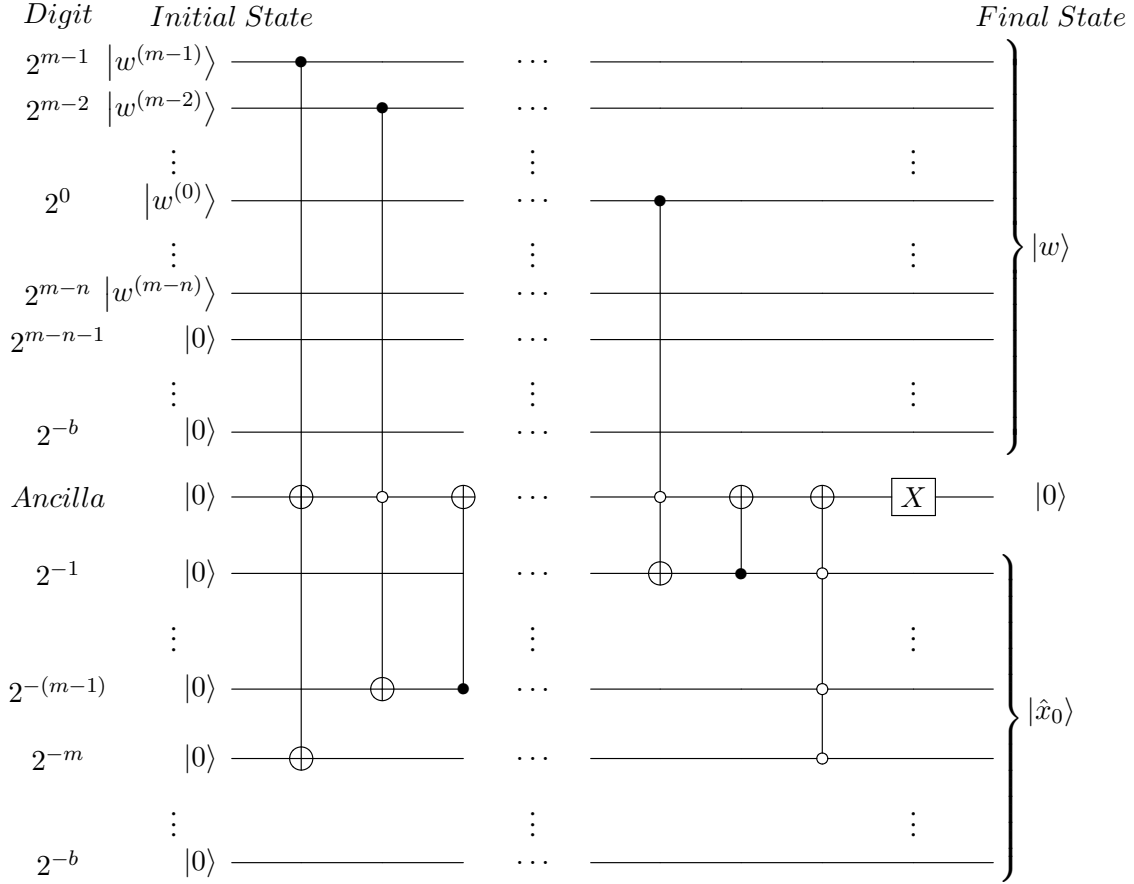

%% file: fig-InitState2.tex
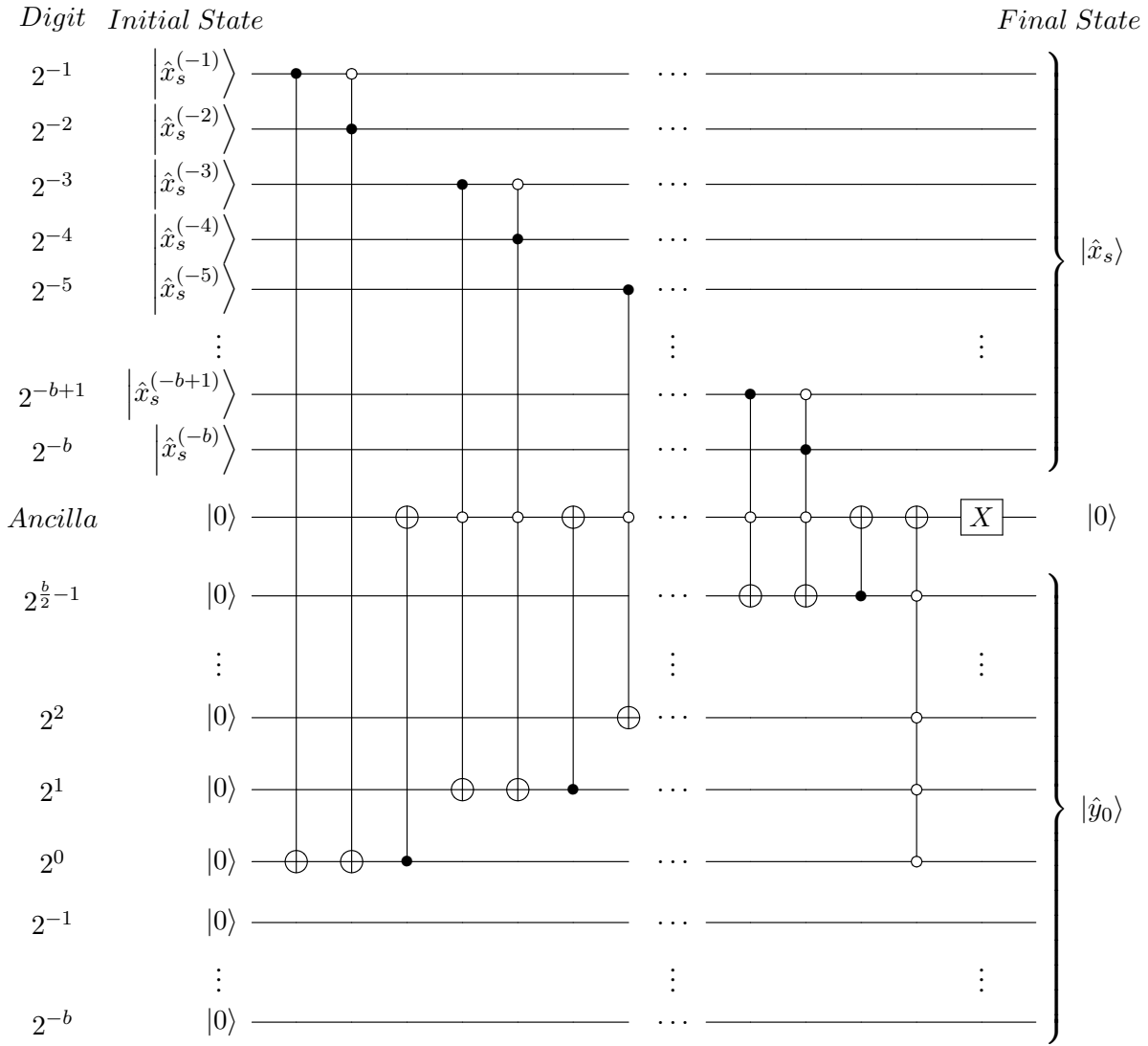
\begin{figure}[H]
\centerline{
\Qcircuit @C = 1.2 em @R = 1.8 em{
Digit & & & & Initial\ State & & & & & & & & & & & & & & & & & & Final\ State \\
2^{-1} & & & & & & \lstick{\ket{\hat{x}^{(-1)}_{s}}} & \ctrl{13} & \ctrlo{1} & \qw & \qw & \qw & \qw & \qw & \dots & & \qw & \qw & \qw & \qw & \qw & \qw \\
2^{-2} & & & & & &  \lstick{\ket{\hat{x}^{(-2)}_{s}}} & \qw & \ctrl{12} & \qw & \qw & \qw & \qw & \qw & \dots & & \qw &  \qw & \qw & \qw & \qw & \qw \\
2^{-3} & & & & & &  \lstick{\ket{\hat{x}^{(-3)}_{s}}} & \qw & \qw & \qw & \ctrl{6} & \ctrlo{1} & \qw & \qw & \dots & & \qw & \qw & \qw & \qw & \qw & \qw \\
2^{-4} & & & & & & \lstick{\ket{\hat{x}^{(-4)}_{s}}} & \qw & \qw & \qw & \qw & \ctrl{5} & \qw &\qw & \dots & & \qw & \qw & \qw & \qw & \qw & \qw \\
2^{-5} & & & & & & \lstick{\ket{\hat{x}^{(-5)}_{s}}} & \qw & \qw & \qw & \qw & \qw & \qw & \ctrl{4} & \dots & & \qw & \qw & \qw & \qw & \qw & \qw & & \raisebox{3 em}{$\ket{\hat{x}_{s}}$} \\
& & & & & \vdots & & & & & & & & & \vdots & & & & & & \vdots \\
2^{-b+1} & & & & & & \lstick{\ket{\hat{x}^{(-b+1)}_{s}}} & \qw & \qw & \qw & \qw & \qw & \qw & \qw & \dots & & \ctrl{2} & \ctrlo{1} & \qw & \qw & \qw & \qw \\
2^{-b} & & & & & & \lstick{\ket{\hat{x}^{(-b)}_{s}}} & \qw & \qw & \qw & \qw & \qw & \qw & \qw & \dots & & \qw & \ctrl{1} & \qw & \qw & \qw & \qw \\
Ancilla & & & & & & \lstick{\ket{0}} &  \qw & \qw & \targ & \ctrlo{4} & \ctrlo{4} & \targ & \ctrlo{3} & \dots & & \ctrlo{1} & \ctrlo{1}  & \targ & \targ & \gate{X} & \qw & & \ket{0} \\
2^{\frac{b}{2} - 1} & & & & & & \lstick{\ket{0}} & \qw & \qw & \qw & \qw & \qw & \qw & \qw & \dots & & \targ & \targ & \ctrl{-1} & \ctrlo{-1} & \qw & \qw \\
& & & & & \vdots & & & & & & & & & \vdots & & & & & & \vdots \\
2^{2} & & & & & & \lstick{\ket{0}} & \qw & \qw & \qw &\qw & \qw & \qw & \targ & \dots & & \qw & \qw & \qw & \ctrlo{-2} & \qw & \qw \\
2^{1} & & & & & & \lstick{\ket{0}} & \qw & \qw & \qw & \targ & \targ & \ctrl{-4} & \qw & \dots & & \qw & \qw & \qw & \ctrlo{-1} & \qw & \qw \\
2^{0} & & & & & & \lstick{\ket{0}} & \targ & \targ & \ctrl{-5} & \qw & \qw & \qw & \qw & \dots & & \qw & \qw & \qw & \ctrlo{-1} & \qw & \qw & & \raisebox{4 em}{$\ket{\hat y_0}$} \\
2^{-1} & & & & & & \lstick{\ket{0}} & \qw & \qw & \qw & \qw & \qw & \qw & \qw & \dots && \qw & \qw & \qw  & \qw & \qw & \qw \\
& & & & & \vdots & & & & & & & & &  \vdots & & & & & & \vdots \\
2^{-b} & & & & & & \lstick{\ket{0}} & \qw & \qw & \qw & \qw & \qw & \qw & \qw & \dots & & \qw & \qw & \qw & \qw & \qw & \qw
\gategroup{2}{22}{9}{22}{1.5 em}{\}} \gategroup{11}{22}{18}{22}{1.5 em}{\}}
}
}
\vspace*{8pt}
\caption{A quantum circuit computing the state $\ket{\hat y_0} = \ket{2^{\lfloor \frac{q-1}{2} \rfloor}}$, for $0 < \hat{x}_s < 1 $ given by $b$ bits, where $q\in \nat$ and $2^{1-q} > \hat{x}_s \geq 2^{-q}$. 
This circuit is used in step 10 of Algorithm 1 SQRT. 
It is for the case of even $b$; a similar circuit follows for odd $b$.}
\label{fig-InitState2}
\end{figure}

%% file: fig-LogOverall.tex
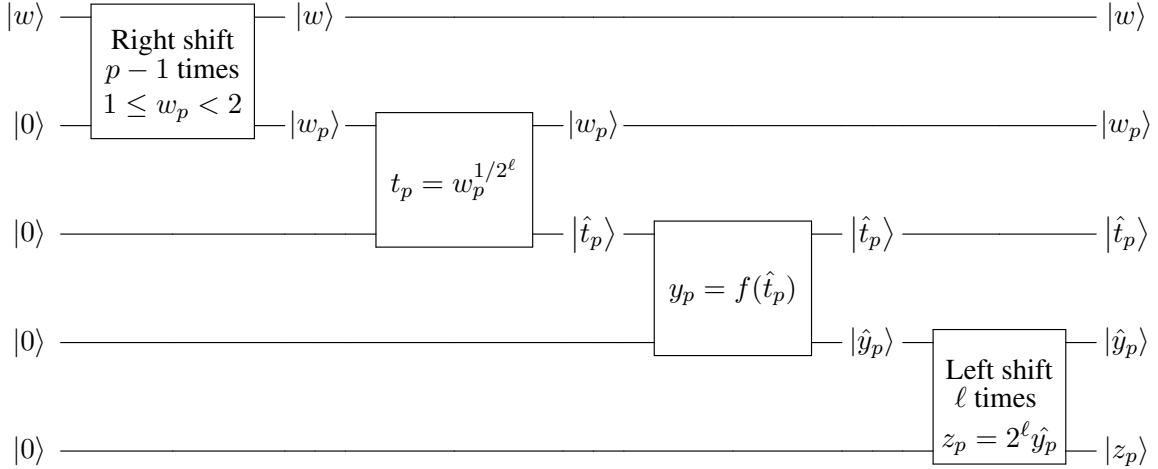
\begin{figure}[H]
\centerline{
\Qcircuit @C = 1.05 em @R = 2.85 em{
\lstick{\ket{w}} & \multigate{1}{p-1\ \text{times}} & \qw & \ket{w} & & \qw & \qw & \qw & \qw & \qw & \qw & \qw & \qw & \qw & \qw & \ket{w} \\
\lstick{\ket{0}} & \ghost{p-1\ \text{times}} & \qw & \ket{w_p} & & \multigate{1}{t_p=w_p^{1/2^\ell}} & \qw & \ket{w_p} & & \qw & \qw & \qw & \qw & \qw \raisebox{-17.5em}{\text{Left shift}} & \qw & \ket{w_p} \\
\lstick{\ket{0}} & \qw \raisebox{9 em}{$1 \leq w_p<2$} & \qw & \qw & \qw & \ghost{t_p=w_p^{1/2^\ell}} & \qw & \ket{\hat{t}_p}  & & \multigate{1}{y_p=f(\hat{t}_p)} & \qw & \ket{\hat{t}_p} & & \qw \raisebox{-15em}{$z_p=2^\ell\hat{y_p}$} & \qw & \ket{\hat{t}_p} \\
\lstick{\ket{0}} & \qw \raisebox{21em}{\text{Right shift}}& \qw & \qw & \qw & \qw & \qw & \qw & \qw & \ghost{y_p=f(\hat{t}_p)} & \qw & \ket{\hat{y}_p} & & \multigate{1}{\ \ell\ \text{times}\ \ } & \qw & \ket{\hat{y}_p} \\
\lstick{\ket{0}} & \qw & \qw & \qw & \qw & \qw & \qw & \qw & \qw & \qw & \qw & \qw & \qw & \ghost{\ \ell\ \text{times}\ \ } & \qw & \ket{z_p} 
}
}
\vspace*{8pt}
\caption{Overall circuit schematic for approximating $\ln w$.  
The gate $f(\hat{t}_p)$ outputs $\hat{y}_p =  (\hat{t}_p - 1) - \frac{1}{2} (\hat{t}_p - 1)^2$. 
Once $z_p$ is obtained the approximation of $\ln w$ is computed by the expression
 $z_p + (p-1)r$,  where $r$ approximates $\ln 2$ with high accuracy and $p-1$ is obtained from a quantum circuit as the one shown in Fig.~\ref{fig-compute_p}.}
\label{fig-LogOverall}
\end{figure}

%% file: fig-compute_p.tex
\begin{figure}[H]

\centerline{
\Qcircuit @C = 1.2 em @R = 1.7 em{
Digit & & & & Initial\ State & & & & & & & & & & Final\ State \\
2^{0} & & & & \lstick{\ket{x^0}}  & \qw & \qw & \qw & \qw & \qw & \qw & \qw & \qw & \qw \\
2^{-1} & & & & \lstick{\ket{x^{(-1)}}}  & \ctrl{11} & \qw  & \qw & \qw & \qw & \qw & \qw & \qw & \qw \\
2^{-2} & & & & \lstick{\ket{x^{(-2)}}} & \qw & \ctrl{9}  & \qw & \qw & \qw & \qw & \qw & \qw & \qw \\
2^{-3} & & & & \lstick{\ket{x^{(-3)}}} & \qw & \qw & \ctrl{9}  & \qw & \qw & \qw & \qw & \qw & \qw \\
2^{-4} & & & & \lstick{\ket{x^{(-4)}}} & \qw & \qw & \qw & \ctrl{6}  & \qw & \qw & \qw & \qw & \qw & & \raisebox{-1 em}{$\ket{x}$} \\
2^{-5} & & & & \lstick{\ket{x^{(-5)}}} & \qw & \qw & \qw & \qw & \ctrl{7} & \qw & \qw & \qw & \qw  \\
2^{-6} & & & & \lstick{\ket{x^{(-6)}}} & \qw & \qw & \qw & \qw & \qw & \ctrl{5} & \qw & \qw & \qw \\
2^{-7} & & & & \lstick{\ket{x^{(-7)}}} & \qw & \qw & \qw & \qw & \qw & \qw & \ctrl{5} & \qw & \qw \\
2^{-8} & & & & \lstick{\ket{x^{(-8)}}} & \qw & \qw & \qw & \qw & \qw & \qw & \qw & \ctrl{1} & \qw \\
2^3 & & & & \lstick{\ket{0}} & \qw & \qw & \qw & \qw & \qw & \qw & \qw & \targ & \qw & & \raisebox{-10 em}{$\ket{p-1}$}\\
2^2 & & & & \lstick{\ket{0}} & \qw & \qw & \qw  & \targ & \targ & \targ  & \targ & \qw & \qw \\
2^1 & & & & \lstick{\ket{0}}  & \qw & \targ  & \targ & \qw & \qw & \targ & \targ & \qw & \qw \\
2^0 & & & & \lstick{\ket{0}} & \targ  & \qw  & \targ  & \qw  & \targ & \qw  & \targ & \qw & \qw \gategroup{2}{14}{10}{14}{1.5 em}{\}} \gategroup{11}{14}{14}{14}{1.2 em}{\}}
}
}
\vspace*{8pt}
\caption{Example of a quantum circuit computing $p-1\ge 0$ required in the last step of Algorithm~3~{LN}. The input to this circuit is
the state $\ket{x}$ computed in Fig.~\ref{fig-RShift} where $x=2^{-(p-1)}$. Recall that $m$ bits are used to store $x$, and clearly $\lceil \log_2 m \rceil$ bits suffice to store $p-1$ exactly.
In this example, $m = 9$. It is straightforward to generalize this circuit to an arbitrary number $m$. }

\label{fig-compute_p}
\end{figure}
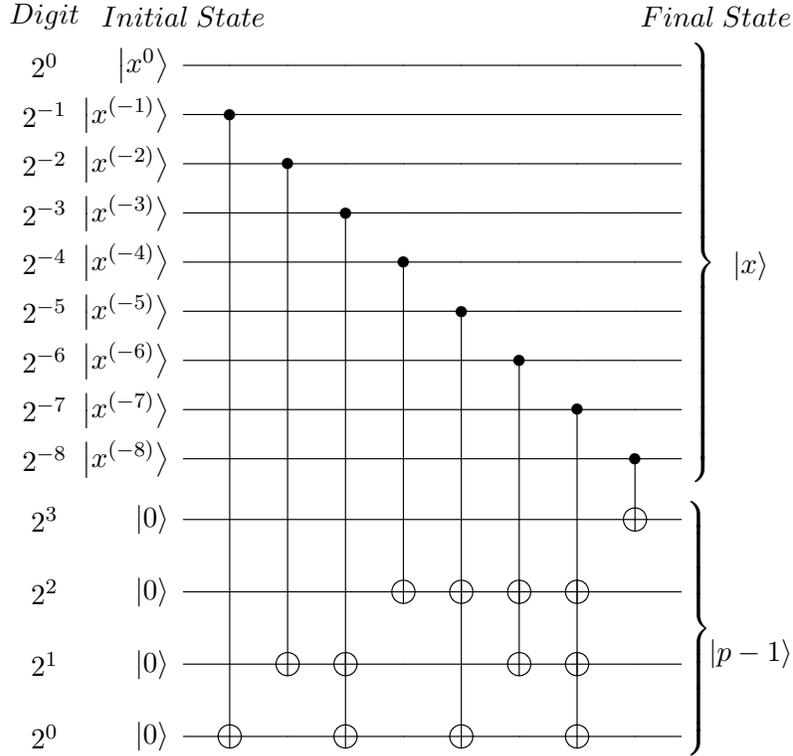

%% file: fig-ShiftInteger.tex
\begin{figure}[H]

\centerline{
\Qcircuit @C = 1.2 em @R = 1.7 em{
Digit & & & & Initial\ State & & & & & & & & & & & & &  &  & & Final\ State \\
2^{m-1} & & & & \lstick{\ket{0}} & \qw & \qw & \qw & \qw & \qw & \qw & \qw & \qw & \qw & \qw & \qw & \qw & \qw & \qw & \qw & \qw \\
& & & \vdots \\
2^{0} & & & & \lstick{\ket{0}} & \qw & \qw & \qw & \qw & \qw & \qw & \qw & \qw & \qw &\qw & \qw & \qw & \qw & \qw & \qw & \qw \\
2^{-1} & & & & \lstick{\ket{w^{(-1)}}} & \ctrl{12} & \qw & \qw & \qw & \qw & \qw & \qw & \qw & \qw & \qw & \qw & \qw & \qw & \qw & \qw & \qw \\
2^{-2} & & & & \lstick{\ket{w^{(-2)}}} & \qw & \ctrl{7} & \qw & \qw & \qw & \qw  & \qw & \qw & \qw & \qw & \qw & \qw & \qw & \qw & \qw & \qw \\
2^{-3} & & & & \lstick{\ket{w^{(-3)}}} & \qw & \qw  & \qw & \ctrl{6} & \qw & \qw & \qw & \qw & \qw & \qw & \qw & \qw & \qw & \qw & \qw & \qw & \raisebox{-1 em}{$\ket{w}$} \\
2^{-4} & & & & \lstick{\ket{w^{(-4)}}} & \qw & \qw & \qw & \qw & \qw & \ctrl{5} & \qw & \qw & \qw & \qw & \qw & \qw & \qw & \qw & \qw & \qw \\
2^{-5} & & & & \lstick{\ket{w^{(-5)}}} & \qw & \qw & \qw & \qw  & \qw & \qw &\qw & \ctrl{4} & \qw  & \qw & \qw & \qw & \qw & \qw & \qw & \qw \\
2^{-6} & & & & \lstick{\ket{w^{(-6)}}} & \qw & \qw & \qw & \qw & \qw & \qw & \qw & \qw & \qw & \ctrl{3} & \qw & \qw & \qw & \qw & \qw & \qw \\
2^{-7} & & & & \lstick{\ket{w^{(-7)}}} & \qw & \qw & \qw & \qw & \qw & \qw & \qw & \qw & \qw & \qw & \qw & \ctrl{2} & \qw & \qw & \qw & \qw \\
2^{-8} & & & & \lstick{\ket{w^{(-8)}}} & \qw & \qw & \qw & \qw & \qw & \qw & \qw & \qw & \qw & \qw & \qw & \qw & \qw & \ctrl{1} & \qw & \qw \\
Ancilla & & & & \lstick{\ket{0}} & \targ & \ctrlo{3} & \targ & \ctrlo{4} & \targ & \ctrlo{2} & \targ & \ctrlo{4} & \targ &\ctrlo{3} & \targ & \ctrlo{4} & \targ & \ctrlo{1} & \targ & \gate{X} & \ket{0} \\
2^3 & & & & \lstick{\ket{0}} & \qw & \qw & \qw & \qw & \qw & \qw & \qw & \qw & \qw & \qw & \qw & \qw & \qw & \targ & \ctrl{-1} & \qw & \raisebox{-10 em}{$\ket{k}$}\\
2^2 & & & & \lstick{\ket{0}} & \qw & \qw & \qw & \qw & \qw & \targ & \ctrl{-2} & \targ & \ctrl{-2} & \targ & \ctrl{-2} & \targ & \ctrl{-2} & \qw & \qw & \qw \\
2^1 & & & & \lstick{\ket{0}} & \qw & \targ & \ctrl{-3} & \targ & \ctrl{-3} & \qw & \qw & \qw & \qw & \targ & \ctrl{-1} & \targ & \ctrl{-1} & \qw & \qw & \qw  \\
2^0 & & & & \lstick{\ket{0}} & \targ & \qw & \qw & \targ & \ctrl{-1} & \qw & \qw & \targ & \ctrl{-2} & \qw & \qw  & \targ & \ctrl{-1} & \qw & \qw & \qw \gategroup{2}{21}{12}{21}{1.5 em}{\}} \gategroup{14}{21}{17}{21}{1.2 em}{\}}
}
}
\vspace*{8pt}
\caption{Example of a quantum circuit computing the positive integer $k$ such that $2^k w \geq 1 > 2^{k-1} w$,
for $0 < w < 1$ with $n-m$ bits after the decimal point. 
In this example, $n-m = 8$. 
It is straightforward to generalize this circuit to arbitrary numbers of bits. 
This circuit is used in Algorithm 5 FractionalPower2. 
}

\label{fig-ShiftInteger}
\end{figure}
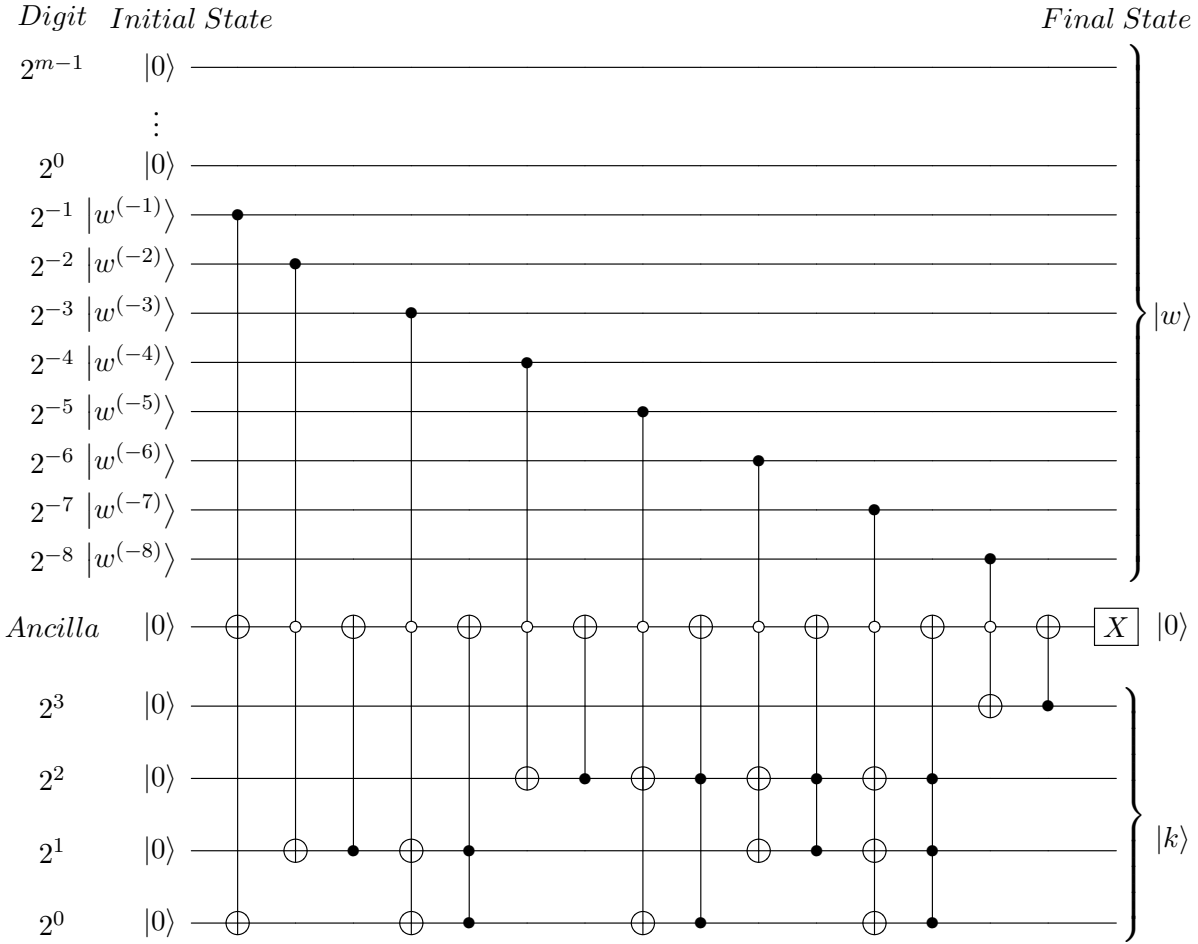

%% file: _ch_ExcitedStates.tex
\chapter{Approximating Ground and Excited State Energies on a Quantum Computer}
\label{ch:ExcStates}

\section{Introduction}
Computing eigenvalues of Hamiltonians with a large number of degrees of freedom is a very challenging problem in computational science and engineering.  Hamiltonian eigenvalues give the system energy levels, corresponding to the \textit{ground} and \textit{excited} states of the system. 
For example, one of the most important tasks in chemistry is to calculate the energy levels of molecules, where the number of degrees of freedom is proportional to the number of particles, 
which are required for computing reaction rates and electronic structure properties that, in particular, depend principally on the low-order energy levels. 
The best classical algorithms known for such problems have cost that grows exponentially in the number of degrees of freedom \cite{lanyon2010towards}. Therefore, efficient quantum algorithms 
for computing Hamiltonian eigenvalues 
would be an extremely powerful tool for new science and technology.

On the other hand, there are a number of recent results in discrete complexity theory suggesting that many eigenvalue problems 
are very hard even for quantum computers because they are QMA-complete 
\cite{KempeLocalHam,intBosons,schuchDFT,childs2013bose, qmaSurvey}. However, discrete complexity theory deals with the worst case over large classes of Hamiltonians. It does not provide methods or necessary conditions determining when an  eigenvalue problem is hard. In fact, there is a dichotomy between theory and practice. As stated in  \cite{love2012back}, \lq\lq complexity theoretic proofs of the advantage of many widely used classical algorithms are few and far between.\rq\rq\
Therefore, it is important to develop new quantum algorithms and
to use them for solving  eigenvalue problems for which quantum computing can be shown to have a significant advantage over classical computing.

Several recent works have made progress in this direction. 
In \cite{GS}, the authors developed an algorithm and proved strong exponential quantum speedup for approximating the ground state energy (i.e., the smallest eigenvaue) of the time-independent Schr\"odinger equation under certain assumptions.
Why this problem is different from the QMA-complete problems of discrete complexity theory was made clear in 
\cite{Qspeedup}. The authors of \cite{Convex} relaxed 
the assumptions of \cite{GS} 
to give a ground state energy approximation algorithm for 
the time-independent Schr\"odinger equation with a convex potential.

An important advance would be to obtain analogous results for approximating excited state energies, under weakened assumptions. The techniques of \cite{GS,Convex,Qspeedup} 
for approximating the ground state energy do not extend to excited state energies. 
In this chapter, 
we present an entirely new approach for approximating a constant number of low-order eigenvalues. 
Our approach applies to a general class of eigenvalue problems \cite{hadfield2015approximating}. 
We illustrate our results by considering the time-independent Schr\"odinger equation 
with a number of degrees of freedom~$d$, 
under weaker assumptions than those of \cite{GS,Convex}. 
More precisely, we consider the eigenvalue problem 
\begin{eqnarray} \label{eqn:TISE1}
\left(-\frac{1}{2}\Delta + V \right) \Psi(x) &=& E \: \Psi(x) \quad x \in I_d = (0,1)^d,\\
\Psi(x) &=& 0 \quad  \quad  \quad \;\;\; x \in \partial I_d,  \label{eqn:TISE2}
\end{eqnarray}
where $\Delta$ denotes the Laplacian and $\Psi$ is a normalized eigenfunction. 
(Here all masses and the normalized Planck constant are set to one, $d$ is proportional to the number of particles, and we assume the potential $V$ is smooth and uniformly bounded as we will explain later.) In general, this problem may have many \textit{degenerate} energy levels 
(eigenvalues with multiplicity greater than one). 

We give a quantum algorithm, and derive cost and success probability bounds, for approximating a constant number 
of low-order eigenvalues, ignoring eigenvalue multiplicities. 
For accuracy $O(\e)$ and success probability at least $3/4$, the cost and the number of qubits of our algorithm are each polynomial in~$d$ and
~$\e^{-1}$, and hence our algorithm is efficient. 
As the best classical algorithms known for this problem have costs that grow exponentially in~$d$, our quantum algorithm gives an exponential speedup. 
The results of this chapter can also be found in \cite{hadfield2015approximating}.

\section{Problem Definition}  \label{sec:ProbDef}
We consider an eigenvalue problem for a self-adjoint operator $L$ with a discrete
spectrum, which generalizes the problem of computing the ground state energy (lowest eigenvalue). 
Let 
\begin{equation}   \label{eqn:exc}
E_{(0)} < E_{(1)} < ... < E_{(i)} < ...
\end{equation}
be 
the eigenvalues of $L$ ignoring multiplicities, which we call the \textit{energy levels} of $L$. 
We refer to energy levels $E$ satisfying $E\leq E_{(j)}$, $j=O(1)$, as \textit{low order}.  
Suppose we want to estimate the lower part of the spectrum with accuracy $O(\e)$.
Since any two distinct eigenvalues of $L$ can be arbitrarily close to each other, any algorithm that approximates the lower part of the spectrum with accuracy $\e$ cannot be expected to distinguish between
all eigenvalues $E_{(k)} \neq E_{(l)}$ with $|E_{(k)} - E_{(l)} | = O(\e)$. We call such eigenvalues $\e$-\textit{degenerate}. So, 
the problem is to obtain an algorithm whose output will be $j$ numbers 
\begin{equation}   \label{eq:tildeE}
\tilde{E}_0 < \tilde{E}_1 < ... < \tilde{E}_{j-1}
\end{equation}
satisfying with high probability the following conditions:
\begin{enumerate}
\item[\bf{C1}] For every $i \neq k\in\{0,\dots,j-1\}$, there exist $E_{(s_i)}\neq E_{(s_k)}$ such that $|E_{(s_i)} - \tilde{E}_{i}|=O(\e)$ and $|E_{(s_k)} - \tilde{E}_{k}|=O(\e)$,  i.e., different outputs are approximations of different eigenvalues with error $O(\e)$, respectively.
\item[\bf{C2}]  If $|\tilde{E}_{i+1} - \tilde{E}_{i}|  = \omega(\e) $, there is no eigenvalue $E$ of $L$ satisfying $\tilde{E}_{i} < E < \tilde{E}_{i+1}$ and $\min(|\tilde{E}_{i+1} -E|, |\tilde{E}_{i} -E|) = \omega(\e)$.\footnote{For functions $f,g\ge 0$ defined on $\reals_+$, the notation $f(\e)=\omega(g(\e))$ means that for any $M>0$, arbitrarily large, we have $f(\e)\ge M g(\e)$ for sufficiently small $\e$.} Thus the algorithm doesn't miss (or skip) any eigenvalues in the lower part of the spectrum unless they are $O(\e)$ apart, i.e.,
{$\e$-degenerate}.
\end{enumerate}
Clearly, if $\e$ is sufficiently small such that the eigenvalues of $L$ are well-separated, then the algorithm produces approximations with error $O(\e)$ of the $j$ smallest distinct eigenvalues. 

We will employ a perturbation-based approach.
Assume that $L^0$ is another self-adjoint operator such that $L=L^0+V$, where $V=L-L^0$, 
with each operator 
acting on the same domain.  
We also assume that $L^0$ and $L$ have discrete spectra and that the eigenspaces associated with each eigenvalue are finite dimensional; see e.g. \cite{Gustafson,Sigal,Titchmarsh}. 
In the general case, selecting the partition of $L$ to $L^0$ and $V$ is not trivial and may significantly affect the problem complexity; we do not deal with this problem here. Our discussion in this section applies equally well to Hermitian matrices.

Let 
\begin{equation}   \label{eqn:sigma}
\sigma \leq E_0 \leq E_1 \leq ... \leq E_i \leq ...
\end{equation}
be the eigenvalues of $L$ indexed in nondecreasing order, where $\sigma$ is a given lower bound. Ignoring possible eigenvalue multiplicities we have the strictly increasing subsequence of eigenvalues~(\ref{eqn:exc}).  
Similarly we denote by 
\begin{equation} \label{eqn:L0eval}
E_0^0 \leq E^0_1 \leq ... \leq E^0_i \leq ...
\end{equation} 
 the eigenvalues of $L^0$ indexed in nondecreasing order, and by 
\begin{equation} \label{eq:l0excited2}
E^0_{(0)} < E^0_{(1)} < ... < E^0_{(i)} < ...,
\end{equation}
the eigenvalues of $L^0$ ignoring multiplicities. 
Assume we know the eigenvalues and eigenvectors of $L^0$. Often this is a reasonable assumption. For example, this is true for the eigenvalues and eigenvectors of the Laplacian $L^0=-\Delta$ defined on the $d$-dimensional unit cube with Dirichlet or Neumann boundary conditions.  

\section{Background} \label{sec:Overview}
We briefly review algorithms for eigenvalue problems. 
Algorithms approximating eigenvalues use a discretization of $L$ to obtain a matrix eigenvalue problem. For example, when $L$ is a differential operator, one can use a finite difference discretization \cite{Leveque}, or a finite element discretization \cite{strangFiniteElement,babuskaOsborn}. In particular, for the time-independent Schr\"odinger equation specified by equations  (\ref{eqn:TISE1}) and (\ref{eqn:TISE2}), a finite difference discretization has been used in \cite{GS,Convex}. Since $L$ is self-adjoint, the resulting matrix is symmetric.
Eigenvalue problems involving 
symmetric matrices are conceptually easy and methods such as the bisection method can be used to solve them with cost proportional to the matrix size, modulo polylog factors \cite{Demmel}. The
difficulty is that the discretization leads to a matrix of size that is exponential in $d$. Hence, the cost for approximating the matrix eigenvalue is prohibitive when $d$ is large. In fact, a stronger result is known, namely the cost of any deterministic classical algorithm approximating the ground state energy 
must be at least exponential in 
the number of degrees of freedom $d$, i.e., the problem suffers from the curse of dimensionality \cite{AP07}. 

More precisely, the discretization must be sufficiently fine so that the eigenvalues of interest are approximated by eigenvalues of the resulting matrix within the specified accuracy $\e$.  This increases the matrix size, which directly impacts the cost.  
Indeed, general matrix eigenvalue problems have been extensively studied in numerical linear algebra, and there are classical algorithms for approximating one, or some, or even all of the eigenvalues and/or the corresponding eigenvectors of a matrix \cite{Demmel,Golub,Parlett,cullum2002lanczos}. Examples of such algorithms include the power method, inverse iteration, the QR algorithm, and the bisection method for symmetric matrices. In particular, the bisection method for symmetric matrices can compute the eigenvalues that lie within a given range. However, the cost of each algorithm scales at least linearly in the size of the matrix. 

It is important to point out that different approaches may lead to different matrix eigenvalue problems that have varying degrees of difficulty. For instance, in quantum chemistry, the first and second quantization approaches for computing energies of the electronic Hamiltonian, as described in \cite{Kassal}, lead to completely different matrices with different notions of degrees of freedom.  Moreover, discretizations of certain problems in physics may lead to eigenvalue problems for stoquastic matrices, that some believe are computationally easier to solve \cite{stoquastic}.

For example, in the estimation of the ground state energy (smallest eigenvalue) of the time-independent Schr\"odinger equation (\ref{eqn:TISE1}), (\ref{eqn:TISE2}) with $V$ uniformly bounded by $1$, the finite difference discretization on a grid will yield a matrix of size 
$m^d \times m^d$, $m= 2^{\lceil  -\log_2 \e \rceil  } -1$ \cite{GS}. This means that the cost of the matrix eigenvalue algorithms mentioned above is bounded from below by a quantity proportional to  
$\e^{-d}$, 
i.e., the cost grows exponentially in $d$.  

To approximate the ground state energy of the problem specified by equations (\ref{eqn:TISE1}) and (\ref{eqn:TISE2}) in the worst case with (relative) error $\e$, assuming $V$ and its first-order partial derivatives are uniformly bounded by $1$, and the function evaluations of $V$ are supplied by an oracle, a much stronger result holds.
The complexity (i.e., the minimum cost of any classical deterministic algorithm, and not just the eigenvalue algorithms mentioned above) is bounded from below by a quantity proportional to 
$\e^{-d}$ as $d\e \rightarrow 0$ \cite{AP07,GS}. So unless $d$ is moderate, the problem is very hard and suffers from the curse of dimensionality. 
The same complexity lower bound applies to the approximation of low-order eigenvalues under the same, or more general, conditions on $V$. Finally, we point out that the complexity of this problem in the classical randomized case is an open question.

In certain cases quantum algorithms may break the curse of dimensionality by computing $\e$-accurate eigenvalue estimates with cost polynomial in $\e^{-1}$ and $d$. This was shown in \cite{GS,Convex} where we saw that for smooth nonnegative potentials that are uniformly bounded by a relatively small constant, or are  convex, 
there exists a quantum algorithm approximating the ground state energy with relative error $O(\e)$ and cost polynomial in $d$ and $\e^{-1}$.
In \cite{GS}, the Laplacian was discretized using a $2d+1$ stencil on a grid, and $V$ was discretized by evaluating it at the grid points. Our results of this chapter continue this line of research, yielding stronger
results under weaker conditions than those of \cite{GS,Convex}.

\subsection*{Quantum Phase Estimation}
There is a well-studied quantum algorithm, quantum phase estimation (QPE) \cite{AbramsLloyd,NC,aspuruGuzik2005yq}, which can be used to approximate eigenvalues of a Hamiltonian $H$. More precisely, the algorithm approximates the phase corresponding to an eigenvalue of a unitary matrix, which in our case is $e^{-iH}$. QPE is efficient if two conditions are met. The first condition is that simulating a system evolving with Hamiltonian $H$ can be done efficiently, i.e., we can approximate $e^{-iHt}$, $t \in \mathbb{R}$, accurately with low cost. The second condition requires that 
we are given a relatively good approximation of an eigenvector corresponding to the eigenvalue of interest. In addition, one should be able to implement this approximation as a quantum state efficiently. The approximate eigenvector is used to form the initial state of QPE. We also remark that QPE uses the (quantum) Fourier transform as a module. The Fourier transform can be implemented efficiently on a quantum computer \cite{NC}. 

We discuss the two required conditions for QPE further. 
Simulating the evolution of a system under a Hamiltonian $H$ appears to be a difficult problem for classical computers when the size of $H$ is large. As proposed by Feynman \cite{Feynman}, quantum computers are able to carry out such simulation more efficiently in certain cases. For example, Lloyd \cite{lloyd1996universal} showed that local Hamiltonians can be simulated efficiently on a quantum computer. About the same time, Zalka \cite{Zalka1,Zalka2} showed that many-particle systems can be also be simulated efficiently on a quantum computer. Later, Aharonov and Ta-Shma \cite{aharonov2003adiabatic} generalized Lloyd's results to sparse Hamiltonians. Berry et al. \cite{berry2007efficient} extended the cost estimates of \cite{aharonov2003adiabatic}. The results of \cite{berry2007efficient} were in turn improved by Papageorgiou and Zhang in \cite{PZ12}. Although there has been more work on quantum Hamiltonian simulation since then, the approach of \cite{berry2007efficient,PZ12} suffices for our discussion. 
These papers assume that $H$ is given by a black-box (or oracle), and that $H$ can be decomposed efficiently by a quantum algorithm, using oracle calls, into a finite sum of Hamiltonians that individually can be simulated efficiently.
In this chapter, where $L=L^0 + V$, we assume that the Hamiltonians resulting from the discretizations of $L^0$ and $V$ can be simulated efficiently on a quantum computer. Their sum, i.e., the Hamiltonian obtained from the discretization of $L$, can be simulated efficiently using splitting formulas such as the Trotter formula, the Strang splitting formula, or Suzuki's high-order formulas; see Appendix \ref{sec:splittingFormulas} for a review. 
(We study Hamiltonian simulation in detail in Chapter \ref{ch:HamSim}.)

We now consider the second requirement of QPE, namely the availability of a good approximate eigenvector. 
QPE will produce an estimate of the eigenvalue $\lambda$ (or more precisely, an estimate of the phase $\phi \in [0,1)$ corresponding to $\lambda$ through $\lambda = e^{2\pi i\phi}$) with success probability proportional to the quality of the approximate eigenvector \cite{NC,AbramsLloyd}. If the eigenvector providing the initial state of QPE 
is known exactly, the parameters of QPE can be set so its success probability is arbitrarily close to 1 \cite{NC}. If, on the other hand, we use an approximate eigenvector, the success probability is reduced proportionally to the square of the magnitude of the projection of the approximate eigenvector onto the actual eigenvector (i.e., the square of the \textit{overlap} between the two vectors)  \cite{AbramsLloyd}.  As long as this overlap is not exponentially small, QPE is efficient. 

We remark that obtaining a good approximate eigenvector required for QPE is a particularly difficult task, in general, when the matrix size is huge. Things are complicated further if one needs a number of different approximate eigenvectors, in order to use QPE to approximate the $j>1$ lowest-order eigenvalues. We overcome this difficulty for $L=L^0 + V$ using the known eigenvalues and eigenvectors of $L^0$, and properties of $V$, as we discuss below.

\section{Algorithm}
\label{sec:Algorithm}  \label{sec:AlgorithmIdea}
We give our algorithm for eigenvalue estimation of a general self-adjoint operator $L=L^0+V$ under the assumptions of Section \ref{sec:ProbDef}. Recall that  $L^0$ and $L$ are self-adjoint operators on a Hilbert Space $\mathcal{H}$ with discrete spectra; e.g., see \cite{Gustafson}. 
The eigenvalue problem for the time-independent Schr\"odinger equation (\ref{eqn:TISE1}) 
is a special case we consider in detail in Section \ref{sec:AlgTISE}.

Our goal is to use QPE to estimate $j$ low-order energy levels of $L$. For this, we need relatively good approximations of the corresponding eigenvectors. Since $L=L^0+V$ (i.e., $L$ and $L^0$ differ by the perturbation $V$), and since we know the eigenvalues and the eigenvectors of $L^0$, we can use them to obtain the necessary approximate eigenvectors. We indicate how this can be done. For simplicity and notational convenience, we do not distinguish between operators and their matrix discretizations in the rest of this section, since it is not important for the moment. Let $E$ denote one of the low-order eigenvalues of $L$ (see equation (\ref{eqn:exc})) that we wish to estimate. Intuitively, we expect a \lq\lq small\rq\rq\ and suitably well-behaved perturbation to have a proportionately \lq\lq small\rq\rq\  effect on the eigenvectors and eigenvalues of $L^0$. Let $u$ be an arbitrary unit vector belonging to the 
eigenspace associated with $E$. Then  there exists an eigenvector $u_k^0$ of $L^0$, similarly corresponding to a low-order eigenvalue, that has an \textit{overlap} (magnitude of projection) with $u$ that is nontrivial, i.e.,  $| \braket{u_k^0}{u}|$ is not extremely small, as we will see later.


One of the keys to our approach is to form a collection $\mathcal{S}$ of the eigenvectors of $L^0$ that correspond to eigenvalues of $L^0$ that satisfy a certain property, which we specify in the next subsection. 
The goal is to have at least one element in $\mathcal{S}$ that has a reasonable overlap with a vector in the eigenspace corresponding to $E_{(i)}$, for each $i=0,1,...,j-1$. We call $\mathcal{S}$ the \textit{set of trial eigenvectors}. We will use each one of the elements of $\mathcal{S}$ repetitively as initial state in QPE, running QPE multiple times, to obtain a sequence of approximations that will lead us to estimates of each $E_{(i)}$.

Let us briefly discuss the idea for constructing $\mathcal{S}$. At one extreme, one could take $\mathcal{S}$ to be all of the eigenvectors of $L^0$, because not all of them have a negligible overlap with the eigenvectors of $L$ corresponding to the eigenvalues of interest. However, then the size of $\mathcal{S}$ can be huge. To limit $|\mathcal{S}|$, we select eigenvectors of $L^0$  that correspond to eigenvalues that do not exceed a certain bound. 
Roughly speaking, we will be excluding eigenvalues of $L^0$ that correspond to energies grossly exceeding the energies of $L$ that we wish to estimate. This idea is made precise in equation ($\ref{eq:upperBound}$) in next section.

The cardinality of $\mathcal{S}$ depends on the eigenvalue distribution of $L^0$.  
If the cardinality of $\mathcal{S}$ is not prohibitively large, and if we can discretize its elements and  efficiently prepare the corresponding quantum states, then we can run QPE repeatedly for the all elements of $\mathcal{S}$ to produce an estimate of $E_{(i)}$ among its different outputs with a sufficiently high probability, $i=0,1,...,j-1$. This probability can be boosted to become arbitrarily close to $1$ using further repetitions of the procedure. We remark that the cardinality of $\mathcal{S}$ depends on the distribution of eigenvalues of $L^0$ and the properties of $V$.
Observe that detecting the desired estimates $\tilde{E}_0, ..., \tilde{E}_{j-1}$ from the outcomes obtained from the different runs of QPE is not a trivial task, and we will show how this is accomplished.

\subsection{Preliminary Analysis} 
\label{sec:PrelimAnal}
Suppose we wish to compute a specific eigenvalue $E$ of $L$. Let $u$ be a unit vector in the (possibly degenerate) subspace associated with $E$, i.e., satisfying $Lu = Eu$.  Then we have
$$ \| L u - L^0 u\|^2  =  \| (L^0 + V) u - L^0 u\|^2 = \|V u \|^2 
\leq  \|V\|^2 .$$
Expanding in the basis of unperturbed eigenvectors, we have $\ket{u} = \sum_i \beta_{i} \ket{u^0_i}$ where $\beta_{i} = \braket{u^0_i}{u}$. Then
$$ \| L u - L^0 u\|^2  = \| E \sum_i \beta_{i} \ket{u^0_i} - \sum_i \beta_{i} E_i^0 \ket{u^0_i}  \|^2 = \| \sum_i \beta_{i} (E - E_i^0)  \ket{u^0_i} \|^2$$
Combining these expressions and using the eigenvector orthonormality gives
\begin{equation} \label{eqn:sum}
 \|V\|^2 \geq \| L u - L^0 u\|^2  = \sum_i |\beta_{i}|^2 (E^0_i - E)^2 
\end{equation}
Assume we know an upper bound $B \geq E$ and a constant $c>1$ such that their exist eigenvalues of 
$L^0$ larger than $c\|V\|+B$, and define the set $\mathcal{I}:=\{i:E^0_i > c\|V\| +B \}$.  
Observe that this is true for instances of the time-independent Schr\"odinger equation \cite{Titchmarsh}. 
From 
(\ref{eqn:sum}), 
we obtain
$$ \|V\|^2 \geq  \sum_{i \in \mathcal{I} } |\beta_{i}|^2 (E_i^0 - E)^2 \geq  \sum_{i \in \mathcal{I}} |\beta_{i}|^2 (c\|V\|)^2 ,$$
which we rearrange to give
\begin{equation}
  \sum_{i \in \mathcal{I}} |\beta_{i}|^2   \leq \frac{1}{c^2},  
\end{equation} 
or, equivalently,  
\begin{equation} \label{eqn:probBound}
 \sum_{i \notin \mathcal{I}} |\beta_{i}|^2  \geq 1 - \frac{1}{c^2  }  =: q .  
\end{equation}
Thus there must exist an index $k \notin \mathcal{I}$ such that $|\beta_k|^2 \geq \frac{q}{|\mathcal{S}|}$. If $|\mathcal{S}|$ is not extremely large, then one of the first 
$|\mathcal{S}|$ eigenvectors of $L^0$ must have a reasonable overlap with $u$. (Here and elsewhere, by reasonable overlap we mean that the magnitude of the projection is not exponentially small in $d$.)

\subsection{Algorithm Description}
\label{sec:AlgorithmDescrip}
Let $V$ be such that $\| V\|:= \sup\{ \|Vu\|: \|u\| = 1\} < \infty$ uniformly in $d$. 
 Assume we are given (or we have derived) $c>1$ 
and $B$ a sufficiently large upper bound on the lower part of the spectrum of~$L$ which is of interest.\footnote{We give an explicit construction for $B$ in equation (\ref{hyp1}).} 
Consider the set of indices 
\begin{equation} \label{eq:upperBound}
\mathcal{I}=\{i:E^0_i-B > c\|V\| \} \neq \emptyset.
\end{equation}

We define $\mathcal{S}$ to be the set of eigenvectors of $L^0$ that correspond to eigenvalues $E_i^0$ with $i \notin \mathcal{I}$;
in the case of degeneracy, it suffices to select any basis of the degenerate subspace. By constructing $\mathcal{S}$ in this way, we are guaranteed that at least one of its elements will  overlap sufficiently with an element of the degenerate eigenspace corresponding to each $E_{(i)}$, for $i=0,1,...,j-1$.
We will show that the magnitude of this overlap is bounded from below by a positive constant. 

We now give our quantum algorithms for approximating $j=O(1)$ low-order eigenvalues of the operator $L=L^0+V$. Algorithm 1 deals with the special case of approximating the ground state energy $E_{(0)}$.
This algorithm illustrates our idea of using a set of trial eigenvectors to approximate an eigenvalue of $L$. 
Algorithm 2 computes the sequence of approximations $\tilde{E}_{1}, \tilde{E}_{2},...,\tilde{E}_{j-1}$, where each $\tilde{E}_{i}$ is computed using the values $\tilde{E}_{0}$ through $\tilde{E}_{i-1}$. Thus the overall procedure consists of iterating Algorithm 2 
until we obtain the $j$ desired estimates of equation $(\ref{eq:tildeE})$.

Let us pretend for the moment that $L^0$ and $L$ are $N\times N$ matrices. 
In the next section we will show how to discretize them and obtain symmetric matrices such that each of the low-order eigenvalues of these matrices approximates the corresponding eigenvalue of the respective operator with error proportional to $\e$. 

Our algorithms are based on QPE and require two quantum registers. 
The first (\textit{top}) register contains sufficiently many qubits $t$ to guarantee the required accuracy $O(\e)$ in the results with a reasonable success probability for QPE. 
The second  (\textit{bottom}) register contains 
the necessary number of qubits to hold an approximate eigenstate.
\vskip 1pc 
\textbf{Algorithm 1. Ground State Energy:}
\begin{enumerate}
     \item Define $\mathcal{S}$, the \textit{set of trial eigenvectors}, to be all eigenvectors of $L^0$ that correspond to eigenvalues $E^0_i$ $i\notin \mathcal{I}$ as defined in equation (\ref{eq:upperBound}). We denote these eigenvectors by~$u_k^0$ for~$k=0,1,.., |\mathcal{S}|-1$.
    \item Set k = 0.
    \item Prepare the initial quantum state $\ket{0}^{\otimes t} \ket{u_k^0}$. The value of $t$ is chosen so that QPE, with relatively high probability, produces outcomes leading to energy estimates with error $O(\e)$. 
   \item Perform QPE with initial state $\ket{0}^{\otimes t} \ket{u_k^0}$ using the unitary matrix $U = e^{ i A/ R}$. $A =  L$ if $L$ is nonnegative definite, and otherwise $A= L - \sigma I $, where $\sigma$ is a lower bound to the minimum eigenvalue of $L$ as we have assumed in the previous section. The parameter $\sigma$ is assumed to be known; see equation (\ref{eqn:sigma}). 
    The parameter $R$ is an upper bound to the spectral norm of $A$, which can be obtained using the eigenvalues of $L^0$ and the properties of $V$. The purpose of $R$ is to ensure that the resulting phases will lie in the interval $[0,1)$.
   \item Measure the first $t$ qubits, which give the result of QPE, and store the resulting value classically. We assume that the measurement outcomes are truncated to $b$ bits and we obtain nonnegative integers in the range $\{0,...,2^b-1 \}$, where $b<t$. The role of the extra qubits $t_0=b-t$ is to increase the success probability of QPE.
    \item $k\leftarrow k+1$.
   \item Repeat steps 3-6 while $k < |\mathcal{S}|$. 
    \item Repeat steps 2-7 $r$ many times, where $r$ is a number precomputed to ensure with high probability that the stored results after $r$ runs contain an estimate of $E_{0}$. 
    \item Take the minimum value of the stored measurement outcomes, mark it as selected, and convert it to an eigenvalue estimate $\tilde{E}_{0}$ of $E_{0}$ using the values of $\sigma$ and $R$ in the definition of $A$.
    \item Output $\tilde{E}_{0}$.
\end{enumerate}
\vskip 1pc
Note, the purpose of step 7 is to run QPE $|\mathcal{S}|$ many times, once with each $\ket{u_k^0}\in \mathcal{S}$ as input, because we do not know which of the elements of $\mathcal{S}$ has the largest overlap with the unknown ground state eigenvector, and the success probability of each run depends on this overlap. Since the largest overlap between the elements of $\mathcal{S}$  and the unknown eigenvector may not be sufficiently large so that the resulting success probability of the algorithm is bounded from below be a constant, say $\frac34$, the purpose of step 8 is to repeat the entire procedure $r$ many times to boost the success probability of computing $\tilde{E}_{0}$ correctly.

\begin{figure*}
\centering
\begin{tikzpicture}[xscale=1.3]
\draw [->](1,1) -- (1,0);
\draw [->](5.5,1) -- (5.5,0);
\draw [->](6.5,1) -- (6.5,0);
\node [above] at (1,1) {$\phi_1$};
\node [above] at (5.5,1) {$\phi_2$};
\node [above] at (6.5,1) {$\phi_3$};
\draw [thick] (-0.5,0) -- (9.5,0);
\draw (0,-.2) -- (0, .2);
\draw (3,-.2) -- (3, .2);
\draw (6,-.2) -- (6, .2);
\draw (9,-.2) -- (9, .2);
\node [below] at (0,-0.2) {$m-1$};
\node [below] at (3,-0.2) {$m$};
\node [below] at (6,-0.2) {$m+1$};
\node [below] at (9,-0.2) {$m+2$};
\end{tikzpicture}
\caption{Example of the selection of the measurement outcomes. Consider three phases $\phi_1$, $\phi_2$ and $\phi_3$ as shown. Assume that the distance between possible outcomes corresponds to error $O(\e)$. 
If $m-1$ is selected to estimate $\phi_1$, the next possible outcome the algorithm selects is $m+1$, which provides an estimate for both $\phi_2$ and $\phi_3$ in this example. Alternatively, if $m$ is selected to provide an estimate of  $\phi_1$, the next possible outcome is $m+2$, which provides an estimate of $\phi_3$, and the algorithm does not care to produce a separate estimate for $\phi_2$. Note that $\phi_2$ and $\phi_3$ are $\e$-degenerate and either can be ignored.}
\label{Fig:phaseEstimation}      
\end{figure*}
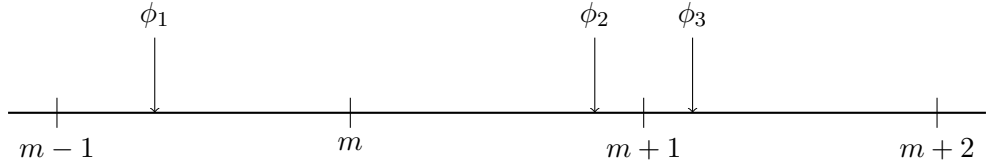

The following iterative algorithm extends Algorithm 1 to compute the sequence of approximations $\tilde{E}_{1}, .. , \tilde{E}_{j-1}$  satisfying the conditions of equation (\ref{eq:tildeE}), respectively. Every term of the computed sequence depends on all of the previously computed terms.
\vskip 1pc 
\textbf{Algorithm 2. Excited State Energies:}
\begin{enumerate}
  \item Consider $A$ as defined in Algorithm 1. Run Algorithm 1 and let $\tilde{E}_{0}$ be its output.
  \item Set $i=1$ and prepare to compute an estimate of $\tilde{E}_{1}$.
  \item Repeat steps 1-8 of Algorithm 1, storing the outcome of every measurement. We assume that the measurement outcomes are truncated to $b$ bits and we obtain nonnegative integers in the range $\{0,...,2^b-1 \}$, where $b<t$. The role of the extra qubits $t_0=b-t$ is to increase the success probability of QPE.
  \item Take the minimum of the measurement outcomes that exceeds by $2$  the last selected outcome and mark it selected. This way, with high probability, for each eigenvalue the error will be $O(\e)$, and the algorithm will not produce two different estimates for the same eigenvalue. 
Note that by  taking the minimum outcome relative to the previously selected outcome implies that the algorithm does not fail to produce estimates for consecutive eigenvalues, unless the eigenvalues differ by $O(\e)$. 
See Figure~\ref{Fig:phaseEstimation}.
  \item Use the values of $\sigma$ and $R$ in the definition of $A$ to rescale and shift the newly selected outcome to obtain the estimate $\tilde{E}_i$.
  \item Set $i\leftarrow i+1$ and prepare to compute the estimate $\tilde{E}_i$. 
  \item Repeat steps 3 through 6 if $i < j$.  
  \item Output $\tilde{E}_{1}, ..., \tilde{E}_{j-1} $.
\end{enumerate}
\vskip 1pc

It is clear that our procedure as outlined by Algorithms 1 and 2 will produce
the $j$ desired estimates (\ref{eq:tildeE}) of equation (\ref{eqn:exc}). However, its cost varies depending on $V$ and the distribution of eigenvalues of $L^0$, which as we already mentioned determine the cardinality of $ \mathcal{S}$.  In the next sections we derive tight estimates for the cost and the success probability of our algorithm for particular choices of $L^0$ and $L$.

We remark that in cases where $L$ and $L^0$ are given explicitly, using their properties one may be able to obtain a set of trial eigenvectors $\mathcal{S}$ with significantly smaller cardinality, 
substantially improving the cost of the algorithm. For example, knowledge of the symmetry groups of $L$, $L^0$, and $V$ could be used to immediately rule out candidate eigenvectors. It is important to observe that different partitionings of the Hamiltonian into $L^0$ and $L-L^0$ may lead to very different sets of trial eigenvectors. Given a Hamiltonian $L$, an important task is to select $L^0$ that will result in a relatively small set of trial eigenvectors which can be computed efficiently.

\section{Application: Time-Independent Schr\"odinger Equation} 
\label{sec:TISE}
Consider the time-independent Schr\"odinger equation on the $d$-dimensional unit cube with Dirichlet boundary conditions,  
\begin{eqnarray}   \label{eqn:eqn}
L u (x) := (-\tfrac 12 \Delta +V)u(x) &=& Eu(x)\quad \mbox{for all}\;\; x\in I_d:=(0,1)^d, \\
u (x) &=& 0 \quad \mbox{for all}\;\; x\in  \partial I_d, \nonumber
\end{eqnarray}
where  $V$ is uniformly bounded by a constant $M$ and has continuous first-order partial derivatives in each direction uniformly bounded by a constant $C$, i.e., $\| \frac{\partial}{\partial x_i} V\| \leq C$. 
Thus, without loss of generality we assume that $V\geq 0$.
We set $$L^0=-\frac{1}{2} \Delta = -\frac{1}{2} \sum_{i=1}^d \frac{\partial^2}{\partial x_i^2}.$$
Assume that the eigenvalues of $L$ and $L^0$ are indexed in nondecreasing order. We want to approximate the first $j$ excited state energies, $E_{(0)},\dots, E_{(j-1)}$ , (i.e., the $j$ smallest eigenvalues ignoring multiplicities) with error proportional to $\e$, modulo $\e$-degenerate eigenvalues as explained previously. 
Thus we are interested in low-order excited state energies because we have assumed that  $j$ is a constant.

\subsection{Set of Trial Eigenvectors}
As we already indicated, the cost of Algorithms 1 and 2 depends on the cardinality of a set of trial eigenvectors $\mathcal{S}$. 
We will now show that $|\mathcal{S}|$ is bounded by a polynomial in $d$
 in the case  of the Schr\"odinger equation we are considering here.

The eigenvalues and eigenvectors of $L^0 = -\frac{1}{2}\Delta$ are known to be
\begin{equation} \label{eqn:eval}
 E_{\vec{k}}^0 =  \frac12 (k_1^2 + k_2^2 + \hdots+ k_d^2)\pi^2 \;\;\;\;  \; \vec{k}=(k_1,...,k_d) \in \mathbb{N}^d
\end{equation}
$$ u_{\vec{k}}^0 (\vec{x})= 2^{d/2} \prod_{i=1}^d \sin(k_i \pi x_i) \;\;\;\;  \vec{x}=(x_1,...,x_d) \in [0,1]^d \;\;\;\; \vec{k}=(k_1,...,k_d) \in \mathbb{N}^d.$$ 
We may reindex them by considering the eigenvalues  in nondecreasing order to obtain $E^0_0\leq E^0_1 \leq ... \leq E_i^0 \leq ...$ as in (\ref{eq:l0excited2}). Thus $E^0_0 = \frac12 d\pi^2 < \frac12 (d+3)\pi^2 = E^0_1 $ and $E^0_1 $ is a degenerate eigenvalue with dimension of its associated degenerate subspace equal to $d$. Similar considerations apply to the rest of the eigenvalues. We remark that the distribution of the eigenvalues of $L^0$ is known \cite{Titchmarsh}.

We will use (\ref{eq:upperBound}) with $c=2$ to derive a set of trial eigenvectors and bound its cardinality. 
In fact, we derive a set of trial eigenvectors that is slightly larger than the set obtained by strictly considering the indices in the complement of $\mathcal{I}$ in (\ref{eq:upperBound}). Yet its size is polynomial in $d$ as we will see, and for the sake of brevity, we also denote this set by $\mathcal{S}$. In particular, we construct the quantity $B$ of equation (\ref{eq:upperBound}) and show a $K=K(j,V)$ such that for $k \geq K$ we have $E_k^0 > 2M + B$. So we obtain an upper bound for the $jth$ largest eigenvalue of $L$.
Clearly the cardinality of $\mathcal{S}$ grows with $B$ because we include eigenvectors of $L^0$ that correspond to increasingly large eigenvalues. The purpose of the construction below is to obtain a crude but helpful in our analysis estimate of the distribution of the eigenvalues of $L$ using the eigenvalues of $L^0$; in particular to cover possible degeneracy of the eigenvalues of $L$.

We select $j+1$ values $E^0_{(s_n)}$ from the strictly increasing sequence of eigenvalues (see  (\ref{eq:l0excited2})) such that 
\begin{equation}  \label{eq:subseq}
E^0_{(s_0)}:= E^0_{(0)}  < E^0_{(0)} + M < E^0_{(s_1)} < E^0_{(s_1)} +M <  \hdots  < E^0_{(s_{j-1})} +M < E^0_{(s_j)} 
\end{equation}
where $E^0_{(s_n)} - E^0_{(s_{n - 1})} > M$, $n=1,\dots,j$. Indeed it is possible to select a subsequence that satisfies these conditions. 
We know that $E^0_{(s_{n-1})} = \frac{1}{2}(k_1^2 + k_2^2 +... k_d^2)\pi^2$ for a certain $\vec{k}$. The inequality 
$$\frac12 ({k'}_1^2 + {k'}_2^2 +... {k'}_m^2)\pi^2 + \frac12 ({k}_{m+1}^2 + {k}_{m+2}^2 +... {k}_d^2)\pi^2  \geq E^0_{(s_{n-1})} + M$$
is satisfied by selecting $m$ to be a suitable constant and then by selecting $k'_i \geq k_i + \gamma_i$, where $\gamma_i$ is a suitable positive integer constant, $i=1,\dots, m$.
For example, after fixing $m$, we can repeatedly increment each of the $k'_i$, $i\in\{1,\dots,m\}$, successively until the desired inequality holds. 
Iteratively, we define $E^0_{(s_n)} = ({k'}_1^2 + {k'}_2^2 +... {k'}_m^2)\pi^2/2 + ({k}_{m+1}^2 + {k}_{m+2}^2 +... {k}_d^2)\pi^2/2$ for $n=1,2,..,j$.

By our construction, the interval $[E^0_{(s_0)},E^0_{(s_j)}]$ contains at least $j$ distinct eigenvalues of $L$, %
since $E^0_{(s_j)} - E^0_{(s_0)} > jM$, and for every $i$, $E_i^0 \leq E_i \leq E_i^0 + M$. Moreover, $E^0_{(s_j)}  = E_0^0 + c'$, where $c'$ is a constant. Thus, we 
take $c=2$ in (\ref{eq:upperBound}) and
define the constant $B$ as
\begin{equation} \label{Bdef}
B:= M + E_{(s_j)}^0 = M + E_{0}^0 + c' .
\end{equation}
From (\ref{eqn:eval}) there exists a $K\in\nat$ such that
\begin{equation} \label{hyp1}
k\ge  K\;\; \Rightarrow \;\;E_k^0 > 2M +B  = 3M + E_{(s_j)}^0 .
\end{equation}
Hence, we construct the \textit{set of trial eigenvectors} $\mathcal{S}$ to be the set of all eigenvectors of $L^0$ that correspond to eigenvalues less than or equal to $3M + E_{(s_j)}^0$. We bound $|\mathcal{S}|$ next. 

The cardinality of  $\mathcal{S}$ is the number of tuples $\vec{k}\in \nat^d$ such that $(k_1^2 + \dots + k_d^2)\pi^2/2 \le 3M + d\pi^2/2 + c' $.
Let $m$ be the number of components $k_{i_1}, ..., k_{i_m}$ of such a $\vec{k}$ that are greater than $1$. Then we have 
$$(d-m)\pi^2/2 +(k_{i_1}^2 +... k_{i_m}^2)\pi^2/2 \le 3M + d\pi^2/2 + c' .$$
Since $k_i \geq 2$ we have
$$3m\pi^2 \leq -m\pi^2 +(k_{i_1}^2 +... k_{i_m}^2)\pi^2 \le  2(3M + c') .$$
Hence, $m$ is $O(1)$.  Therefore, in order to construct $\mathcal{S}$ one needs to consider tuples $\vec{k}'\in\nat^d$ where at most a constant number of components are greater than $1$.
The number of such tuples depends on the number of possible combinations by which  one can select a constant number of components of $\vec{k}'$ to be greater than or equal to~$2$. Therefore, this number is polynomial in $d$.\footnote{This follows immediately for $m=O(1)$ from the bound $\binom{d}{m}\leq \frac{d^m}{m!} = {\rm poly}(d)$.}

Table \ref{tab:tab1} 
shows the eigenvalues of the Laplacian by considering tuples where a constant number $m$ of components exceed $1$, assuming that these components are each bounded by a constant~$N$.
Observe that in all cases, since $m=O(1)$, the multiplicity of the eigenvalues 
grows as a polynomial function of~$d$.

\begin{table}
\centering 
\caption{Distribution of eigenvalues of $L^0 = -\frac12 \Delta$ with respect to the number $m$ of indices $k_i \geq 2$. }
\label{tab:tab1}   
\begin{tabular}{c c c}  
\hline\noalign{\smallskip} 
$m$ & Combinations & Eigenvalue \\ [0.5ex] 
\hline\noalign{\smallskip} 
$0$ & $\binom{d}{0}$ & $d\pi^2/2$ \\
$1$ & $\binom{d}{1}$ & $(d-1)\pi^2/2 + k_{i_1}^2\pi^2/2$ \\
$2$ & $\binom{d}{2}$ &  $ (d-2)\pi^2/2 + (k_{i_1}^2+k_{i_2}^2)\pi^2/2$  \\
$...$\\
$l\leq d$ & $\binom{d}{l}$ & $(d-l)\pi^2/2 + (k_{i_1}^2+..+k_{i_l}^2)\pi^2/2$ \\
[1ex] 
\hline\noalign{\smallskip} 
\end{tabular}
\end{table}

Therefore, the cardinality of $\mathcal{S}$ is polynomial in $d$. As shown in Section \ref{sec:PrelimAnal}, for every eigenvector of $L$ that corresponds to an eigenvalue less than or equal to $B$, there exists an eigenvector of~$L^0$ in~$\mathcal{S}$ such that the two eigenvectors have a nontrivial overlap and it follows from (\ref{eqn:probBound}) with $c=2$ that the magnitude squared of this projection of the one onto the other will be at least~
~$\frac{3}{4}\frac{1}{|\mathcal{S}|} = \frac1{{\rm poly}(d)}$, i.e., at worst polynomially small with respect to $d$, as desired.   


\subsection{Finite Difference Discretization}  
\label{sec:FiniteDiff}
We obtain a matrix eigenvalue problem by discretizing  (\ref{eqn:eqn}) on a grid with mesh size $h= \frac{1}{N+1}$, $N\in \nat$, using finite differences \cite{Leveque,GS}.
This yields a matrix $M_h:= -\tfrac{1}{2}\Delta_h + V_h$ with size $N^d \times N^d$. The matrix $-\frac12\Delta_h$ is obtained using a $2d+1$ stencil for the Laplacian \cite[p.60]{Leveque}. 
It is known that the low-order eigenvalues of $M_h$ approximate the corresponding eigenvalues of $L$. The eigenvalues and eigenvectors of $-\tfrac{1}{2} \Delta_h$ are known and are given by
\begin{equation}
E^0_{h,\vec{k}} = \frac{2}{h^2} \sum_{i=1}^d \sin^2(\pi h k_i /2) \;\;\;\;  \;\vec{k}=(k_1,...,k_d) \;\;\;\; 1\leq k_i \leq N
\end{equation}
\begin{equation}
u^0_{h,\vec{k}} = \bigotimes_{i=1}^d v_{k_i},
\end{equation} 
where the vectors $v_{k_i}\in \mathbb{R}^d$ have coordinates
\begin{equation} \label{eq:discreteEvecs}
v_{k_i, \ell} = \sqrt{2h} \sin(k_i \ell \pi h ) \;\;\;\;  \;\ell = 1,2,...,N \;\;\;\;  \;i= 1,2,...,d.
\end{equation}
Similarly to (\ref{eq:l0excited2}), we index the eigenvalues of $-\tfrac{1}{2} \Delta_h$ in increasing order ignoring multiplicities to obtain
\begin{equation} \label{eq:l0excited}
E^0_{h,(0)} < E^0_{h,(1)} < ... < E^0_{h,(i)} < ...
\end{equation}
Then from \cite{Weinberger1} we have
\begin{equation}  \label{eq:cdh2}
|E^0_{h,(k)}-E^0_{(k)}| \leq Cdh^2    \;\;\;\;  \;\ \text{ for } k = O(1)
\end{equation}
where $C>0$ is a constant.

$V_h$ is an $N^d \times N^d$ diagonal matrix which contains evaluations of $V$ at the grid points truncated to $\lceil  \text{log}_2 h^{-1} \rceil $ bits of accuracy. Thus $M_h$ is symmetric, positive definite, and sparse. This matrix has been extensively studied in the literature \cite{Demmel,FW,Leveque}. 
For $V$ that has bounded first-order partial derivatives and $k=O(1)$, using the results of \cite{Weinberger1,Weinberger2} we have that 
there exists a matrix eigenvalue $E_{h,k'}$ such that
\begin{equation} \label{eqn:err}
|E_{(k)} - E_{h,k'}| = O( d h)  
\end{equation}
 as $dh \rightarrow 0$, where $E_{(k)}$ is defined in (\ref{eqn:exc}). We will use the algorithms of Section \ref{sec:Algorithm} to approximate the low-order eigenvalues of $M_h$, which as we have seen approximate the low-order eigenvalues of $L$. For this, we need to construct the set of trial eigenvectors $\mathcal{S}$, and estimate its cardinality.
Recall that for the continuous operator, the set of candidate eigenvectors is derived using equation (\ref{hyp1}), and in particular by selecting the eigenvectors of $L^0$ that correspond to eigenvalues less or equal to $2M + B = 3M + E_{(s_j)}^0$. So for the discretized case we select the eigenvectors of $-\frac12 \Delta_h$ that correspond to eigenvalues less than or equal to $3M + E_{(s_j)}^0 + O(dh^2)$ due to equation (\ref{eq:cdh2}). Since $dh \rightarrow 0$, without loss of generality we slightly modify equation (\ref{hyp1}), to select the eigenvectors of $M_h$ that correspond to eigenvalues less than or equal to
\begin{equation} \label{eq:defB}
2M+ B = 3M + E_{(s_j)}^0 + 1
\end{equation}
 for sufficiently small  $h$, where this equation effectively redefines $B$ by increasing its value by $1$. Thus the cardinality of $\mathcal{S}$ in the case of the matrix $M_h$ follows from the continuous case and remains polynomial in $d$.

Specifically, we define
\begin{equation}   \label{eq:SB}
\mathcal{S}:= \{ u^0_{h,k} : E^0_{h,k} \leq 3M + E_{(s_j)}^0 + 1   \}
\end{equation}

\subsection{Algorithm for Excited State Energies}
\label{sec:AlgTISE} 
We now give the details of Algorithms  1 \& 2 
applied to the time-independent Schr\"odinger equation (\ref{eqn:eqn}). Given $\e$, the algorithms produce the $j$ eigenvalue estimates $\tilde{E}_0 < ... < \tilde{E}_{j-1}$ of equation (\ref{eq:tildeE}).
Algorithm 1 computes $\tilde{E}_0$. For this, QPE \cite{NC} is applied repeatedly with its initial state taken to be every single element of the set of trial eigenvectors $\mathcal{S}$. We use  repetitions of the procedure to boost the success probability. 
We remark that our Algorithm 1 computes the ground state energy in a way similar to \cite{GS,Convex}, but under weakened assumptions.
Algorithm 2 iterates $j-1$ times the procedure of Algorithm 1, at each iteration producing the next estimate $\tilde{E}_i$ by taking into account all the previously produced estimates as we will explain below.

Both algorithms use QPE as the main module. The purpose is to compute approximations of the eigenvalues of the matrix $M_h$ of the previous section. Setting $N=2^{\lceil  2\log_2 (d/\e)  \rceil}$, we discretize (\ref{eqn:eqn}) with mesh size $h=\frac{1}{N+1} < \frac{\e^2}{d^2}$ to obtain the $N^d\times N^d$ matrix $M_h$, where we have $N^d = O((\frac{d}{\e})^{2d})$.  From (\ref{eqn:err}), we obtain that the low-order matrix eigenvalues approximate the low-order eigenvalues of the continuous operator with error proportional to $dh = O(\frac{\e^2}{d})$. 
The reason we have taken very small $h$ is because we want to ensure that $\e$-degenerate eigenvalues of the continuous operator will be approximated by tightly clustered eigenvalues of $M_h$. As $M$ is a constant, without loss of generality we may assume that $\e^{-1} \gg M$. Since the largest eigenvalue of $-\frac12\Delta_h$ is bounded from above by $2dh^{-2}$, and $V$ is uniformly bounded by $M$, we obtain that $\|M_h\|$ is bounded from above by $2dh^{-2}+M \ll 3dh^{-2}$, in the sense that $\frac{M}{dh^{-2}}=o(1)$. 

Let $R=3dh^{-2}$ and consider the matrix $W=e^{i M_h /R}$. Its eigenvalues are $e^{i E_h /R} = e^{ \frac{2\pi i E_h }{ 2 \pi R} } = e^{2\pi i \phi}$, where $E_h$ is an eigenvalue of $M_h$ and $\phi:= \frac{E_h}{2\pi R}$ denotes the corresponding phase.
 
QPE is used to compute an approximation $\hat{\phi}$ of $\phi$ with $b = 5\lceil \log_2 \frac{d}{\e} \rceil + 7$ bits of accuracy, and from this we get $\tilde{E} = 2\pi R\hat{\phi}$ so that 
\begin{equation}  \label{eq:totErr}
|E-\tilde{E}| \leq |E- E_h  | + |E_h - \tilde{E}| = O(\e) 
\end{equation}
where $E$ denotes the eigenvalue of $L$ that $E_h$ approximates according to (\ref{eqn:eqn}). QPE uses two registers, the top and the bottom. The size of the top register is related to the accuracy of QPE and its success probability. Recall that QPE succeeds when it produces an estimate with accuracy $2^{-b}$. The bottom register is used to hold an (approximate) eigenvector of $M_h$ corresponding to the phase of interest, and therefore has size $d\log_2 N = d\cdot O(\log \frac{d}{\e})$. The number of qubits in the top register is $t=b+t_0$, so that QPE has accuracy $2^{-b}$ with probability at least $1-\frac{1}{2(2^{t_0}-2)}$, assuming that an exact eigenvector is provided as initial state in the bottom register \cite[Sec. 5.2]{NC}. QPE uses powers of $W$, namely $W^{2^0}, W^{2^1},...,W^{2^t-1}$. We will approximate these powers using a splitting formula with error, as we will see below. This reduces the success probability of QPE to at least $p :=  1 - \frac{1}{2^{t_0} -2}$. We will set $t_0$ to be logarithmic in $d$, and will give all the details later on when dealing with the cost of our algorithm.

Consider an eigenvalue $E_h \leq B + 2M$ (see equations (\ref{eq:upperBound}) and (\ref{eq:defB})) of the matrix $M_h$ and let $u_h$ denote an eigenvector corresponding to $E_h$. Then QPE with initial state some $u^0_{h,i} \in \mathcal{S}$ succeeds with probability  at least $p_{u_h}(i):=| u_h^T u_{h,i}^0  |^2 \cdot p    = | u_h^T u_{h,i}^0  |^2    \cdot (1-\frac{1}{2^{t_0}-2})$  \cite{AbramsLloyd}.

Recall that $\mathcal{S}$ contains eigenvectors of $L^{0}$ that correspond to eigenvalues $E^0_h\leq B +2M$ as defined in (\ref{eq:SB}), and that the cardinality of $\mathcal{S}$ is polynomial in $d$. 
Applying the same approach of Section \ref{sec:PrelimAnal} for the eigenvectors of $M_h$, we conclude that 
for every eigenvector $u_h$ of the matrix $M_h$ that corresponds to an eigenvalue less than $B$, there exists a vector $u^0_{h,k} \in \mathcal{S}$ such that $| u_h^T u_{h,k}^0  |^2 \geq \frac{3}{4|\mathcal{S}|}$, where we have used equation (\ref{eqn:probBound}) with $c=2$ (since the value of $B$ we are using here leads to $c=2$ in this case too). 
Thus, after we run QPE with each element of $\mathcal{S}$ as initial state, the probability that at least one of the outcomes (in principle we do not know which one) will give a good estimate of  $E_{h}$ is at least $p_{u_h}(k) \geq \frac{3}{4|\mathcal{S}|}  p$.

We repeat the whole procedure $r$ times to boost the success probability of obtaining an estimate of $E_h$ with accuracy $\e$. Indeed, the probability that QPE fails with all initial states taken from $\mathcal{S}$ and in all its $r|\mathcal{S}|$ repetitions is
\begin{equation}
\left(  \prod_{i=1}^{|\mathcal{S}|} (1-p_{u_h}(i)) \right)^r \leq (1-p_{u_h}(k) )^r   \leq e^{-r p_{u_h}(k)} \leq e^{-r \frac{3}{4|\mathcal{S}|}p}
\end{equation}
Thus, the probability that at least one of the $r|\mathcal{S}|$ outcomes will lead to an approximation of $E_h$ with accuracy $O(\e)$ is at least 
\begin{equation}
1 -e^{-r \frac{3}{4|\mathcal{S}|}p}
= 1- e^{-r \frac{3}{4|\mathcal{S}|}   \cdot (1-\frac{1}{2^{t_0}-2})}
\end{equation}
We can boost this probability to be arbitrarily close to 1 by taking $r={\rm poly}(d)$, since $|\mathcal{S}|$ is polynomial in $d$.

Observe that Algorithm 1 selects the minimum measurement outcome from all the runs of QPE, and uses it to obtain $\tilde{E}_0$. Let this outcome be $m'\in\{ 0,\dots, 2^{t}-1\}$. The algorithm converts $m'$ to $m_0 =  \lfloor m' 2^{-t_0} \rfloor \in\{ 0,\dots, 2^{b}-1\}$ and uses it to obtain $\tilde{E}_0$, according to the formula 
\begin{equation}  \label{eq:Ephi}
\tilde{E}_0 = 2\pi R \hat{\phi}_0  = 2\pi R \frac{m_0}{2^b}.
\end{equation}
Since $|m'/2^t - \phi_0|\le 2^{-b}$ and $\phi_0, m'/2^t\in[ m_0/2^{b} , (m_0 +1 ) / 2^ b]  $, it follows that 
$ |2\pi R \phi_0 - \tilde{E}_0|  =      2\pi R \cdot |\phi_0 - m_0 / 2^ b| \leq 2\pi R / 2^b \leq \e$, which together with (\ref{eqn:err}) gives (\ref{eq:totErr}).

Let $G_0=\{ m : |\frac{m}{2^{b}}-\phi_0| \leq \frac{1}{2^b}  \}$.
Algorithm 1 fails either if none of the converted outcomes is an element of $G_0$, or at least one of the  converted outcomes is an element of $G_0$, but there is another converted outcome (produced by a failure of QPE) smaller than the minimum element of $G_0$. Thus, we can bound the total probability of failure by
\begin{eqnarray} \nonumber
\text{Pr(Algorithm 1 fails) } &=&\text{Pr(none of the outcomes leads to an element of } G_0\text{) } \\ \nonumber &+& \text{Pr(one of outcomes leads to an element of } G_0 \\
\nonumber &\ & \text{ but there is at least one smaller converted outcome)} \\
\nonumber &\le& e^{-r \frac{3}{4|\mathcal{S}|}p} 
+ \text{Pr(QPE failed in at least one of the } |\mathcal{S}|\text{ runs)} 
\\
\nonumber &\le& e^{-r \frac{3}{4|\mathcal{S}|}p} 
+(1 - \text{Pr(every run of QPE approximates }\\
\nonumber &\ &  \qquad \qquad  \qquad  \qquad \text{one of the phases with error }  2^{-b}) ) \\
\nonumber &\le& e^{-r \frac{3}{4|\mathcal{S}|}p} + (1- p^{r|\mathcal{S}|}  ) \\
\label{eq:probBound}&\le& e^{-r \frac{3}{4|\mathcal{S}|}(1-\frac{1}{2^{t_0}-2})} + \left( 1- \left(1- \frac{1}{2^{t_0}-2}\right)^{r|\mathcal{S}|} \right)  \\
\nonumber &\le& e^{-r \frac{3}{4|\mathcal{S}|}(1-\frac{1}{2^{t_0}-2})} +  \frac{r|\mathcal{S}|}{2^{t_0}-2} , 
\end{eqnarray} 
where the third from last inequality follows from equation  (\ref{eq:probBound2}) below.
Observe that this bound can be made arbitrarily close to $0$ by selecting the number of repetitions $r$ to be a suitable polynomial in $d$, 
since $|\mathcal{S}|$ is polynomial in $d$, and by taking $t_0 = \beta \log d$, where $\beta$ is an appropriately chosen constant. We have used the fact that if a measurement outcome $\ell$ fails to estimate any of the phases, i.e., $|\frac{\ell}{2^b} - \phi_s| > 2^{-b}$ for all phases $\phi_s$ corresponding to eigenvalues of $M_h$, then 
\begin{equation}   \label{eq:probBound2}
\text{Pr}(\ell)=\sum_{s=0}^{N^d-1} |c_s|^2 |\alpha(\ell,\phi_s)|^2 \leq \sum_{s=0}^{N^d-1} |c_s|^2 \frac{1}{2^{t_0}-2} = \frac{1}{2^{t_0}-2}     = (1- p).
\end{equation}
Here, the $c_s$ denote the projections of the initial state onto each of the eigenvectors of $M_h$, and
the $|\alpha(\ell,\phi_s)|^2$ denote the probability to get outcome $\ell$ given the exact eigenvector $u_{h,s}$ as input.
We have upper bounds for these quantities from \cite[Eq. 5.34]{NC}.
Therefore, the probability the measurement outcome estimates at least one (or, some) phase is $1-\text{Pr}(\ell) \ge p$. 

Recall that the set $\mathcal{S}$ has been constructed using an upper bound for $E_{(j-1)}$; see equations (\ref{eq:subseq}) and (\ref{eq:SB}). Algorithm 2 essentially repeats Algorithm 1 $(j-1)$ times, but selects the converted measurement outcome in a different way by considering the already selected outcomes.  At repetition $i$, it selects the minimum converted outcome $m_i$ that exceeds the outcome selected at the previous iteration by at least 2, i.e., $m_i \geq m_{i-1} +2$, where $m_i = \lfloor m'_i 2^{-t_0} \rfloor$ and $m'_i$ is a measurement outcome at the $i$th run, $i=1,2,3, j-1$; see also equation (\ref{eq:Ephi}).
The success probability for both Algorithm 1 and Algorithm 2 follows from (\ref{eq:probBound}) and is at least
\begin{equation} \label{eq:TotalSuccProb}
\left(1-\left(e^{-r \frac{3}{4|\mathcal{S}|}(1-\frac{1}{2^{t_0}-2})} +  \frac{r|\mathcal{S}|}{2^{t_0}-2} \right)\;\right)^j, 
\end{equation}
which can be made arbitrarily close to 1 by selecting $r$ to be a suitable polynomial in $d$ and taking $t_0$ to be sufficiently large.

Note that Algorithm 2 computes $\tilde{E}_i = 2\pi R \frac{m_i}{2^b}$, $i=1,2,\dots, j-1$, as estimates of the eigenvalues according to equation (\ref{eq:tildeE}) and the conditions \textbf{C1} and \textbf{C2} that follow it.
If both algorithms are successful with high probability, at the $i$th run we have that there exists a phase $\phi_i$ corresponding to an eigenvalue of $M_h$ such that $|2\pi R \phi_i - \tilde{E}_i| \leq \frac{2\pi R}{2^b} \leq  \e$. The condition $m_i \geq m_{i-1} +2$ in the selection of measurement outcomes guarantees that for any two  
$i_1\neq i_2$, the computed matrix eigenvalue approximations satisfy $\tilde{E}_{i_1}\neq \tilde{E}_{i_2}$ and 
$|\tilde{E}_{i_1} -\tilde{E}_{i_2}| = \Omega(\e)$
because for the corresponding phases we have $\phi_{i_1} \neq \phi_{i_2}$ as belonging to different intervals;
see Figure~\ref{Fig:phaseEstimation}. Moreover, the $\tilde E_{i_1}$ and $\tilde E_{i_2}$ also approximate different eigenvalues $E_{i_1}\ne E_{i_2}$ of the continuous operator because we have used a very fine discretization.
Finally, the algorithm does not fail to produce consecutive eigenvalues unless they differ by less than $O(\e)$ because we always select the minimum outcome that satisfies $m_i \geq m_{i-1} +2$.

\subsubsection{Cost of Quantum Phase Estimation}
\label{sec:Cost}

Algorithms 1 \& 2 use QPE as a module. The cost of QPE depends on the cost to prepare its initial state, and on the cost to implement the matrix exponentials $W^{2^0},W^{2^1},..W^{2^{t-1}}$, where $W= e^{i M_h/R}$. We approximate these exponentials below using Suzuki-Trotter splitting, the analysis of which proceeds similarly to that of \cite{PZ12,GS}; see also the details given in Chapter \ref{ch:HamSim}. 

The initial states are taken from $\mathcal{S}$ which contains eigenvectors of $-\frac12 \Delta_h$ according to (\ref{eq:SB}). Each eigenvector can be prepared efficiently using the quantum Fourier transform, which diagonalizes the Laplacian, with a number of quantum operations proportional to $d \cdot \log^2 \frac{d}{\e}$ and using number of qubits $\log_2 N^d = d \cdot O(\log \frac{d}{\e})$. We remark that from the tensor product structure of the eigenvectors of $-\frac12 \Delta_h$, it suffices to prepare eigenvectors of the one-dimensional Laplacian; see e.g. \cite{Poisson,Klapp,Wavelet}.

Now let us turn to the approximation of the matrix exponentials. We simulate the evolution of the Hamiltonian $H = M_h/R$ for times $2^\tau$, 
 $\tau=0,1,\dots, t-1$, where we have set $t=b+t_0$. Let $H = H_1 + H_2$
where $H_1 = -\Delta_h / 2R $ and $H_2 = V_h/ R$, where we assume $V$ is given by an oracle.

To simulate quantum evolution by $H_1$, assuming the known eigenvalues of $-\frac12 \Delta_h$ are given by a quantum query oracle with $O(\log \frac{1}{\e})$ bits of accuracy, we again use the quantum Fourier transform to diagonalize $H_1$ with cost (i.e., a number of quantum operations) bounded by $d \cdot O(\log^2 \frac{d}{\e})$, and requiring  a number
of qubits proportional to $d \log \frac{d}{\e}$. Alternatively, if the eigenvalues of $-\frac12 \Delta_h$ are implemented explicitly (without an oracle) by the quantum algorithm, then the number of quantum operations required is a low-order polynomial in $d$ and $\log_2 \frac{1}{\e}$, and so is the number of qubits \cite{Poisson}. For simplicity, we will not pursue this alternative here.  
The evolution of a system with Hamiltonian $H_2$ can be implemented using two quantum queries returning the values of $V$ at the grid points, and phase kickback. The queries are similar to those in Grover's algorithm \cite{NC} and the function evaluations of $V$ are truncated to $O(\log\frac 1\e)$ bits. 

We use a splitting formula $S_{2k}$ of order $2k+1$, $k\ge 1$, to
approximate $W^{2^t}=e^{i (H_1+H_2)2^t}$ by a product of the form
\begin{equation}\label{eq:r50}
\prod_{\ell=1}^{N_t} e^{i A_\ell z_\ell},
\end{equation}
where $A_\ell \in\{H_1,H_2\}$ and suitable $z_\ell$ that depends on $t$ and $k$.

The splitting formula $S_{2k}$ is due to Suzuki \cite{Suzuki90,Suzuki91}. It is used to approximate 
$e^{i(B+C)\Delta t}$, where $B$ and $C$ are Hermitian matrices. This formula is defined recursively by
\begin{eqnarray*}
S_2(B,C,\Delta t) &=& e^{iB\Delta t/2} e^{iC\Delta t} e^{iB\Delta t/2} \\
S_{2k}(B,C,\Delta t) &=& [S_{2k-2}(B,C, p_k\Delta t) ]^2 S_{2k-2}(B,C, (1-4p_k) \Delta t) \\
&& \hspace{10pc} \times [S_{2k-2}(B,C, p_k\Delta t) ]^2,
\end{eqnarray*}
where $p_k= (4 - 4^{1/(2k-1)})^{-1}$, $k=2,3,\dots$. 

Unfolding the recurrence above and combining it with \cite[Thm. 1]{PZ12} we obtain that the approximation of $W^{2^\tau}$ has the form
\begin{equation}
\label{eq:unfold}
\widetilde W^{2^\tau} = e^{iH_1 a_{\tau,0}} e^{iH_2 b_{\tau,1}} e^{i H_1 a_{\tau,1}} \cdots e^{iH_2 b_{\tau,L_\tau}} e^{i H_1 a_{\tau,L_\tau}},
\end{equation}
where
$a_{\tau,0},\dots, a_{\tau,L_\tau}$ and $b_{\tau,1},\dots,b_{\tau,L_\tau}$ and $L_\tau$ are parameters, $\tau=0,\dots, t_0+b-1$. 
The number of exponentials involving $H_1$ and $H_2$ in the expression above is $N_\tau=2L_\tau+1$. An explicit algorithm for computing each $\widetilde W^{2^\tau}$ is given in \cite{GS}.

Let $\| \cdot\|$ be the matrix norm induced by the Euclidean vector norm.
From \cite[Thm. 1 \& Cor. 1]{PZ12} the number $N_t$ of exponentials needed to approximate $W^{2^t}$
by a splitting formula of order $2k+1$ with error $\e_\tau$, $\tau=0,\dots, t_0+b-1$, is
\begin{equation*}
N_\tau \leq  16e \|H_1\| 2^\tau\, \left(\frac {25}3\right)^{k-1}
\left(\frac{8e\,2^\tau\|H_2\|}{\e_\tau}\right)^{1/(2k)}, 
\end{equation*}
for any $k \geq 1$. 
Since we want to approximate all the $W^{2^\tau}$, $\tau=0,1,...,t_0+b-1$, we sum the number of exponentials required to approximate each one of them. 
Thus the total number of matrix exponentials required by Algorithm 2, $\mathcal{N}_{tot}$, is bounded from above by
\begin{eqnarray}
\mathcal{N}_{tot}&=&jr|\mathcal{S}| \sum_{\tau=0}^{t_0+b-1} N_\tau \label{eq:numexp}  \nonumber \\
&\le&  jr|\mathcal{S}|  \left( 16e \| H_1\| \left(\frac {25}3\right)^{k-1}
 \left(8e\|H_2\|\right)^{1/(2k)}  \sum_{\tau=0}^{t_0+b-1}2^\tau
\left( \frac {2^\tau}{\e_\tau}\right)^{1/(2k)} \right). 
\end{eqnarray}
The factor $jr|\mathcal{S}|$ is the number of executions of QPE performed by our algorithms, and the second factor is the cost of a single QPE. Note that $j$ is the number of eigenvalues we wish to estimate, $|\mathcal{S}|$ is the number of eigenvectors we use as initial states, and $r$ is the number of times we repeat QPE per initial state to boost the success probability of getting the desired outcome.
We select a polynomial $g(d)$ such that the product $r|\mathcal{S}|/g(d) = o(1)$ (as $d\rightarrow \infty$). We then select the error of each exponential to be $\e_\tau = \tfrac {2^{\tau+1 - (b + t_0)}}{40 g(d)}$,
$\tau=0,\dots,t_0+b-1$. It is easy to check that $\sum_{\tau=0}^{t_0+b-1}\e_\tau \le \tfrac 1{20g(d)}$.
Thus the success probability of QPE is reduced by at most twice this amount \cite[p. 195]{NC}, giving $1- \frac{1}{2(2^{t_0}-2)} - \frac{1}{10g(d)}$. Next we set $t_0 = \lfloor \log_2 (5g(d) + 2)    \rfloor$, to get $p = 1 - \frac{1}{2^{t_0} -2}$ that we used above in deriving equation (\ref{eq:probBound}). Our choice of $g(d)$ and $t_0$ aims to make the bound of equation (\ref{eq:TotalSuccProb}) arbitrarily close to 1.

The largest eigenvalue of $-\Delta_h$ is $4dh^{-2} \sin^2(\pi N h/2) < 4dh^{-2}$. Since $R=3dh^{-2}$,  $H_1 = - \frac{1}{2}\Delta_h/ R = - \frac12\frac{1}{3dh^{-2}}\Delta_h$ and we have
$\norm{H_1} \leq \frac{2dh^{-2}}{3dh^{-2}} = \frac23$. 
Since $V$ is uniformly bounded by $M$ and $H_2 = V_h/ R$ we 
have $\norm{H_2} \leq M/3 d h^{-2}$. 
Substituting the value of $\e_\tau$ in (\ref{eq:numexp}), yields that the algorithm uses a number of
exponentials of $H_1$ and $H_2$ that satisfies
\begin{eqnarray*}
\mathcal{N}_{tot}
&\le&  jr|\mathcal{S}|  \left( 16e \| H_1\| \left(\frac {25}3\right)^{k-1}
 \left(8e\|H_2\|\right)^{1/(2k)}  \right) \sum_{\tau=0}^{t_0+b-1}2^\tau
\left( \frac {40 g(d) 2^\tau}{ 2^{\tau+1 - (b + t_0)}}\right)^{1/(2k)}  \\
&\leq&  jr|\mathcal{S}|  \left( 16e \| H_1\| \left(\frac {25}3\right)^{k-1}
 \left(8e\|H_2\|\right)^{1/(2k)}  \right)
\left( 20 g(d) 2^{t_0+b}\right)^{1/(2k)}  \\
&\leq&  jr|\mathcal{S}| \left( 16e \| H_1\| 2^{t_0 + b} \left(\frac {25}3\right)^{k-1}
\left( 160e\,2^{t_0 +b}\|H_2\| g(d) \right)^{1/(2k)} \right) .
\end{eqnarray*}
Using the bounds on $\|H_1\|$ and $\|H_2\|$, we obtain
$$
  \mathcal{N}_{tot} \leq  jr|\mathcal{S}| \left(  \frac{32e}{3}     2^{t_0 + b} \left(\frac {25}3\right)^{k-1}
\left( 160e\,2^{t_0 +b}   \frac{Mh^2}{3d}    g(d) \right)^{1/(2k)} \right).
$$
%
From $b = 5 \lceil \log_2 \tfrac{d}{\e} \rceil +7$, we have $2^b = 2^{5\lceil log_2 \frac{d}{\e}  \rceil  +7} \leq 2^{12} \left( \frac{d}{\e} \right)^5 = O(\frac{d^5}{\e^5})$.
Since $h < \frac{\e^2}{d^2}$, we have $2^b h^2 \leq 2^{12} \frac{d}{\e} =   O(\frac{d}{\e})$. 
Also, $2^{t_0} = 2^{\lfloor \log_2 (5g(d) + 2)    \rfloor} \leq 5g(d) +2$. 
We obtain
\begin{eqnarray} \label{eq:numexp2}
 \mathcal{N}_{tot} &\leq&  jr|\mathcal{S}| \left(  \frac{32e}{3}   \left(5g(d) + 2\right) \left( 2^{12} \left( \frac{d}{\e} \right)^5 \right)  \left(\frac {25}3\right)^{k-1} \right)\nonumber \\  
 &\qquad&  \qquad \times  
 \left( 160e\, \left(5g(d) + 2\right)  \left(  2^{12} \frac{d}{\e} \right)  \frac{M}{3d}    g(d) \right)^{1/(2k)} \nonumber \\
 &\leq&  jr|\mathcal{S}|\left( \widetilde C\; \frac{d^5 g(d)}{\e^5} \left(\frac {25}3\right)^{k-1} \left(\widehat{C}\;
\frac{g^2(d)}{\e}        \right)^{1/(2k)} \right) ,
%
\end{eqnarray}
for any $k>0$, where $\widetilde C$ and $\widehat{C}$ are suitable constants.

The {\it optimal} $k^*$, i.e., the one minimizing the upper bound for $\mathcal{N}_{tot} $
in (\ref{eq:numexp2}), 
is obtained in \cite[Sec. 5]{PZ12} and is given by
\[ k^* = \left\lfloor \sqrt{\frac{1}{2}  \log_{25/3}   \left(\widehat{C}\;
\frac{g^2(d)}{\e}        \right)     } + \frac 12 \right\rfloor = 
  \bar{C} \sqrt{\ln \frac {d}{\e}},
\]
for a suitable constant $\bar{C}$, 
since $g(d)$ is a polynomial in $d$ and we are taking its logarithm. 
With  $k^*$ and using again \cite[Sec. 5]{PZ12}, equation (\ref{eq:numexp2}) yields 
\begin{equation}\label{eq:numqueries}
\mathcal{N}_{tot}^{*} \leq \widetilde{C}   jr|\mathcal{S}| \frac{d^5}{\e^5}\; g(d)\;  e^{ 2\bar{C} \sqrt{ \ln \frac{25}{3}   \ln \frac{d}{\e}}}
=O \left(  g^2(d) \left(\frac{d}{\e}\right)^{5+\eta} \right) \quad {\rm as\ } d\e\to 0,
\end{equation}
where we have used $j=O(1)$ and $r|\mathcal{S}|=o(g(d))$, and where the equality above holds asymptotically for arbitrarily small $\eta>0$. 

We remark that of the $N_{tot}^*$ matrix exponentials roughly half involve $H_1$ and the remaining involve $H_2$; see (\ref{eq:unfold}). Since each exponential involving $H_2$ requires two queries the total number of queries is also of order $N_{tot}^*$. The cost to prepare the initial state, to diagonalize $-\frac12 \Delta_h$, and to implement the inverse Fourier transform that is applied prior to measurement in QPE, is proportional to 
$$ d\log^2 \frac{d}{\e} + (t_0+b)^2  =  O\left(d\log^2 \frac{d}{\e} \right),$$
since $t_0+b = O(\log \frac{d}{\e})$. 
Hence, the total number of quantum operations, excluding queries, is proportional to 
\begin{equation}\label{eq:numops}
\mathcal{N}_{tot}^* \cdot d\log^2 \frac{d}{\e}.
\end{equation}
Equations 
(\ref{eq:numqueries}) and (\ref{eq:numops}) yield 
that the total cost of the algorithm, including the number of
queries and the number of all other quantum operations, is proportional to
$$d\: g^2(d)  \left( \frac{d}{\e}\right)^{5 + \delta} ,$$
where $\delta >0$ is arbitrarily small. 

Finally, using equation (\ref{eq:TotalSuccProb}) we can select $r$ to be polynomial in $d$ and obtain success probability at least $\frac{3}{4}$, and the cost remains polynomial in $\frac{1}{\e}$ and $d$. We summarize our results in the following theorem. 
\vskip 2em
\begin{theorem} 
\label{EXCthm1}
Consider the time-independent Schr\"odinger equation  (\ref{eqn:eqn})  on the $d$-dimensional unit cube with Dirichlet boundary conditions and where the potential $V$ and its first-order derivatives are uniformly bounded. 
Algorithms 1 \& 2 of Section \ref{sec:Algorithm} compute approximations of  $j=O(1)$ low-order eigenvalues 
as in equation (\ref{eq:tildeE}),
each with error $O(\e)$ and satisfying conditions {\bf{C1}} and {\bf{C2}} of Section \ref{sec:ProbDef}, with overall success probability at least
$$\left(1-\left(e^{-r \frac{3}{4|\mathcal{S}|}(1-\frac{1}{2^{t_0}-2})} +  \frac{r|\mathcal{S}|}{2^{t_0}-2} \right)\;\right)^j \geq \frac34, $$
where $r$ and $|\mathcal{S}|$ are polynomial in $d$, 
$t_0 = \lfloor \log_2 (5g(d) + 2)    \rfloor$, 
and $g(d)$ is a polynomial in $d$ selected such that $r|\mathcal{S}| = o(g(d))$.
The algorithms apply QPE with initial state each element of a set of trial eigenvectors $\mathcal{S}$, and repeat this
procedure $r$ times. 
They use a number of queries proportional to
$$ \left( \frac{d}{\e} \right)^{5+\delta} \; g^2(d) 
\quad {\rm as\ } d\e\to 0, $$
and a number of quantum operations excluding queries proportional to
$$\left( \frac{d}{\e} \right)^{5+\delta} \; d\;g^2(d)\;       \quad {\rm as\ } d\e\to 0,$$
where $\delta >0$ is arbitrarily small. The algorithms use a number of qubits proportional to
$$ d\; \log \frac{d}{\e} +  \log g(d).$$
\end{theorem}
\vskip 2em
\begin{rem}
The $5$ in the exponent of $\frac{d}{\e}$ is due to the fact that for simplicity we have taken $N=2^{\lceil2\log_2 \frac{d}{\e} \rceil}$ in the discretization of the continuous operator and our consequent choice of $b$, the number of bits of accuracy of QPE. As explained, a fine discretization is needed to ensure that degenerate eigenvalues of the continuous problem are approximated by tightly clustered eigenvalues of the matrix. By taking slightly coarser discretization, 
it is possible to reduce this exponent. As our goal was to establish
an algorithm with cost polynomial in $d$ and $\e^{-1}$ we do not pursue this further. 
\end{rem}

\begin{rem}
The classical complexity of approximating a constant number of low-order eigenvalues with error $\epsilon$ grows as $\left(\frac{1}{\epsilon}\right)^d$  in the deterministic worst case \cite{Qspeedup}. 
Since our quantum algorithm for this problem has cost polynomial in $d$ and $\frac{1}{\epsilon}$, it vanquishes the curse of dimensionality.
\end{rem}

\section{Discussion}
There are a number of recent results 
suggesting that certain eigenvalue problems are very hard, 
even for quantum computers \cite{KempeLocalHam,intBosons,schuchDFT,childs2013bose, qmaSurvey}. On the other hand, obtaining positive results for eigenvalue problems showing where quantum algorithms give advantages over classical algorithms is particularly important towards understanding the power of quantum computers. 

We show such a positive result for the approximation of ground and excited state energies on a quantum computer. 
In summary, general conditions for the efficient approximation of 
a constant number of low-order excited state energies 
follow by combining conditions for efficient quantum simulation and for deriving a relatively small set of trial eigenvectors that can be implemented efficiently as quantum states. 
For quantum algorithms, previous approaches for computing the ground state energy require stronger conditions on $V$ than those we consider here, and these approaches do not extend to computing excited state energies. We have developed an entirely new approach to approximate not only the ground state energy, but also excited state energies, with cost polynomial in~$d$ and~$\e^{-1}$. For the special case of the time-independent Schr\"odinger equation with~$d$ degrees of freedom we study, our quantum algorithm vanquishes the curse of dimensionality.

We remark on several open problems. 
We have assumed that $L=L^0 +V$, but such a partition need not be unique, and different partitions may result in algorithms with significantly different costs. It is possible, that with additional assumptions, one would be able to determine suitable partitions leading to fast algorithms. 
 Such a characterization is an open problem.  We have provided a condition for constructing a set of trial eigenvectors~$\mathcal{S}$. Improving this condition to minimize the size of the resulting set~$\mathcal{S}$ is another open problem. Finally, in the initial investigation of quantum algorithms for eigenvalue problems, 
 strong assumptions on~$V$ were considered in order to obtain efficient algorithms. Progressively, culminating with our work, these assumptions have been weakened. It is important to continue working in this direction to further extend the scope of our algorithm, in particular to first-quantized approaches for  important problems in physics and chemistry.

%% file: _ch_HamSim.tex
\chapter{Divide and Conquer Approach to Hamiltonian Simulation}
\label{ch:HamSim}

\section{Introduction}
Simulating quantum mechanical systems is a very important yet very difficult problem. 
The computational cost of the best classical deterministic algorithm known grows 
exponentially with the system size. In some cases classical randomized 
algorithms, such as quantum Monte Carlo, 
have been used to overcome the difficulties, but these algorithms also have
limitations.
On the other hand, as Feynman proposed \cite{Feynman}, quantum computers may be able to 
carry out the simulation more efficiently than classical computers. 
This led to a large body of research dealing with quantum algorithms for 
Hamiltonian simulation 
\cite{lloyd1996universal,boghosian1998simulating,Zalka1,Zalka2,somma2002simulating,aharonov2003adiabatic,childs2004quantum,berry2007efficient,kassal2008polynomial,wiebe2010higher,childs2009limitations,childs2010simulating,poulin2011quantum,whitfield2011simulation,wiebe2011simulating,PZ12,childs2012hamiltonian,berry2012black,raeisi2012quantum,berry2013exponential,berry2014exponential,berry2015simulating,berry2015hamiltonian,poulin2014trotter,babbush2015chemical,babbush2016exponentially},
with efficient algorithms found for simulating many classes of important Hamiltonians. 
In particular, 
these 
algorithms have 
numerous applications to problems in physics and chemistry 
\cite{byrnes2006simulating,Kassal,jordan2014quantum,jordan2014quantum2,wecker2015solving,vivsvnak2015quantum}. 

Nevertheless, for certain problems, despite being \lq\lq efficient,\rq\rq\ the cost of implementing these quantum algorithms appears prohibitive; i.e., even though they scale polynomially with respect to the problem size, 
the resources required are formidable for problems of interest in practice. 
A prototypical example which we will 
consider in this chapter is the \textit{electronic Hamiltonian}, 
which 
encodes the energy level structure of a given molecule, 
although we emphasize that our results are general and apply to other problems.  
The (Born-Oppenheimer approximate) electronic Hamiltonian in the second quantized form~\cite{Szabo,helgaker2014molecular} is given by 
 \begin{equation}  \label{eq:elecHam}
H \: = \: \sum_{p,q=1}^\mathcal{N} h_{pq} a^{\dagger}_p a_q  +\:  \frac12 \sum_{p,q,r,s=1}^\mathcal{N} h_{pqrs} a^{\dagger}_p a^{\dagger}_q a_r a_s.
\end{equation} 
Here $a^{\dagger}$ and $a$ are fermionic creation and annihilation operators, respectively, 
and the coefficients $h_{pq}, h_{pqrs}\in\reals$ for $p,q,r,s=1,\dots,\mathcal{N}$ are provided as input. Combining adjoint pairs of terms as Hamiltonians $H_j$ we may write
$$ H=\sum_{j=1}^m H_j,$$
where each $H_j$ may be simulated efficiently \cite{whitfield2011simulation}, and their number 
is $m=\Theta(\mathcal{N}^4)$. The parameter $\mathcal{N}$ is at least the number of electrons, and should be taken larger than this to increase the accuracy in the underlying problem. 
For important problems 
in chemistry that are believed to be beyond the capabilities of classical algorithms, 
where, say, $\mathcal{N} \simeq 100$, 
the number of Hamiltonian terms $m$ is proportional to $10^8$. Hence, even for a quantum algorithm with relatively low polynomial cost dependence on $m$, say $m^2$, the simulation cost is immediately prohibitive, independent of its dependence on the other simulation problem parameters (time, accuracy, etc.).  

It is thus 
critical to 
derive simulation algorithms with reduced cost dependence on $m$. 
In this chapter, we show a general 
\textit{divide and conquer} 
approach which takes advantage of Hamiltonian structure. 
By partitioning a Hamiltonian $H=\sum_{j=1}^m H_j$ into a number of partial sums, simulating each sum separately, and then recombining the partial results, we obtain refined cost bounds which can lead to faster simulation. 
In particular, for simulating the electronic Hamiltonian (\ref{eq:elecHam}) we show that, under reasonable assumptions taken from the literature, our algorithm reduces the simulation cost dependence from $\mathcal{N}^8$ to $\mathcal{N}^9$, to between $\mathcal{N}^5$ to $\mathcal{N}^7$, with exponent depending on the particular problem representation and details. 
Previously, a sequence of papers has argued for the possibility of similar cost improvements 
based on heuristics or empirical evidence  \cite{wecker2014gate,poulin2014trotter,mcclean2014exploiting,babbush2015chemical}. 
In contrast, we provide rigorous cost and error bounds for our algorithms. 
The following table summarizes our cost estimates, which we explain in detail 
in Section~\ref{sec:QChem}. 

\begin{table}[h!]
\begin{center}
\begin{tabular}{| c | c |}
	\hline
	Method & Cost dependence on $\mathcal{N}$ \\
	\hline
	Splitting Formulas \cite{PZ12,wecker2014gate} & $\mathcal{N}^8 - \mathcal{N}^{9}$  \\
	\hline
	Truncated Taylor Series \cite[Eq. 46]{babbush2016exponentially}$^*$
	& ${\mathcal{N}^8} $  \\
		\hline
	Our algorithms with local basis functions & $\mathcal{N}^{5}-\mathcal{N}^{7}$  \\
		\hline
\end{tabular}
\end{center}
\caption{Summary of the cost estimates, expressed in terms of $\mathcal{N}$, for different algorithms for the simulation of the electronic Hamiltonian (\ref{eq:elecHam}). 
$(^*)$ The cost scaling is improved to $\mathcal{N}^5$ under strong assumptions 
on the basis functions and the computation of the $h_{pq}$, $h_{pqrs}$  \protect\cite{babbush2016exponentially}.}
\label{tab:qChemSummary}
\end{table}

Although the derivation and analysis of our algorithms is complicated, 
we emphasize that 
their implementation is relatively straightforward, and they yield an easily computable sequence of basic quantum operations that is similar in form to commonly used Suzuki-Trotter splitting formula approaches. 
We remark that a recent paper \cite{babbush2016exponentially}, specific to the electronic structure problem, has 
given a sophisticated algorithm with cost scaling nearly as $\mathcal{N}^5$, 
but under even stronger assumptions.  
Our approach seeks to give simple-to-implement quantum algorithms that are generally applicable, and can achieve similar performance improvements 
for important problems.

We emphasize that our 
approach is general and applies to simulation problems beyond the electronic Hamiltonian. 
We derive our algorithms, analysis, and results in terms of a general Hamiltonian simulation problem, and revisit 
quantum chemistry as an application at the end of the chapter.

\subsection{Problem Definition and Background}   \label{sec:probDefHamSim}
In the Hamiltonian simulation problem one is given a Hamiltonian $H$ acting on $q$ qubits, 
a time~$t\in\reals$, and an accuracy demand~$\e$, 
and the goal is to derive a quantum algorithm 
that constructs an operator~$\widetilde{U}$ 
which approximates the unitary operator~$e^{-iHt}$ with error $\|\widetilde{U}-e^{-iHt}\| \leq \e$ measured in the spectral norm.%
\footnote{The spectral-, or two-norm, of a Hermitian matrix $H$ is 
the magnitude of the maximum eigenvalue of $H$, and typically written~$\|H\|_2$.} 
When the Hamiltonian is given explicitly, the size of the quantum circuit realizing the algorithm is its cost. 
In particular, the cost depends on the complexity parameters $q$, $t$ and $\e^{-1}$. 
On the other hand, when the Hamiltonian is given 
by an oracle, the number of queries (oracle calls) used by the algorithm plays a major role in  its cost, in addition to the number of qubits and the other necessary quantum operations. 
Different types of queries have been considered in the literature. 

Many papers study only the query complexity.
For example, \cite{berry2007efficient,PZ12} use splitting formulas%
\footnote{Recall that high-order splitting formulas \cite{Suzuki90,Suzuki91} were used for Hamiltonian simulation in Algorithms $1$ and $2$ of Chapter \ref{ch:ExcStates}. 
We provide a review of splitting formulas in Appendix \ref{sec:splittingFormulas}. }
of order $2k+1$ 
to simulate  $H=\sum_{j=1}^m H_j$, $\|H_1\|\geq \|H_2\| \geq \dots \geq \|H_m\|$. 
They approximate $e^{-iHt}$ with error $\e$ by an ordered product of $N$ unitary operators of the form $e^{-iH_{j_\ell}t_{\ell}}$, $j_\ell \in \{1,\dots, m\}$, $|t_\ell | \leq t$, $\ell=1,\dots,N$. 
It is assumed that the Hamiltonian $H$ is given by an oracle (a \lq\lq black-box\rq\rq), and that 
$H$ can be decomposed efficiently by a quantum algorithm using oracle calls
into a sum of Hamiltonians $H_j$, $j=1,\dots,m$, that individually can be simulated efficiently.
This kind of query has been considered in numerous other papers, see e.g., \cite{lloyd1996universal,aharonov2003adiabatic,wiebe2010higher,childs2012hamiltonian}. 
The cost of the simulation is the total number of oracle calls,\footnote{Since the $H_j$ are obtained by decomposing $H$ by the algorithm, an oracle call to any $H_j$ is simulated by making oracle calls to $H$; see \cite[Sec. 5]{berry2007efficient} for details.} which is proportional to the number $N$ of exponentials $e^{-iH_{j_\ell}t_{\ell}}$. 
Then \cite{PZ12} shows the number of 
exponentials is bounded from above by
\begin{equation} \label{eq:PZ}
 N\le 
 m^2 \norm{H_1}t \left( \frac { 4em \norm{H_2} t}{\e }
\right)^{1/(2k)} \frac {16e}{3} \left( \frac {25} 3 \right)^{k-1},
\end{equation}
where $\|\cdot \|$ is the spectral norm. 
We shall also use the number $N$ as our measure of simulation cost for the algorithms we give later 
in this chapter.

On the other hand, \cite{berry2014exponential} uses a different type of query to simulate $d$-sparse Hamiltonians. In particular,  one is given access to a $d$-sparse Hamiltonian $H$
acting on $q$ qubits via a black box that accepts a row index $i$ and a number $j$ between $1$ and $d$, and
returns the position and value of the $j$th nonzero entry of $H$ in row $i$.  The paper shows a clever technique applied in combination with oblivious amplitude amplification to derive an algorithm simulating $d$-sparse Hamiltonians with a number of queries
\begin{equation} \label{eq:costLog}
N= O\left( \frac{\tau\log \tau/ \e}{\log \log \tau/\e}\right), \;\;\;\; \tau = d^2 \|H\|_{max} t,
\end{equation}
where $\|\cdot \|_{max}$ is the maximum norm.  
This is an important result. 
The  dependence of the cost on~$\e^{-1}$  is exponentially better in the latter case. However, this fact is not sufficient to conclude that the algorithm is exponentially faster than previously known simulation algorithms, because the size of the other complexity parameters,  $\tau$ and particularly the Hamiltonian norm $\| H\|_{\max}$,  needs to be taken into account as well.

For instance, the spectral and maximum norms are proportional to $d h^{-2}$ in the case where $H = -\Delta_h + V_h$ is a  matrix obtained from the discretization of the $d$-variate Laplacian $\Delta$ and a uniformly bounded $d$-variate potential function $V$ on a grid with mesh size $h$; see \cite{Demmel} for details. 
In this case, the sparsity of $H$ is $\Theta(d)$. 
Thus, for univariate functions 
the sparsity is constant. 
If we set $h=\e$,\footnote{When dealing with partial differential equations, the mesh size $h$ determines the discretization error, which subject to smoothness conditions often is $O(h^\alpha)$, for some $\alpha >0$. In terms of the partial differential equation, the combination of the discretization error and the simulation error  determines the accuracy of the final result. 
In this sense $h$ and $\e$ are related.} 
both cost estimates (\ref{eq:PZ}, \ref{eq:costLog}) become polynomial in $\e^{-1}$ and there is no exponential speedup. 
It is easy to extend this argument to $d$-variate functions and the situation is more interesting. In this case $H$ is a matrix of size $\e^{-d}\times\e^{-d}$.
For $k=1$ the bound (\ref{eq:PZ}) is proportional to
$$d \e^{-2.5} t^{1.5}$$  
while that of (\ref{eq:costLog}), modulo polylogarithmic factors, is proportional to
$$d^3 \e^{-2} t.$$
Both query estimates are \textit{low degree polynomials} in each of the complexity parameters.
Moreover, polynomial improvements, 
such as reducing the exponent of $d$ in (\ref{eq:costLog}) by one, 
as in \cite{berry2015hamiltonian},  hardly make a difference.
This situation is typical for matrices obtained from the discretization of ordinary and partial differential equations. 
We may have an exponential speedup 
when $\tau$ 
is at most polylogarithmic in $\e^{-1}$, but
 this is not typically the case in practice. 
Indeed \cite{berry2014exponential} does not mention any practical situation where an exponential speedup is realized.
These considerations apply to other recent papers also showing polylogarithmic dependence on $\e^{-1}$ of the query complexity \cite{berry2015simulating,berry2015hamiltonian}.

It is interesting to observe that the query complexity might be low and depend on $\e^{-1}$ polylogarithmically as in \cite{berry2015simulating}, yet when one considers the total gate count the picture may be quite different. An example can be found in 
\cite[Table~1 \& Table~2]{babbush2016exponentially} which applies \cite{berry2015simulating} to the simulation of the Born-Oppenheimer second-quantized electronic Hamiltonian (\ref{eq:elecHam}). 
 In particular the query complexity is proportional to $t\, \mathcal{N}^4$ times a quantity
 polylogarithmic in $t$, $\mathcal{N}$ and $\e^{-1}$, while the total gate count is proportional to 
 $t\, \mathcal{N}^8$ times a quantity polylogarithmic in $t$, $\mathcal{N}$ and $\e^{-1}$.
 Improvements of the cost are possible under significant assumptions on the 
 class of basis functions used and assumptions about 
 the cost and accuracy in computing the $h_{pq}$ and $h_{pqrs}$ by the quantum algorithm. 
Moreover, in chemistry the desired accuracy is not arbitrarily small~\cite{babbush2015chemical} and thus 
 it may impact the cost only by a constant factor. The important parameter is $\mathcal{N}$ which is the number of single-particle basis functions used in the approximation of the Born-Oppenheimer electronic Hamiltonian. 
 Larger values of $\mathcal{N}$ give more accurate approximations of the Hamiltonian operator.

Although improving exponentially the dependence of the simulation cost on $\e^{-1}$ is very significant, it is not a panacea.  We already mentioned that the other complexity parameters may be dominant.
There are other issues as well to consider. Queries that require one to have precomputed and stored the positions and values of all nonzero matrix entries of the Hamiltonian, or else have an efficient routine to generate these, can be restrictive. 
Similar concerns are discussed in \cite{aaronson2015read}. 
Moreover, simulation algorithms relying on oblivious amplitude amplification are probabilistic. This means that for applications where numerous Hamiltonian simulations need to be carried out, such as in phase estimation, the overall success probability must be further boosted. Making such an algorithm deterministic in practice is a numerical stability consideration.

On the other hand, without being advocates of splitting methods, we cannot avoid recognizing they have some very appealing features. Splitting methods \lq\lq conserve an important symmetry of the system in problems of quantum dynamics and Hamiltonian dynamics\rq\rq,  a \lq\lq remarkable advantage\rq\rq\ according to \cite{hatano2005finding}. Suzuki also remarks that splitting methods are particularly useful for studying quantum coherence \cite{Suzuki90}. 
 Simulation using splitting methods is deterministic in the sense that 
any repetition produces the same output with exactly the same accuracy. 
The simulation methods   \cite{berry2014exponential,berry2015simulating,berry2015hamiltonian} do not have these properties. 

\subsection{Divide and Conquer Simulation}
We give a new approach
for simulating Hamiltonians of the form 
\begin{equation}  \label{eq:HamDef}
H=\sum_{j=1}^m H_j.
\end{equation}
Our approach is especially useful when
the number $m$ of Hamiltonians $H_j$ is large, and many of the $H_j$  have relatively small norm. 
Such Hamiltonians are common in physics and chemistry \cite{ortiz2001quantum,suzuki1976generalized,suzuki2012quantum,peruzzo2014variational,Kassal,wecker2015solving,vivsvnak2015quantum}. 
For example, a system of interacting bodies or particles is described typically by a Hamiltonian of the above form. 

Without loss of generality, assume that 
the $H_j$ are indexed as 
\begin{equation}  \label{eqn:HamOrdering}
 \|  H_1 \| \geq \| H_2 \| \geq \dots  \geq \|H_m\|.
\end{equation}
For many problems, the norms $\|H_j\|$ vary substantially, and many Hamiltonians may have norm $\|H_j\| \ll \|H_1\| $.  
Then one can take advantage of the discrepancy between the norm sizes to derive fast simulation algorithms.  
The main idea is as follows: 
\begin{enumerate}
\setcounter{enumi}{-1}
\item  
Partition the Hamiltonians $H_1,H_2,\dots,H_m$ into groups using the magnitude of their norms. 
Ideally, Hamiltonians of proportionate norms are grouped together.  
\item  Approximate $e^{-iHt}$ by a splitting formula applied with respect to the partition into groups, 
pretending that the sums of Hamiltonians in each group can be simulated exactly. 
\item  Simulate the sum of the Hamiltonians in each group separately with sufficient accuracy. 
\item  Combine all the group simulation results 
to give the overall simulation of $H$.  
\end{enumerate} 
A 
top-level description of the procedure above applied to two groups and utilizing splitting formulas is shown in Figure \ref{fig:flowchart} below. Nevertheless, our approach is not limited to splitting formulas. 
\vskip 1pc
\begin{figure}[h!]
\centering
\includegraphics[width=12cm]{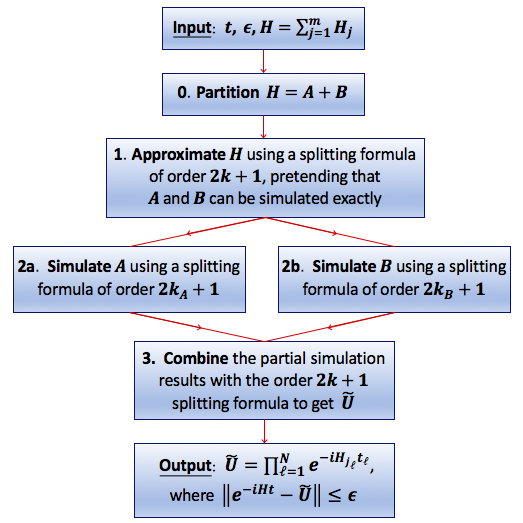}
\caption{Divide and conquer simulation using splitting formulas.}
\label{fig:flowchart}
\end{figure}

To motivate this idea consider the bound (\ref{eq:PZ}) which depends particularly on $m$, $\| H_1\|$ and $\| H_2\|$, and not on $\| H_3\|,\dots, \| H_m\|$.  
For the sake of argument, suppose $m$ is huge 
and $\| H_2\| \gg \| H_3\|$. 
Then we can
split the Hamiltonians in two groups $\{H_1, H_2\}$ and $\{H_3,\dots,H_m\}$,
simulate $A:=H_1+H_2$ and $B:= H_3+\dots+H_m$ independently, and then combine the partial simulation results using a splitting formula. Observe that $e^{-iHt} \to e^{-iAt}$ 
as $\| H_3\| \to 0$ and in the limit the total simulation cost becomes independent of $m$. 
Thus in the limit the bound (\ref{eq:PZ}) holds with $m$ replaced by $2$.  
This suggests that when
many Hamiltonians are relatively small in norm one should be able to 
improve the cost estimate~(\ref{eq:PZ}) 
by partitioning them into groups and, for instance, using splitting formulas of different orders as we indicate in Figure \ref{fig:flowchart}, and we will explain in detail later. 
 
An example application with these properties is the simulation of the electronic Hamiltonian~(\ref{eq:elecHam}). 
This problem has been well-studied in the literature; see e.g. \cite{lanyon2010towards,Kassal,whitfield2011simulation,wecker2014gate,poulin2014trotter,mcclean2014exploiting,babbush2015chemical,hastings2015improving,babbush2016exponentially}. 
Recall the number of single-particle basis functions $\mathcal{N}$ is typically chosen to be 
proportional to the number of particles in a given problem, and the number of Hamiltonians is
$m=\Theta(\mathcal{N}^4)$.  
The best classical algorithms can reasonably solve problem instances with $\mathcal{N}$ in the range $50-70$, and it is believed that a quantum computer able to simulate problem instances with $\mathcal{N} \simeq 100$ will solve many important applications ranging from chemical engineering to biology \cite{wecker2014gate}.  
In these cases, the required modest number of qubits (typically $\Theta(\mathcal{N})$ \cite{aspuruGuzik2005yq}) makes these very attractive applications for early quantum computers. 
As explained, despite the many recent advances in quantum simulation algorithms, 
the cost of Hamiltonian simulation remains the primary bottleneck to solving this problem on a quantum computer, 
Indeed, reducing the simulation cost dependence on $m$ (i.e., on~$\mathcal{N}$) for this problem has been the subject of considerable recent effort \cite{wecker2014gate,poulin2014trotter,mcclean2014exploiting,hastings2015improving,babbush2015chemical,babbush2016exponentially}. 
 %
Furthermore, in many situations,  
it has been observed that the Hamiltonian norms vary significantly, and many of them are relatively small \cite{helgaker2014molecular,jones2012faster}. 
It has been suggested that this could be used in some way to potentially reduce the simulation cost, without any rigorous analysis \cite{jones2012faster,mcclean2014exploiting,hastings2015improving,babbush2015chemical}. 
In contrast, in this chapter we develop algorithms that use the discrepancy between sizes of Hamiltonian norms to speedup Hamiltonian simulation and we derive their cost in full detail.

\subsection{Overview of Main Results}
For simplicity, 
we consider here the case where we partition the Hamiltonians in two groups, but the idea extends to many groups, as we show in Section \ref{sec:GeneralParadigm}. Let  $H=A+B$,
with $A=\sum_{i=1}^{m'} H_i$ and $B=\sum_{i=m'+1}^m H_i$, with $m' \ll m$, where 
again $\|H_1\| \geq \|H_2\| \geq \dots \geq \|H_m\| $. 
The bound~(\ref{eq:PZ}) for the number of queries 
$N$ scales with $m$ as $m^{2+1/2k}$, and our goal is to improve that.

\begin{enumerate}
\item Suppose we have two \textit{arbitrary} algorithms 
$\widetilde{U}_A(\tau) \simeq e^{-iA\tau}$ and $\widetilde{U}_B(\tau) \simeq e^{-iB\tau}$ 
for approximately simulating the Hamiltonians $A$ and $B$, respectively, for time $\tau\in\reals$. 
We show how  splitting formulas may be used to combine $\widetilde{U}_A$ and $\widetilde{U}_B$ such that an approximation $\widetilde{U}(t)$ to $U(t)=e^{-iHt}$ is achieved. 
For example, dividing the time $t$ into $n$ intervals of length $\tau := t/n$ and 
using the Strang splitting formula \cite{PZ12} we get the overall approximation
\begin{equation}  \label{eq:mainApprox}
 \widetilde{U}(t) := \left(\widetilde{U}_A(\tau/2)\; \widetilde{U}_B(\tau)\;
\widetilde{U}_A(\tau/2)\right)^n  
 \simeq \left(  e^{-iA\tau/2} e^{-iB\tau} e^{-iA\tau/2}   \right)^n
 \simeq e^{-iHt}. 
\end{equation}
Then to obtain 
 $\|U(t) - \widetilde{U}(t)\|= O( \e)$ it suffices that $\| e^{-iA\tau/2}- \widetilde{U}_A(\tau/2)\|$ and $\| e^{-iB\tau}- \widetilde{U}_B(\tau)\|$ are each of order $\e/n$. 
Higher order splitting formulas may be used instead of the Strang splitting formula 
such that the error and resulting cost are further reduced. 

In the following items we use splitting formulas to derive $\widetilde{U}_A$ and
$\widetilde{U}_B$; however, in principle, different 
applicable simulation algorithms could be used for each of $\widetilde{U}_A$ and
$\widetilde{U}_B$. 
Moreover, 
we use the ordering $\|H_1\|\geq \|H_2\|\dots$ to partition the $H_j$, 
though in practice criteria other than the norms may be used to 
group the Hamiltonians, such as sparsity, commutativity, or unitarity or any other property which may allow one to use an advantageous algorithm for simulating the Hamiltonians in that group.

\item  We use splitting formulas (of orders $2k_A+1$ and $2k_B+1$, respectively) to obtain the approximations
 $\widetilde{U}_A(t)$ and $\widetilde{U}_B(t)$, which we combine
 with an order $2k+1$ splitting formula. The resulting total number of queries $N$ for simulating $H=A+B$ satisfies
\begin{equation}   \label{eq:costBound}
 N \leq  8m' 5^{k+k_A-2} \; \max\{ n_A , n \}  +
4(m-m')5^{k+k_B-2}  \; \max\{n_B,n \}  ,
\end{equation}
where 
\begin{itemize}   \label{eqn:setn2}
\item $n \geq  \|A\|t  \left(  16e \|B\|t / \e  \right)^{1/2k}  \frac{8e}{5} \left( \frac53 \right)^{k}          \;\;\;\;$ for $\|A\|\geq \|B\|$, 
\item $n_A = m' \|H_1\| t  \left(\frac{64e}{5} \: m'  \|H_2\| t /\e  \right)^{1/2k_A}  7e \left( \frac53 \right)^{k_A - k} $,
\item $n_B =  (m-m') \|H_{m'+1}\| t   \left( \frac{64e}{5} \: (m-m') \|H_{m'+2}\| t /\e  \right)^{1/2k_B}  14e \left( \frac53 \right)^{k_B-k} $.
\end{itemize}

The form of the cost bound (\ref{eq:costBound}) is similar to that of (\ref{eq:PZ}), but with refined cost dependence. 
Roughly speaking, the two terms of the cost bound above correspond to the cost of simulating the Hamiltonians in the two groups forming $A$ and $B$, respectively, 
plus some partitioning/recombining overhead that is captured by the maximum function.

The novelty of the algorithm is that it uses  
substantially fewer exponentials to simulate Hamiltonians of small norm, relative to the number of exponentials required for Hamiltonians of much larger norm, 
while maintaining the desired accuracy. 
In this respect, different time slices are chosen adaptively 
to simulate Hamiltonians in different groups. 
As a result, it is possible to use few exponentials to simulate 
a large number of Hamiltonians $H_j$ of relatively small norm
for longer time slices, and this reduces the overall simulation cost. 

We emphasize that even though the cost bound (\ref{eq:costBound}) appears complicated, implementing the algorithms achieving this bound is straightforward, 
with similar implementation details to those of splitting formulas; see e.g. (\ref{eq:mainApprox}). 

Items $3$ and $4$ below illustrate the impact of the divide and conquer approach, relative to earlier work, 
as the number  $m$ of terms 
grows and becomes huge. 
Item $5$ estimates the practical advantage of the divide and conquer approach for simulating the electronic Hamiltonian. 

\item For the case $k=k_A=k_B=O(1)$, 
and assuming that a large number of Hamiltonians have very small norm such that 
$(m-m')\|H_{m'+1}\| \leq m' \|H_2\|$, 
we can select $n$ so that 
$n_A \geq n \geq n_B$ and  
\begin{eqnarray*}
N &=& O\left(m'^{\, 2+1/2k} \;\|H_1\|t \; (\|H_2\|t/\e )^{1/2k} \right) \\
&+& O\left((m-m')^{1+1/2k}\, m'  \;\|H_1\|t \; (\|H_{m'+1}\|t/\e )^{1/2k} \right). 
\end{eqnarray*}
In particular, when
a relatively small number of $H_j$ form $A$ so that   
$m'=O(m^a)$, and 
the remaining $H_j$ have small norms in the sense that 
$(m-m')\|H_{m'+1}\| / \|H_2\|=O(m^b)$, for $0\leq b \leq a < 1$, 
we have a speedup  over the number of queries in (\ref{eq:PZ}) 
given by 
$$  \frac{N}{N_{prev}} = O\left( \frac{1}{m^{(1-a)+(1-b)/2k}} \right) ,$$
independently of $t,\e$, where $N_{prev}$ denotes the upper bound shown in (\ref{eq:PZ}) with the same $k$.  
Observe that this quantity goes to $0$ as $m\rightarrow \infty$.

\item In \cite{berry2007efficient,PZ12,childs2012hamiltonian}, 
it is shown how for splitting methods, the order of the splitting formula may be selected \lq\lq optimally\rq\rq\ such that the respective cost bounds are minimized. 
We show how optimal parameters $k^*$, $k^*_A$, and $k^*_B$ may be similarly selected for our algorithms. 
Let $N^{*}_{prev}$ and $N^{*}$ be the resulting numbers of queries for the algorithm in \cite{PZ12} and for our algorithm, respectively. 
We show conditions for a strong speedup over \cite{PZ12} in the sense that
$$\frac{N^{*}}{N^{*}_{prev}} \; \xrightarrow[m\rightarrow \infty]{}  0  \;\;\;\;\;\;\; \text{for fixed } t,\e.$$

\item We apply our algorithm to the approximate electronic Hamiltonian (\ref{eq:elecHam}) of quantum chemistry. Let~$\mathcal{N}$ be the number of single-particle basis functions. The number of Hamiltonians in (\ref{eq:elecHam}) is $\Theta(\mathcal{N}^4)$. We can assume that the largest Hamiltonian norm in the sum is constant. 
It is known that in practical cases a large number of  terms have very small norm \cite{jones2012faster,mcclean2014exploiting,babbush2015chemical}. This allows us to dramatically improve the simulation cost. 
Table \ref{tab:qChemSummary} 
presented 
at the beginning of this chapter 
illustrates this point by comparing our technique to others. 
Recall that  
the important complexity parameter is  $\mathcal{N}$; 
we express the cost with respect to  $\mathcal{N}$ in the table, 
assuming $t,\e$ are fixed. 
 
Our cost estimates of $\mathcal{N}^{5}-\mathcal{N}^{7}$ are consistent with empirical studies 
indicating that previous cost and error estimates may be overly conservative for practical applications~\cite{poulin2014trotter}.  
 
We emphasize that standard circuits implementing the evolution under the individual terms in (\ref{eq:elecHam}) can be incorporated into our algorithm directly to yield its gate level implementation. For example, one can use the circuits in~\cite{whitfield2011simulation}.   
 \end{enumerate}

In the remainder of this chapter, we give our approach and algorithms in detail 
and derive the above results. 
Several of the more involved proofs are deferred to Appendix \ref{app:HamSim}. 
The results of this chapter can also be found in \cite{hadfield2017hamsim}.

\section{Preliminary Analysis}   \label{sec:GeneralParadigm}
Our goal is to take the Hamiltonian simulation problem  and partition it into a number of smaller and simpler Hamiltonian simulation problems, then solve each one of them, and combine the results. The splitting should be customized to take advantage of the properties of each of the subproblems, yielding refined bounds for the overall simulation cost.

In certain applications, for instance in chemistry, Hamiltonians with extremely small norm can be discarded from the sum (\ref{eq:HamDef}) as a preprocessing step, to the extent that this does not affect the desired accuracy. We formalize this idea in the following subsection. 

%
\subsection{Discarding Small Hamiltonians}
Hamiltonians of very small norm relative to the accuracy $\e$ may be discarded, and it suffices to consider the simulation problem for the remaining Hamiltonians. This may substantially reduce the cost, 
particularly for problems where $\e$ is not arbitrarily small.

\begin{prop}   \label{prop:discardHamsNew}
Let $H= A+B$ where $H$, $A$, $B$ are Hamiltonians, $t>0$, and $\e>0$. 
If 
\begin{equation}   \label{eqn:smallHamCondNew}
  \|B\|t \leq \e/2,
\end{equation}
and $\widetilde{U}$ is such that $\|e^{-iAt} - \widetilde{U}\| \leq \e/2$ then  $\|e^{-iHt} - \widetilde{U}\| \leq \e$. 
\end{prop}
%
The proof of the proposition is shown in Appendix \ref{app:HamSim}. 
Thus, when the conditions of the proposition are satisfied, simulating $A$ with error $\e/2$ implies the simulation of $H$ with error $\e$.

\begin{rem}    
Equation (\ref{eqn:smallHamCondNew}) implies that the aggregate norm of the discarded Hamiltonians must be small, not just the norms of the discarded Hamiltonians themselves. 
Generally, Hamiltonians cannot be discarded without considering how many they are and the magnitudes of the other problem parameters.
\end{rem}  
In practical applications a large number of \lq\lq negligible\rq\rq\ Hamiltonians are sometimes 
discarded, often using heuristics.  
For example, in quantum chemistry, an ad hoc fixed cut-off parameter, say $10^{-10}$, is used \cite{helgaker2014molecular}. (For applications such as eigenvalue estimation, a relatively large error can be tolerated for Hamiltonian simulation \cite{NC}.) However, in general the effect of discarding terms must be accounted for in the error analysis.

We will assume that possible discarding of Hamiltonians according to Proposition \ref{prop:discardHamsNew} may have happened as a preprocessing step. 
Our results and proof techniques do not depend on whether Hamiltonians have been discarded or not. Thus, from this point on $m$ will refer to the total number of Hamiltonians that we 
consider as input for our algorithms.

\subsection{Recursive Lie-Trotter Formulas}
Suppose the number $m$ of Hamiltonians is large, and we 
are given a partition as 
\begin{equation}  \label{eqn:defAB}  
H=A+B:=(H_1+\dots +H_{m'}) + (H_{m'+1}+\dots+H_m).
\end{equation}
We consider partitions into two groups to make the ideas of this section clear; it is straightforward to extend to 
an arbitrary number of groups $\mu$. 
 As $A$ and $B$ are themselves Hamiltonians, we may apply the Lie-Trotter formula 
(see equation (\ref{eqn:LieTrotterFormula}) in Appendix \ref{app:HamSim}) 
 with respect to them to give
\begin{equation}  \label{eqn:GeneralizedLieTrotterFormula}
\lim_{n \rightarrow \infty} (e^{-iAt/n} e^{-iBt/n} )^n = e^{-iHt}.
\end{equation}
Thus, we see that if we are able to approximate $e^{-iAt/n}$ and $e^{-iBt/n}$
then we should be able to combine the approximations as in (\ref{eqn:GeneralizedLieTrotterFormula}) to 
approximate $e^{-iHt}$. 

Indeed, we can 
 again apply the Lie-Trotter formula (\ref{eqn:LieTrotterFormula}) to each $e^{-iAt/n}$ and $e^{-iBt/n}$ 
to yield the 
\textit{Recursed Lie-Trotter formula}
\begin{equation}  \label{eqn:GeneralizedLieTrotterFormula2}
\lim_{\alpha, \beta, n \rightarrow \infty} \left(
(e^{-iH_1t/\alpha n} \dots e^{-iH_{m'}t/\alpha n} )^\alpha
 (e^{-iH_{m'+1}t/\beta n} \dots e^{-iH_{m}t/\beta n} )^\beta
\right)^n = e^{-iHt},
\end{equation}
where the limits may been taken 
in any order; see Appendix \ref{app:HamSim} for the proof. 

Compared to (\ref{eqn:LieTrotterFormula}), there are now three parameters $n,\alpha,\beta$ in (\ref{eqn:GeneralizedLieTrotterFormula2}) which reduce the error of the truncated product approximation as they are increased. 
Suppose $\|H_1\| \gg \|H_\ell \|$ for some $1\leq \ell \ll m$; then, grouping the largest Hamiltonians in $A$ and the remaining Hamiltonians in $B$, it follows that we may want to take $\alpha > \beta$ as to reduce the overall error, while keeping $\beta$ relatively small to reduce the overall cost. 
We will shortly derive divide and conquer simulation algorithms based on splitting formulas which will take  
 three parameters $k, k_A, k_B$ specifying the order of each formula. Thus we may use a high order splitting formula for $A$ and a lower order splitting formula (and also larger time slices) for $B$, without compromising the error and such that the overall cost is reduced.

We remark that generalizing (\ref{eqn:GeneralizedLieTrotterFormula2}) to more than two groups of Hamiltonians gives a Trotter step parameter $\alpha_i$ for each group. Alternatively, this formula could be recursed deeper 
by further decomposing $A$ and $B$ into subgroups of Hamiltonians and again applying (\ref{eqn:LieTrotterFormula}). 
Finally, the ideas of this subsection are easily generalized from the Trotter approximation to higher order formulas.

\subsection{Combining Different Simulation Methods} \label{sec:GeneralParadigmDetails}

We now describe our approach generally. 
Consider a Hamiltonian $H$ as in (\ref{eq:HamDef}, \ref{eqn:HamOrdering}), and let $U=e^{-iHt}$. 
Assume the $H_j$ have been partitioned into $\mu=O(1)$ disjoint groups, where we denote by $A_1,\dots,A_\mu$ the sums of the Hamiltonians in the respective groups. 
We are not concerned with how the partitioning is done at this point. As we will see later, the partitioning can be done adaptively and follows from general cost estimates.  In practice, small values of $\mu$ may suffice and we'll see such an example in Section \ref{sec:QChem}.

Then $H=A_1+\dots+A_\mu$.   
Assume the $A_j$ have been indexed so that $\|A_1\| \geq \|A_2\| \geq \dots \|A_\mu\|$. 
Suppose we divide the simulation time $t$ into intervals $\Delta t = t/n$, 
$n \in \nat$; we will show how to select $n$ later.  Applying a 
splitting formula of order $2k+1$ with respect to this partition yields the operator 
\begin{equation}  \label{eqn:UhatGen}
\widehat{U}:= (S_{2k}(A_1,\dots,A_\mu,t/n))^n
=  \left( \prod_{\ell=1}^{N_{k,\mu}} e^{-i  A_{j_\ell} t_{\ell} /n } \right)^n,  \;\;\;\;\;\;  j_\ell \in \{1,\dots,\mu \}, \;\;\;\;  \sum_{{\ell=1}}^{N_{k,\mu}} t_{\ell}= \mu t,   
\end{equation}
where $N_{k,\mu}=(2\mu - 1)5^{k-1}$, $|t_{\ell}|\leq t/n$, 
and $S_{2k}(A_1,\dots,A_\mu,t/n)$ is given in (\ref{eqn:defS2k}).
Then, if we have algorithms $\widetilde{U}_{A_{j}}(\tau)$ to simulate (approximately) each exponential $e^{-iA_j\tau}$ in the right-hand side above, we can
substitute them into (\ref{eqn:UhatGen}) and obtain the approximation  
\begin{equation}   \label{eqn:UtildeGen}
\widetilde{U}:= (\widetilde{S}_{2k}(A_1,\dots,A_\mu,t/n))^n = \left( \prod^{N_{k,\mu}}_{\ell=1} \widetilde{U}_{A_{j_\ell}}(t_{\ell}/n)\right)^n,  \;\;\;\;\;\;  j_\ell \in \{1,\dots,\mu \}, \;\;\;\;  \sum^{N_{k,\mu}}_{{\ell=1}} t_{\ell}= \mu t. 
\end{equation}
We emphasize $\widetilde{S}_{2k}(A_1,\dots,A_\mu,t/n)$ is constructed by expanding $S_{2k}(A_1,\dots,A_\mu,t/n)$ as an ordered product of exponentials $e^{-i A_{j_\ell}t_{\ell} }$ and replacing each $e^{-iA_{j_\ell}t_{\ell}  }$ with $\widetilde{U}_{A_{j_\ell}}(t_{\ell})$. 
The precise ordering of the product and the values $t_{\ell}$ are obtained from the particular choice of the splitting formula of order $2k+1$; see \cite{Suzuki90,Suzuki91}. 
For example, for $k=1$, this gives
\begin{equation}   \label{eqn:S2general}
\widetilde{S}_{2}(A_1,\dots,A_\mu,t/n) = \widetilde{U}_{A_1}(t/2n)  \dots  \widetilde{U}_{A_{\mu-1}}(t/2n)   \widetilde{U}_{A_\mu}(t/n) \widetilde{U}_{A_{\mu-1}}(t/2n)   \dots \widetilde{U}_{A_1}(t/2n).
\end{equation}
In principle, any available method may be used to implement the approximations $\widetilde{U}_{A_j}$, with the possibility of using different subroutines for different $j$.  

We bound the overall error by
\begin{equation}  \label{eqn:totalErrorTriangleGeneral}
\| U - \widetilde{U} \| \leq \| U- \widehat{U}\| + \| \widehat{U} - \widetilde{U} \|.
\end{equation}
We refer to $\| U- \widehat{U}\|$ and $\| \widehat{U} - \widetilde{U} \|$ as the \textbf{first-step error} and \textbf{second-step error}, respectively. 
Clearly, if both error terms are $O(\e)$, then so is the overall error $\| U - \widetilde{U} \|$.

The first-step error depends only on the splitting formula used at the first step, and is independent of the 
subroutines used to simulate each group at the second step. 
We have
\begin{equation}   \label{eqn:topLevelErrorGeneralParadigm}
\| U- \widehat{U}\| = \|(e^{-iHt/n})^n -  (S_{2k}(A_1,\dots,A_\mu,t/n))^n\|
\leq n  \|e^{-iHt/n} -  S_{2k}(A_1,\dots,A_\mu,t/n)\|,
\end{equation}
where $\|e^{-iHt/n} -  S_{2k}(A_1,\dots,A_\mu,t/n)\|$ is the error of $S_{2k}$ over a single time slice.  
Following the approach of \cite{PZ12} (see equation (\ref{eqn:def0Mnew})) for the simulation of a sum of $\mu$-many Hamiltonians with accuracy $\e/2$ at the first step, we define the quantity 
\begin{equation}   \label{eqn:defMgeneral}
M=  \left(   \frac{4e \mu t \|A_2\| }{\e/2}  \right)^{1/2k}  \frac{4e\mu}{3}  \left( \frac53 \right)^{k-1},
\end{equation}
which gives the first-step time slice size as $\Delta t := (M\|A_1\|)^{-1}$. 
The number of first-step time slices
is $n= \lceil M\|A_1\|t \rceil$. Observe that the final time slice may be smaller than $\Delta t$. With this in mind, for simplicity we assume here that $M\|A_1\|t$ is an integer. 

The second-step error is
\begin{eqnarray}   \label{eqn:errorSecondTermGeneralCase}
\| \widehat{U} - \widetilde{U}\| &=& \| S_{2k}(A_1,\dots,A_\mu,t/n)^{n} - \widetilde{S}_{2k}(A_1,\dots,A_\mu,t/n)^{n}  \| \nonumber \\
&\leq& n \left\|    \prod^{N_{k,\mu}}_{\ell=1} e^{-i A_{j_\ell} t_{\ell}/n  } -  \prod^{N_{k,\mu}}_{\ell=1} \widetilde{U}_{A_{j_\ell}}(t_{\ell}/n) \right\|  \nonumber\\
&\leq& n \sum^{N_{k,\mu}}_{\ell=1}    \| e^{-iA_{j_\ell} t_{\ell}/n} - \widetilde{U}_{A_{j_\ell}} (t_{\ell}/n)  \|
\end{eqnarray}
Hence, a sufficient condition for $\| \widehat{U} - \widetilde{U}\| \leq  \e/2$ is that the error of each stage satisfies 
$$\| e^{-iA_{j_\ell} t_{\ell}/n} - \widetilde{U}_{A_{j_\ell}} (t_{\ell}/n)  \| \leq \frac{\e}{2N_{k,\mu} n} .$$ 

Assume the cost 
$N_{j_{\ell}}=N(A_{j_\ell},t_{\ell}/n)$
of each simulation subroutine $\widetilde{U}_{A_{j_\ell}} ( t_{\ell}/n )$ is expressed in terms of the number of exponentials of the form $e^{-iH_{j} z}$, where the $H_j$ belong to the group forming $A_{j_\ell}$, for suitable values $z\in \reals$. 
The total simulation cost is the number of time slices $n$ times the cost per time slice. The latter is 
$\sum_{\ell=1}^{N_{k,\mu}} N_{j_{\ell}},$ and 
therefore the total simulation cost is
\begin{equation}  \label{eqn:overallCostGen}
N = n \cdot \left( \sum_{\ell =1}^{N_{k,\mu}} N_{j_{\ell}}  \right).
\end{equation}

\section{Divide and Conquer Splitting Formulas}  \label{sec:DCAlgorithms}
Consider again a Hamiltonian $H=\sum_{j=1}^m H_j$ as in (\ref{eq:HamDef}, \ref{eqn:HamOrdering}), 
partitioned into two groups $H=A+B$ as in~(\ref{eqn:defAB}).     
The two algorithms we present are illustrated in Figure \ref{fig:flowchart}. 
Algorithm $1$ is a special case of Algorithm $2$. 
Both algorithms, like splitting formulas, result in an ordered product of exponentials $\widetilde{U}=e^{-iH_{j_1}t_1}e^{-iH_{j_2}t_2}\dots e^{-iH_{j_N}t_N}$, $|t_\ell |\leq t$. 
The difference between our algorithms is that Algorithm $1$ uses $k=1$ in the first step, while Algorithm $2$ considers arbitrary $k$. 
Even though this difference might appear minor, the analysis of Algorithm $2$ is much more complicated. Algorithm $1$ is simpler to understand and implement, while Algorithm $2$ is more general, offering one the possibility to reduce the number of exponentials by selecting $k>1$. 

\subsection{Algorithm 1}   \label{sec:AlgStrangTop}
Algorithm $1$ follows the construction of Section \ref{sec:GeneralParadigmDetails} for the general case, applied to the partition $H=A+B$. 
At the \textbf{first step}, applying the Strang splitting formula ($k=1$) gives the operators
\begin{equation}  \label{eqn:Uhat}
\widehat{U}:= (S_2(A,B,\Delta t))^n   = ( e^{-iA\Delta t/2} e^{-iB\Delta t}e^{-iA\Delta t/2} )^n,
\end{equation}
where $\Delta t = t/n$ and we will define $n$ below. 
For the \textbf{second step}, Algorithm $1$ 
approximates the operators $e^{-iA\Delta t/2}$ and $e^{-iB\Delta t}$ 
using different high-order splitting formulas $\widetilde{U}_A(\Delta t/2)$ and $\widetilde{U}_B(\Delta t)$, of orders $2k_A+1$ and $2k_B+1$, respectively. 
This yields the overall approximation $\widetilde{U}$ of $U=e^{-iHt}$ which is defined by 
\begin{eqnarray}  \label{eqn:Utilde}
\widetilde{U} &:=& (\widetilde{S}_2(A,B,\Delta t) )^n 
= \left(\widetilde{U}_A(\Delta t/2)\widetilde{U}_B(\Delta t)\widetilde{U}_A(\Delta t/2)\right)^n.
\end{eqnarray}
Note that in general 
$\widetilde{U}_A(\Delta t/2) \widetilde{U}_A(\Delta t/2) \neq \widetilde{U}_A(\Delta t)$.

As in \cite{PZ12} let
$$\mathcal{H}_j :=\begin{cases}
    H_j / \| H_1\|         & (1 \leq j \leq m')\\
    H_j / \| H_{m'+1}\| & (m' < j \leq m) . \end{cases}  
$$
For splitting formulas, such a rescaling of the Hamiltonian norms is equivalent to a rescaling of the respective group 
simulation times, i.e.,  
$S_{2k_A}(\mathcal{H}_{1},\dots, \mathcal{H}_{m'}, \|H_1\|\tau)=S_{2k_A}(H_{1},\dots, H_{m'}, \tau )$
and 
$S_{2k_B}(\mathcal{H}_{m'+1},\dots, \mathcal{H}_{m}, \|H_{m'+1}\|\tau)=S_{2k_B}(H_{m'1},\dots, H_{m}, \tau )$. 
Observe that the Hamiltonians in $A$ and $B$ are rescaled by different quantities, 
which leads to different simulation times for each.   

The 
time slice sizes for simulating $\widetilde{U}_A(\Delta t/2)$ and $\widetilde{U}_B(\Delta t)$ are $1/M_A$ and $1/M_B$, respectively, where $M_A$ and $M_B$ are defined below. 
Thus, applying splitting formulas of orders $2k_A +1$ and $2k_B+1$  for $U_A$ and $U_B$, respectively,  gives 
\begin{equation}   \label{eqn:UtildeA}
\widetilde{U}_A(\Delta t/2) := S_{2k_A}(\mathcal{H}_{1},\dots, \mathcal{H}_{m'}, 1/M_A)^{ \lfloor M_A\|H_{1}\|\Delta t/2 \rfloor }  S_{2k_A}(\mathcal{H}_{1},\dots, \mathcal{H}_{m'}, \delta_A/M_A)  ,
\end{equation} 
\begin{equation}   \label{eqn:UtildeB}
\widetilde{U}_B (\Delta t) := S_{2k_B}(\mathcal{H}_{m'+1},\dots, \mathcal{H}_{m}, 1/M_B)^{ \lfloor M_B\|H_{m'+1}\| \Delta t \rfloor } S_{2k_B}(\mathcal{H}_{m'+1},\dots, \mathcal{H}_{m}, \delta_B/M_B).
\end{equation}
Since we have effectively rescaled the simulation times by dividing by the respective largest Hamiltonian norms, we are actually subdividing an interval of size $\|H_1\|\Delta t/2$ into $\lceil M_A\|H_1\|\Delta t/2 \rceil$ intervals of length at most $1/M_A$ for the simulation of $\widetilde{U}_A(\Delta t/2)$, and 
into $\lceil M_B\|H_{m'+1}\|\Delta t \rceil$ intervals of length at most $1/M_B$ for $\widetilde{U}_B(\Delta t)$. 
Clearly, the last of these subintervals in either case may have length less than $1/M_A$ or $1/M_B$, respectively.
In such a case, the length of the last subinterval is equal to 
$\delta_A/M_A$  or $\delta_B/M_B$, with 
$\delta_A:=M_A\|H_1\|\Delta t/2 - \lfloor M_A\|H_1\|\Delta t/2 \rfloor$ and $\delta_B:=M_B\|H_{m'+1}\|\Delta t - \lfloor M_B\|H_{m'+1}\|\Delta t \rfloor$, respectively.  
That is the reason why we have taken the floors of the exponents in the first factors of (\ref{eqn:UtildeA}) and (\ref{eqn:UtildeB}).

For the simulation error, from (\ref{eqn:totalErrorTriangleGeneral}) we have $\| U - \widetilde{U} \| \leq \| U- \widehat{U}\| + \| \widehat{U} - \widetilde{U} \|$. 
Thus, to guarantee $\| U - \widetilde{U} \|\leq \e$, we require $\| U- \widehat{U}\| \leq \e/2$ and $\| \widehat{U} - \widetilde{U} \| \leq \e/2$.

We consider each error term separately.  
The first error term 
 is independent of 
 the algorithms used for $\widetilde{U}_A$ and $\widetilde{U}_B$, and results only from the 
first-step Strang splitting and time slice size ${\Delta t = t/n}$, $n\in \nat$. 
From Lemma \ref{lem:wecker2}, shown in Appendix \ref{app:HamSim}, we have   
\begin{equation} \label{eqn:topError}
\| U- \widehat{U}\|  \leq \frac23 t \Delta t^2 \|A\| \|B\| \cdot \max\{ \|A\|,\|B\| \}.
\end{equation} 
From (\ref{eqn:overallCostGen}) the cost of our algorithm is proportional to $n$, and therefore we would like to select it to be as small as possible. 
Setting the right hand side of the equation above to $\e/2$ we obtain
\begin{equation} 
n \geq  \sqrt{ 4 t^3\; \|A\| \|B\| \max\{ \|A\|,\|B\| \}  / 3\e } .
\end{equation}
For instance, when $\|A\| \geq \|B\|$, 
from the triangle inequality  bounds 
$\|A\| \leq m' \|H_1\|$ and $\|B\| \leq (m-m')\|H_{m'+1} \|$, 
to obtain $\| U- \widehat{U}\| \leq \e/2$ 
it therefore suffices to select $n$ as 
\begin{equation}  \label{eqn:setn}
n:= \bigg \lceil  \sqrt{4/3} \;  m'\|H_1\| t \sqrt{ (m-m')\|H_{m'+1} \|  t/ \e}  \bigg\rceil .
\end{equation} 

Now consider the second error term. 
As $\| S_2\|=\|\widetilde{S}_2\|=1$, 
we have (cf. eq. (\ref{eqn:errorSecondTermGeneralCase}))
\begin{eqnarray*}
\| \widehat{U} - \widetilde{U}\| &=& \| S_2(A,B,\Delta t)^{n} - \widetilde{S}_2(A,B,\Delta t)^{n}  \| 
\leq n \| S_2(A,B,\Delta t) - \widetilde{S}_2(A,B,\Delta t) \|  \\
&\leq& n \| e^{-iA\Delta t/2} e^{-iB\Delta t}e^{-iA\Delta t/2} - \widetilde{U}_A(\Delta t/2)\widetilde{U}_B(\Delta t)\widetilde{U}_A(\Delta t/2) \|  \\
&\leq& n \left(2 \| e^{-iA\Delta t/2} - \widetilde{U}_A (\Delta t/2)  \| + \| e^{-iB\Delta t} - \widetilde{U}_B (\Delta t)  \|  \right), 
\end{eqnarray*}
where the terms $\| e^{-iA\Delta t/2} - \widetilde{U}_A (\Delta t/2)  \| $ and $\| e^{-iB\Delta t} - \widetilde{U}_B (\Delta t)  \|$ bound the error of each $\widetilde{U}_A(\Delta t/2)$ and $\widetilde{U}_B(\Delta t)$. 
Hence, to ensure $\| \widehat{U} - \widetilde{U}\| \leq \e/2$, we require
\begin{equation}
\| e^{-iA\Delta t/2} - \widetilde{U}_A (\Delta t/2)  \|  \leq \e/8n  \;\;\;\;
\text{ and } \;\;\;\;
 \| e^{-iB\Delta t} - \widetilde{U}_B (\Delta t)  \| \leq \e/4n.
\end{equation}
The quantity $M_A=M_A(k_A)$ is defined by applying (\ref{eqn:def0Mnew}) to the simulation of $A$ with time $t/2n$ and error at most $\e/8n$, to obtain 
$$M_A :=  \left(   \frac{4em' (t/2n) \|H_2\| }{(\e/8n)}  \right)^{1/2k_A}  \frac{4em'}{3} \left( \frac53 \right)^{k_A-1} = \left(   \frac{16em' t \|H_2\| }{\e}  \right)^{1/2k_A}  \frac{4em'}{3} \left( \frac53 \right)^{k_A-1}.$$ 
Remarkably, observe that the factors of $n$ have canceled, 
i.e., the time interval size for 
each application of $\widetilde{U}_A(\Delta t/2)$ depends only on the original problem time and error parameters and not on the number of time slices $n$ we subdivided $t$ into. 
Further note that when $16em' t \|H_2\| \leq \e$, then $M_A$ is bounded from above independently of~$\e$. This means that we are dealing with an easy problem for the simulation of $A$, 
so the interesting case is when $16em' t \|H_2\| > \e$, and we will consider this case from now on. Similar considerations apply to the simulation of~$B$.  

To bound the 
cost of each $\widetilde{U}_A(\Delta t/2)$, we apply \cite[Thm. 1]{PZ12}. 
Thus, 
 the number of exponentials $N_A$ 
required for each application of $\widetilde{U}_A(\Delta t/2)$ 
satisfies
\begin{eqnarray*}
N_A &\leq& (2m'-1)5^{k_A-1}  \lceil M_A \|H_1\| \Delta t/2  \rceil . 
\end{eqnarray*}
We have already mentioned that the quantity 
$\lceil M_A \|H_1\| \Delta t/2  \rceil$ 
gives the number of subintervals of length at most $1/M_A$ 
 that each time slice $\|H_1\| \Delta t/2$ is subdivided. 
 When the ceiling function argument is at most one, no sub-division 
is necessary. Then it may be possible to reduce the cost further by decreasing $k_A$. 

Now consider $\widetilde{U}_B(\Delta t)$. 
For the simulation of $B$ for time $\Delta t= t/n$ and error at most $\e/4n$, we set $M_B = M_B(k_B)$ as in 
(\ref{eqn:def0Mnew}) to obtain
\begin{eqnarray*}
M_B &:=&  \left(   \frac{4e(m-m') (t/n) \|H_{m'+2}\| }{(\e/4n)}  \right)^{1/2k_B}  \frac{4e(m-m')}{3} \left( \frac53 \right)^{k_B-1} \\
&=& \left(   \frac{16e(m-m') t \|H_{m'+2}\| }{\e}  \right)^{1/2k_B}  \frac{4e(m-m')}{3} \left( \frac53 \right)^{k_B-1}.
\end{eqnarray*}
Once again, $M_B$ is independent of $n$. 
As above, the interesting case is 
$16e (m-m') \|H_{m'+2}\|t > \e$, because otherwise 
$M_B$ would be independent of $\e$ and the problem would be easy. 
Applying again \cite[Thm. 1]{PZ12}, 
the number $N_B$ of exponentials for each application of $\widetilde{U}_B(\Delta t)$ satisfies 
\begin{eqnarray*}
N_B &\leq& (2(m-m')-1)5^{k_B-1}  \lceil M_B \|H_{m'+1}\| \Delta t \rceil.
\end{eqnarray*}
In this case the quantity 
$\lceil M_B \|H_{m'+1}\| \Delta t  \rceil$ 
gives the number of subintervals of length at most $1/M_B$ that each time interval of size $\|H_{m'+1}\|\Delta t$ is subdivided. 

We may now bound the total cost of our algorithm, 
i.e. bound the number $N$ of exponentials of the form $e^{-H_jt_{\ell}}$, $j\in\{1,\dots,m\}$,
that are used to construct $\widetilde{U}$. 
From (\ref{eqn:overallCostGen}), we have 
\begin{eqnarray}   \label{eqn:totalCost1}
N &=& n\cdot (2N_A + N_B) \nonumber \\
&\leq&  n\cdot \left( 4m'5^{k_A-1} \bigg \lceil M_A \|H_1\| \frac{t}{2n}\bigg \rceil +
2(m-m')5^{k_B-1} \bigg \lceil M_B \|H_{m'+1}\| \frac{t}{n} \bigg \rceil  \right).
\end{eqnarray}
We summarize the results for Algorithm $1$ in the following proposition.
\begin{prop} \label{prop:alg1}
Let $H=\sum_{i=1}^m H_i$, $\|H_1\|\geq \|H_2\| \geq \dots \geq \|H_m\|$, $m\geq 2$, with given partition
$H=A+B$, $A=\sum_{i=1}^{m'} H_i$ and $B=\sum_{i=m'+1}^m H_i$. 
Let $t>0$ and $1\geq \e >0$, 
and assume $16em'\|H_2\|t \geq \e$ and  $16e(m-m')\|H_{m'+2}\|t \geq \e$.  
Let $n\in \nat$ such that
\begin{equation}   \label{eqn:nPropAlg1}
n \geq  \sqrt{ 4 \, t^3\, \|A\| \|B\|  \|C\|  / 3\e }  ,
\end{equation}
where $\|C\|=\max\{\|A\|,\|B\|\}$. For any $k_A,k_B \in \nat$, 
define the quantities 
$$M_A=\left(   \frac{16em'  \|H_2\| t}{\e}  \right)^{1/2k_A}  \frac{4em'}{3} \left( \frac53 \right)^{k_A-1}$$
$$M_B = \left(   \frac{16e(m-m') \|H_{m'+2}\| t}{\e}  \right)^{1/2k_B}  \frac{4e(m-m')}{3} \left( \frac53 \right)^{k_B-1},$$
and let $\widetilde{U}$ be defined by (\ref{eqn:Utilde}). 
Then the number $N$ of exponentials for the approximation of $e^{-iHt}$ by $\widetilde{U}$ with accuracy $\e$ is at most 
\begin{equation} \label{eqn:thmTotalCostAlg1}
N \leq  4m'5^{k_A-1} n  \bigg \lceil M_A \|H_1\| \frac{t}{2n}\bigg \rceil  +
2(m-m')5^{k_B-1} n \bigg \lceil M_B \|H_{m'+1}\| \frac{t}{n} \bigg \rceil  .             
\end{equation}
\end{prop}
\vskip 1pc
For $x,y > 0$, it is easy to show $x \lceil y/x \rceil \leq \max \{ x, 2y \}$. 
Thus we have the following corollary. 
\begin{cor} \label{cor:alg1WithMax}
Let $n_A = M_A \|H_1\| t$ and $n_B = 2M_B \|H_{m'+1}\| t$. 
The bound to the number of exponentials of Proposition  \ref{prop:alg1}, equation (\ref{eqn:thmTotalCostAlg1}), may be expressed as 
\begin{equation}   \label{eqn:costBoundCor}
 N \leq  4m'5^{k_A-1} \cdot \max\{  n_A, n\}  \;\;+\;\;
2(m-m')5^{k_B-1} \cdot \max\{n_B , n\}  .
\end{equation}
\end{cor}

\begin{rem}
Observe that if $n_A, n_B \geq n$, then modulo constants (\ref{eqn:costBoundCor}) implies that the cost for simulating $H=A+B$ is 
upper bounded by the sum of the cost upper bounds for simulating $A$ and $B$ independently. 
\end{rem}

\begin{rem}   
The simulation cost bound (\ref{eqn:costBoundCor}) is 
minimized with respect to $k_A$ and $k_B$ by 
selecting 
optimal values~$k^*_A,k^*_B\:$ such that 
 $1\leq k^*_A \leq k^{(max)}_A$ and $1\leq k^*_B \leq k^{(max)}_B$, where 
\begin{eqnarray*}
k^{(max)}_A &=&  \bigg \lceil \; \sqrt{   \tfrac12   \log_{25/3} ( 16e\: m'\|H_2\| t / \e }) \;  \bigg \rceil,\\
k^{(max)}_B &=&  \bigg \lceil \; \sqrt{   \tfrac12   \log_{25/3} ( 16e\: (m-m') \|H_{m'+2}\| t / \e }) \;  \bigg \rceil.
\end{eqnarray*}
If $n \geq M_A(1)\|H_1\|t$, 
then $k_A$ is optimally selected as $k^*_A=1$.  
Alternatively, if ${M_A(k_A^*)\|H_1\|t \geq n }$, then $k^*_A=k_A^{(max)}$. 
Similar remarks apply for 
$k_B$, where instead of $m'$, $\|H_2\|$, and $M_A$ we use $(m-m')$, $\|H_{m'+2}\|$, and $M_B$. 
We formalize how to optimally select the splitting formula orders for the general case in Section  \ref{sec:splitFormOrders}.   
\end{rem}
It is relatively straightforward to extend Algorithm $1$ to a partition of $H$ into $\mu \geq 2$ groups $H=A_1 + \dots + A_\mu$. We consider this for the more general Algorithm $2$ in the next section.

\subsection{Algorithm 2} 
Algorithm $2$ generalizes Algorithm $1$ by applying an arbitrary splitting formulas at its \textbf{first step} instead of 
specifically the Strang splitting formula; see Figure \ref{fig:flowchart}. 
The details and analysis of Algorithm $2$ are similar to, but more complicated than, those of Algorithm $1$. 
We state the main results, and provide the proofs in Appendix \ref{app:HamSim}.  

We again consider the simulation of a partitioned Hamiltonian $H=A+B$, with  $A=H_1+\dots+H_{m'}$, $B=H_{m'+1}+\dots+H_m$. Just like in Algorithm $1$, the \textbf{second step} of Algorithm $2$ uses splitting formulas of orders $2k_A+1$ and  $2k_B+1$ for the simulations of $A$ and $B$, respectively, and combines the partial results using a splitting formula of order $2k+1$.

\begin{prop} \label{thm:alg2}
Let $H=\sum_{i=1}^m H_i$, $\|H_1\|\geq \|H_2\| \geq \dots \geq \|H_m\|$, $m\geq 2$, with given partition
$H=A+B$, $A=\sum_{i=1}^{m'} H_i$ and $B=\sum_{i=m'+1}^m H_i$.  
Let $\|C\| := \max\{\|A\|,\|B\|\}$ and $ \|D\| := \min\{\|A\|,\|B\|\}$. 
Assume $\|C\|t\geq 1$, $16e m' \|H_2\| t \geq \e$, $16e (m-m') \|H_{m'+2}\| t \geq \e$, and $16e \|D\|  t \geq \e$. 
For $k, k_A, k_B \in \nat$, 
define the quantities 
\begin{itemize}   \label{eqn:setn2b}
\item $n \geq  \|C\|t  \left( 16e \|D\| t /\e  \right)^{1/2k}  \frac{8e}{5} \left( \frac53 \right)^{k}   $,
\item $n_A = m' \|H_1\| t  \left(\frac{64e}{5} \: m' \|H_2\| t /\e  \right)^{1/2k_A}  7e \left( \frac53 \right)^{k_A - k} $,
\item $n_B =  (m-m') \|H_{m'+1}\| t   \left( \frac{64e}{5} \: (m-m') \|H_{m'+2}\| t /\e  \right)^{1/2k_B}  14e \left( \frac53 \right)^{k_B-k} $.
\end{itemize}
and let $\widetilde{U}$ be defined by (\ref{eqn:UtildeGen}). 
Then the number $N$ of exponentials for the approximation of $e^{-iHt}$ by $\widetilde{U}$ with accuracy~$\e$ is at most 
\begin{equation} \label{eqn:thmTotalCostAlg2}
 N \leq  8m' 5^{k+k_A-2} \; \max\{ n_A , n \}  +
4(m-m')5^{k+k_B-2}  \; \max\{n_B,n \} =: \eta(k,k_A,k_B).
\end{equation}
\end{prop}

\begin{rem}
If $k=1$, we recover the cost bound (\ref{eqn:costBoundCor}) of Algorithm $1$, up to 
 constant factors. 
Note that in some cases, e.g. when $\|D\|t/\e$ is large, even though we could use $k=1$, selecting a value $k>1$ may yield $n$ that is substantially smaller than 
that shown in (\ref{eqn:nPropAlg1}) in 
Proposition~\ref{prop:alg1}. 
\end{rem}

\begin{rem}
If any of the conditions  $\|C\|t\geq 1$, $16e m' \|H_2\| t \geq \e$, $16e (m-m') \|H_{m'+2}\| t \geq \e$, or $16e \|D\|  t \geq \e$ are violated, then we end up with an easier simulation problem as 
$\e$ is relatively large. 
So, in this sense, these conditions specify the interesting case. 
\end{rem}
%
Algorithm $2$ extends to the case where $H$ is partitioned into $\mu \geq 2$ groups $H=A_1 + \dots + A_\mu$. 
The algorithm is now specified by $\mu+1$ parameters $\underline{k}=\{k, k_1,\dots, k_\mu\} \in \nat^{\mu +1}$. 
The overall approximation $\widetilde{U}$ of $U=e^{-iHt}$  becomes 
\begin{eqnarray}  \label{eqn:UtildeGen2b}
\widetilde{U} &:=& \left(\widetilde{S}_{2k}(A_1,A_2, \dots A_\mu, \Delta t) \right)^n,
\end{eqnarray}
where $\widetilde{S}_{2k}(A_1,A_2, \dots A_\mu, \Delta t)$ is constructed as in (\ref{eqn:UtildeGen}). 

We summarize the results for this case in the following theorem, whose proof can be found in Appendix \ref{app:HamSim}.  
 A partition of $H=\sum_{i=1}^m H_i$ to $H=\sum_{j=1}^\mu A_j$, $2\leq \mu \ll m$,
where each $A_j$ is a sum of a subset of the $H_i$,  
  is \textit{disjoint} if each $H_i$ is contained in a single $A_j$. 
 Let each $A_j$ contain $m_j$ of the $H_i$, such that $\sum_{j=1}^\mu m_j = m$.  
  We use 
 $H_{(j,1)}$ to denote the largest Hamiltonian norm in a group
 $$ H_{(j,1)} = \max_{H_i \in A_j} \|H_i\|,$$ 
 and similarly $H_{(j,2)}$ denotes the second largest Hamiltonian norm. 

\begin{theorem} \label{thm:alg2generalPartition}
Let $H=\sum_{i=1}^m H_i$ be disjointly partitioned as $H=\sum_{j=1}^\mu A_j$, labeled such that 
$\|A_1\|\geq \|A_2\| \geq  \dots \geq \|A_j\|$. 
Let $t>0$ and $1\geq \e >0$.
Suppose $\mu \|A_1\|t\geq 1$,  $\|A_2\|  t \geq \e$, and $\mu m_j \|H_{(j,2)}\| t \geq \e$.  
For 
$k, k_1, \dots, k_\mu \in \nat$, 
let $n\in \nat$ be such that
\begin{equation}   \label{eqn:nThmAlg2}
n(k) \geq  \mu  \|A_1\| t \left( \frac{ 8 e \mu \|A_2\|t }{\e}  \right)^{1/2k}  \frac{4e}{5}  \left(\frac{5}{3} \right)^{k}  . 
\end{equation}
and define the quantities 
\begin{equation}
n_{A_j} (k,k_j) = m_j \|H_{(j,1)}\| t 
\left(  \frac{32e}{5} \frac{ \mu m_j   \|H_{(j,2)}\| t }{\e}  \right)^{1/2k_j}  7e \left( \frac53 \right)^{k_j-k},
\;\;\;\;\;\; j=1,\dots, \mu.
\end{equation}
Consider $\widetilde{U}$ to be defined by (\ref{eqn:UtildeGen2b}). 
The number $N$ of exponentials for the approximation of $e^{-iHt}$ by $\widetilde{U}$ with accuracy $\e$ is at most 
\begin{equation} \label{eqn:thmTotalCostAlg2gen}
N  \leq 8 \sum_{j=1}^\mu  5^{k+k_j -2 } m_j  \max\{ n(k), n_{A_j} (k,k_j) \} =: \eta (\underline{k})  . 
\end{equation}

\end{theorem}

\begin{rem}   \label{rem:alg1}
The way the Hamiltonians are grouped will influence the upper bound (\ref{eqn:thmTotalCostAlg2gen}). 
Ideally, the formation of the groups should minimize this upper bound. 
Roughly speaking, Hamiltonians of relatively large norm should be put in groups of relatively small cardinality. 
\end{rem}

\begin{rem}   \label{rem:alg1b}
The bound 
for the number of exponentials 
holds under general conditions and does not depend on how the partitioning of the Hamiltonians into groups is performed. 
Finding parameters that minimize  (\ref{eqn:thmTotalCostAlg2gen}) is a separate task, which is to be carried out on a classical computer. 
\end{rem}

\subsection{Selecting the Order of the Splitting Formulas}  \label{sec:splitFormOrders}
For any partitioning of the Hamiltonians into $\mu$ groups we 
show how to determine the order of the splitting formulas. 
The parameters $k_1,\dots,k_\mu$ allow splitting formulas of different orders to be used for the Hamiltonians in each group. 
The parameter $k$ determines the order of the splitting formula used 
in the first algorithm step. 
Ideally we want to find the optimal parameters $k^*,k^*_1,\dots,k^*_\mu$ that minimize the simulation cost bound (\ref{eqn:thmTotalCostAlg2gen}), which takes the value $\eta^* := \eta (k^*,k^*_1,\dots,k^*_\mu)$. 
This expression is complicated  and to simplify matters
 we provide sharp upper bounds $k^{(max)}$, $k_1^{(max)},\dots, k_\mu^{(max)}$ 
 to the optimal values.  
The upper bounds turn out to be turn out to be very slowly growing functions (sublogarithmic in the problem parameters) which means that for all practical instances the parameters  $k^*,k^*_1,\dots,k^*_\mu$ yielding the lowest cost upper bound can be found quickly by exhaustive search.

\begin{prop}   \label{prop:optks}
The simulation cost bound (\ref{eqn:thmTotalCostAlg2gen}) of Theorem \ref{thm:alg2generalPartition} is minimized by integers 
$k^*,k^*_1,\dots,k^*_\mu$ satisfying
\begin{equation}
1 \leq k^* \leq k^{(max)},\;\;\;  1 \leq k_j^* \leq k_j^{(max)}\;\;\; j=1,\dots,\mu,
\end{equation}
where 
\begin{equation}  \label{eqn:koptn2}
k^{(max)} := \max \bigg\{ {\rm round} \left( \sqrt{   \frac12   \log_{25/3}  \frac{8e \mu \| A_2 \| t }{ \e} } \; \right), 1  \bigg\}
\end{equation}
and 
\begin{equation}  \label{eqn:koptA2}
k^{(max)}_j := \max \bigg\{ {\rm round} \left( \sqrt{   \frac12   \log_{25/3} \frac{32e \mu m_j \|H_{(j,2)}\|t }{ 5\e} } \; \right)  , 1  \bigg\},  \;\;\;\;\; j=1,\dots,\mu.
\end{equation}
\end{prop}
\begin{proof}
%
Let the functions $g(k):=5^k n(k)$ and $h_j(k_j):=5^{k+k_j} n_{A_j}(k,k_j)/ 3^k$. Note that $k$ cancels out in the latter case so $h_j(k_j)$ is a univariate function. 
Consider minimizing $g(\cdot)$ and $h_j(\cdot)$ independently. For $g(k)$,  
setting its derivative to zero gives  
$$ 2k^2 \ln \frac{25}{3} - \ln  \frac{ 8 e \mu \|A_2\|t }{\e}  = 0, $$
which 
gives $k^{(max)}$ as in (\ref{eqn:koptn2}).
Repeating this argument for $h_j(k_j)$ gives  
$k^{(max)}_j$ as in (\ref{eqn:koptA2}). 
Since $g(\cdot)$ and $h_j(\cdot)$ are log-convex functions \cite{boyd2004convex}, the values $k^{(max)}$ and $k_j^{(max)}$ give the respective minima. 
%
Next, observe that we may rewrite the right-hand side of  (\ref{eqn:thmTotalCostAlg2gen}) as 
\begin{equation}   
\eta (\underline{k})  = 
\frac{8}{25} \sum_{j=1}^\mu  m_j  \max\{5^{k+k_j } n(k), 5^{k+k_j }  n_{A_j} (k,k_j) \}  \; 
= \frac{8}{25} \sum_{j=1}^\mu  m_j  \max\{5^{k_j } g(k), 3^k h(k_j) \} .
\end{equation}

Now assume $k_1,\dots,k_\mu$ are arbitrary but fixed. 
Then $\eta(k,k_1,\dots,k_\mu)\ge \eta(k^{(\max)},k_1\dots,k_\mu)$ for $k> k^{(\max)}$
since the arguments of the maximum function cannot decrease.
By a similar argument, for $k_j> k_j^{(max)}$ 
$\eta(k,k_1,\dots, k_j,\dots k_\mu)\ge \eta(k,k_1\dots, k_j^{(\max)},\dots k_\mu)$.
Therefore, the minimizers $k^*, k^*_1,\dots,k^*_\mu$ of 
(\ref{eqn:thmTotalCostAlg2gen}) satisfy $k^*\leq k^{(max)}$ and $k^*_j \leq k_j^{(max)} $, for $j=1,\dots,\mu$.
\end{proof}

\begin{rem} Equation (\ref{eqn:koptA2}) shows that small group cardinality and small Hamiltonian norms  
reduce the order of the splitting formula that suffices for the simulation of a given group. 

Indeed, from the arguments of the maximum function in (\ref{eqn:thmTotalCostAlg2gen}), we have that if $n(k) \ge n_{A_j}(k, k_j)$ for all $k$ and some $j$, then $k_j^*=1$. On the other hand, if $n(k) < n_{A_j}(k,k_j)$ for $k\leq k^{(max)}$ and some $j$, then $k_j^*=k_j^{(\max)}$. 
Thus, roughly speaking, Hamiltonians of small norm may be grouped and  
simulated with a low-order splitting formula (e.g., $k_j=1$), whereas groups of Hamiltonians of large norm in general require higher order formulas to achieve the lowest simulation cost. 
\end{rem}

\subsection{Speedup}   \label{sec:speedup}
We illustrate our results by showing the speedup of our algorithms relative to those in \cite{PZ12} for a number of cases. 
Generally, our approach is preferable when there is a disparity in the Hamiltonian norms and many of them are very small.
 
From \cite{PZ12}, we have the number of exponentials is bounded as 
\begin{equation}   \label{eq:Nprev}
N_{prev}(k) = O \left(  m^2 \|H_1\|t  \left( \frac{mt\|H_2\|}{\e}\right)^\frac{1}{2k} \left(\frac{25}{3}\right)^{k}\right),
\end{equation}
where $k$ is the order of the splitting formula. 
Selecting $k=k^*$ as in (\ref{eqn:optk}) this becomes  
\begin{equation}    \label{eq:NprevStar}
N_{prev}^* = O\left( m^2 \|H_1\|t \right) \cdot e^{2\sqrt{\frac12 \ln \frac{25}{3} \ln \frac{4emt\|H_2\|}{\e}} }.
\end{equation}
Note that the second factor $e^{2\sqrt{\frac12 \ln \frac{25}{3} \ln \frac{4emt\|H_2\|}{\e}} } = O((mt\|H_2\|/\e)^\delta)$ for any $\delta >0$. The explicit expressions for 
(\ref{eq:Nprev}, \ref{eq:NprevStar})
are shown in (\ref{eqn:PZ12boundexpoForm}, \ref{eqn:PZ12boundOptimalk}). 

For simplicity, we consider $\mu = 2$, i.e., $H=A+B$ with 
$A=H_1+H_2+\dots + H_{m'}$, $m' < m$, 
and $B$ equal to the sum of the remaining Hamiltonians. 
The number of exponentials for Algorithm $2$ is then 
shown in (\ref{eqn:thmTotalCostAlg2}) in Proposition~\ref{thm:alg2}. 
Assume that $\|A\| \geq \|B\|$ and in addition that  
\begin{equation}  \label{assump:AgreaterThanB}
(m-m')\|H_{m'+1}\| \leq  m' \|H_2\|, 
\end{equation}
Note that the left-hand side of the inequality above is an upper bound to $\|B\|$, and the inequality relates that to the number of Hamiltonians forming $A$ times the overall second largest Hamiltonian norm. This condition is easy to check in principle, and it holds especially in cases where the original Hamiltonians have quite disproportionate norms and have been partitioned accordingly. 

Clearly, we have $\|A\| \leq m' \|H_1\|$ and $\|B\|\leq (m-m')\|H_{m'+1}\|$, and hence we may select the parameter~$n$ of Proposition~\ref{thm:alg2} as
\begin{equation}   \label{eq:localn}
n(k) = \bigg\lceil m' \|H_1\|t  \left(\frac{ 16e (m-m')\|H_{m'+1}\| t}{\e}  \right)^{1/2k}  \frac{8e}{5} \left( \frac53 \right)^{k}  \bigg\rceil . 
\end{equation}
Recall the quantities $k^*, k_A^*, k_B^*$ and $k^{(max)}, k_A^{(max)},k_B^{(max)}$ shown in Proposition \ref{prop:optks}. 
For any algorithm parameters $k,k_A,k_B$, the cost bound of Proposition \ref{thm:alg2} 
satisfies $N\leq  \eta^* \leq \eta(k,k_A,k_B)$, 
where $\eta^*$ 
denotes the optimize cost bound $\eta(k^*,k^*_A,k^*_B)$ of Algorithm~$2$. 

For different 
cases of the algorithm parameters we have the following speedups.
\begin{enumerate}
\item Comparison when all splitting formulas have the same order, i.e.,  $k=k_A=k_B$, $k=O(1)$:

The cost bound (\ref{eqn:thmTotalCostAlg2}) 
has the same dependence on $t$ and $\e$ 
as that of  (\ref{eq:Nprev}), 
so when we divide the two cost bounds to obtain the speedup the parameters $t$ and $\e$ cancel out. 
From Proposition \ref{thm:alg2}, (\ref{assump:AgreaterThanB}), and (\ref{eq:localn}), we have 
$\max\{n_A,n \} = c_1 n_A$ and $\max\{n_B,n \} = c_2 n$, where $c_1,c_2 \geq 1$ are constants. 
%
Hence, (\ref{eqn:thmTotalCostAlg2}) gives cost 
\begin{eqnarray*}
N \leq \eta(k,k,k) &\leq&  8m' 5^{2k-2} c_1  n_A +
4(m-m')5^{2k-2} c_2 n \\
&\leq&  C \cdot \big( m'^2 \|H_1\|t \; (m'\|H_2\| t/\e)^{1/2k}  \\ 
&+& (m-m')m'\|H_1\|t \; ((m-m')\|H_{m'+1}\| t/\e)^{1/2k}   \big) ,
\end{eqnarray*}
where $C$ is a constant, 
and hence the speedup over \cite{PZ12} (with the same $k$) is 
\begin{eqnarray}   \label{eqn:speedupCase1}
\frac{N}{N_{prev}(k)} 
&=& O \left(\left(\frac{m'}{m}\right)^{2+1/2k} \right)+ O \left( \frac{m'}{m }  \left( \frac{(m-m')\|H_{m'+1}\|}{m\|H_2\|}\right)^{1/2k} \right) 
\end{eqnarray}
for all $\e$, $t$. 
Therefore, the algorithm in \cite{PZ12} is slower than the one in this chapter by a factor proportional to a polynomial in $m^\prime/m$, the degree of which is in the range $[1,2.5]$. 
This is particularly important when $m^\prime\ll m$.

\item 
Comparison to the cost of \cite{PZ12} with optimally chosen parameter: 

We use the previous case to derive a rough estimate.
Observe that, for 
fixed $k$ we have 
$$\frac{N_{prev}(k)}{N^*_{prev}} = O(m\|H_2\|t/\e)^{1/2k}. $$
Thus, 
again considering 
$k=k_A=k_B$, $k=O(1)$, 
we have 
\begin{eqnarray}   \label{eqn:speedupCase1b}
\frac{N}{N^*_{prev}} &\le&
\frac{\eta(k,k,k)}{N_{prev}(k)} \; \frac{N_{prev} (k)}{N^*_{prev}} \\
&=& 
O \left(\left(\frac{m'}{m}  \right)^{2}    \left(\frac{m'\|H_2\|t}{\e}\right)^{1/2k}       \right)
+ O \left( \frac{m'}{m }  \left(\frac{(m-m')\|H_{m'+1}\| t}{\e}\right)^{1/2k}  \right),\nonumber 
\end{eqnarray}
which follows from 
 (\ref{eq:NprevStar}) and (\ref{eqn:speedupCase1}). 
Therefore, for fixed $\|H_2\|$, $t$ and $\e$, the algorithm
in \cite{PZ12} with optimally chosen parameters remains slower than Algorithm $2$ with arbitrary $k=k_A=k_B$. 
The speedup 
depends on a polynomial in $m^\prime/m$, the degree of which is in the range $[1,2]$. 

Clearly, optimally selecting $k$, $k_A$, and $k_B$ 
can only improve the speedup. 

\item Comparison among optimal methods, i.e., using the respective optimal splitting formulas:

Assuming that all complexity parameters are fixed, 
with the exceptions of $m$ and $m'=O(m^{5/6})$, 
we have 
\begin{equation}  \label{eqn:speedupLimit}
\frac{N}{N^*_{prev}} \leq \frac{\eta^{*}}{N^*_{prev}}  \; \xrightarrow[m\rightarrow \infty]{} \;  0,
\end{equation}
where $\eta^*$ is 
the optimized cost bound of Algorithm $2$. 
The proof is given in Appendix~\ref{app:HamSim}. 

In this sense we achieve a strong speedup over \cite{PZ12}.  

\item 
Comparison when a significant number of Hamiltonians have very small norm relative to~$\|H_2\|$:

Recall that we are 
interested in simulation problems where a significant number of the Hamiltonians $H_j$ are relatively small in norm, 
where existing simulation methods do not take advantage of this structure. 

We use two parameters 
$0\leq b \leq a < 1$ to describe the relationship of $\|B\|$ and $\|A\|$. 
This approach has applications to problems such as the simulation of the electronic Hamiltonian, as we will see in the following section. 
Suppose 
\begin{equation}  \label{assump:normBp}
\|B\|\leq (m-m')\|H_{m'+1}\|/\|H_2\|=O((m-m')^b)  \;\;\;\;\;\; \text{ for a given } b \in [0,1),
\end{equation}
i.e., not only do the $H_j$, $j>m'$, that form $B$ have small norm individually, but $\|B\|$ is relatively small also. 
For example, 
we could have $(m-m')=10^6$ and the Hamiltonians 
comprising $B$ to have norms at most $10^{-4}$, so that $\|B\| \leq 10^6\cdot 10^{-4} = 10^{2}$, i.e., $b \simeq 1/3$. 

Recall $1 \leq m' < m$ because $m'$ is the number of Hamiltonians forming $A$. 
Further suppose 
\begin{equation} \label{assump:m'q}
m' = O(m^a) \;\;\;\;\;\; \text{ for some } a \in [0,1).
\end{equation}

For the case $k=k_A=k_B=O(1)$ above (where the speedup is independent of $\e$ and $t$), using these assumptions in (\ref{eqn:speedupCase1}) we obtain 
\begin{eqnarray*}  
\frac{\eta(k,k,k)}{N_{prev}(k)}
&=& O \left(\left(\frac{m^a}{m}\right)^{2+1/2k} \right)+ O \left( \frac{m^a}{m }  \left(\frac{ (m-m')^b }{m}\right)^{1/2k} \right) \\
&=& O \left(\frac{1}{m^{(1-a) + (1-b)/2k}}  \right). 
\end{eqnarray*}

Similarly, for the case above where we obtain (\ref{eqn:speedupCase1b}), using the new assumptions we obtain  
$$\frac{N}{N^*_{prev}} = O\left( \frac{1}{m^{1-a - b/2k}}  \right)  .  $$
Therefore, selecting $k$ such that the exponent of the denominator is positive yields a speedup the grows with $m$. 
In the next section we will use the parameters $a$ and $b$ to estimate the cost of our algorithms for practical instances of the electronic Hamiltonian. 
\end{enumerate}
We emphasize that for the speedup estimates derived  in this section 
we have made many simplifications, and they are hence 
quite conservative. 
For practical problem instances, the speedups may be 
much greater than those indicated here.

\section{Application to Quantum Chemistry}    \label{sec:QChem}
Solving difficult problems in quantum chemistry is viewed as a 
primary application of quantum computers. 
We apply our algorithms to simulate  
the electronic Hamiltonian, which describes molecular systems. 
Quantum algorithms for 
simulating the electronic Hamiltonian have applications to 
the calculation of electronic energies (i.e., the \textit{electronic structure problem}), and also reaction rates, and other chemical properties \cite{aspuruGuzik2005yq,kassal2008polynomial,wang2008quantum,Kassal,whitfield2011simulation}. 

Robust classical algorithms for 
this simulation exist (e.g., diagonalization), but in general they are 
intractable as their cost 
grows exponentially with the number of particles. 
Thus, large molecules 
are 
out of reach for classical computers \cite{whitfield2011simulation}.  
On the other hand, 
quantum algorithms 
\cite{aspuruGuzik2005yq,whitfield2011simulation} can efficiently simulate the second-quantized 
electronic Hamiltonian (\ref{eq:elecHam}). 
There exist quantum algorithms for this simulation problem with cost polynomial in the number $m$ of Hamiltonian terms. 
Unfortunately, 
the combination of the size of $m$ and the 
%
degree of the polynomial 
makes the algorithms 
impractical in many cases of interest 
 \cite{whitfield2011simulation,toloui2013quantum,wecker2014gate,mcclean2014exploiting,peruzzo2014variational}. 
Hence, reducing the cost of Hamiltonian simulation will have a significant impact in chemistry. 

\subsection{Electronic Hamiltonian} 
Recall the second-quantized 
Born-Oppenheimer electronic Hamiltonian (\ref{eq:elecHam}), i.e.,\footnote{Using \textit{atomic units}, where the electron mass, electron charge, Coulomb's constant, and reduced Planck constant $\hbar$ are unity, the electronic Hamiltonian can be written in the given \textit{dimensionless} form; see, e.g.,  \cite[Sec. 2.1.1]{Szabo} for details.} 
\begin{equation*}   
H: = \sum_{p,q=1}^\mathcal{N} h_{pq} a^\dag_p a_q + \frac12 \sum_{p,q,r,s=1}^\mathcal{N} h_{pqrs} a^\dag_p a^\dag_q a_r a_s.
\end{equation*}
The quantities $h_{pq}$ and $h_{pqrs}$ are obtained by considering $\mathcal{N}$ single-particle basis functions (spin orbitals) taken from a given family of such functions. 
Particularly, the $h_{pq}$ and $h_{pqrs}$ are one-electron and two-electron integrals, respectively, as defined in 
\cite[Sec. 3]{whitfield2011simulation}. 
For our purposes the $h_{pq}$ and $h_{pqrs}$ 
are problem inputs. 
The $a^\dag_p$ and $a_p$ are the creation and annihilation operators 
for the $p$th orbital, which encode the fermionic exchange antisymmetry of the problem.  
The general Hamiltonian form 
is the same for all molecules with the same number of single particle orbitals $\mathcal{N}$. 
Therefore, 
the Hamiltonian of a particular molecule  
is defined by $\mathcal{N}$ and the $h_{pq}$ and $h_{pqrs}$.  

The Hamiltonian above can be written in the form 
\begin{equation}   \label{eq:elecHamNew}
H= \sum_{j=1}^m H_j , 
\end{equation}
where $m= \Theta(\mathcal{N}^4)$, and the $H_j$ are Hamiltonians obtained from the terms of (\ref{eq:elecHam})
by combining adjoint pairs; 
see e.g. 
\cite{whitfield2011simulation,jones2012faster}. 
Thus, modulo constant factors, the norms $\|H_j\|$ are given by 
the $|h_{pq}|$ and $|h_{pqrs}|$. 
These quantities 
depend on molecular geometry and the chosen set of basis functions \cite{Szabo,helgaker2014molecular}. 
For basis functions that are spatially localized, which are called local basis sets, many of the $|h_{pq}|$ and $|h_{pqrs}|$ are small or 
very small relative to their largest magnitude 
 \cite{clementi1972computation,christoffersen1972ab,helgaker2014molecular}. 
We use this disparity to partition the Hamiltonians into groups for our algorithms.

For example, 
\cite{jones2012faster} considers the quantum simulation of the lithium hydride (LiH) molecule with different choices of basis sets. 
The authors of \cite{jones2012faster} consider 
Slater-type (STO-3G) \cite{hehre1969self} and triple-zeta (TZVP) \cite{dunning1971gaussian} basis sets and 
in both cases they find that a substantial fraction of the $H_j$ have quite small norm. 
In Table~\ref{table:chemNorms}, we illustrate how one can partition the Hamiltonian using the $h_{pq}$, $h_{pqrs}$ values shown in 
\cite{jones2012faster} to obtain $H=A+B$, where the Hamiltonian $B$ is the sum of the terms for which the corresponding $|h_{pq}|$ and $|h_{pqrs}|$ are less than or equal to different \lq\lq cutoffs\rq\rq\, and $A$ is the sum of the remaining terms. 
Clearly, different partitions lead to different bounds for the norm of each group, 
which will be reflected in the cost bounds of the 
algorithm as shown in Theorem 
\ref{thm:alg2generalPartition}. 
Extending this idea to $\mu>2$ groups is straightforward. 


\begin{table}[h!]    
\begin{center}
\begin{tabular}{| c | c | c | c || c  | c | c |}
	\hline
	Basis Set  & Cutoff  & $m$ & $m'$ & $\|A\|$ & $\|B\|$ \\
	\hline
	STO-3G & $10^{-10}$ &  $231$ & $99$ &$10^2$ & $10^{-8}$ \\
	\hline
	TZVP & $10^{-10}$ & $22155$ & $10315$ & $10^{4}$ &  $10^{-6}$ \\
	\hline
	TZVP & $10^{-4}$ & $22155$ & $9000$ & $10^{4}$ &  1\\
		\hline
			TZVP & $10^{-3}$ & $22155$ & $6000$ & $10^{4}$ &  10 \\
		\hline
			TZVP & $10^{-2}$ & $22155$ & $2000$ & $10^3$ &  $10^2$ \\
		\hline
\end{tabular}  
\end{center}
\caption{
For simulating LiH with bond distance $1.63$ \AA, in \cite{jones2012faster} they approximate the Born-Oppenheimer electronic Hamiltonian in two ways. The first approximation uses a minimal basis set (STO-3G) and has $m=231$. The second approximation uses a more accurate basis set (TZVP) and has $m=22155$. In 
each case $m$ is the number of non-zero $h_{pq}$ and $h_{pqrs}$. 
The quantity $m'$ is the number of $|h_{pq}|$ and $|h_{pqrs}|$ that are larger than the different cutoff values.  
The quantities of the leftmost four columns in the first two rows are 
 are taken from \cite[Sec. 3.2]{jones2012faster}. The quantities of the leftmost four columns in the remaining rows are estimated from \cite[Fig. 10]{jones2012faster}. 
 We consider  $H= A+B$, where $A$ is the sum of the $m'$ terms corresponding to $|h_{pq}|$, $|h_{pqrs}|$ larger than the cutoff, and $B$ is the sum of the remaining $m-m'$ terms. The norms of $A$ and $B$ are estimated 
 using the triangle inequality, i.e.,  
 $\|A\| \leq m' \|H_1\|$ and $\|B\| \leq (m-m')\|H_{m'+1}\|$, where 
$\|H_1\|\geq \dots \geq \|H_{m'}\| \geq \dots \|H_m\|$,
 and for simplicity we assume $\|H_1\| \simeq 1$. 
 Thus, we estimate  $\|A\|  \simeq m'$ and $\|B\| \simeq (m-m')\times$cutoff,  rounded to the nearest power of $10$ for simplicity. 
 }    \label{table:chemNorms}
\end{table}

We digress for a moment to remark that in practical applications of quantum chemistry, the computational cost is often reduced by discarding terms of $H$ that have  \lq\lq negligible\rq\rq\ norm relative to some cut-off, say $10^{-10}$ \cite{clementi1972computation,jones2012faster,helgaker2014molecular}, but this cannot be done in an ad hoc way.  
Recall that Proposition \ref{prop:discardHamsNew} shows that we may possibly, 
depending on the particular problem instance, 
discard certain terms from $H$,   
subject to the relationship between the cutoff,  $t$, $\e$, and the number of terms~$(m-m')$ below the cutoff.  
On the other hand, when the product of the cutoff with~$(m-m')$ exceeds~$\e/t$, we cannot arbitrarily discard~$(m-m')$ terms, even though individually they may be tiny, because this could introduce truncation error that would exceed the desired simulation accuracy. 
This is also made particularly clear in the last three rows of Table~\ref{table:chemNorms}, where excluding the terms below the cutoff may introduce error exceeding any reasonable accuracy as evidenced by the respective estimates of $\|B\|$. 

\subsection{Simulation Cost} 
In chemistry problems the desired simulation accuracy is not arbitrarily small, while $\mathcal{N}$ can be quite large so that (\ref{eq:elecHamNew}) 
adequately represents the system of interest~\cite{mcclean2014exploiting,babbush2015chemical}. Therefore, the important parameters affecting the simulation cost are the number of single-particle basis functions $\mathcal{N}$, and the magnitudes of the $h_{pq}$ and $h_{pqrs}$.  

In second quantization, i.e., the occupation number representation,  states are given by linear combinations of  $\mathcal{N}$-bit strings, where a $1$/$0$ indicates which orbitals are occupied/unoccupied by electrons, respectively \cite{Szabo,helgaker2014molecular}. 
 Thus $H$ acts on $\mathcal{N}$ qubits. 
Each Hamiltonian $H_j$ in (\ref{eq:elecHamNew}) can be represented by  tensor products of Pauli matrices 
through the
Jordan-Wigner transformation, and can be simulated efficiently using
$O(\mathcal{N})$ standard 
quantum gates \cite{whitfield2011simulation}. 
Alternatives to the Jordan-Wigner transformation have been proposed,
such as the Bravyi-Kitaev transformation  \cite{bravyi2002fermionic,seeley2012bravyi} 
which improves the gate count for simulating the individual $H_j$ 
to $O(\log \mathcal{N})$. Hence, our cost bounds 
for the number of queries (exponentials)  
immediately translate to bounds for the total gate count through  multiplication. 
Thus, using \cite{bravyi2002fermionic,seeley2012bravyi}, modulo polylogarithmic factors, the total gate count is proportional to the number of queries. This is what we will consider for our algorithms. 

Now consider the simulation of $H$ using splitting formulas. 
For $\mathcal{N}$ spin-orbitals, the number of terms in (\ref{eq:elecHamNew}) is $m=\Theta(\mathcal{N}^4)$.  
Naively applying an order $2k+1$ splitting formula (\ref{eqn:PZ12bound}) 
yields a number of queries 
(i.e., a number of exponentials)
$$O\left(\mathcal{N}^{8+2/k} \|H_1\|t \; (\|H_2\|t/\e)^{1/2k} (25/3)^k \right).$$ 
Thus, for arbitrary $k$ the cost grows with $\mathcal{N}$ at least as 
$\mathcal{N}^{8}$. 
In particular, if we use the Strang splitting formula $(k=1)$, 
the number of queries is proportional to $\mathcal{N}^{10}$. 
Hence, a straightforward application of splitting formulas yields a number of queries in the range $\mathcal{N}^8 - \mathcal{N}^{10}$, which clearly becomes impractical even for moderate $\mathcal{N}$ (e.g., $\mathcal{N}=100$). 

Improving this cost bound is critical for quantum computers to have an impact in quantum chemistry applications. 
A sequence of papers \cite{wecker2014gate,poulin2014trotter,mcclean2014exploiting,hastings2015improving,babbush2015chemical,babbush2016exponentially} 
describe the recent progress. 
They show both analytic and empirical results. 
Some of them perform gate-level optimizations across queries, 
and are thereby specific to the particular problem instance.  
In \cite[Table I]{wecker2014gate}, the number of queries using the Strang splitting formula is shown to be proportional to $\mathcal{N}^{10}$, which corresponds to the one that follows from \cite{PZ12} shown above. 
It is also shown in \cite[App. B]{wecker2014gate} that the number of queries can be reduced to become $\mathcal{N}^{9}$, and it is conjectured that the proof leading to this reduction in the case $k=1$ could be extended to high-order splitting formulas $(k>1)$. 
The paper also considers the implementation of the queries using the 
Jordan-Wigner transformation. Thus, the total gate count becomes proportional to $\mathcal{N}^{10}$, but allowing parallel gate execution the circuit depth becomes proportional to $\mathcal{N}^{9}$. 
Moreover, the authors of the paper carried out 
numerical tests of molecules from a random ensemble 
suggesting a number of queries proportional to $\mathcal{N}^{8}$ as shown in \cite[Table I]{wecker2014gate}. 
Gate-level optimizations on the entire circuit are considered in \cite{hastings2015improving}. In particular, using the Jordan-Wigner transformation for implementing the queries, the authors of that paper conclude that their optimizations make the total gate count proportional to the total number of queries. 
Therefore, for the Strang formula 
as presented in \cite{wecker2014gate}, the total gate count  
is proportional to $\mathcal{N}^{9}$, and allowing parallel execution in conjunction with gate-level optimization leads to a circuit with depth proportional to $\mathcal{N}^{7}$. 
In \cite{mcclean2014exploiting, babbush2015chemical}, it was 
argued using empirical evidence 
that similar improvements on the number of queries are possible for certain restricted but commonly used basis function sets, 
and this may lead to a number of queries proportional to $\mathcal{N}^{5.5}-\mathcal{N}^{7}$, while \cite{poulin2014trotter} reports even better empirical query estimates 
in the range $\mathcal{N}^{5.5}-\mathcal{N}^{6.5}$. 
Finally, a recent paper \cite{babbush2016exponentially} 
that uses the simulation method of \cite{berry2015simulating} 
with different queries than the matrix exponentials used in splitting formulas, obtains a total gate count proportional to 
$\mathcal{N}^{8}$, modulo polylogarithmic factors. 
Furthermore, in a special case they are able to obtain a total gate count proportional to $\mathcal{N}^5$ (up to polylogarithmic factors), under strong assumptions on the basis functions and the computation of the $h_{pq}$, $h_{pqrs}$ and the resulting accuracy and cost. However, we point out that other authors consider the computation of these quantities to be \lq\lq complicated business\rq\rq\ in general \cite[sec. 9.9.5]{helgaker2014molecular}. 

Further note that the possibility of using problem specific information in quantum chemistry (e.g., simulating different Hamiltonians for different amounts of time, or simulating them in a certain order) to improve the simulation cost was suggested in \cite{jones2012faster,wecker2014gate,poulin2014trotter,hastings2015improving,babbush2015chemical}
without presenting an algorithm or a rigorous analysis exhibiting error and cost bounds. 
Our goal is obtain rigorous simulation cost improvements under fairly general conditions. 

\subsection*{Divide and conquer simulation}
It is well known 
that the locality of physical interactions can be exploited 
to give substantial advantages 
for classical algorithms \cite{goedecker1999linear}, yet only recently considered in detail for quantum algorithms \cite{mcclean2014exploiting}. 
By utilizing local basis functions, which are 
localized near atomic centers and 
have mutual overlap which is 
typically 
exponentially decaying with their separation \cite{mcclean2014exploiting}, 
there generally exists a characteristic distance between atomic centers beyond which the corresponding integrals will be negligible. 
In particular, molecules with large \textit{physical} size, for which a large fraction of atomic orbitals are sufficiently distant from each other, will have many Hamiltonian terms with very small norms. 
Similarly, another example is the Hartree-Fock basis (which is accurate and commonly used for small systems), where many off-diagonal Hamiltonians are very small \cite{wecker2014gate}.  
Therefore, in many applications of interest we expect the number of 
Hamiltonian terms with substantial norm to scale with $\mathcal{N}$ much lower than $\mathcal{N}^4$  \cite{clementi1972computation,almlof1982principles,strout1995quantitative,helgaker2014molecular}. We take advantage of this property to derive faster quantum algorithms. 

Consider a fixed set of local basis functions. 
In general, 
the number of \lq\lq non-negligible\rq\rq\ $|h_{pq}|$ and $|h_{pqrs}|$ 
is significantly less than $\mathcal{N}^4$. 
In \cite[Sec. 9.12.2]{helgaker2014molecular}, it is argued that for sufficiently large molecules this number is  
of order $\mathcal{N}^2$. 
In \cite{mcclean2014exploiting}, the authors claim that this number can scale even as $\mathcal{N}$ using local basis functions. 
Also,  
\cite{babbush2015chemical} has found this number to be $O(\mathcal{N})$ modulo logarithmic factors. 
Using these estimates, we obtain $a=1/2$ 
or $a=1/4$,  
where we have assumed 
$A =  \sum_{j=1}^{m'}  H_j $ with  $m'=O(m^a)$ as in (\ref{assump:m'q}), and 
$B =  \sum_{j=m'+1}^{m}  H_j $ with $\|B\|\leq (m-m')\|H_{m'+1}\|=O(m^b)$ as in (\ref{assump:normBp}), 
with $0\le b\le a$. 
Taking $\|H_1\|=O(1)$ then further implies $\|H_\mathcal{N}\| = O(m^a)$. 
Observe that our assumptions are consistent with the situation depicted in Table \ref{table:chemNorms}. 

Now consider Algorithm $2$ with $H=A+B$. 
Assume $t$ and $\e$ are arbitrary but fixed, and 
let us study the simulation cost with respect to $\mathcal{N}$. 
Even if we do not select the optimal values for  $k$, $k_A$, and $k_B$, and we simply assume they are $O(1)$, we obtain a simulation cost improvement. 
The quantities of Proposition  \ref{thm:alg2} become $\|C\|=O(m^a)$, $\|D\|=O(m^b)$, and hence
$n = O(m^{a+b/2k})$, $n_A = O(m^{a+a/2k_A}) $, and $n_B = O(m^{b+b/2k_B}) $. 
Using these quantities and (\ref{eqn:thmTotalCostAlg2})  the simulation cost is bounded from above by
$$c_1 m^{2a} \max\{ m^{b/2k}, m^{a/2k_A} \} +
c_2 m \max\{ m^{a+b/2k}, m^{b+b/2k_B}\},$$
where $c_1,c_2 > 0$ are constants and $t,\e$ are fixed.  
Since $0\le b\le a\le 1/2$ the previous expression is bounded by a quantity proportional to
$$m \: \max\{ m^{a+b/2k}, m^{b+b/2k_B} \}. $$  
Taking $k\le k_B$ yields that the simulation cost is proportional to
\begin{equation}
 \label{eqn:qChemCostbound1}
m^{1 +a+b/2k}=O(\mathcal{N}^{4+4a + 2b/k}).
\end{equation}

Recall that using the Bravyi-Kitaev representation  \cite{bravyi2002fermionic}  for implementing
the terms of (\ref{eq:elecHamNew}), the number of queries of our algorithms, modulo polylogarithmic factors, is proportional to the total gate count. 
We compare our results  to those from the literature in Table \ref{tab:qchemMain}
(as summarized in Table~\ref{tab:qChemSummary} previously).

\begin{rem}
Using the estimates $a=1/2$ and $a=1/4$ 
for local basis functions from \cite{helgaker2014molecular,mcclean2014exploiting,babbush2015chemical} 
Table \ref{tab:qchemMain} shows that the number of queries of 
Algorithm $2$ scales as $\mathcal{N}^5 - \mathcal{N}^7$,  
$0 \leq b \leq a $. 
This is consistent with the empirical results 
in \cite{poulin2014trotter,babbush2015chemical}. 
\end{rem}

\vskip 1pc
\begin{table}[h!]   
\begin{center}
\begin{tabular}{| c | c |}     
	\hline
	Method & Cost Dependence on $\mathcal{N}$ \\
	\hline
	Suzuki-Trotter splitting formulas \cite{PZ12} & $\mathcal{N}^8 - \mathcal{N}^{10}$  \\
	\hline
	 Improved Strang splitting for the electronic Hamiltonian 
	  \cite{wecker2014gate} & $\mathcal{N}^9$  \\
	\hline
	Empirical scaling of random \lq\lq real\rq\rq\ molecules \cite{wecker2014gate} &  $\mathcal{N}^8$ \\  
	\hline 
		Truncated Taylor series \cite{babbush2016exponentially} $\;\;$ (\# gates) & $ \mathcal{N}^8 $ \\
	\hline
	Improved empirical scaling
	\cite{poulin2014trotter,mcclean2014exploiting,babbush2015chemical} & $\mathcal{N}^{5.5}-\mathcal{N}^{7}$  \\
		\hline
			On-the-fly algorithm \cite{babbush2016exponentially} $\;\;\;$ (\# gates) & $\mathcal{N}^5$  \\
		\hhline{|=|=|}
			Algorithm $2$ :  $(a,b)=(3/4,0)$ & $\mathcal{N}^{7}$  \\
		\hline
	Algorithm $2$ :  $(a,b)=(1/2,1/2)$ & $\mathcal{N}^{6}-\mathcal{N}^{7}$  \\
		\hline
		Algorithm $2$ :  $(a,b)=(1/2,1/4)$ & $\mathcal{N}^{6}-\mathcal{N}^{6.5}$  \\
		\hline
		Algorithm $2$ :  $(a,b)=(1/2,0)$ & $\mathcal{N}^{6}$  \\
		\hline
		Algorithm $2$ :  $(a,b)=(1/4,1/4)$ & $\mathcal{N}^{5}-\mathcal{N}^{5.5}$  \\
		\hline
		Algorithm $2$ :  $(a,b)=(1/4,0)$ & $\mathcal{N}^{5}$  \\
		\hline
\end{tabular}
\end{center}
\caption{Comparison of empirical and analytic cost 
bounds with respect to the number $\mathcal{N}$ of single-particle basis functions for the simulation of the electronic Hamiltonian. 
The top half of the table are estimates taken from the literature, ignoring any polylogarithmic factors.
The bottom half of the table is the estimated scaling for Algorithm~$2$, where the parameters~$a$ and~$b$ have been estimated; see the text for details. 
We give a range for the cost dependence in cases where it 
varies with some of the algorithm parameters, or when 
the cost is obtained empirically. 
All cost estimates refer to the number of queries, except in the case of 
\cite{babbush2016exponentially} which does not use splitting formula and presents the total gate count. 
For all estimates concerning queries, the transition from queries to gate counts
involves a multiplication by a 
$O(\log \mathcal{N})$ factor in the most favorable case. }
\label{tab:qchemMain}
\end{table}

\begin{rem}
If $a,b \rightarrow 0$, 
the cost of Algorithm $2$ tends to  $O(\mathcal{N}^4)$, which is a lower bound to the simulation cost since the input size is $\Theta(\mathcal{N}^4)$. 
In contrast, a naive application of
an order $2k+1$ splitting formula without partitioning the Hamiltonian would still have cost proportional to $\mathcal{N}^{8+2/k}$. 
\end{rem}

Thus, our algorithms exhibit speedup for simulation of the electronic Hamiltonian comparable to the empirical predictions discussed above. Characterizing classes of basis functions and molecules where tighter bounds on the distribution of Hamiltonians of large norm and their number is an important open problem. 

Our algorithms take advantage of problem structure in terms of the Hamiltonian norms, without relying on other domain-specific information or implementation-level assumptions. 
As part of future work, it would be interesting to study how 
gate-level optimizations and other information specific to chemistry 
 could further improve the performance of our algorithms. Furthermore, partitioning the Hamiltonian into $\mu>2$ groups may lead to further cost improvements in applications. 

We conclude this section 
by emphasizing that 
the advantages of our approach may extend to problems beyond quantum chemistry.

\section{Discussion}    \label{sec:Discussion}
Splitting formulas and similar approaches have many advantages for Hamiltonian simulation, and are especially important for near-term quantum computing. 
Our algorithms take advantage of the problem structure without relying on heavy assumptions and are as simple to implement as standard splitting formulas, but can lead to significantly lower cost. 
Just like splitting formulas, our algorithms succeed 
deterministically and therefore they can be used as subroutines that are called numerous times in other  quantum algorithms without this affecting the overall success probability. 
The reduced cost of our algorithms may make them especially suitable for applications in near-term quantum computing devices which will likely have limited resources available.

We emphasize that our results are general, and may be improved given further structural information for a given problem. 
In particular, our error and cost estimates are worst-case, and hence may be overly pessimistic for application to real-world problem instances. 
Nevertheless, the rigorous cost and error bounds we derive, 
allow our algorithms to be used as well-characterized subroutines, which is critically important for the development and deployment of future quantum algorithms.  

Finally, we remark that it is possible 
to extend our algorithms 
to the case of time-dependent Hamiltonians $H=H(t)=\sum_{j=1}^m H_j(t)$, using similar techniques as  
those applied for 
splitting formulas in \cite{wiebe2010higher,wiebe2011simulating,poulin2011quantum}; we leave this as an  area for future investigation. 

%% file: _ch_QAOA1.tex
\chapter{Quantifying the Performance of Low-Depth QAOA}
\label{ch:QAOAperformance}

\section{Introduction}
As small quantum computers begin to emerge 
in the near future, 
practitioners will 
be empowered to experiment with and analyze 
a new frontier of quantum algorithms. One promising area of application is to approximately 
solve  
challenging optimization problems. Indeed, for many such problems, 
classical algorithms finding the optimal solution 
require a number of steps that is exponential in the input size in the worst case, so we often must settle for  algorithms that produce approximate solutions (in a polynomial number of steps). Hence, an important  natural question to explore is whether or not quantum computers offer advantages for 
approximate optimization.  
It is thus important to 
develop new algorithmic techniques and derive new methods of analysis towards resolving this question. 

Recently, Farhi et al.~\cite{Farhi2014}
proposed a new class of quantum algorithms and heuristics, the Quantum Approximate
Optimization Algorithm (QAOA), to tackle
challenging approximate optimization problems on gate model
quantum computers. 
In QAOA, the problem Hamiltonian, which encodes the
objective function of the given optimization problem, and the mixing Hamiltonian, 
which 
transfers probability amplitude between different basis states encoding problem solutions, 
are applied in alternation $p$ times each to a suitable initial state. Then a computational basis measurement is performed, which returns an approximate  solution. The process is repeated a number of times and the best outcome kept. 
A handful of recent papers suggest such circuits may be powerful for 
different types of computational problems
~\cite{Farhi2014b,Farhi2016,Shabani16,Jiang17,Wecker2016training,hadfield2017quantum}. 

The QAOA algorithm and resulting quantum circuits are parameterized by the times (angles) for which the
problem Hamiltonian and the mixing Hamiltonian are applied at each iteration. 
A level-$p$ (depth-$p$) algorithm has $2p$ parameters. 
The success of QAOA relies on being able to find a good 
set of parameters. 
For QAOA$_p$ of a fixed depth $p$, straightforward sampling of the search space was
proposed~\cite{Farhi2014}, but this is practical in general only for small $p$;
as the level increases the parameter optimization
becomes inefficient due to the curse of dimensionality \cite{guerreschi2017practical,wang2017quantum}. 
Elegant analytical tools 
for specific problems 
can provide
parameter values for $p\gg 1$ that give near optimal performance, e.g., 
for searching an unstructured database~\cite{Jiang17}, but for general problems 
practically efficient search strategies are needed. 
Here, we analytically 
study the performance of 
QAOA applied to 
the Maximum Cut (MaxCut) problem, and characterize the optimal algorithm parameters. 

In Ref.~\cite{Farhi2014}, Farhi et al.~investigated low-depth QAOA for MaxCut 
for specific (bounded-degree) graphs, and provided numerical results
for a few special cases. 
We extend the known results for MaxCut by 
deriving a 
sequence of 
analytic expressions for the expected performance of the algorithm for $p=1$ on both arbitrary and 
special classes of graphs, 
which can be solved to obtain the
optimal algorithm parameters. 
We apply these results to bound the approximation ratio achieved on certain classes of graphs.
In particular, for MaxCut on a triangle-free graph with maximum vertex degree~$D$, we show that QAOA$_1$ 
achieves a solution within a factor of 
$$\frac{1}{2}+ \Omega \left(\frac{1}{\sqrt{D}}\right)$$
of the optimal value. A similar expressions with a correction term proportional to $1/D$ is obtained for general graphs. 

The proofs of our results rely on the algebraic properties of the Pauli matrices.  
A similar analysis quickly becomes cumbersome as the number $p$ of QAOA iterations increases, 
as the amount of terms involved in the analysis grows exponentially with $p$. 
(Indeed, optimizing QAOA in general appears to suffer from the curse of dimensionality, i.e., the number possible combinations of angles grows exponentially with $p$.) 
Nevertheless, for $p=2$ we solve for the performance of QAOA on a toy problem, called the ring of disagrees \cite{Farhi2014}, 
producing a particularly complicated intermediate result which exemplifies the difficulty of deriving similar 
expressions for higher $p$ or for more general graphs.  
Similarly, we apply our technique to the Directed MaxCut problem, which even for $p=1$ results in complicated expressions for the expected algorithm output. 

We use MaxCut as a case study to demonstrate our approach to quantifying the performance of QAOA. 
In Section \ref{sec:PauliSolver}, we propose the \textit{Pauli Solver} algorithm, which is a classical method for computing the expected output of a QAOA circuit. 
The primary advantage of the Pauli solver is that it doesn't require computing the QAOA state explicitly, 
avoiding any large matrix-vector multiplications. 
Applying this algorithm as an analytic procedure is the primary technique we use for proving the main results of this chapter, though it may also be implemented numerically for instance-wise optimization of the QAOA parameters.    
Moreover, this algorithm is general, and we explain how it may be applied to 
QAOA applications beyond MaxCut. 

Many of the results of this chapter can also be found in \cite{wang2017quantum}. 

\section{Background}   \label{eq:QAOAprelim} 
We briefly review the rich field of approximation algorithms, and provide 
the details of the 
Quantum Approximate Optimization Algorithm (QAOA). This material will also be utilized in Chapter~\ref{ch:QACOA}.

\subsection{Approximate Optimization}
In a \textit{combinatorial optimization problem}, we 
seek to maximize (or minimize) 
an \textit{objective function}\footnote{In Chapter \ref{ch:QACOA}, we consider more general domains than $\{0,1\}^n$, which result from feasibility constraints and choice of problem encoding.}
 $f:\{0,1\}^n \rightarrow \reals$. , 
We are interested in problems where finding the optimal 
value is computationally difficult, i.e., the best classical algorithms known have costs that scale exponentially with the input size $n$. 
Hence, for such problems we must settle for an \textit{approximation algorithm} or \textit{heuristic}, a 
procedure that for all $n$ terminates in an amount of time polynomially bounded in $n$ 
and produces as good a solution as possible. This has led to rich theories of approximation algorithms, hardness of approximation, and approximation complexity, 
in addition to a variety of heuristic approaches for different problems. 
We provide a brief outline of some important results;   
see, e.g., \cite{Vazirani,ShmoysBook,hromkovic2002algorithmics,trevisan2004inapproximability,Ausiello2012complexity,AroraBarak} for comprehensive overviews of these subjects. 

For some problems, we have results showing that no polynomial time classical algorithm 
exists that always produces an approximation within some factor of the optimal value (unless something believed to be unlikely such as P=NP is true), and, moreover, 
in some cases we have algorithms that achieve these optimal bounds. 
We emphasize that such results are worst-case. Indeed, in some cases, heuristics are known that perform much better than any such algorithm on \textit{practical instances}, i.e., return a solution much closer to optimal than the worst-case hardness bounds would suggest. 
These heuristics are typically evaluated empirically, and may or may not have general performance guarantees, or may require exponentially large running time on some problem instances. On the other hand, for some problems there are significant gaps between the performance of the best algorithm known and the best hardness result. Such problems are natural targets for heuristics and improved algorithms, and in particular quantum approaches, to provide better performance. 
Hence, it is important for scientists and engineers dealing with such problems to have at their disposal a variety of tools and algorithms. 

We are primarily interested in 
\textit{NP-optimization} (NPO) problems, where, informally, the corresponding decision problem (Given $k$, does there exist an $x$ such that $f(x) \geq k$ ?) is in NP
(typically, NP-complete). Hence, given a witness solution $x'$, we can efficiently compute $f(x')$ to verify that $f(x')\geq k$. In this chapter we study an important example, the MaxCut problem, and we will see many more examples of such problems in Chapter \ref{ch:QACOA}. 

For a maximization problem, let $f^*= \max_x f(x)$ denote the optimal value, with optimal solution  
$x^*= \arg\max_x f(x)$. Note that we are not concerned whether $x^*$ is unique. 
A solution $x$ is an $R$-\textit{approximation} if 
$f(x)/f^* \geq R$,  with $0\leq R \leq 1$.  We call $R$ the \textit{approximation ratio}. 
Similarly, an algorithm $\mathcal{A}$ achieves approximation ratio $R$ if it satisfies $f(\mathcal{A}(y)) / f^* \geq R$ for all problem instances $y$ 
(i.e., in the worst case), where  $A(y) \in \{0,1\}^n$ denotes the solution returned by~$\mathcal{A}$. If $\mathcal{A}$ runs in time polynomial in the problem 
size, 
then we call it an 
$R$-\textit{approximation}.  

For a minimization problem, the approximation ratio for a solution $x$ may be similarly defined, but with $R\geq 1$ and $f(x) / f^* \leq R$, where $f^*$ denotes the minimal value, and again similarly defined for an approximation algorithm $\mathcal{A}$. Unfortunately, there are multiple conventions in the literature, 
with 
the approximation ratio defined to be either $R\geq 1$ or $0\leq R \leq 1$, for both maximization or  minimization problems. In either convention, the ratio is simply the inverse $1/R$ of the other convention,  
so for a fixed problem and given $R$ there is no ambiguity as to which convention is being used. 
For convenience, we will primarily consider approximation ratios $R \leq 1$. 

We remark that approximation schemes utilizing randomness are also common in the literature. 
As is standard practice \cite{goemans1995improved}, we use $R$-\textit{approximate algorithm} to describe randomized polynomial time algorithms that output solutions with expected value at least $R$ times the optimal value.

It turns out that different problems may possess very different properties with respect to efficient approximation. As with decision problems, this has led to rich theories of approximation complexity, and natural grouping of problems into approximation complexity classes. We 
mention three of the most important classes, though there naturally exists others.   
The complexity class APX contains problems which can be efficiently approximated (on a classical computer) to within some constant factor. 
 Similarly, PTAS is the subset of problems which can be efficiently approximated (with respect to the input size $n$) to within any fixed $\e$, and FPTAS is the further subset of problems where the approximation is also efficient with respect to $\e^{-1}$ (as $\e \rightarrow 0$). 
As problems in FPTAS can be efficiently solved to arbitrary accuracy and in this sense are effectively solvable, 
primary targets for 
quantum algorithms 
are problems in PTAS or APX, 
in particular problems where the best classical algorithms are not known to be optimal and hence new algorithms could offer significant improvement. (We will see various examples of problems in these classes in Chapter~\ref{ch:QACOA}.)

Moreover, \textit{complete} problems can be also defined for many approximation complexity classes, which gives a notion of the hardest problems in each class. 
This requires computational reductions between problems, in particular, reductions preserving approximate solutions in some sense, of which a richer variety exists than as compared to decision problems \cite{crescenzi1997short}.  
See, e.g., \cite{Ausiello2012complexity} for details.  

Finally, 
many important results have been obtained showing that particular problems cannot be approximated better than some factor, 
known as the \textit{hardness of approximation}. 
Typically, such results show an approximation ratio for a given problem such that an efficient algorithm beating this ratio could be used to solve an NP-hard problem, and hence such an algorithm 
is impossible unless P=NP. 
For certain problems, these results imply the best approximation algorithms known are essentially optimal. 
For example, as we discuss below, if P$\neq$NP then MaxCut cannot be approximated better than $0.941$, whereas, under a weaker complexity theoretic conjecture, we similarly have that the $0.8785$ ratio achieved by the Goemans-Williamson algorithm is 
optimal.  

These difficulties in   
 solving optimization problems efficiently 
 have led to much excitement about the possibility of using quantum information processing devices for approximation. 
Indeed, currently available quantum devices such as the D-WAVE quantum annealers may be appropriately viewed as heuristic solvers, since generally we do not have performance guarantees, and these devices must be characterized empirically; see e.g. \cite{mcgeoch2014adiabatic}. It remains an important open problem whether or not such devices can provide an advantage for real-world optimization problems. 
We will instead focus here on quantum gate model algorithms, which 
offer greater design flexibility and several 
implementation advantages as we will discuss further in Chapter \ref{ch:QACOA}.   

\subsection{Quantum Approximate Optimization Algorithm (QAOA)}  \label{sec:QAOAdetails}
The Quantum Approximate Optimization Algorithm \cite{Farhi2014} seeks to approximately solve hard optimization problems on a quantum computer. This family of parameterized algorithms builds off of both 
quantum annealing 
and  
earlier proposals for gate model quantum optimization \cite{hogg2000quantum}. 
We give an overview of the details of QAOA and some important related results. 

Consider the problem of maximizing an \textit{objective function} 
$f(x)$ acting on $n$-bit strings $x$.  
We may always map such a function to a Hamiltonian 
$H_f$ with eigenvalues that encode the $2^n$ values $f(x)$ takes over all possible 
inputs. 
Thus, finding the optimal value of $f$ corresponds to a special case of 
the problem of finding extremal eigenvalues for $H_f$. 

We  say a Hamiltonian~$H_f$ \textit{represents} a function 
$f:\{0,1\}^n \rightarrow \reals$ if
its eigenvalues satisfy 
$$ H_f \ket{x} = f(x) \ket{x} \;\;\;\;\;\; \text{ for all } \; x\in\{0,1\}^n.$$
Given such a Hamiltonian, if we can compute its expansion in terms of Pauli $Z$ operators
\begin{equation}   \label{eq:diagHam}
H_f = c_0 I + \sum_{j=1}^n c_1 Z_j + \sum_{j<k} c_{jk}Z_jZ_k + \dots \;\;\;\;\;\;\;\; c_\alpha \in 
\reals,
\end{equation}
then we can 
easily simulate $H_f$,
i.e., implement the operation $U_f(t) =e^{-iH_ft}$, $t\in [0,2\pi]$, 
 using 
controlled-NOT and $Z$-rotation gates to simulate each term in the sum independently, and without requiring any additional ancilla qubits; 
such a quantum circuit is shown in Figure \ref{fig:RZZcircuit}. 
If the number of terms in the sum is not too many, i.e., bounded by a polynomial in the number of variables $n$, then this simulation is efficient. 
On the other hand, if $f$ is efficiently computable classically, then it is well known that we may efficiently simulate $H_f$ 
using ancilla qubits (to compute and store $f(x)$) and controlled $Z$-rotation gates to then apply the proportional phase shift; see e.g. \cite{NC,childs2004quantum}. 
We provide a design toolkit elaborating on 
the construction and simulation of Hamiltonians of the form (\ref{eq:diagHam}) in Section \ref{sec:QAOtoolkit}, along with several related results, with the details given in Appendix~\ref{ch:QAOA0}. 

The primary application of QAOA we consider is to solving \textit{constraint satisfaction problems}, where the objective function~$f$ 
to maximize 
is given by a (possibly weighted) sum of $m$ many  \textit{soft constraints} (Boolean clauses)~$f_j$ to be satisfied,   
$f(x)=\sum_{j=1}^m f_j(x)$.  
In many important cases such problems are NP-hard to solve optimally, and therefore we must settle for approximate solutions. When the constraints involve a constant number of variables, e.g. as in Max-$k$-SAT, then the number of terms in the resulting Hamiltonian is bounded by a constant times $m$, so such Hamiltonians can be efficiently constructed and simulated with polynomial cost. 

For a given problem instance on $n$ binary variables $x\in \{0,1\}^n$ with objective function $f(x)$, a QAOA mapping 
consists of three components:
\begin{enumerate}
\item An \textbf{initial state}, which we take to be the equal superposition state 
$$\ket{s}=\ket{+}^{\otimes n} = \frac{1}{\sqrt{2^n}}\sum_{x=0}^{2^n -1} \ket{x}.$$
\item A family of \textbf{phase separation operators} 
$$U_P(\gamma) = e^{-i\gamma H_f},$$ where $\gamma\in [0,2\pi)$ and $H_f$ acts on basis states $\ket{x}$ as $H_f\ket{x}=f(x)\ket{x}$. 
\item A famly of \textbf{mixing operators} 
$$\;U_M(\beta)= e^{-i\beta B},$$ where $\beta\in [0,2\pi)$ and $B=\sum_{j=1}^n X_j$. 
\end{enumerate}
The circuit depth parameter $p\in \nat$ and the $2p$ angles 
$\gamma_1,\dots,\gamma_p, \beta_1,\dots,\beta_p$ 
then suffice to specify a QAOA circuit, which we denote QAOA$_p$. 
More precisely, a QAOA mapping for a given problem is a classical procedure which uniformly and efficiently maps each problem instance to a corresponding quantum circuit that depends on the parameter
values selected for $\gamma$ and $\beta$. Note that the same mapping applies whether the goal is maximization or minimization of the objective function $f$, up to the sign of $H_f$. 
We consider more general operators and initial states for QAOA in Chapter  \ref{ch:QACOA}. 

The interval $[0,2\pi)$ for the angles $\gamma,\beta$ is chosen arbitrarily, and different intervals may be used. 
Thus, the angles for QAOA are real numbers. 
To avoid complicating details, we do not deal here  with issues of precision regarding how such angles may be represented. Generally, the constructions we give in the remainder of this chapter and in Chapter \ref{ch:QACOA} will result in quantum circuits consisting of fixed quantum gates such as Hadamard and CNOT gates, and single-qubit rotation gates that depend on $\gamma,\beta$. Standard techniques are known for efficiently approximating such rotations with 
a number of quantum gates taken from a universal set 
that scales polylogarithmically with the desired accuracy; see the discussion in Appendix \ref{ch:QC} and the related details in \cite{NC}.

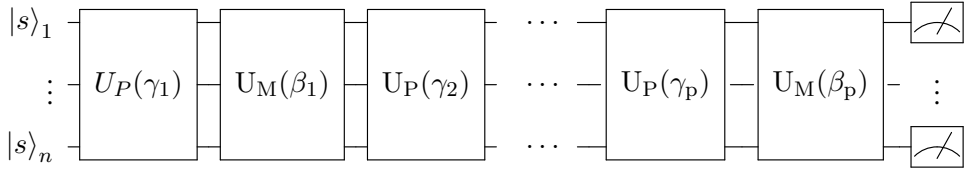
\begin{figure}[H]
\centerline{
\Qcircuit @C=0.4em @R=1em {
\lstick{\ket{s}_1}  & \multigate{2}{U_P(\gamma_1)} & \qw & \multigate{2}{\rm U_M(\beta_1)} & \qw & \multigate{2}{\rm U_P(\gamma_2)} & \qw & & & &  \hdots & & & & \qw &  \multigate{2}{\rm U_P(\gamma_p)} & \qw & \multigate{2}{\rm U_M(\beta_p)} & \qw & \meter \\
\lstick{\vdots} & \ghost{U_P(\gamma_1)}& \qw & \ghost{\rm U_M(\beta_1)} & \qw & \ghost{\rm U_P(\gamma_2)} & \qw & & & &  \hdots & & & & \qw & \ghost{\rm U_P(\gamma_p))} & \qw & \ghost{\rm Mix(\beta_P)}  & \qw & \vdots  \\
\lstick{\ket{s}_n}  & \ghost{{U_P(\gamma_1)}} & \qw & \ghost{\rm U_M(\beta_1)}& \qw & \ghost{\rm U_P(\gamma_2)}& \qw & & & &  \hdots & & & & \qw & \ghost{\rm U_P(\gamma_p)} & \qw &\ghost{\rm U_M(\beta_p)} & \qw & \meter
}
}
\caption{The Quantum Approximate Optimization Algorithm (QAOA$_p$).} 
\label{fig:QAOAcircuit}
\end{figure}

Given a QAOA mapping and set of angles, the algorithm consists of 
applying the phase and mixing operators 
in alternation to create the QAOA$_p$ state
\begin{equation}  \label{eq:QAOApState} 
\ket{\boldsymbol{\gamma \beta}} =    U_M(\beta_p)    U_P(\gamma_p) \dots U_M(\beta_1)    U_P(\gamma_1) \ket{s}, 
\end{equation}
followed by a computational basis measurement. This process is repeated a number of times, and the best solution kept after evaluating $f$ at each of the measurement outcomes. 
A high-level QAOA$_p$ circuit is shown in Figure \ref{fig:QAOAcircuit}. 
We remark that the description of the problem instance is encoded in the QAOA operators themselves, and is not otherwise input into the quantum algorithm; therefore, each QAOA circuit is problem instance dependent.
\vskip 1pc
Then, if somehow we can determine 
\lq\lq good\rq\rq\ angles 
$\gamma_j,\beta_j$ such that the state $\ket{\boldsymbol{\gamma \beta}}$  
has significant 
projection (i.e., probability amplitude) onto the subspace of states corresponding to good approximate solutions, then repeated preparation and measurement of $\ket{\boldsymbol{\gamma \beta}}$ will yield such a solution with high probability. Indeed, if 
such a QAOA$_p$ state can be found with probability amplitude 
at least $1/{\rm poly}(n)$ 
for a given solution state $\ket{x'}$, then repeating this process a polynomial number of times will yield the solution $x'$ with high probability. 

Clearly, the success of this approach depends on the number $p$ of QAOA rounds, and the angles 
$\gamma_1,\dots,\gamma_p, \beta_1,\dots,\beta_p$. 
When $p=O(1)$, 
QAOA is hoped to be especially suitable for small quantum computers \cite{Farhi2017}. 
Indeed, promising results have been shown even for $p=1$ \cite{Farhi2014b}. 
The angles 
can either be determined in advance through analysis and numerical testing, 
or searched for as part of the algorithm. 
In general, $\ket{\boldsymbol{\gamma \beta}}$ is a vector 
exponentially large in $n$, so deriving or finding parameters producing a \lq\lq good\rq\rq\ state is a difficult task. Moreover, knowing which solutions are \lq\lq good\rq\rq\ may not be meaningful without knowing the optimal solution itself. 
Hence, we follow the approach of \cite{Farhi2014} and 
consider the value of the objective function output by the algorithm in expectation (i.e., we take the average of the function evaluation on the measurement outcomes). Bounds on this quantity will lead to useful performance bounds to the approximation ratio achieved by QAOA. 

For a fixed set of angles, each preparation and measurement of $\ket{\boldsymbol{\gamma \beta}}$ returns  a solution $x$ with some probability.
Repeating this process,  
the expected value of the objective function is $$\langle f \rangle_{\text{QAOA}_p} = \bra{\boldsymbol{\gamma \beta}} H_f\ket{\boldsymbol{\gamma \beta}}  =: \langle H_f \rangle.$$ 
Thus, 
for fixed $p$ the best the algorithm can do with respect to the expected solution output is 
\begin{equation}  \label{eq:Fp}
F_p:= \max_{\gamma_1,\dots,\beta_p} \bra{\boldsymbol{\gamma \beta}} H_f\ket{\boldsymbol{\gamma \beta}}.
\end{equation}
We refer to angles $\gamma_1^*, \dots, \beta_p^*$ maximizing the right-hand side of (\ref{eq:Fp}) as \textit{optimal}. 
Selecting such angles and 
repeating the algorithm
a polynomial number of times
produces with high probability\footnote{Fom the Chebyshev inequality, an outcome of at least $F_p - 1$ will be obtained with probability at least $1 - 1/m$ after $O(m^2)$ repetitions 
In \cite{Farhi2014}, 
it is argued 
that $O(m \log m)$ repetitions suffice for MaxCut. } 
 a solution $x$ such that $f(x)$ is close to $F_p$. %
For a constraint satisfaction problem with $m$ constraints, i.e., $0\leq f(x) \leq f^* \leq m$ for optimal value $f^*$, 
the expected approximation ratio with optimally selected angles then satisfies 
\begin{equation}
\max_{\gamma \beta} \langle R \rangle \geq F_p / m.
\end{equation}
Note that for some problems, a better upper bound than $m$ to the optimal solution $f^*$ is known, which given $F_p$ yields a sharper lower bound to $\langle R \rangle$.

We typically consider QAOA$_p$ for fixed $p$. Clearly, the set of QAOA$_p$ states contains the  QAOA$_q$ states for $q<p$, so as $p$ increases the performance of QAOA$_p$, 
with respect to optimally selected angles, can only improve. The trade-off of course is increased implementation cost, and increased difficulty to find such optimal or near-optimal angles. 
Clearly, the low $p$ implementations are the most suitable for early quantum computers. 
We show below that for the MaxCut problem, the QAOA$_p$ state can be implemented using $O(p(n+m))$ basic quantum gates, 
and $O(pn^2)$ basic gates suffice for more general quadratic unconstrained optimization problems. 

Henceforth, by \textit{basic quantum gates} we mean controlled-not (CNOT) gates and arbitrary single-qubit gates; see Appendix \ref{ch:QC} for a review of some important quantum gates.

We remark that there are close connections between QAOA and quantum annealing. 
In particular, standard techniques (i.e., Suzuki-Trotter splitting formulas) to approximate the latter with discrete steps naturally yields a quantum state of the form (\ref{eq:QAOApState}). Hence, from the adiabatic theorem, it follows that
\begin{equation}
\lim_{p \rightarrow \infty} F_p = f^*,
\end{equation}
i.e., for $p$ sufficiently large there exists optimal angles with expected value arbitrarily close to the optimal solution. This was shown in \cite{Farhi2014} for the MaxCut problem under minor assumptions, but is easily seen to also apply to many other 
optimization problems. 
Thus, the QAOA method is sound in the sense that it in principle can solve optimization problems optimally. 
Unfortunately, this result does not tell us how large $p$ must be selected to obtain a certain quality of approximation, or how the algorithm behaves for constant or polynomially bounded $p$, 
and how to select the optimal angles. 

\begin{rem}  \label{rem:QAOA1}
For a given class of problem instances, if we can show for QAOA$_p$ 
there exists angles such that $\langle H_f \rangle \geq \alpha$, then from standard probabilistic arguments, there must exist a solution $x$ such that 
$\alpha \leq f(x) \leq f^*$, where $f^*$ is the optimal value. 
\end{rem}

Since their initial proposal \cite{Farhi2014}, QAOA circuits have sparked considerable interest in the research community. As mentioned, 
a QAOA$_1$ algorithm was shown to beat the 
best 
classical approximation algorithm known for 
the problem Max-E3Lin2, 
only to subsequently inspire a slightly better classical
algorithm \cite{Farhi2014b,barak2015beating}. 
The performance of QAOA$_p$ for Max-E3Lin2 with $p > 1$ has yet to
be determined, which is the truly interesting case.  

Generally, 
characterizing the power of QAOA$_p$ circuits remains the most important open problem. 
The lowest depth version QAOA$_1$ has provable performance guarantees for certain problems, 
though there exist better classical approximation algorithms. 
However, the class of QAOA$_1$ circuits turns out to be surprisingly powerful. 
In a recent paper, Farhi and Harrow~\cite{Farhi2016} proved that, 
under reasonable complexity assumptions, the output distribution of
 QAOA$_1$ circuits cannot be efficiently 
approximated by a classical algorithm.\footnote{More precisely, it was shown that if one could efficiently classically sample from the output of a QAOA$_1$ circuit, then the Polynomial Hierarchy would collapse to its third level, which is believed to be unlikely; see \cite{Farhi2016} for details.}  
QAOA circuits are therefore
among the most promising candidates for early demonstrations of
``quantum supremacy''~\cite{preskill_quantum_2012,boixo_characterizing_2016}, i.e., the physical demonstration on a quantum computer of a computational task which cannot be reproduced classically without an exponentially scaling amount of resources.    
However, it remains an open question whether QAOA circuits provide a quantum
advantage for approximate optimization, whether in terms of better approximation algorithms, or heuristics that empirically outperform existing methods on practical problem instances. 

Ultimately, the success of the QAOA approach will depend on finding effective
parameter-setting strategies.
For fixed $p$, and problems where the objective function is given by a sum of locally acting terms, 
the optimal angles 
can be computed
in time polynomial in the number 
of qubits~\cite{Farhi2014}. 
With increasing $p$, however, exhaustive search of the
QAOA parameters becomes inefficient due to the
curse of dimensionality; if we discretize so that each angle can take on $L$ different values, 
searching to find the optimal angles (without any further structural information) 
takes a number of steps at least $\Omega(L^p)$ in the worst case. 
Hence, we will 
study approaches to bounding the expected performance of QAOA, but with an eye to techniques for parameter setting.  
We will focus on the MaxCut problem, originally considered in \cite{Farhi2014}, as a prototype for the application of QAOA to other problems

Finally, we remark that QAOA circuits 
have also been considered for applications to exact optimization
\cite{Jiang17,Wecker2016training} and sampling \cite{Farhi2016}. In particular, the authors of \cite{Jiang17} show that QAOA circuits are powerful enough to achieve the $O(\sqrt{2^n})$ query complexity 
of Grover's quantum algorithm for unstructured search problems, 
showing the first definitive quantum advantage for QAOA circuits with finite $p$. 
Nevertheless, in this chapter we restrict our attention to applications of QAOA to approximate optimization problems. 

\section{Unconstrained Optimization} \label{sec:QUBO}
We first consider \textit{unconstrained} optimization problems, where 
all strings $x\in \{0,1\}^n$ encode valid solutions. 
With the natural mapping to $n$ qubits, 
where the~$j$th qubit encodes the~$j$th binary variable~$x_j$, 
such problems 
are especially well-suited for QAOA, typically resulting in simple 
initial states, phase operators, and mixing operators, and yielding 
especially  
low-resource constructions. These implementation advantages make unconstrained problems particularly  promising target applications for 
quantum approximation on early quantum computers. 

For 
unconstrained problems, it is relatively straightforward to implement QAOA$_p$ using basic quantum gates. 
For the initial state, 
we may 
use the equal superposition state 
$\ket{s}=\ket{+}^{\otimes n}=\frac{1}{\sqrt{2^n}} \sum_{x=0}^{2^n -1}  \ket{x}$. 
This state may be easily prepared from $\ket{0}^{\otimes n}=\ket{0^n}$ using $n$ Hadamard~$({\textrm H})$ gates in depth $1$ (i.e., applied in parallel), as follows from the identity $\ket{s}={\textrm H}^{\otimes n} \ket{0^n}$. Similarly, the standard mixing operator $U_M(\beta)=e^{-i\beta \sum_{j=1}^n X_j}=\prod_{j=1}^n  e^{-i\beta X_i}$ may be used, and implemented with $n$ $X$-rotation gates $R_X(\beta/2)$ in depth $1$. Hence, for problems where the phase operators can also be implemented relatively inexpensively, QAOA$_p$ can be applied with relatively low resources. 

We next quantify the required resources for the general class of \textit{quadratic} unconstrained 
binary optimization problems. In the following section we consider 
in detail a prototypical problem from this class, the MaxCut problem. 

\subsection{Quadratic Unconstrained Binary Optimization}
A general and important class of pseudo-Boolean optimization problems are 
\textit{quadratic unconstrained binary optimization} (QUBO) problems \cite{mcgeoch2014adiabatic}, where we seek to minimize a function  
\begin{equation}  \label{eq:QUBO}
f(x) = a + \sum_{j=1}^n c_j x_j + \sum_{j<k} d_{jk} x_{j} x_k,
\end{equation}
with $a, c_j, d_{jk}\in \reals$, $x_j \in \{0,1\}$. 
Indeed, this is the class of problems 
(ideally) implementable on a quantum annealing device such as, for example, the D-WAVE 2X, where the qubit interactions are themselves quadratic \cite{bian2010ising,mcgeoch2014adiabatic}.  
The QUBO class also contains many problems, via polynomial reductions, which at first sight are not quadratic, sometimes requiring extra variables; 
indeed, 
the 
natural QUBO decision problem 
is 
NP-complete~\cite{GareyJohnson}. A variety of exact and 
approximate classical algorithms and heuristics have been developed for 
QUBO problems; 
see e.g. \cite{boros2007local,tavares2008new} and the references therein. 

We remark that although (\ref{eq:QUBO}) contains only positive variables $x_j$, 
as $\overline{x}_j=1-x_j$,  
it is without loss of  generality. For example, 
the clause $x_1\oplus x_2$ may be equivalently written 
 as $x_1+x_2-2x_1x_2$, which is of the same form as (\ref{eq:QUBO}). Similarly, other common Boolean  clauses may be mapped to Hamiltonians using the rules given explicitly in Section \ref{sec:QAOtoolkit} (and their derivations in Appendix \ref{ch:QAOA0}).  Applying these results, 
we have the following lemma.

\begin{lem}   \label{lem:QUBOcost}
The QUBO 
function (\ref{eq:QUBO}) maps to a Hamiltonian given as a quadratic 
sum of Pauli $Z$ operators, with size (number of terms) at most $1 + n/2 + n^2/2$. 
Explicitly, we have
\begin{equation} \label{eq:QUBOHam}
H_f = (a + c + d ) I - \frac12 \sum_{j=1}^n (c_j + d_j) Z_j + \frac14 \sum_{j<k} d_{jk} Z_{j} Z_k,
\end{equation}
where we have defined $c = \frac12 \sum_{j=1}^n c_j, d = \frac14 \sum_{j<k} d_{jk}$, and $d_j = \frac12 \sum_{k\neq j} d_{jk}$ with $d_{jk}=d_{kj}$.
\end{lem}
\begin{proof}
The proof 
follows directly from applying Theorem \ref{thm:pseudoBool} of Appendix \ref{ch:QAOA0} to (\ref{eq:QUBO}). 
\end{proof}

The terms in (\ref{eq:QUBOHam}) mutually commute. 
Thus, we can simulate $H_f$, i.e., implement the QAOA phase operator $U_P(\gamma)=e^{-i\gamma H_f}$, using at most $n$-many $R_Z$ gates and $\binom{n}{2}$-many $R_{ZZ}$ gates. 
As shown in Figure \ref{fig:RZZcircuit}, 
each $R_{ZZ}$ can be simulated with $2$ CNOT gates and a $R_Z$ gate 
 (up to an irrelevant global phase), hence $U_P(\gamma)$ can be implemented with 
$\frac32 n^2 - \frac32 n$ basic quantum gates.  
For the initial state $\ket{s} = \ket{+}^{\otimes n}$ 
and the mixing operator $U_M(\beta) = e^{-i\beta \sum_{j=1}^n X_j}$, 
we obtain the following bound to the total gate cost. 

\begin{theorem}
For a QUBO problem on $n$ variables, we can prepare the QAOA$_p$ state using at most 
\begin{equation}
\frac32 p n^2 -\frac12pn + n
\end{equation}
CNOT and single qubit gates ($R_X$, $R_Z$, ${\rm H}$), with depth $O(pn)$, and $n$ qubits. 
\end{theorem}
\begin{proof}
The theorem follows 
from Lemma \ref{lem:QUBOcost} and the discussion of the general case above. 
\end{proof}
We remark that for a given instance of a particular problem, the implementation cost may be much lower. For example, in the next section we will see that the QAOA$_p$ construction for the MaxCut problem on bounded degree graphs requires a number of gates scaling as~$pn$ rather than~$pn^2$. 

We may similarly construct QAOA circuits for objective functions of order higher than quadratic. For maximum degree $d=O(1)$, 
the phase operator can be implemented with $O(n^d)$ basic gates, and the QAOA$_p$ state can be implemented with $O(p n^d)$ basic gates. It is straightforward to construct such phase operators and to explicitly bound their implementation costs using the results of 
Appendix~\ref{ch:QAOA0}, so we do not explore this in detail here. In the remainder of this chapter we focus on strategies for analyzing the performance of QAOA and finding good algorithm parameters.

\section{Maximum Cut} \label{sec:MaxCut}
We consider QAOA applied to the 
MaxCut optimization problem, as studied in \cite{Farhi2014,wang2017quantum}. 
In an instance of MaxCut, we are given a graph, and we seek to \lq\lq cut\rq\rq\ as many edges as possible by dividing the vertices into two sets.  
This problem may be naturally represented as a QUBO. 
We summarize the QAOA implementation of MaxCut \cite{Farhi2014}. 
We then derive analytic formulas 
for the performance of QAOA$_1$ for MaxCut, 
reproducing 
and significantly extending 
the results 
of \cite{Farhi2014}, where numerical results were found for limited classes of graphs. 
\paragraph{Problem:} Given a graph $G=(V,E)$, 
with $|V|=n$ vertices and $|E|=m$ edges, 
partition $V$ into two sets~$V'$ and $V\setminus V'$ such that the number of edges crossing the cut, i.e., edges with one endpoint in $V$ and the other in $V\setminus V'$, is maximized. 

The MaxCut problem has many applications in various fields, including circuit layout design and statistical physics \cite{barahona1988application,deza1994applications}. 
The corresponding decision problem, deciding if there exists a cut of size at least $k$, is NP-complete \cite{GareyJohnson}, so we cannot hope to efficiently find the optimal solution in general unless P=NP. 
For approximation, MaxCut is 
APX-complete \cite{papadimitriou1991,khanna1998syntactic}, 
which means it has no polynomial-time approximation scheme (PTAS) unless P=NP. Thus, the best we can do is a constant-factor approximation. 
Indeed, it is NP-hard to approximate MaxCut better than $0.941$ \cite{haastad2001some}. 
Using semidefinite programming and randomization, the Goemans-Williamson algorithm  \cite{goemans1995improved} achieves an approximation ratio of $0.8785$, 
the best classical algorithm known. It has been shown that if the unique games conjecture is true (a weaker complexity theoretic conjecture than P$\neq$NP), then this value is optimal \cite{khot2007optimal}. 

The problem remains hard to approximate on graphs of bounded degree.  
For a graph with maximum vertex degree 
$D_G \geq 3$, MaxCut can be efficiently approximated to within
$0.8785 + O (D_G)$~\cite{feige2002improved}. In particular, for $D_G=3$ this gives a $0.921$ approximation, 
yet this case remains APX-complete~\cite{papadimitriou1991}. 
%
Indeed, the MaxCut decision problem remains NP-complete for $D_G=3$ \cite{yannakakis1978node}.   
On the other hand, MaxCut is known to be solvable in polynomial time when restricted to certain simple classes of graphs, 
for example, planar graphs, toroidal graphs, or graphs not contractable to the complete graph with five vertices $K_5$ \cite{hadlock1975finding,barahona1983max}. 
For the application of QAOA to early quantum computers, we are particularly interested in subclasses of problems, such as MaxCut on bounded degree graphs, where fewer implementation resources are required than the general case, yet the problem remains hard to approximate. 

We emphasize that the stated complexity and hardness results concern the worst-case performance of efficient algorithms approximately solving MaxCut. 
Different randomized heuristics are known for MaxCut, which may 
produce much better solutions for practical instances \cite{festa2002randomized}.

\paragraph{Construction.}
The QAOA construction for MaxCut follows directly from that of the general 
QUBO case discussed above, resulting in the same construction as the one given in \cite{Farhi2014}.   
For an $n$-vertex graph, states are represented with $n$ qubits, with the $2^n$ computational basis states encoding 
all $2^n$ possible partitions of $V$. 
Each solution $x$ specifies a unique $S' \subset V$. 
Explicitly, basis states encode the $n$ binary indicator variables $x_1,\dots,x_n$, with $x_j =1$ indicating that vertex $j$ is included in $S'$. 
All states are feasible, 
so the problem is unconstrained. 
We consider the 
the initial state $\ket{s}=\ket{+}^{\otimes n}$ and mixing Hamiltonian $B= \sum_{v\in V} X_v$ 
as in \cite{Farhi2014}, though other initial states or mixing operators are possible. 
As explained, $\ket{s}$ can be prepared, and the mixing operator $U_M(\beta)=e^{-i\beta B}$ can be implemented, with $n$ single-qubit gates each.  

The objective function counts the number of edges crossing the partition, and  
is represented by the quadratic Hamiltonian 
\begin{equation} \label{eq:MaxCutObjHam}
C = \sum_{(uv)\in E} C_{uv},  \;\;\;\;\;\;\;\;\;\;\;\ C_{uv}= \frac12(I-Z_uZ_v),
\end{equation}
where each $C_{uv}$ encodes the predicate $x_u \oplus x_v$, which is true when the edge $(uv)$ is properly colored. 
For convenience, 
here we use $C$ for the objective Hamiltonian instead of $H_f$, as well as for the objective function $C(x)$, which is consistent with the notation of \cite{Farhi2014}. 
Thus, 
ignoring the global phase term, the phase operator $U_P(\gamma) = e^{-i\gamma C}$ can be implemented with at most $3m$ basic gates. 

Hence, the QAOA$_p$ state $\ket{{\bf \gamma \beta}}$ for MaxCut can be prepared with $n+p(n+3m)$ basic quantum gates. 
In particular, for bounded degree graphs $\left(D_G=O(1)\right)$, only $O(pn)$ basic gates are required.

\subsection{Performance}
We turn to the performance of QAOA for MaxCut. 
For $p=1$, \cite{Farhi2014} derives approximation ratio bounds for $2$-regular and $3$-regular graphs based on numerical results. We generalize these bounds to arbitrary graphs by deriving an exact formula for $\langle C \rangle=\bra{\gamma\beta}C\ket{\gamma\beta}$, which  
is used to bound the expected approximation ratio $\langle R \rangle$. 
Unfortunately, deriving similar results for higher $p$, which is the true question of interest, appears to be a difficult problem. We derive a complicated expression for $p=2$ on a particularly simple family of graphs  which exemplifies this difficulty.

Consider an arbitrary edge $(uv)\in E$, and 
let $d=d_u = \deg(u) - 1$ and $e = e_u = \deg(v) -1$. 
Let $f=f_{uv}$ be the number of triangles in the graph containing $(uv)$, i.e., 
$f=|\mathrm{nbhd}(u)\cap \mathrm{nbhd}(v) |$,
where the \textit{neighbourhood} function $\mathrm{nbhd}(v)$ gives the set of vertices adjacent to $v$. 
The following theorem shows that the $p=1$ expectation value $\langle C_{uv} \rangle$ depends only on the angles $(\gamma,\beta)$, 
and the \textit{neighbourhood parameters} $(d,e,f)$ of the edge $(uv)$, i.e.,
the local structure of the subgraph containing $(uv)$ and its adjacent vertices.  
Hence, the overall expectation value $\langle C \rangle$ 
reduces to a sum over triplets $(d,e,f)$, weighted by the number of times an edge with these parameters appears in $G$. 

For MaxCut, the optimal solution $C^*=C(x^*)$ is at most $m$, 
so the expected approximation ratio satisfies $\langle R \rangle \geq \langle C \rangle / C^* \geq \langle C \rangle / m$. The case $C^*=m$ occurs if $G$ is a bipartite graph, which can be checked (and if so, MaxCut solved) in linear time \cite{cormen2009introduction}. Hence, we emphasize that the lower bound $\langle R \rangle \geq \langle C \rangle / m$ is quite conservative, 
and the algorithm may perform much better on instances occurring in practice. 
(For fixed angles, $\langle R \rangle$ may be significantly larger than $\langle C \rangle / m$ if $C^*<m$.) 

\begin{theorem}   \label{thm:generalMaxCut}
Consider the QAOA$_1$ state $\ket{\gamma \beta}$ for MaxCut on a graph $G$. 
\begin{itemize}
\item For each edge $(uv)$, 
\begin{eqnarray}  \label{eq:costBoundMaxCutEdge}
\bra{\gamma \beta }  C_{uv}  \ket{\gamma \beta} \; 
&=& \frac12 +
\frac14 \sin (4\beta) \sin \gamma \; (\cos^d \gamma + cos^e \gamma ) \\ 
 &-&\frac14 \sin^2 (2\beta) \cos^{d+e-2f}\gamma \; (1-\cos^f (2\gamma)) \;\;  =: \; \langle C_{uv}\rangle (d,e,f)
 \nonumber 
\end{eqnarray}
where $d = deg(u) - 1$,  $e = deg(v) - 1$, and $f$ is the number of triangles in the graph containing $(uv).$  

\item The overall expectation value is 
\begin{equation}  \label{eq:costBoundMaxCut}
 \langle C \rangle =\bra{\gamma \beta} C \ket{\gamma \beta} = \sum_{(d,e,f)} \langle C_{uv}\rangle (d,e,f)\;  \chi (d,e,f)  ,
\end{equation}
where $\chi(d,e,f)$ gives the number of edges in $G$ with neighbourhood parameters $(d,e,f)$.
\end{itemize}
\end{theorem}
We prove the theorem using the Pauli Solver algorithm which we introduce in the next section. 

For a \textit{fixed} arbitrary graph, 
the expectation value $\langle C \rangle$ for QAOA$_1$ 
may thus be efficiently classically computed, \textit{analytically in closed form}, for any angles $\gamma,\beta$, and hence efficiently optimized. 

Note that, recalling Remark \ref{rem:QAOA1}, these results imply that there exists a partition that cuts at least 
$\lceil \max_{\gamma,\beta} \langle C \rangle \rceil$ edges. 
However, QAOA (or some other algorithm) 
is still required to 
\textit{find} a 
bit string realizing such an approximation or better.

For graphs with structure or symmetry, we may significantly simplify the result (\ref{eq:costBoundMaxCut}). 
To prove the theorem, we first show a general procedure for computing expectation values for QAOA$_p$.  
We then consider the special cases of  
triangle free and regular graphs 
as corollaries. 
 
\subsection{Pauli Solver Algorithm}  \label{sec:PauliSolver}     
We describe a high-level 
classical procedure for deriving $\langle C \rangle$ using the 
properties of the Pauli matrices and their exponentials. (See Appendix \ref{ch:QC} for a review of the most important properties.) 
We then apply this approach explicitly to prove Theorem \ref{thm:generalMaxCut}. 
This 
procedure is general and may be applied to different mappings, initial states, or to other problems beyond MaxCut. Moreover, this approach works in principle for arbitrary QAOA depth~$p$. 

Observe that, for a general Hamiltonian $H$ expanded in the basis of tensor products of Pauli matrices (see equation (\ref{eq:PauliExpansion})), only the terms in the sum that are strictly composed of $X$ and $I$ operators will have non-zero expectation value 
for the state $\ket{s}=\ket{+}^{\otimes n}$, and hence only these terms will contribute to $\bra{s}H\ket{s}$. 
Thus, in the spirit of the \textit{Heisenberg representation} of quantum mechanics \cite{sakurai1995modern}, 
instead of evolving the state $\ket{s}$, 
to compute $\langle C \rangle$
we (equivalently) evolve the 
observable $C$ itself. 
For a unitary evolution $U$, this corresponds to the transformation $C \rightarrow U^{\dagger} C U$, called \textit{conjugation} of $C$ by $U$. 

Hence, 
for a general objective function $C=\sum_{\ell=1}^m C_\ell$
(and corresponding Hamiltonians $C$, $C_\ell$),  
define the operator $Q=U_M(\beta)U_P(\gamma)$ for QAOA$_1$. 
Suppose we can compute, somehow,  
the Pauli expansions of each~$C_{\ell}$ conjugated by~$Q$, 
\begin{equation}   \label{eq:PauliSumMaxCut}
Q^\dagger C_{\ell} Q =  a_0 I + \sum_{j=1}^n \sum_{\sigma = X,Y,Z} a_{j\sigma} \sigma_j 
+ \sum_{j \neq k } \sum_{\sigma,\lambda = X,Y,Z} a_{j\sigma\lambda} \sigma_j  \lambda_k + \dots, 
\end{equation}
$a_\alpha \in \reals$. 
Then, using $\bra{+}I\ket{+}=\bra{+}X\ket{+}=1$ and $\bra{+}Y\ket{+}=\bra{+}Z\ket{+}=0$, 
$\langle C_{\ell}\rangle$ is given by 
$$ 
 \bra{\gamma \beta }  C_{\ell} \ket{\gamma \beta} = \bra{s} (Q^\dagger C_{\ell} Q)  \ket{s}
= a_0  + \sum_{j=1}^n a_{jX} 
+ \sum_{j \neq k }  a_{jkXX}  + \dots,
$$
i.e., the sum of the coefficients of the 
terms containing only $X$ or $I$ factors in (\ref{eq:PauliSumMaxCut}). 
Hence, computing the Pauli coefficients~$a_\alpha$ of~$Q^\dagger C_{\ell} Q$ gives~$\langle C_{\ell} \rangle$ directly. 
This suggests the following general (classical) procedure for computing $\langle C \rangle$. 

\paragraph{Algorithm computing $\langle C \rangle$:} 
\begin{enumerate}
\item Compute $Q^{\dagger} C_{\ell} Q$ as a sum of Pauli operators as in (\ref{eq:PauliSumMaxCut}).
\item Discard all terms in the sum containing a $Y$ or a $Z$ factor (i.e., keep the terms containing strictly $X$ and $I$ factors).
\item Set $\langle C_{\ell} \rangle$ as the sum of the remaining coefficients.
\item Apply Steps $1-3$ for each edge 
$\ell=1,\dots,m$ and return the overall sum $\langle C \rangle = \sum_\ell \langle C_{\ell} \rangle$. 
\end{enumerate}
The same steps apply for higher $p$ by replacing $Q$ with $Q_p = U_M(\beta_p)U_P(\gamma_p)\dots U_M(\beta_1)U_P(\gamma_1)$. 
However, the difficulty is that, in general, the number of terms to deal with in the sum (\ref{eq:PauliSumMaxCut}) grows exponentially with $p$ in this case. 
By implementing the 
algorithm in Python, in Sec.~\ref{sec:Ringp2} we derive an expression for $\langle C_{\ell}\rangle$ for $p=2$  
in a particular case. For $p=1$, the number of terms may be few enough that it is possible to derive general results such as those of 
Theorem \ref{thm:generalMaxCut}, 
whereas this becomes cumbersome even for $p=2$. However, $\langle C\rangle$ 
can always be solved for using the algorithm, and the angles optimized, on an instance-by-instance basis. 

The advantage of the Pauli Solver algorithm is that $\langle C\rangle$ can often be computed much more efficiently by conjugating $C$ to $Q^{\dagger} C Q$ and computing the coefficients than by computing (and storing) the state $\ket{\gamma \beta}$ explicitly. 
This, in general, depends on the given problem and the form of the phase and mixing operators. 
In particular,~$Q^{\dagger} C_{\ell} Q$ can often be computed directly using the algebraic properties of the Pauli matrices, avoiding any large matrix-vector multiplications. 
For MaxCut, using the 
locality of the clause Hamiltonians $C_{uv}$, and the product structure of the phase and mixing operators, in computing $\langle C_{uv}\rangle$ we can ignore many of the terms in $U_M(\beta)$ and $U_P(\gamma)$ which substantially simplifies the computation. These ideas will be made clear in the proof of Theorem \ref{thm:generalMaxCut} below. 

The Pauli Solver algorithm is also suitable for different initial states $\ket{s}$. In this case, Steps $2$ and $3$ must be appropriately modified to keep only the terms in the Pauli expansion that contribute to $\bra{s} \cdot \ket{s}$, which will depend on the particular $\ket{s}$. 

We now use this procedure to prove Theorem \ref{thm:generalMaxCut}. We show the steps explicitly, with the goal that similar techniques may be used to characterize the performance of QAOA for other problems. 
 
\begin{proof}[Proof of Theorem \ref{thm:generalMaxCut}]
Consider QAOA$_1$ applied to MaxCut with $Q=e^{-i \beta B} e^{-i \gamma C}$. 
Observe that $\langle C \rangle = \bra{s}Q^{\dagger} C Q \ket{s} 
=\sum_{(uv)\in E} \langle C_{uv} \rangle 
= \frac{m}2  -  \frac12 \sum_{(uv)\in E}\bra{s}Q^{\dagger} Z_u Z_v Q \ket{s}$. 
Hence, 
to compute~$\langle C \rangle$ 
it suffices to compute the 
quantities~$\langle Z_uZ_v \rangle$. 

Fix an edge $(uv)\in E$. Recall that each Pauli matrix $\Lambda$ satisfies $e^{-i\theta \Lambda} = \cos(\theta)I-i \sin(\theta) \Lambda$. Let~$c=\cos 2\beta$ and $s=\sin 2\beta$. 
Consider 
the action of 
$Q$ on $C_{uv}$ as first a conjugation by the mixing operator, 
followed by a conjugation by the phase operator. 
From the commutation properties of the Pauli matrices, 
most of the terms in the mixing operator~$e^{-i \beta B}=\prod_{j=1}^n e^{-i\beta X_j}$ 
will 
commute through and cancel, and 
we have 
\begin{equation}   \label{eq:mixOpConj}
e^{i \beta B} Z_uZ_v e^{-i \beta B} = e^{2i \beta X_u} e^{2i \beta X_v} Z_uZ_v
 =c^2 Z_uZ_v + sc (Y_uZ_v + Z_uY_v) + s^2 Y_uY_v.
\end{equation}
 Similarly, 
for the subsequent application of the phase operator,  
terms corresponding to 
 edges not containing $u$ or $v$ also commute through and cancel. 
The first term on the right, $c^2Z_uZ_v$, commutes with $e^{-i \gamma C}$ and thus does not  contribute to the expectation value. 
 We conjugate each remaining term in (\ref{eq:mixOpConj}) separately by $e^{-i \gamma C}$. Let $c' = \cos \gamma$ and $s' = \sin \gamma$. We have
 \begin{eqnarray*}
 \bra{s} e^{i\gamma C} Y_{u}Z_v e^{-i\gamma C} \ket{s} 
 &=&  \bra{s}e^{2i\gamma C_{uv}} e^{2i\gamma C_{u}} Y_{u}Z_v  \ket{s} \\ 
 &=&\bra{s} e^{-i \gamma Z_u Z_v }e^{-i \gamma \sum_{w\in \mathrm{nbhd}(u)\setminus v} Z_u Z_w } Y_{u}Z_v  \ket{s}\\
&=& \bra{s} (I c' - i s' Z_uZ_v) \prod_{i=1}^d (I c' - i s' Z_uZ_{w_i})  Y_{u}Z_v \ket{s},
\end{eqnarray*}
where  $C_u=\sum_{w\in \mathrm{nbhd}(u)\setminus v} C_{uw}$ and $C_v=\sum_{w\in \mathrm{nbhd}(v)\setminus w} C_{vw}$. Recall that $d=|\mathrm{nbhd}(u)\setminus v|$ and $e=|\mathrm{nbhd}(v)\setminus u|$. 
Expanding the product on the right hand side above gives a sum of tensor products of Pauli operators. 
Clearly, the only term that can contribute to the expectation value is the one proportional to $Z_uZ_v * I^{\otimes d} * Y_uZ_v = -i X_u$. Thus, we have
$$ \bra{s} e^{i\gamma C} Y_{u}Z_v e^{-i\gamma C}  \ket{s}  = \bra{s} -i s' c'^d (-i X_u)  \ket{s} =  - s' c'^d.$$
By symmetry, this implies 
$ \bra{s} e^{i\gamma C} Z_{u}Y_v e^{-i\gamma C}  \ket{s}   =  - s' c'^e.$
Observe that these terms depend only on the numbers of neighbours $d$ and $e$. 

The last term in (\ref{eq:mixOpConj}) becomes  
  \begin{eqnarray*}
 \bra{s} e^{i\gamma C} Y_{u}Y_v e^{-i\gamma C}  \ket{s} 
  &=&  \bra{s}e^{2i\gamma C_{u}} e^{2i\gamma C_{v}} Y_{u}Y_v  \ket{s} \\ 
  &=& \bra{s}  \prod_{i=1}^d (c' I - i s' Z_uZ_{w_i}) \prod_{j=1}^e (c'I - i s' Z_vZ_{w_j})  Y_{u}Y_v   \ket{s}.
 \end{eqnarray*}
In this case, there are many terms in the product which can contribute to the expectation value. 
The simplest terms that contribute 
are $\bra{s} (c'I)^{d+e-2} (-is'Z_uZ_w)(-is'Z_vZ_w)Y_u Y_v \ket{s}= c'^{d+e-2}  s'^2$, of which 
there are $f$ many, 
corresponding to the $f$ triangles containing $(uv)$.  
 As $Z_uZ_{w_i}*Z_uZ_{w_i}=I$, if $f>2$ then higher order terms 
 will contribute. 
 The next higher-order terms result from three different pairs $(Z_uZ_{w_i},Z_vZ_{w_i})$ in the product, and hence their contribution is proportional to~$s'^6$. There are $\binom{f}{3}$ many such terms. 
 Thus, for the general case, we have 
    \begin{eqnarray}    \label{eqn:YYtermThm1proof}
 \bra{s} e^{i\gamma C} Y_{u}Y_v e^{-i\gamma C}  \ket{s} 
 &=& \binom{f}{1}  c'^{d+e-2}  s'^2 +  \binom{f}{3}  c'^{d+e-6}  s'^6 +  \binom{f}{5}  c'^{d+e-10}  s'^{10}+ \dots \nonumber \\
 &=& c'^{d+e-2f} \sum_{i=1,3,5,\dots}^f \binom{f}{i}  (c'^{2})^{f-i}(s'^2)^i.  
 \end{eqnarray}
To sum this series, 
we twice apply the binomial theorem to yield  
$$  \sum_{i=1,3,\dots}^f \binom{f}{i} a^{f-i} b^{i} = \frac12 ( (a+b)^f - (a-b)^f) .$$
Thus the above sum becomes 
$$  \sum_{i=1,3,\dots}^f \binom{f}{i}  (c'^{2})^{f-i}(s'^2)^i = \frac12( c'^2 + s'^2 )^f - (c'^2 - s'^2)^f = \frac12 (1 - \cos^f 2\gamma),$$
which gives
\begin{equation}  \label{eqn:YYtermSumThm1proof}
 \bra{s} e^{i\gamma C} Y_{u}Y_v e^{-i\gamma C}  \ket{s}  = \frac12  c'^{d+e-2f} (1 - \cos^f 2\gamma).
\end{equation} 
Combining the above results with basic trigonometric identities gives (\ref{eq:costBoundMaxCutEdge}). 

Finally, summing~$\langle C_{uv}\rangle$ over the set of edges gives (\ref{eq:costBoundMaxCut}). 
\end{proof}

\subsection{Triangle-Free Graphs}
For \text{triangle-free graphs}, Theorem \ref{thm:generalMaxCut} yields  
simple results for $\langle C\rangle$ that are 
particularly amenable to further analysis. 
We first consider the case of graphs of fixed vertex degree. 

\begin{cor}   \label{cor:MaxCutTriFreeGraphs}
For a $D$-regular triangle-free graph, 
for QAOA$_1$ we have 
\begin{equation}    \label{eq:triangleFreeExpec}
\langle C \rangle =  \frac{m}{2} + \frac{m}{2} \sin 4\beta \sin \gamma \cos^{D-1} \gamma, 
\end{equation}
with maximum value
\begin{equation}    \label{eq:triangleFreeExpec2} 
  \max_{\gamma, \beta}  \langle C \rangle = \frac{m}{2}  +   \frac{m}{2} \frac{1}{\sqrt{D}} \left(\frac{D-1}{D}\right)^{(D-1)/2} =: C^{reg}_{max} (D),
\end{equation}
and approximation ratio satisfying 
\begin{equation}  \label{eq:firstApproxRatioBound}
  \max_{\gamma, \beta} \; \langle R \rangle > \frac12+ \frac{1}{2\sqrt{e}} \frac{1}{\sqrt{D}}  = \frac12 + \Omega(\frac{1}{\sqrt{D}}).
\end{equation}
\end{cor}
\begin{proof}
Theorem \ref{thm:generalMaxCut} gives the first equation directly, and the second equation then follows applying simple calculus. 
(The optimal angles are given below in Theorem \ref{thm:maxCutAngles}.) 
For the third equation we use the bounds $\langle R \rangle \geq \langle C \rangle /m$ and $\left(\frac{d}{d+1}\right)^d > 1/e$. 
\end{proof}
In particular, (\ref{eq:triangleFreeExpec2}) gives $  \max_{\gamma, \beta}  \langle R \rangle \geq 0.75$, $0.69245$, $0.66238, 0.64310$ for $D = 2,3,4,5$, respectively. The cases $D=2,3$ reproduce results found by numerical simulation in \cite{Farhi2014}. 

We next consider arbitrary triangle-free graphs. 
As any edges with a degree one vertex can always be trivially cut, we may assume 
the minimal vertex degree is at least two. 
Observing that (\ref{eq:triangleFreeExpec}) can only decrease with $D$ for fixed $\gamma,\beta$, we have the following corollary. 

\begin{cor}  
For an arbitrary triangle-free graph with maximum degree $D_G$, 
and $n_D$ vertices of degree $D$, $D=2,3,\dots,D_G$,   
for QAOA$_1$ we have 
\begin{equation}    \label{eq:triangleFreeExpec3}
\langle C\rangle =   \frac{m}{2} + \frac{1}{4} \sin 4\beta \sin \gamma \; \sum_{D} D \; n_{D}  \cos^{D-1} \gamma  , 
\end{equation}
for which the previous corollary 
gives the lower bound 
\begin{equation}    \label{eq:triangleFreeExpec4}
\max_{\alpha, \beta} \langle C \rangle \geq C^{reg}_{max} (D_G) > \frac{m}{2} + \frac{m}{2\sqrt{eD_G}},
\end{equation}
and hence the expected approximation ratio satisfies
\begin{equation}  \label{eq:triFreeBoundR}
\max_{\gamma,\beta} \langle R \rangle > \frac12 + \frac{1}{2\sqrt{e}} \frac{1}{\sqrt{D_G}}.
\end{equation}
\end{cor}
Thus, 
$\max_{\alpha, \beta} \langle C \rangle > \frac{m}{2}$, 
so QAOA always beats random guessing on triangle-free graphs, i.e., there always exist angles $\gamma,\beta$ with expected approximation ratio strictly greater than $\frac12$. 
The result (\ref{eq:triFreeBoundR}) follows 
from (\ref{eq:triangleFreeExpec4}) 
using $ \langle R \rangle \geq \langle C \rangle /m$, 
which is a relatively crude lower bound to the expected approximate ratio. 
Indeed, there exist families of triangle-free graphs with $m$ edges such that the best possible cut contains 
only $C^* = \frac{m}{2} + \Theta(m^{4/5})$ edges \cite{alon1996bipartite}; clearly, the approximation ratio will be much higher on such instances. (On the other hand, an $m$-edge triangle-free graph could be bipartite or nearly bipartite, so we can have graphs with $C^* \simeq m$ in the worst case.)  

Using the 
corollaries, we 
classify the optimal QAOA$_1$ angles for MaxCut on triangle free graphs.

\subsubsection*{Optimal Angles}
Recall a pair of angles $(\gamma^*,\beta^*)$ is \textit{optimal} if they maximize the lower bound to the expected approximation ratio $\langle R \rangle$. 
For our purposes here we consider angles optimal if they maximize $\langle C \rangle/m $. 

\begin{theorem}   \label{thm:maxCutAngles}
For QAOA$_1$ applied to MaxCut on any triangle-free graph, the optimal angles 
maximizing $\langle C\rangle$ (or, equivalently,  maximizing $\langle C\rangle/m$) satisfy the following:
\begin{itemize}
\item For a $D$-regular triangle-free graph, 
the unique smallest positive optimal pair of angles is 
\begin{equation}  \label{eq:minOptAngles}
(\gamma^*,\; \beta^*) := ( \arctan \frac{1}{\sqrt{D-1}}, \pi/8),
\end{equation}
for $D\geq 2$ (i.e., no other optimal pair $(\gamma,\beta)$ exists with $0\leq \gamma \leq \gamma^*$ or $0\leq \beta \leq \beta^*$). 

All optimal angles are periodic in $\gamma,\beta$ with periodicity depending on $D$:
\begin{itemize}
\item 
If $D$ is even, there is a second independent pair of optimal angles given by $(-\gamma^*,-\beta^*)$, 
independent in the sense that 
all optimal angles are generated from these two pairs as 
\begin{equation}  \label{eq:angles}
(\gamma^* + a\pi , \; \beta^* + b\frac{\pi}{2}),   \;\;\;\;\;
(- \gamma^*+ c \pi , \; - \beta^* + d\frac{\pi}{2}),
        \;\;\;\;\; \;\;\;\;\;  a,b,c,d \in \integers.
\end{equation}
\item 
Else if $D$ is odd, there are four independent pairs of optimal angles 
$(\gamma^*,\beta^*)$, 
$( -\gamma^*,-\beta^*)$, 
$(\pi -\gamma^*,\beta^*)$, and  
$(\pi + \gamma^*,-\beta^*)$, 
and all optimal angles are generated from one of these pairs, denoted $(\gamma',\beta')$, as
\begin{equation}  \label{eq:angles2}
( \gamma' + a2\pi , \;  \beta' + b\frac{\pi}{2}),    \;\;\;\;\; \;\;\;\;\;   a,b \in \integers.
\end{equation}
\end{itemize}

\item For an arbitrary triangle-free graph with maximum vertex degree $D_{G}$ and minimum vertex degree $D_{min}$, the smallest positive optimal angles
 $\gamma^*,\beta^* \in [0, \pi/2]$ 
 satisfy
\begin{equation}    \label{eq:optAnglesTriFree}
\arctan \frac{1}{\sqrt{D_{G}-1}} \; \leq \;  \gamma^* \; \leq \;  \arctan \frac{1}{\sqrt{D_{min}-1}} , \;\;\;\;\;\;\;\; \;\;\;   \beta^* =  \frac{\pi}{8} . 
\end{equation} 
Given such a pair, the angles $(-\gamma^*, -\beta^*)$ are also optimal, and both pairs are $2\pi$-periodic in the first argument and $\pi/2$-periodic in the second, with respect to optimality. 

\end{itemize} 
\end{theorem}
The proof 
is given in Appendix \ref{app:QAOA}. 
The theorem implies the smallest optimal angles are 
$(\pi/4,\pi/8)$ and $(0.6155,\pi/8)$ for $2$-regular and $3$-regular triangle-free graphs, respectively, 
which reproduces results found numerically in \cite{Farhi2014}. 
%
Plugging the angles (\ref{eq:minOptAngles}) into (\ref{eq:triangleFreeExpec}) gives~(\ref{eq:triangleFreeExpec2}).

\subsection{General Graphs}
For a general graph with bounded degree, 
combining previous results we have the following lemma. 
\begin{lem}   \label{lem:genGraphsBound}
For QAOA$_1$ applied to MaxCut on a graph $G$ with maximum vertex degree $D_G$, and containing $F$ triangles, in addition to $\: \max_{\gamma, \beta} \langle C \rangle \geq \frac{m}2$ 
we have
\begin{equation}  \label{eq:genGraphsBound2}
\max_{\gamma, \beta} \langle C \rangle \geq \frac{m}{2} + \frac{m}{2\sqrt{e}} \frac{1}{\sqrt{D_G}} - O\left(\frac{F}{D_G}\right).
\end{equation}
\end{lem} 
\begin{proof} 
Clearly, $\max_{\gamma, \beta} \langle C \rangle \geq \bra{ \gamma^* \beta^*} C \ket{\gamma^* \beta^*}$,  
where $\gamma^* = \arctan\frac{1}{D_G-1},\beta^* = \pi/8$ are taken from (\ref{eq:optAnglesTriFree}), i.e., pretending the graph had no triangles. 
Plugging $\gamma^*,\beta^*$ into (\ref{eq:costBoundMaxCutEdge}), it is straightforward to derive  
(\ref{eq:genGraphsBound2}). 
Note that the right-hand side may become less than $m/2$ as the number of triangles $F$ becomes large; this shows that for such cases the angles $\gamma^*,\beta^*$ are no longer good choices. Indeed, setting $\gamma=0=\beta$, we can always obtain $ \langle C \rangle = \frac{m}{2}$ for any graph.
\end{proof}
We remark that there exist families of $m$-edge graphs such that the best possible cut contains $C^*= \frac{m}{2} + \sqrt{\frac{m}{8}} + O(m^{1/4})$ edges \cite{alon1996bipartite,alon1998bipartite}.  
Clearly, the expected approximation ratio will be much higher on such instances than $\langle C\rangle / m$. 

Thus, for general graphs, we have shown that the performance of QAOA depends strongly on the graph topology of a given instance. 
Using Lemma \ref{lem:genGraphsBound} it is straightforward to derive the following 
lower bound on the performance of QAOA on general graphs. 

\begin{theorem}   \label{thm:genGraphsMaxCutIntro}
For QAOA$_1$ applied to MaxCut on a graph $G$ with 
bounded maximum vertex degree $D_G=O(1)$, 
we have
\begin{equation}  \label{eq:genGraphsBound}
\max_{\gamma, \beta} \langle R \rangle \geq \frac{1}{2} + \frac{1}{2\sqrt{e}} \frac{1}{\sqrt{D_G}} - O\left(\frac{1}{D_G}\right).
\end{equation}
\end{theorem} 
\begin{proof}
The proof follows from Lemma \ref{lem:genGraphsBound} using the bound $\langle R \rangle \geq \langle C \rangle /m$ and the fact that for graphs with bounded degree $D_G=O(1)$, the number of triangles $F$ in the graph is $O(m)$. 
\end{proof}

\begin{rem}
It is 
worthwhile to elaborate on the expected approximation ratio lower bounds 
(\ref{eq:firstApproxRatioBound}), (\ref{eq:triFreeBoundR}), and 
(\ref{eq:genGraphsBound}). 
For graphs with large maximum degree $D_G$, these bounds become close to $1/2$, which is the approximation ratio obtained for MaxCut by random guessing. 
These bounds are not competitive with the best classical algorithm known, the Goemans-Williamson algorithm~\cite{goemans1995improved} based on semidefinite programming, which achieves an approximation ratio of $0.8785$, independently of $D_G$.  
Moreover, under a plausible conjecture from computational complexity theory, no polynomial-time classical algorithm can do better than this in general. 

Indeed, the Goemans-Williamson algorithm, published in 1995, 
was a huge breakthrough for the MaxCut problem, after 20 years of relative standstill. Previously, a (deterministic) $1/2$-approximation had been found \cite{sahni1976p}, which despite much effort, led to a sequence of algorithms with relatively minor improvements (i.e., no improvement to the $1/2$ factor), yielding approximation ratios 
$R=\frac12 +\frac{1}{2m}$, $R=\frac12 +\frac{1}{2n}$, $R=\frac12 +\frac{n-1}{4m}$, and 
$R=\frac12 + \frac{1}{D_G}$, 
for graphs with $n$ vertices and $m$~edges  \cite{vitanyi1981well,poljak1982polynomial,haglin1991approximation,hofmeister1996combinatorial}. (See \cite{goemans1995improved} for an insightful discussion on the history of approximation algorithms for MaxCut.) Thus, we take the results (\ref{eq:firstApproxRatioBound}), (\ref{eq:triFreeBoundR}) and (\ref{eq:genGraphsBound})  
as important positive indicators that quantum computers may be useful for approximating hard optimization problems such as MaxCut. Whether QAOA, or another quantum approximation algorithm, can 
improve upon 
the result of \cite{goemans1995improved} remains a tantalizing 
open problem. 
\end{rem}

\begin{rem}
For each class of graphs studied, the obtained lower bounds to $\langle R \rangle$ decrease as the maximum vertex degree $D_G$ increases. This suggests that 
as $D_G$ increases, we 
may need to take higher~$p$ 
for QAOA$_p$ to obtain the same performance as on graphs with smaller $D_G$. 
\end{rem}

\subsection{Depth-Two QAOA for the Ring of Disagrees}  \label{sec:Ringp2}
While it is in principle straightforward to extend our results for MaxCut to QAOA$_p$ with $p>1$, 
the number
of terms in the analysis quickly becomes prohibitive for direct calculation. 
The expectation value $\langle C_{uv}\rangle$ for each edge $(uv)$ will now depend on its $p$-local graph topology (i.e., the subgraph induced from the set of vertices within edge-distance $p$ of $u$ or $v$), which becomes difficult to succinctly characterize as $p$ increases. 
We show here how even for $p=2$, and for one of the simplest possible graphs, the number of terms is daunting. 

Consider MaxCut on a $2$-regular connected graph, called the \textit{ring of disagrees}, which is a useful toy problem for QAOA studied in \cite{Farhi2014,wang2017quantum}. 
In physics, it is equivalently described as the closed
one-dimensional chain of spin-1/2 particles with nearest-neighbour antiferromagnetic
couplings.  
Assume $n$ is even, so the optimal cut size is trivially seen to be $C^*=m$. 
For a given choice of angles~$\gamma,\beta$, the expected approximation ratio is given by $\langle R\rangle = \langle C\rangle/m$. 
In \cite{Farhi2014}, the optimal expected approximation ratios for this problem 
were found numerically to be 
$3/4$ and $5/6$ 
for QAOA$_1$ and QAOA$_2$, respectively. 

In \cite{wang2017quantum}, we study MaxCut on the ring of disagrees using a different approach inspired from physics. 
By reformulating the QAOA mapping for this problem to a fermionic representation using the Jordan-Wigner transformation \cite{lieb2004two}, we show that the 
parameterized QAOA evolution translates equivalently into the quantum control of an ensemble of independent spins, significantly simplifying the analysis. This, in principle, allows for the optimal expected approximation ratio to be computed for arbitrary $p$; 
see  \cite{wang2017quantum} for details. 
Furthermore, we show how symmetries satisfied by the optimal angles can make finding such angles much easier. 
Unfortunately, it is unclear whether or not this fermionic approach can be extended to derive performance bounds 
or parameter setting strategies 
for MaxCut on general graphs. 
Here, we again apply our 
Pauli Solver algorithm, but with $p=2$. 

For QAOA$_1$, Corollary \ref{cor:MaxCutTriFreeGraphs} and Theorem \ref{thm:maxCutAngles} reproduce the optimal expected approximation ratio of~$3/4$ for the ring of disagrees, 
with the optimal angle pairs for 
$\gamma, \beta \in [0,\pi)$ 
given by
$$ (\gamma^*,\beta^*) = (\frac{\pi}{4},\frac{\pi}{8}), \;  (\frac{\pi}{4},\frac{5\pi}{8}),\;  (\frac{3\pi}{4},\frac{3\pi}{8}), \;  (\frac{3\pi}{4}, \frac{7\pi}{8}).$$

For QAOA$_2$ on the ring of disagrees, 
we consider maximization of the quantity 
$$ \langle C\rangle:=\bra{\boldsymbol{\gamma}, \boldsymbol{\beta}} C \ket{\boldsymbol{\gamma} \boldsymbol{\beta}}= \sum_{(uv)\in E}   \bra{s} e^{i\gamma_1 C} e^{i \beta_1 B}  e^{i\gamma_2 C} e^{i \beta_2 B} C_{uv}  e^{-i \beta_2 B}  e^{-i\gamma_2 C}   e^{-i \beta_1 B}  e^{-i\gamma_1 C}  \ket{s} .$$
Using the Pauli Solver algorithm and some 
trigonometric simplifications,   
the expected approximation ratio is found to be 
\begin{equation}   \label{eq:ApproxRatioRing}
\langle R \rangle = \frac12 + f(\gamma_1,\beta_1,\gamma_2,\beta_2),
\end{equation}
where 
the function $f(\gamma_1,\beta_1,\gamma_2,\beta_2)$ is given by 
\begin{eqnarray*}
f(\gamma_1,\beta_1,\gamma_2,\beta_2) 
&=& \frac{1}{128} *\bigg( 4 \cos (2 (\gamma_1 - 2 \beta_1)) + 4 \cos (2 (\gamma_2 - 2 \beta_1)) - 
4 \cos (2 (\gamma_1 + 2 \beta_1))     \\
&-& 4 \cos (2 (\gamma_1 + \gamma_2 - 2 \beta_1))  - 4 \cos (2 (\gamma_2 + 2 \beta_1)) + 4 \cos (2 (\gamma_1 + \gamma_2 + 2 \beta_1)) \\
&-& 6 \cos (2 (\gamma_1 - \gamma_2 - 2 \beta_2)) +  4 \cos (2 (\gamma_2 - 2 \beta_2)) 
    +     6 \cos (2 (\gamma_1 + \gamma_2 - 2 \beta_2)) \\ 
&-& 6 \cos (2 (\gamma_1 + 2 \beta_1 - 2 \beta_2)) + 3 \cos (2 (\gamma_1 - \gamma_2 + 2 \beta_1 - 2 \beta_2)) \\
&+& 3 \cos (2 (\gamma_1 + \gamma_2 + 2 \beta_1 - 2 \beta_2))+ 6 \cos (2 (\gamma_1 - \gamma_2 + 2 \beta_2))\\
 &-& 4 \cos (2 (\gamma_2 + 2 \beta_2)) - 6 \cos (2 (\gamma_1 + \gamma_2 + 2 \beta_2)) 
   + 6 \cos (2 (\gamma_1 - 2 \beta_1 + 2 \beta_2))\\
    &-&  3 \cos (2 (\gamma_1 - \gamma_2 - 2 \beta_1 + 2 \beta_2)) 
    -3 \cos (2 (\gamma_1 + \gamma_2 - 2 \beta_1 + 2 \beta_2)) \\
    &+&  6 \cos (2 (\gamma_1 - 2 (\beta_1 + \beta_2))) + 3 \cos (2 (\gamma_1 - \gamma_2 - 2 (\beta_1 + \beta_2))) \\
    &-& 4 \cos (2 (\gamma_2 - 2 (\beta_1 + \beta_2))) +  7 \cos (2 (\gamma_1 + \gamma_2 - 2 (\beta_1 + \beta_2)))\\
    &-& 6 \cos (2 (\gamma_1 + 2 (\beta_1 + \beta_2))) -  3 \cos (2 (\gamma_1 - \gamma_2 + 2 (\beta_1 + \beta_2)))  \\
    &+&4 \cos (2 (\gamma_2 + 2 (\beta_1 + \beta_2))) -7 \cos (2 (\gamma_1 + \gamma_2 + 2 (\beta_1 + \beta_2))) \bigg).
\end{eqnarray*}
Thus. 
even for a simple graph such as the ring of disagrees, the expected output of QAOA$_2$ depends on the angles $\gamma_1,\gamma_2,\beta_1,\beta_2$ in a 
relatively complicated way. 
Optimizing this function gives 
$$\max_{\gamma_1,\gamma_2,\beta_1,\beta_2} \langle R \rangle = 0.8333,$$  confirming the value
found numerically in \cite{Farhi2014}. 
A particular set of optimal angles is%
~$(\gamma_1,\gamma_2,\beta_1,\beta_2) =$ 
$(0.655871, 0.62143, 1.24286,  0.327935),$
~with this choice satisfying $\gamma_2 = 2*\beta_1$, $\gamma_1=2*\beta_2$, and $\gamma_2 + \beta_2 = \pi/2$ (or, equivalently, $\gamma_1 + 4 \beta_1 = \pi $). These symmetry conditions may be explicitly enforced to yield a simpler expression for $f(\gamma_1,\beta_1,\gamma_2,\beta_2)$. 
 
The optimal angles in general can again be seen to obey periodicity conditions.
Suppose $(\gamma_1^*, \beta_1^*,\gamma_2^*,\beta_2^*)$ is a set of optimal angles. Then 
the angles $(\gamma_1^*  +  a\pi, \beta_1^* + b\pi/2,\gamma_2^* + c\pi,\beta_2^* + d\pi/2)$ are also optimal for $a,c,b,d \in \integers$. 

Clearly, the result (\ref{eq:ApproxRatioRing}) indicates that bounding the performance of QAOA$_2$ for MaxCut on general graphs, or even relatively simple subclasses of graphs, is a nontrivial task. 
We leave this important question as an open 
problem for future research.

\subsection{Weighted Maximum Cut}
A natural generalization 
is \textit{Weighted MaxCut} where we are given a weight $w_{uv} $ for each edge and seek a partition such that the sum of the weights of edges crossing the partition is maximized. 
The corresponding weighted decision problem was one of Karp's original NP-complete problems \cite{karp1972reducibility} (shown via a reduction from the number partitioning problem). 

The QAOA construction is the same as MaxCut, up to the multiplicative weights in the objective Hamiltonians, 
which become
$$C = \sum_{(uv)\in E} C_{uv},  \;\;\;\;\;\;\;\;\;\;\;\ C_{uv}= \frac{w_{uv}}{2}(I-Z_uZ_v).$$ 
The case $w_{uv}=1$ gives MaxCut. Generalizing our previous approach 
gives the following result. 

\begin{theorem}   \label{thm:weightedMaxCut}
Consider QAOA$_1$ applied to Weighted MaxCut on a triangle-free graph.
Letting $\gamma_{uv} := \gamma w_{uv}$
and $W := \sum_{(uv)\in E} w_{uv}$, 
the overall expectation value 
$\bra{\gamma \beta }  C \ket{\gamma \beta}$
is 
\begin{equation}  \label{eqn:costBoundWeightedMaxSat}
 \frac{W}{2} +   \frac{\sin 4\beta  }{4}   \sum_{(uv)\in E}    w_{uv} \sin \gamma_{uv} \left( \prod_{t \in \mathrm{nbhd}(u)\setminus \{v\}}   \cos \gamma_{ut} \;\; + 
 \prod_{w \in \mathrm{nbhd}(v)\setminus \{u\}}   \cos \gamma_{vw} \right)   .
\end{equation}
\end{theorem}
The proof of Theorem \ref{thm:weightedMaxCut} 
follows 
similarly to the proof of 
Theorem \ref{thm:generalMaxCut}, so we omit its details. 
%
We remark that it is also possible to extend Theorem \ref{thm:weightedMaxCut} to general graphs. However, in this case, the term in $\langle C_{uv}\rangle$ which is non-zero for triangles (i.e., analogous to (\ref{eqn:YYtermThm1proof}) in the proof of Theorem \ref{thm:generalMaxCut})  
will now depend on the graph weights, and can no longer be collapsed in general to a succinct result after summing over the edges as in (\ref{eqn:YYtermSumThm1proof}). This results in a substantially more complicated formula for $\langle C \rangle$. As a simple example, we consider the ring of disagrees with weights. 

\begin{cor}
For the weighted ring of disagrees, for an edge $(uv)$ 
with adjacent edges $(tu)$ and $(vx)$ we have 
 $$ 
\bra{\gamma \beta }  C_{uv}  \ket{\gamma \beta}  
=    \frac{w_{uv}}{2} + \frac{w_{uv}}{2} \sin 4\beta \sin \gamma w_{uv}  \cos \frac{\gamma(w_{tu} + w_{vx})}{2}
\cos \frac{\gamma(w_{tu} - w_{vx})}{2}  .$$
\end{cor}

Importantly, for a given a problem instance, (\ref{eqn:costBoundWeightedMaxSat}) may be efficiently maximized classically, with optimal value and angles depending on the particular problem graph and weights.

Finally, given further information such as the distribution of weights $w_{uv}$ over a given class of problem instances, it may be possible to use Theorem \ref{thm:weightedMaxCut} to obtain further quantities of interest, e.g., the expected value $\langle C\rangle$ taken with respect to this distribution.

\subsection{Directed Maximum Cut}
Another natural generalization is MaxCut on directed graphs.  
 In Directed MaxCut (MaxDiCut), we are given a \textit{directed} graph $G=(V,E)$ and we 
seek to find a 
subset of the vertices $L\subset V$, $R=V\setminus L$,  
such that the number of directed edges $|uv)$ with $u\in L$ and $v \in R$ is maximized. Here we use $|uv)$ to indicate the directed edge from $u$ to $v$. 

We emphasize that only directed edges from $L$ to $R$, and not those from $R$ to $L$, are counted by the objective function. Furthermore, the usual problem formulation considers weighted edges, and we seek to maximize the total edge weight from $L$ to $R$. For simplicity, for the remainder of the section we take all weights to be $1$. It is relatively straightforward to obtain a weighted version of the theorem below, similar to (undirected) Weighted MaxCut. 

Suppose each vertex $u\in V$ is the left endpoint of $\ell_u$-many edges and the right endpoint of $r_u$-many edges, and define $k_u = \ell_u - r_u$. 
Then $\sum_u \ell_u = \sum_u r_u =m$, and $\sum_u k_u = 0$.

Let the indicator variable $x_u$ be $1$ if vertex $u$ is assigned to $R$. 
The possible vertex partitions 
are again encoded with $n$ qubits. 
Recall that for 
MaxCut, the objective function for each edge $x_u \oplus x_v = \bar{x}_u x_v + x_u\bar{x}_v$ was encoded as the Hamiltonian $\frac12(I- Z_u Z_v)$. 
For MaxDiCut, 
the objective function counts the number of edges strictly leaving $L$, which for an edge $|uv)$ becomes
$$C_{uv} = \bar{x}_u x_v= \frac14(I+Z_u - Z_v - Z_uZ_v).$$
Observe that if both $|uv)$ and $|vu)$ are in the graph, then $C_{uv} + C_{vu} = \frac12 (I - Z_uZ_v)$, i.e., the combination $C_{uv} + C_{vu}$ acts exactly as an undirected edge in the sense of (\ref{eq:MaxCutObjHam}). 
(However, in this case at most one of $|uv)$ or $|vu)$ can be cut.)
We use this to simplify the objective Hamiltonian. 

Partition the edge set $E$ into the sets we call the \textit{directed} and \textit{undirected} edges
$$ D :=  \{|uv) \; \in E : \;\; |vu) \; \notin E  \}, \;\;\;\;\;  U := \{ (uv) : \;\;  |uv) \; \in E  \text{ and } |vu) \in E   \}, $$
where a pair of edges $|uv),|vu)$ is included as a single element $(uv)\in U$. 
Note that $|D|+2|U|=|E|=m$. 
Then 
the objective Hamiltonian $C=\sum_{|uv)\in E} C_{uv}$ becomes 
\begin{equation}   \label{eqn:CMaxDiCut}
C = \frac{m}{4}I + \frac14 \sum_{u\in V} k_u Z_u  -\frac14 \sum_{|uv)\in D} Z_uZ_v - \frac12  \sum_{(uv)\in U} Z_uZ_v. 
\end{equation}
Observe that the $ZZ$ terms are symmetric with respect to flipping the edge directions, and all information about the direction of each edge lies in the $ \frac14 \sum_{u\in V} k_u Z_u$ term. 

The analysis of QAOA$_1$ for MaxDiCut is similar to that of MaxCut, but considerably more complicated. 
We give results for oriented graphs and triangle-free directed graphs, which are two more manageable cases; it is relatively straightforward but complicated to extend the proof of the theorem below 
to arbitrary graphs.  
A graph is \textit{oriented} if it contains no symmetric edge pairs $|uv)$ and $|vu)$, i.e.,  $U=\emptyset$ and $D=E$.
A \textit{triangle} in a directed graph is defined to be any 
subset of three edges forming a cycle when the direction of each edge is ignored. 

For each $u \in V$, let $D_u \subset D$ be the edges in $D$ containing $u$, let $U_u \subset U$ be the edges in $U$ containing $u$, 
and define $d_u=|D_u|$ and $e_u=|U_u|$.

\begin{theorem}   \label{thm:MaxDiCut}
Consider QAOA$_1$ applied to 
MaxDiCut, 
with the quantities 
$d_u$, $e_u$, and $k_u$ for each vertex $u$ defined as above. 
\begin{itemize}
\item For an oriented graph, let $f_{uv}$ be the number of triangles containing an edge $|uv)$. 
Then 
\begin{equation}  \label{eqn:costBoundMaxDiCut}
 \bra{\gamma \beta} C \ket{\gamma \beta} 
 = \frac{m}{4} + \frac14 \sum_{u\in V} k_u \mathcal{K}_u 
+\frac18 \sum_{|uv)\in D} (\mathcal{D}_{u} + \mathcal{D}_{v} -\mathcal{T}_{uv} )  ,
\end{equation}
where 
we have the quantities 
\begin{eqnarray*}
\mathcal{K}_u &=& \sin (2\beta) \sin (\gamma k_u/2) \cos^{d_u} (\gamma/2),\\
\mathcal{D}_{u} &=& \sin( 4\beta)  \sin (\gamma/2)  \cos(\gamma k_u/2)  \cos^{d_u -1} (\gamma/2 ),\\
  \mathcal{T}_{uv}  &=&   \sin^2 (2\beta)  \cos^{d_u + d_v -2 - 2f_{uv}}\left(\tfrac{\gamma}{2}\right) 
\left(\cos\left(\tfrac{\gamma}{2}(k_u - k_v)\right)   - \cos^{f_{uv}} (\gamma)  \cos\left(\tfrac{\gamma}{2}(k_u + k_v)\right)  \right) .
\end{eqnarray*}

\item For a triangle-free directed graph, the overall expectation value is  
\begin{equation}  \label{eqn:expecCDirMaxSat}
 \bra{\gamma \beta} C \ket{\gamma \beta} 
 = \frac{m}{4} + \frac14 \sum_{u\in V} k_u K'_u 
+\frac18 \sum_{|uv)\in D} \mathcal{D}_{uv} 
+ \frac14  \sum_{(uv)\in U} \mathcal{U}_{uv} ,
\end{equation}
where 
we now have the quantities 
$\; K'_u = \cos^{e_u} (\gamma) K_u$, 
\begin{eqnarray*}
\mathcal{D}_{uv} &=& \mathcal{D}_u \cos^{e_u} (\gamma)  + \mathcal{D}_v \cos^{e_v} (\gamma) \\
&-& \sin^2 (2\beta) \sin( \tfrac{\gamma k_u}{2})   \sin( \tfrac{\gamma k_v}{2}) \cos^{d_u +d_v -2}(\tfrac{\gamma}{2}) \cos^{e_u + e_v} (\gamma), \\
U_{uv} &=& \sin( 4\beta) \sin(\gamma)   \left( \cos( \tfrac{\gamma k_u}{2})  \cos^{d_u} (\tfrac{\gamma}{2} ) \cos^{e_u-1} (\gamma)
+ \cos( \tfrac{\gamma k_v}{2})  \cos^{d_v} (\tfrac{\gamma}{2}) \cos^{e_v-1}( \gamma)
\right)  \\
&-& \sin^2 (2\beta) \sin( \tfrac{\gamma k_u}{2})   \sin( \tfrac{\gamma k_v}{2}) \cos^{d_u +d_v}(\tfrac{\gamma}{2}) \cos^{e_u + e_v - 2} (\gamma).
\end{eqnarray*}
\end{itemize}
\end{theorem}
%
The proof is again based on the Pauli Solver approach and is similar to but more involved than that of Theorem \ref{thm:generalMaxCut}. The proof details are given in Appendix \ref{app:QAOA}. 

In principle, further results for MaxDiCut may be derived using the 
theorem, as was done for MaxCut. However, the same difficulties apply in analyzing the performance for $p>1$. 
We leave further investigation of the performance of QAOA for MaxDiCut as a direction of future research.

\section{Discussion}
Using quantum computers to  
approximately solve hard optimization problems 
is an exciting new theoretical direction. Unfortunately, the most important problems remain open, and, as we have demonstrated, appear to elude simple resolutions.   

The results in this chapter are important first steps towards a fuller understanding of the QAOA algorithm. 
While we have not yet answered the most critical question, namely to characterize the performance of QAOA$_p$ generally, 
we have 
made important progress, improving significantly on known results for QAOA$_1$ applied to MaxCut. In particular, 
 we have shown exact analytic formulas for several results given in \cite{Farhi2014} that were found numerically. Moreover, we have presented the \textit{Pauli Solver} algorithm, which can in principle be used to derive performance bounds for a wider variety of problems, and can potentially assist in finding good QAOA parameters. 

As gate model quantum computers begin to 
come online over the next several years, we expect 
experimentation and empirical analysis to enable a significant expansion of our knowledge of quantum approximation algorithms and heuristics. 
In particular, smaller quantum computers may be useful for characterizing the performance of QAOA on larger quantum computers. 
For example, for problems consisting of 
clauses $C_\ell$  
each acting on a bounded number of variables $k \ll n$, such as MaxCut, 
a relatively small quantum computer (requiring much fewer than $n$ qubits) could be used to compute the quantities $\langle C_\ell \rangle$ for QAOA$_p$ with fixed $p$. 
These quantities then 
in turn could be used to characterize the expected QAOA output 
for the objective function $C=\sum_{\ell} C_\ell$ on $n$ variables, i.e., the performance of QAOA$_p$ on a much larger problem instance (to be executed on a larger $n$ qubit quantum computer). 

Generally, the performance of QAOA in practice will be highly dependent on the ability to find good angles. These may be found in advance through analysis similar to our demonstrated techniques, or found on an instance-by-instance basis by incorporating searching over angles as part of the QAOA algorithm. Finding further techniques for reducing the cost of this search is important towards improving the efficacy of the algorithm. 
The most important question is how QAOA performs (i.e., how the optimized approximation ratio scales) for $p>1$. In particular, results for $p=O(1)$ are enticing for application to early quantum computers, and results for, say, $p=O(\log(n))$ or $p=\textrm{poly}(n)$ are important for characterizing the power of QAOA itself. Unfortunately, we give evidence in this chapter that deriving performance bounds for $p>1$ is a difficult problem. 
We are, however, optimistic that the techniques of this chapter may be used to study MaxCut further, or to analyze the application of QAOA to other problems. 

In the next chapter, we give a generalization of QAOA that is particularly suitable to low-resource implementations for \textit{constrained} optimization problems. The techniques presented in 
this chapter similarly apply to the analysis of constrained problems; however, for these cases, the QAOA constructions themselves, and likewise the details of the analysis, become further complicated due to the additional constraints. 
A major breakthrough we leave for future investigation is to find 
new approaches to analyzing the performance of QAOA that are generally applicable.

%% file: _ch_QAOA2.tex
\input{macros2}

\chapter{Quantum Approximate Optimization with Hard and Soft Constraints}
\label{ch:QACOA}

\section{Introduction}
While some small-scale exploration of quantum 
algorithms and heuristics for 
approximate optimization 
beyond quantum annealing has been possible through 
classical simulation, 
the exponential overhead in such
simulations has greatly limited their usefulness. 
The next decade will see a blossoming of quantum algorithms 
as a broader and more flexible array of quantum computational hardware becomes
available.
The immediate question is: 
which 
algorithms 
should we prioritize
that will give us insight into the power and utility of quantum computers? 
One leading candidate is
the 
Quantum Approximate Optimization Algorithm (QAOA), 
for which a number of
tantalizing related results have been 
obtained~\cite{Farhi2014b,Farhi2016,Shabani16,Jiang17,Wecker2016training,wang2017quantum,Venturelli17}. 
As discussed in Chapter \ref{ch:QAOAperformance},  
QAOA facilitates low-resource implementations for 
unconstrained optimization problems, although the performance of QAOA for these problems remains open.  
It is important to 
derive 
constructions for even more general classes of problems, where we may not have good classical approximation algorithms at all, that in particular also exhibit low or modest resource requirements. 
Indeed, 
implementing such QAOA constructions 
to find approximate solutions may lead to the first examples of experimental quantum computers performing truly \textit{practically useful} computations.

In this chapter, we formally describe the Quantum Alternating Operator Ansatz (QAOA), 
extending the approach of Farhi \ea~\cite{Farhi2014} to encompass alternation
between more general families of operators.\footnote{As the Quantum Alternating Operator Ansatz generalizes the Quantum Approximate Optimization Algorithm, by design we use the same acronym \textit{QAOA} for both.} 
The essence of this extension is to consider the alternation of operators drawn from 
general parameterized families of unitaries, rather than only those 
that correspond to the time-evolution of a fixed local Hamiltonian with the time specified by the parameter. 
Thus, this ansatz supports the representation of a larger
and potentially more useful set of states than the original formulation.  
For cases that call for mixing only within a feasible subspace, 
refocusing on unitary operators rather than Hamiltonians leads to
a variety of possible 
mixing operators, 
many of which 
are much simpler and can be implemented more efficiently 
than 
those of the original framework. 
Such mixers are particularly useful for optimization problems with
\textit{hard constraints} that must always be satisfied, defining
a feasible subset of solutions, and \textit{soft constraints} which we seek to satisfy as many of as possible. 
Simple and efficient implementations are especially important towards enabling earlier experimental
exploration of quantum alternating operator approaches to a wide variety of 
potential applications, including 
approximate
optimization, exact optimization, and sampling problems.

We 
 specify a framework for this ansatz, laying out design criteria 
for constructing initial states, phase operators, and mixing operators. 
We then detail QAOA mappings of several important
optimization problems, including Maximum Independent Set, 
three graph coloring optimization problems, 
and the Traveling Salesman problem. 
The constructions described 
serve as prototypes for many other optimization problems; a compendium of mappings is included in \cite{hadfield2017quantum}. 

For each problem we show how a variety of different mixing operators may be constructed by 
combining local Hamiltonians and unitaries. In particular, we describe \textit{sequential} (ordered product) mixers, which are similar in form to a Trotterization step, 
and we show explicit circuits and bound their costs. 
Our constructions utilizing   
sequential mixers are simple by design and 
exhibit relatively low resource scaling, as desired for early quantum hardware. 
We summarize 
these implementation 
results in 
Table~\ref{tab:QAOAconstructions} below. 
We emphasize that our ansatz encompasses even more general mixing operators, generally requiring higher implementation cost; 
investigating the trade-off between increased resource requirements and algorithm performance is an important direction for future work. 

It is worthwhile to remark on the relation between these mappings and
those for quantum annealing and adiabatic quantum optimization. 
Because current quantum annealers have a fixed driver Hamiltonian
(which is similar to the mixing Hamiltonian in the QAOA setting),  
all problem dependence must be captured in the objective Hamiltonian on such devices.
Hence, to deal with hard constraints, the 
typical strategy is to add extra terms to the objective Hamiltonian which penalize states encoding infeasible solutions such that these states are avoided;~
see, e.g.,~\cite{biswas2017nasa,rieffel2015case,LucasIsingNP}. 
But this approach means that the algorithm must search a much larger
space than would be necessary if the evolution was somehow restricted to feasible configurations. 
This issue, and other drawbacks, led 
Hen \& Spedalieri~\cite{Hen2016quantum} and Hen \& Sarandy~\cite{Hen2016driver}
to suggest a different approach for adiabatic quantum optimization 
in which the standard driver Hamiltonian is replaced by an alternative one 
that, given a feasible initial state, confines the evolution to the feasible subspace. 
We apply similar ideas to 
the quantum circuit model, leading to a much more general approach with wider applicability. 
Many of 
the results of this chapter can also be found in \cite{hadfield2017qaoaPMES,hadfield2017quantum,hadfield2018representation}. 

\vskip 2pc
\begin{table}[h]
\begin{center}
\begin{tabular}{| c  || c | c |}
	\hline
	QAOA$_p$ Problem & \# of Qubits & \# of Basic Gates \\ 
	\hhline{|=||=|=|}
	Quadratic Unconstrained Binary Optimization  & $n $ & $O(p(m+n))$ \\	
	\hline
	Max Independent Set  & $n + 1$ & $O(p(m+n))$ \\
	\hline
	Max $k$-Colorability (Max $k$-Cut)  & $kn$ & $O(pk(m+n))$ \\
	\hline
	Max $k$-Colorable Induced Subgraph  & $(k+1)n+1$ & $O(p(km+n))$ \\
	\hline
	Min Chromatic Number $\;$($k=D_G+O(1)$)   & $(n+1)k+1$ & $O(p(k^2m + kn))$ \\
	\hline
	Traveling Salesman  & $n^2$ & $O(pn^3)$ \\
	\hline
	Single Machine Scheduling (Min Total Tardiness)  & $nP$ & $O(pn^2P)$ \\  
	\hline
\end{tabular}
\end{center}
\caption{Summary of implementation costs for creating a QAOA$_p$ state for the indicated problems with $n$ variables. 
The initial state $\ket{s}$, phase operator $U_P(\gamma)$, and mixing operator $U_M(\beta)$ for each problem are given in the following sections. 
In each case, the mixing operator may be replaced by $U_M^r(\beta)$, $r=O(1)$, without affecting the scaling of the cost estimates. Quadratic Unconstrained Binary Optimization is addressed in Chapter \ref{ch:QAOAperformance}, with $m$ specifying the number of quadratic terms. For the last problem, $P$ is the sum of the processing times for each job. Basic gates are defined to consist of  
CNOT and arbitrary 
single-qubit gates.}
\label{tab:QAOAconstructions}
\end{table}

\section{The Quantum Alternating Operator Ansatz}
We formally describe the \textit{Quantum Alternating Operator Ansatz} (QAOA),  
generalizing the approach of Farhi \ea~\cite{Farhi2014}.
QAOA, in our sense, encompasses a more general class of quantum states that may be
algorithmically accessible and useful. 

We consider here QAOA for approximate optimization problems, though it may also have other applications,  
such as, for example, exact optimization or sampling problems  \cite{Jiang17,Wecker2016training,Farhi2016}.

An instance of an \emph{optimization problem} is a pair $(\domain, \objFunc)$,
where $\domain$ is the \emph{domain} (set of valid solutions) and 
$\objFunc: F \rightarrow \mathbb R$ 
is the \emph{objective function} to be optimized. 
%
Recall 
from Section \ref{sec:QAOAdetails} that 
a QAOA mapping is the same between the maximization or minimization versions of a given problem, up to trivial sign flips, 
and possibly a different choice of initial state. Hence, in this chapter we we will generally not be concerned with which is the case and we consider optimization problems generally.

Earlier in Chapter \ref{ch:QAOAperformance} we considered 
\textit{Hamiltonian-based QAOA} (H-QAOA) \cite{Farhi2014}, the subclass of QAOA circuits in
which both the phase operators 
 $\phaseUnitary(\gamma) =  e^{-i \gamma
\phaseHam}$ and the mixing operators 
$\mixUnitary(\beta) = e^{-i \beta
\mixHam}$ correspond to time-evolution under Hamiltonians $\phaseHam$ and
$\mixHam$, respectively. 
%
The quantum approximate optimization algorithm as originally proposed fits 
within this paradigm, whereas 
our construction below encompasses much more general operators; we will see 
a variety of explicit examples in the subsequent sections.

Let $\domainQ$ be the Hilbert space of dimension $|\domain|$, whose standard
basis we take to be 
$\left\{\ket{\mathbf{x}} : \mathbf{x} \in \domain\right\}$.
A general $\QAOA$ circuit is defined by  
two parameterized families of operators on $\domainQ$: 
\begin{itemize}
\item a family of 
\emph{phase separation operators} $\phaseUnitary(\gamma)$ that depends on the
\emph{objective function} $\objFunc$, and 
\item a family of \emph{mixing operators} $\mixUnitary(\beta)$ that depends on the domain and its structure, 
\end{itemize}
where $\gamma$ and $\beta$ are real parameters.
A $\QAOA_p$ circuit consists of $p$ alternating applications of operators from these two families,  
\begin{equation}  \label{eq:QAOApOp}
\QAOAcirc_p(\boldsymbol \gamma, \boldsymbol \beta)
=
\mixUnitary(\beta_p) \phaseUnitary(\gamma_p)
\cdots 
\mixUnitary(\beta_1) \phaseUnitary(\gamma_1). 
\end{equation}
The Quantum Alternating Operator Ansatz 
then consists of the states 
that can be represented (i.e., generated) by the application of such a circuit to a suitably simple initial state $\ket{\initial}$, 
\begin{equation}
\ket{\boldsymbol \gamma \boldsymbol \beta}
=
\QAOAcirc_p(\boldsymbol \gamma, \boldsymbol \beta)
\ket{\initial}.
\end{equation}

Hence, for a given optimization problem, a {\it QAOA mapping\/} 
consists of 
a family of phase separation operators, a family of mixing operators, and a
starting state. Once a problem encoding onto qubits is selected, the QAOA mapping can be used to compile directly to a gate-level quantum circuit. 
Note that different problem encodings lead to different gate and qubit costs. 

Constrained optimization problems require optimization over \textit{feasible} solutions,  
generally a subset of a \emph{configuration space} (such as $\{0,1\}^n$) that is 
often specified by a set of Boolean 
predicates called 
\textit{feasibility} (or \textit{hard}) \textit{constraints}, 
which are satisfied by feasible solutions. 
Hard constraints result both from the problem itself and how it is encoded, 
often specifying nontrivial subsets. 
For implementation on 
quantum hardware, 
it is typically easier to encode the entire configuration space onto qubits, 
with the problem domain subsumed by the (more general) 
\emph{feasible subspace}, 
which results from both the natural structure of the domain and 
how the configuration space is encoded. 
For convenience, we will use the terms domain and feasible subspace interchangeably. 
Quantum states (generally, superpositions) lying entirely in the feasible subspace are called \emph{feasible states}. 

For a given problem, our goal is to design 
families of mixing operators that preserve feasibility; 
then, given a feasible initial state, 
the QAOA state will remain feasible always (for all possible algorithm parameters and $p$), and, in particular, the final QAOA state will be feasible, 
so any computational basis measurement performed 
is guaranteed to produce a feasible solution. 
This avoids all difficulties related to dealing with infeasible states directly; in particular, without this property many measurement outcomes could yield infeasible solutions, which 
would have to be carefully accounted for in analyzing the success probability and performance of the algorithm, or dealt with by some other means. We formalize this idea as design criteria in Section \ref{sec:designCrit} below. 

For an objective function $f$ 
we define
$\objHam$ to be the Hamiltonian that acts as $f$ on basis states 
\begin{equation}     \label{eq:functionHam}
    \objHam \ket{\mathbf{x}} = \objFunc(\mathbf{x}) \ket{\mathbf{x}}.
\end{equation}
In prior work, the domain $\domain$ is
the set of all $n$-bit strings, 
$\phaseUnitary(\gamma) = e^{-i\gamma\objHam}$, and
$\mixUnitary(\beta) = e^{-i\gamma\stdDriver}$, where, with just one exception, the mixing Hamiltonian is $\stdDriver = \sum_{j=1}^n X_j$.
Recall $X_j$ denotes the Pauli matrix $X$ acting on the $j$th qubit,  
and similarly for $Y_j$ and $Z_j$.
The one exception is Section~VIII of~\cite{Farhi2014}, which discusses a variant for 
the Max Independent Set problem, in which $\domain$ is the set of
bit strings 
encoding independent sets, 
and the mixing operator is 
$\mixUnitary(\beta) = e^{-i\gamma H_M}$ where
\begin{equation}
\braket{\x}{H_M | \y} = 
\begin{cases}
1 & \quad \x, \y \in \domain \text{ and } \hamDist(\x,\y) = 1\\
0 & \quad \text{otherwise}.
\end{cases}
\end{equation}
The Hamiltonian $H_M$ 
connects feasible states with Hamming distance one 
(i.e., independent sets differing only by a single vertex). 
However, Section~VIII of~\cite{Farhi2014} does not discuss the implementability
of $\mixUnitary(\beta)$. We consider this problem in Section \ref{sec:QAOAbitstring} and derive the implementation cost shown in Table \ref{tab:QAOAconstructions} using a closely related mixing operator. 

We extend 
this approach to applying QAOA to constrained optimization problems, 
with a focus towards implementability, both in the short and long terms. 
We also build on the 
ideas developed for adiabatic quantum optimization (AQO) by 
Hen and Spedalieri~\cite{Hen2016quantum} and 
Hen and Sarandy~\cite{Hen2016driver}, 
though the gate-model setting of QAOA leads to different implementation considerations than those for AQO.\@
For example, Hen \ea~identified Hamiltonians of the form
$\mixHam = \sum H_{j,k}$, where $H_{j,k} = X_j X_k + Y_j Y_k$, 
as useful in the AQO setting 
for restricting state evolution to the feasible subspace for certain problems. 
By incorporating this restriction directly into the mixing operator itself, it was found that the resources required for implementation could be substantially reduced as compared to different approaches for ensuring feasibility, such as the standard approach of including extra terms in the objective Hamiltonian to penalize infeasible states. 
Analogously, the mixing unitary 
$\mixUnitary(\beta) = e^{-i\beta\mixHam}$
meets our design criteria, specified in the next section, 
for applications of QAOA to a number of 
optimization problems including many of those 
considered in~\cite{Hen2016quantum,Hen2016driver}.
Since the Hamiltonians $H_{j,k}$ and $H_{i,l}$ do not commute in general, 
compiling $\mixUnitary(\beta)$ to two-qubit gates is nontrivial. 
One could 
Trotterize, or use higher-order splitting 
formulas, to approximately implement $\mixUnitary(\beta)$ 
in terms of exponentials of individual $H_{j,k}$, which are each efficiently simulatable; 
recall the discussion of Hamiltonian simulation in Chapter \ref{ch:HamSim}. 
Alternatively, $\mixUnitary(\beta)$ can be implemented efficiently using techniques related to the quantum Fourier transform~\cite{verstraete2009quantum}.  
Instead, in particular, we will propose 
alternative mixing operators such as
$$\mixUnitary(\beta) = e^{-i\beta H_{S_\ell}}\cdots e^{-i\beta H_{S_2}}e^{-i\beta H_{S_1}},$$ where 
the 
$H_{j,k}$ have been partitioned into $\ell$ subsets (partial sums) $S_1,\dots,S_\ell$ containing only 
 mutually commuting pairs. 
Such mixing operators may often be selected by design to be 
much simpler to implement than~$e^{-i\beta H_M}$, 
motivating in part our more general ansatz.

We remark that, clearly, there are 
obvious further generalizations in which $\phaseUnitary$ and $\mixUnitary$ are
taken from families parameterized by more than a single parameter.  
Such operator families could be designed to take advantage of specific quantum hardware. 
For example, in~\cite{Farhi2017} a different free parameter for every term in the mixing Hamiltonian is considered. 
In this chapter, we 
consider 
one-dimensional families of operators, 
given that this is already a 
rich area of study, 
with the task of finding good parameters $\gamma_1, \ldots, \gamma_p$
and $\beta_1, \ldots, \beta_p$ already challenging enough due to the 
curse of dimensionality~\cite{wang2017quantum}.
A larger parameter space may support more effective circuits, but further 
increases the difficulty of finding good parameters. 

\subsection{Design Criteria}\label{sec:designCrit}
Here, we 
specify design criteria for the three components of
a QAOA mapping of a problem, namely, the initial state, the phase operators, and the mixing operators. 
\paragraph{Initial State.} We require that the initial state $\ket{\initial}$ be feasible, and moreover it must be 
\emph{trivial} to implement, by which we mean that it can be created by
a constant-depth 
quantum circuit from the $\ket{0\dots 0}$ state. 
The standard initial state $\ket{+\cdots+}$ from \cite{Farhi2014} 
may be obtained from the $\ket{0\dots 0}$ state by a depth-$1$ circuit applying 
a Hadamard gate $\textrm{H}$ to each qubit.  
For our purposes, it is often convenient to select the initial state to be a single feasible solution $\ket{\x}$, $\x\in F \subset \{0,1\}^n$, which 
can be prepared by a depth-$1$ circuit consisting of up to $n$ single-qubit bit-flip operations $X$. In such cases, 
the initial phase operator applies a global phase and can be discarded,  
and hence we may reindex 
to consider QAOA
as starting with 
a single mixing operator $\mixUnitary(\beta_0)$ applied to 
the basis state $\ket{\x}$.
This $0th$ round of QAOA creates a superposition state 
\begin{equation}
 \ket{s} = \mixUnitary(\beta_0)\ket{\x}.
\end{equation}

We remark that the constant-depth criterion could be relaxed to logarithmic depth if needed.
It should not be relaxed too much:
relaxing the criterion to polynomial depth would 
obviate the usefulness of the
ansatz as a model for a strict subset of states producible via
polynomially-sized quantum circuits.  
Algorithms with more 
complicated initial states 
may be considered
hybrid algorithms, with an initialization part and a QAOA part. 
\paragraph{Mixing unitaries (``Mixers'').}  We require the family of 
mixing operators $\mixUnitary(\beta)$ to  
\begin{itemize}
\item \textit{preserve the feasible subspace}: for all values of the parameter $\beta$ the resulting unitary
takes feasible states to feasible states, and
\item \textit{explore the feasible subspace}: provide possible transitions between all feasible solutions. 
More concretely, for any pair of feasible 
computational basis 
states $\x, \y \in \domain$,
there is some parameter value $\beta^*$ 
and some positive integer $r$
such that the corresponding mixer
\textit{connects} those two states: 
$\left|\Braket{\x | \mixUnitary^r(\beta^*) }{ \y}\right| > 0$. $\;$ (Note that $\mixUnitary^r(\beta)$ denotes  $(\mixUnitary(\beta))^r$.)
\end{itemize}

We remark that these criteria are 
intentionally not overly restrictive, facilitating the design of a variety of mixing operators with different trade-offs. In particular, given a general mixing operator~$\mixUnitary(\beta)$, 
applying it $r$ times gives the operator $\mixUnitary^r(\beta) \neq \mixUnitary(r\beta)$, which may provide transitions between states not connected by any $\mixUnitary(\beta)$ alone. If $r=O(1)$, then the increased overhead to implement $\mixUnitary^r(\beta)$ is relatively small. Note that for the special case of H-QAOA, 
the mixing operator $\mixUnitary(\beta)=e^{-i\beta H_M}$ satisfies 
 $\mixUnitary^r(\beta)=\mixUnitary(r\beta)$, so  
repetitions of the mixing operator do not give any advantages in this case. 
\paragraph{Phase separation unitaries.} We require 
 the family of 
phase separation operators~$\phaseUnitary(\gamma)$ 
to be diagonal in the computational basis. 
We take 
\begin{equation}  \label{eq:PhaseOpGeneral}
\phaseUnitary(\gamma) = e^{-i \gamma \objHam},
\end{equation}
up to trivial global phase terms which act as $e^{-i \gamma a}I$, $a \in \reals$, 
and may be ignored. 
In the constructions of this chapter we consider only phase separators where 
$H_f$ represents the classical objective function $f$, though more general types of phase separators may be considered (e.g., using $H_{\tilde{f}}$ where $\tilde{f}$ is a simpler to implement approximation of $f$). \\

Together, these criteria \textit{restrict state evolution to the feasible subspace}. 
In particular, all computational basis measurements are guaranteed to return a feasible string. 
We remark that 
the restriction to feasible states 
often allows for substantial simplification of the phase separation operator, reducing its implementation cost, as we shall see in several of the problem constructions we study. 

\subsection{Simultaneous and Sequential Mixers}  \label{sec:SimSeqMixers}
The implementation of diagonal 
phase operators 
of the form (\ref{eq:PhaseOpGeneral}) 
was addressed in Chapter  
\ref{ch:QAOAperformance}, and we give several more general results 
in Section \ref{sec:QAOtoolkit} below. 
Here, we consider the construction of mixing operators. 
By deriving 
simple transformations between states that preserve feasibility, 
we can map these transformations to Hamiltonians $B_j$, and then combine the Hamiltonians and their corresponding unitaries ${\rm exp}(-i\beta B_j)$ to yield mixing operators satisfying the design criteria.   

The original formulation of QAOA 
considered the domain $\{0,1\}^n$ of all bit strings and 
used the mixing operator 
$U_M(\beta)={\rm exp}(-i\beta B)$, with $B= \sum_{j=1}^n X_j$. (For the remainder of this chapter it will be convenient to use $B$ instead of $H_M$ to denote mixing Hamiltonians.) 
As 
the $X_j$ mutually commute and each acts on a single qubit, we have 
$U_M(\beta) = \prod_{j=1}^n e^{-i\beta X_j}$ which 
may be implemented with $n$ many $X$-rotation ($R_X$) gates and depth $1$. 
In the subsequent sections, we give constructions for problems 
with nontrivial domains. 
This will result, generally, in mixing Hamiltonians of the form 
$$B= \sum_{j=1}^\ell B_j,$$ 
where each $B_j$ acts on 
a 
subset of the qubits and $\|B_j\|=O(1)$ (typically, $\|B_j\|=1$). 
However, $[B_j, B_k] \neq 0$ in general,  
so $e^{-i\beta B} \neq \prod_{j=1}^\ell e^{-i\beta B_j}$, 
and more sophisticated 
Hamiltonian simulation
techniques are required to implement ${\rm exp}(-i\beta B)$. 
We refer to Hamiltonian-based mixers of the form ${\rm exp}(-i\beta B)$ as \textit{simultaneous mixers}. 

Indeed, 
suppose the decomposition $B= \sum_{j=1}^\ell B_j$ satisfies the following properties:
\begin{itemize}
\item for each $j$ and for any $\beta$, 
the exponential ${\rm exp}(-i\beta B_j)$ can be efficiently implemented, and 
\item for each $j$, $B_j$ maps feasible states to feasible states. 
\end{itemize}
Then, 
for any permutation $j_1,\dots,j_\ell$ of $1,\dots,\ell$, 
the \textit{sequential mixer} defined as the ordered product
\begin{equation}  \label{eq:seqMixer}
U_M (\beta) =  e^{-i\beta B_{j_\ell}} \dots e^{-i\beta B_{j_2}} e^{-i\beta B_{j_1}}
\end{equation}
also preserves feasibility.  
The cost of implementing $U_M(\beta)$ is the cost of implementing the~$\ell$ many~$e^{-i\beta B_j}$ operations. 

Importantly, different orderings of the exponentials $e^{-i\beta B_j}$ in the product 
(\ref{eq:seqMixer}) result in inequivalent operators. 
We may associate a sequential mixer to each of the $\ell!$ possible orderings of the~$e^{-i\beta B_j}$, some of which result in equivalent operators. If two operators $e^{-i\beta B_j}$ and $e^{-i\beta B_{k}}$, $j\neq k$, act on disjoint sets of qubits, they may be implemented 
in parallel, and moreover $e^{-i\beta B_j}e^{-i\beta B_k}=e^{-i\beta (B_j + B_k)}$. 
Therefore, selecting an ordering where many such pairs are adjacent can significantly reduce the resulting circuit depth. Generally, given a disjoint partition $P=\{P_1,P_2, \dots, P_\alpha\}$, $\alpha \leq \ell$, of $[\ell]=\{1,\dots,\ell\}$ (i.e., $\cup_j P_j = [\ell]$ and $P_i \cap P_j = \emptyset$ for $i \neq j$), we define the 
the \textit{partitioned sequential} mixer to be 
\begin{equation}  \label{eq:partMixer}
U^{(P)}_M = \prod_{i=1}^\alpha \prod_{j \in P_i} {\rm exp}(-i\beta B_j).
\end{equation}
Then, if each $P_i$ 
contains $B_j$   
that act on disjoint sets of qubits, it may be possible to implement $U^{(P)}_M$ with much lower circuit depth than that of an arbitrary partition (i.e., an arbitrary ordering). 

On the other hand, 
it is easy to see that the implementation costs of the sequential and simultaneous mixers are polynomially related. 
Indeed, using the Strang ($k=1$) splitting formula, from \cite{PZ12} (cf. equation (\ref{eq:PZ})) the cost of simulating ${\rm exp}(-i\beta B)$, 
with $B=\sum_{j=1}^\ell B_j$ and $\|B_j\|=O(1)$, 
is at most 
$N$ times the maximal cost of 
any ${\rm exp}(-i\beta B_j)$, where the number $N$ of such exponentials is at most 
$$ N = O(\ell^2 \beta  (\ell \beta /\e)^{1/2}   ). $$
Using higher order splitting formulas ($k>1$) reduces the exponent above from 
$1/2$ to $1/2k$. 
Thus for $\ell = {\rm poly}(n)$, $\beta=O(1)$, and accuracy $\e=1/{\rm poly}(n)$, the 
cost is polynomial in $n$. 
Still more sophisticated Hamiltonian simulation algorithms could be used to reduce the cost dependence on $\e$ to $O(\log(1/\e))$, though potentially requiring more complicated implementations. 

As we are particularly interested in applications to early quantum computers, we will 
focus on 
sequential mixers similar to (\ref{eq:seqMixer}) and (\ref{eq:partMixer}) for deriving implementation cost estimates in our constructions to follow. Generally, these mixers will be defined up to the order of exponentials in the product, or equivalently, a corresponding partition. 
As partitions may be selected on an instance-by-instance basis, 
and, moreover, optimized for compilation to specific gate sets, 
we will not consider them in detail here; 
possible partitions and their selection are discussed in 
 \cite{hadfield2017quantum}. Moreover, as mentioned, 
 each mixer $U_M(\beta)$ we propose can always be replaced by $U_M^r(\beta)$, $r=O(1)$, with $r$ times the implementation cost, for potentially more rapid mixing.  
 With these caveats in mind, it suffices for each of our constructions to specify a single sequential mixer.

Our main technique will be to construct mixers based on \textit{local mixing rules}, which correspond  to Hamiltonians $B_j$ acting on a small number of qubits, possibly controlled by a number of other qubits. In each case, the $B_j$ will themselves preserve feasibility, and the operators ${\rm exp}(-i\beta B_j)$ will be efficiently implementable. In the remainder of this chapter 
we demonstrate the ideas of this section by deriving explicit constructions for a variety of problems. 

In future applications, as more powerful quantum computing devices come online, simultaneous (i.e., Hamiltonian-based) mixers may be more appealing. It is an important future research problem to quantify the performance of QAOA for given problems with respect to simultaneous or the various possible sequential mixers, and the trade-offs between performance and implementation cost.

\subsection{Constraint Satisfaction via  Commuting Operators} 
We show here how mixing operators that preserve feasibility may be derived from the commutation properties of the Hamiltonians for a problem and its QAOA construction. 
 
Suppose that, for a given problem, the feasible subspace is specified exactly as the ground state (minimal eigenvalue) eigenspace of a Hamiltonian $A$, 
which typically encodes a suitable function, i.e., the hard constraints.   
For example, we may have $H_A = \sum_j H_{g_j}$, where the $H_{g_j}$ encode Boolean functions $g_j$ such that $\bigvee_j g_j (x) = 0$ if and only if $x$ is feasible. 
As $H_A$ is diagonal, it trivially commutes with the objective Hamiltonian $[H_A,H_f]=0$. 

Now consider a Hamiltonian-based mixing operator 
$U_M (\beta) = e^{-i\beta B}$. 
We require $[B,H_f]\neq 0$, else, clearly, the QAOA dynamics will be trivial. 
However, if we can select, somehow, $B$ such that $[B,H_A]=0$, then evolution under linear combinations of $B$ and $H_f$ is guaranteed to not mix between the eigenspaces of $H_A$. 
Thus, if the initial state is feasible, the QAOA evolution using such Hamiltonians $B$ and $H_f$ preserves feasibility at all times. 

This was observed in \cite{Hen2016quantum,Hen2016driver} 
for adiabatic quantum optimization, 
where it was shown that mixing Hamiltonians satisfying these properties could be used instead of penalty Hamiltonian terms, with several advantages in that setting including reduced resource requirements for implementation. We extend these ideas to the quantum gate model with more general unitaries, which provides a useful  
tool for finding mixing operators satisfying our design criteria. 

For a Hamiltonian $H_A$ on $n$ qubits, 
we say that a unitary operator $U$ \textit{preserves eigenspaces of} $H_A$ if for any eigenvector $\psi$ of $H_A$ with eigenvalue $a$, 
$U\psi$ is also an eigenvector of $H_A$ with eigenvalue~$a$. 
Clearly, this is a stronger condition than 
preserving feasibility. 

\begin{prop}  \label{prop:commutingHams1}
A unitary operator $U$ preserves eigenspaces of $H_A$ if and only if $[U,H_A]=0$. 
\end{prop}
\begin{proof}
Suppose $[U,H_A]=0$, which is trivially equivalent to $UH_AU^{-1} = H_A$. 
Then for $\psi$ such that $H_A\psi=a\psi$, we have $H_A( U \psi)= U H_A \psi = a (U \psi)$ as desired.
For the other direction, suppose $U$ preserves eigenspaces and 
consider an arbitrary eigenvector $\psi$ of $H_A$. Then 
$[U,H_A]\psi = UH_A\psi - H_AU\psi = aU\psi - aU\psi = 0$. As the eigenvectors (including degeneracy) of a Hermitian operator $A$ give a basis for the Hilbert space of $n$ qubits, this suffices to show  $[U,H_A]=0$.
\end{proof}

\begin{prop}    \label{prop:commutingHams2}
Consider unitary operators $U=U_\ell U_{\ell-1} \dots U_1$, with $U_j = e^{-i \alpha_j H_j}$, $\alpha_j \in \reals$. Then $[H_j,H_A]=0$ for  $j=1,2,\dots, \ell$ is a sufficient but not necessary condition for $U$ to preserve eigenspaces of $H_A$ for all $\alpha_j$.
\end{prop}
\begin{proof}
The condition $[H_j,H_A]=0$ implies $[U_j,H_A]=0$ 
from the series expansion of $e^{-i \alpha_j H_j}$, and hence $[U,H_A]=0$, so $U$ preserves eigenvalues by Proposition \ref{prop:commutingHams1}. This argument holds for any~$\alpha_j $. 
To see that the conditions 
are not necessary, consider 
a single qubit with $H_A=Z$ and $B=X$. Then $U = e^{-i 2\pi B} = I$ satisfies $[U,H_A]=0$, but $[B,H_A]=[X,Z]=-2iY\neq 0$. 
\end{proof}
These propositions are general and apply to quantum algorithms beyond QAOA or quantum annealing. 
For QAOA, Proposition \ref{prop:commutingHams2} implies that 
for a mixer 
$U_M(\beta)= e^{-i\beta H_1}e^{-i\beta H_2}\dots e^{-i\beta H_m} $, a sufficient condition for the quantum state to remain feasible is that the initial state is feasible and $[H_j,H_A]=0$ for $j=1,\dots,m$.  
Thus 
it suffices to consider Hamiltonian  
commutators to ensure our mixing operators satisfy the desired design criteria. 

We elaborate on how to select and construct such Hamiltonians and give several examples in the remainder of the chapter.

\section{Design Toolkit for Quantum Optimization}   \label{sec:QAOtoolkit}
\input{_QAOtoolkit}

\section{Mappings on Bits}  \label{sec:QAOAbitstring}
We first consider problems where the configuration space is naturally expressed as the set of $n$-bit strings. 
Recall that \textit{unconstrained} problems, where every string is feasible, were considered in 
Section \ref{sec:QUBO}, in particular for the case of quadratic objective functions. 
Here we consider \textit{constrained} binary optimization, and show how a suitable generalization of the mixing operator facilitates the application of QAOA to such problems.

\subsection{Max Independent Set}

\paragraph{Problem:} Given a graph $G=(V,E)$, with $|V|=\numVars$ and $|E|=\numConstraints$, 
find the largest cardinality subset $V' \subset V$ of mutually nonadjacent vertices. 

No 
polynomial-time classical algorithm exists for Max Independent Set unless P=NP~\cite{trevisan2004inapproximability}, and the best algorithms known for general graphs give approximations within a polynomial factor.\footnote{Max Independent set is in fact complete for Poly-APX~\cite{bazgan2005completeness}, the class of problems efficiently approximable to within a ${\rm poly}(n)$ factor.}  
On bounded degree graphs with maximum degree $D_G \geq 3$, Max Independent Set can be approximated to $(D_G +2)/3$~\cite{bazgan2005completeness}, but remains APX-complete~\cite{papadimitriou1991}.

Our construction generalizes that of Sec.~VII of~\cite{Farhi2014}. 
The configuration space is the set of $n$-bit strings $x=x_1x_2\dots x_n$ representing subsets of vertices $V' \subset V$, 
where $i \in V'$ if and only if the indicator variable $x_i = 1$. The domain is the subset of $n$-bit strings corresponding to independent sets of $G$. Note that the domain is dependent on the problem instance.

The objective function may be written $\objFunc(x) = \sum_{j=1}^n x_j $, which counts the number of vertices in $V'$, and maps to the Hamiltonian 
\begin{equation}
H_f =  \sum_{u\in V} \frac12 (I-Z_{u}) = \frac{n}{2}I -  \frac12 \sum_{u\in V} Z_{u} .
\end{equation}
Dropping the constant (identity matrix) term, which affects the algorithm dynamics only trivially via a global phase, the phase operator $e^{-iH_f}$ becomes 
\begin{equation}
\phaseUnitary(\gamma) = e^{i  \frac{\gamma}{2} \sum_{u\in V} Z_{u}} = \prod_{u\in V} e^{i \frac{\gamma}{2} Z_{u}} = 
\prod_{u\in V} R_{Z_u} (-\gamma). 
\end{equation}
Clearly $\phaseUnitary(\gamma)$ can be implemented with $n$ single-qubit ($Z$-rotation) $R_Z$ gates and depth $1$ (the gates do not overlap so can be implemented simultaneously).

We remark that $H_f$ gives the correct value of $f(x)$ on the feasible subspace of states representing independent sets, but gives erroneous values on infeasible states.    
Typical methods \cite{LucasIsingNP,Hen2016quantum,Hen2016driver} for dealing with infeasible states require additional complicated Hamiltonian terms to be added to~$H_f$, whereas our approach avoids this. 
Hence, restricting state evolution to the feasible subspace allows for simple low-cost phase operators. 

As an initial state, we may take $\ket{\initial} = \ket{0}^{\otimes n}$ which encodes the empty set and is assumed to be trivial to prepare. 
Alternatively, suppose we used a classical algorithm or heuristic to find an approximate solution $y$. 
Then the initial state $\ket{\initial} = \ket{y}$ could be used, with cost at most $n$ many $X$-gates and depth $1$, in addition to the cost of the classical preprocessing. Both of these states are clearly feasible, and the latter is problem instance dependent. 

Following our design criteria of Sections~\ref{sec:designCrit} and~\ref{sec:SimSeqMixers}, we define two mixing operators. In order to preserve feasibility, we will utilize controlled quantum operations. 

Observe that given an independent set $V'$, adding a vertex $w \notin V'$ to $V'$ preserves feasibility only if none of the neighbours (adjacent vertices) of $w'$ are already in $V'$. 
On the other hand, we can always remove any vertex $w \in V'$ without affecting feasibility. 
Combining these properties 
in a reversible way gives the following feasibility-preserving transformation rule.  

{\bf Mixing Rule:} flip the bit $x_w$ if and only if $\bar{x}_{v_1  } \bar{x}_{v_2 }\dots \bar{x}_{v_\ell }=1$, where $v_1,\dots,v_\ell$ are the 
vertices adjacent to $w$. 
Using the results of Section \ref{sec:QAOtoolkit}, 
we 
encode this rule as the Hamiltonian
\begin{equation}  \label{eqn:driverIndepSet}
B_{u} =   (\bar{x}_{v_1  } \bar{x}_{v_2 }\dots \bar{x}_{v_\ell }) \cdot X_u 
 = \frac{1}{2^{\ell}} X_u \; 
\prod_{j=1}^{\ell} (I+Z_{v_j }). 
\end{equation}
Exponentials of such Hamiltonians correspond to controlled unitaries. 
 Writing the control predicate as $f_u = \prod_{v \in \neighborFunc(u)} \bar{x}_{v}$, we define  
 the operator 
\begin{equation}
 U_{M,u}(\beta) = e^{-i\beta B_u} = \Lambda_{f_u} ( e^{-i\beta X_u}),
\end{equation} 
which is a multiqubit controlled rotation that applies an $X_u$-rotation to a basis state only if the control condition $f_u$ is true. Explicitly, for a basis state $\ket{x}$, $x\in\{0,1\}^n$, $U_{M,u}$ acts as 
%
\begin{equation}  \label{eq:ctrlMixerOnBasisStates}
U_{M,u}(\beta)\ket{x} =  ( f_u(x) \cos\beta + \overline{f_u}(x) )\ket{x}    -i  f_u(x) \sin \beta  \ket{x_1 \dots x_{u-1}\overline{x}_u x_{u+1} \dots x_n},
\end{equation}
%
so clearly each $U_{M,u}(\beta)$ preserves feasibility. Furthermore, clearly a sequence of such transformations connects every independent set to the empty set, and vice versa, so there exists a sequence connecting any two feasible states. 

 From the Hamiltonian $B=\sum_u B_u$, we define two related but inequivalent mixers:
 \begin{itemize}
\item the \emph{simultaneous} (Hamiltonian-based) 
controlled-$X$ mixer 
$\;\;\; U_M^{(H)} (\beta) = \timeEv{\beta B},$ 
\item the \emph{sequential} (partitioned) controlled-$X$ mixer 
$\;\;\; \mixUnitary(\beta) =  \prod_{u\in V}  U_{M,u}(\beta). $ 
\end{itemize}

The simultaneous mixer, while consistent with the original proposal of \cite{Farhi2014}, 
is nontrivial to implement because the Hamiltonians $B_u$ in $B$ do not mutually commute in general. 
Indeed, 
the primary advantage of the sequential mixer is that it 
results in much simpler quantum circuits. 
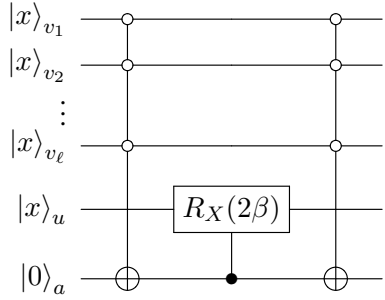
\begin{figure}[H]
\centerline{
\Qcircuit @C = 1.2 em @R = 1.2 em {
\lstick{\ket{x}_{v_1}}   & \ctrlo{1} & \qw      & \ctrlo{1}   & \qw       \\
\lstick{\ket{x}_{v_2}}    & \ctrlo{2} & \qw      & \ctrlo{2}  & \qw   \\
\lstick{\vdots} &  & \\
\lstick{\ket{x}_{v_\ell}} & \ctrlo{2} & \qw      & \ctrlo{2}  & \qw   \\
\lstick{\ket{x}_{u}}        & \qw    & \gate{R_X (2\beta) } & \qw  & \qw \\ 
\lstick{\ket{0}_a}     &  \targ     & \ctrl{-1}    & \targ & \qw   \\ 
}
}
\vspace*{8pt}
\caption{Quantum circuit implementing the mixing operator $U_{M,u}(\beta)=\Lambda_{\overline{x}_{v_1}\overline{x}_{v_2}\dots\overline{x}_{v_\ell}} (X_u)$  
for Maximum Independent Set. 
The circuit consists of two $(\ell+1)$-bit Toffoli gates,   
for $v_1,\dots,v_\ell$ the neighbours of vertex $u$, and a controlled $X$-rotation gate.  
The ancilla qubit is initialized and returned to $\ket{0}$. 
The circuit can be implemented with $O(\ell)$ basic gates.} 
\label{fig:MaxIndepSetcircuit}
\end{figure}

As explained, there is freedom to select the ordering of the product defining the sequential mixer,  with some orderings yielding 
implementation advantages.  
In particular, for bounded-degree graphs with maximum degree $D_G=O(1)$, many of the $U_{M,u}$ will act on disjoint sets of qubits and hence can be implemented in parallel. Thus, 
by partitioning the $U_{M,u}$ into groups of terms acting on disjoint qubits, 
$U_{M}$ may be implemented 
with depth $O(D_G) = O(1)$, which is significantly less than $n$. 
We further elaborate on 
possible 
partitions 
for mixing operators
in~\cite{hadfield2017quantum}. 

The sequential mixer $\mixUnitary(\beta)$  
consists of 
multiqubit controlled $X$-%
rotations $\Lambda_{f_u} ( e^{-i\beta X_u})$, which may each be implemented efficiently
using basic quantum gates, as we now explain. 
Appending an ancilla qubit labeled $a$ and initialized to $\ket{0}$, which
we use to store the value $f_u(x)$, 
we may implement the action of $ \Lambda_{f_u} ( e^{-i\beta X_u})$ using the operator
\begin{equation}  \label{eq:controlledRX}
\Lambda_{f_u} (X_a) \;
\Lambda_{x_a} ( e^{-i\beta X_u}) \; 
 \Lambda_{f_u} (X_a),
\end{equation}
shown in Figure \ref{fig:MaxIndepSetcircuit}. 
 The first operator $\Lambda_{f_u} ( X_a )$ is a multi-controlled $X$ gate, i.e., a multiqubit Toffoli gate, which computes $f_u$ in the ancilla register, 
taking each basis state 
 $\ket{x}\ket{0}$ to $\ket{x}\ket{0\oplus f_u(x)}$. 
 An important result for our purposes is that 
any such multi-controlled Toffoli $\Lambda_{f_u} (X_a)$ acting on $\ell+1$ qubits can be implemented with $O(\ell)$ basic gates and using one additional ancilla qubit \cite{VBE96,NC}. 
(Note that the white circles in Fig.  \ref{fig:MaxIndepSetcircuit}, which indicate negated control variables, are immaterial; such a Toffoli can be implemented from a regular Toffoli with $2$ additional $X$ gates per control line.) 
 The second operator in (\ref{eq:controlledRX}) is an $X$-rotation controlled by the value of $f_u(x)$ in the ancilla qubits, and the final operator $\Lambda_{f_u} (X_a )$ 
 uncomputes (clears) the ancilla qubit $a$ for reuse.

Using the constructions of \cite{VBE96}, we may implement each $\Lambda_{f_u} ( e^{-i\beta X_u})$ using $O(D_u)$ CNOT and single-qubit gates, 
i.e., \textit{basic gates}.  
Here $D_u$ is the degree of the vertex $u$, and hence the number of control bits in $f_u$. Therefore,  
$\mixUnitary(\beta)$ may be implemented using at most 
$O(\sum_u \degree_u)=O(\numConstraints)$ basic gates, and a single ancilla qubit. 
It is tedious but straightforward to derive estimates of the constants in the implementation cost, but this is not our concern here. Similarly, different compilations, in particular to different gate sets, are possible.

{\bf Implementation Cost:} The initial state $\ket{0\dots 0}$ is trivial to prepare. Each application of the QAOA operator
 $Q = \mixUnitary(\beta) U_P(\gamma) $ requires at most $O(m+n)$ basic quantum gates, and $n+1$ qubits. Thus the QAOA$_p$ state can be created with 
 $O(p(m+n))$ basic quantum gates.

We emphasize that different gate sets, compilation choices, and optimizations are possible, with varying cost trade-offs and suitability for a given architecture. In particular, the mixing operator 
$U_M(\beta)$ may be replaced with $U_M^r(\beta)$, $r=O(1)$, 
affecting the cost estimates only by a constant.  

\subsubsection{Applications to Other Problems}
Our construction for Max Independent Set 
extends to the 
related problems Max Clique 
and Min Vertex Cover, which we now summarize. 
It is interesting to observe that while these three problems have very similar QAOA constructions, 
each has quite different properties concerning the best classical algorithms known and hardness of approximation.  
Exploring these connections is an interesting future research direction. 

Note that it is straightforward to extend all three problem constructions 
to vertex-weighted 
problem variants 
via the modification $H_f= \sum_{i=1}^n w_ix_i.$

\subsubsection*{Max Clique}
\paragraph{Problem:} Given a graph $G=(V,E)$, find the largest cardinality clique (a subset $V' \subset V$ of mutually-adjacent vertices). 

The Maximum Clique decision problem is NP-hard \cite{karp1972reducibility}, and the optimization problem cannot be approximated better than $O(n^{1-\e})$ for any $\e$ (as the graph becomes large) unless P=NP \cite{zuckerman2006linear}. 
The best algorithm for general graphs achieves a $O(n \frac{ (\log \log n)^2}{(\log n)^3})$-approximation \cite{feige2004approximating}. The problem of finding cliques of fixed size was considered for adiabatic 
quantum optimization in \cite{childs2002finding}.

Observe that every clique in $G=(V,E)$ 
is an independent set in the complement graph $G^c = (V, \binom{V}{2} \setminus E)$.
Thus, Maximum Clique can be approximated with the above construction for Maximum Independent Set applied to $G^c$. Note that is this case, feasible states now encode valid cliques of $G$. 
The details follow as for Max Independent Set, 
and the same compilation costs apply (with the parameters of $G^c$).  
The details follow as above using the parameters of $G^c$.

\subsubsection*{Min Vertex Cover}\label{sec:MinVertexCover}
{\bf Problem:} Given $G=(V,E)$, 
minimize the size of a subset $V' \subset V$ that covers $V$ (i.e.,
for every $(uv)\in E$, $u\in V'$ or $v\in V'$). 

The Minimum Vertex Cover problem is APX-complete~\cite{papadimitriou1991}.  It has a $(2-\Theta(1/\sqrt{\log n}))$-approximation~\cite{karakostas2009better}, but cannot be approximated better than $1.3606$ unless P=NP~\cite{dinur2005hardness}. 

We again reduce the problem to Maximum Independent Set, though as approximation problems they are not equivalent~\cite{trevisan2004inapproximability}. 
A subset $V' \subset V$ is a vertex cover if and only if $V \setminus V'$ is
an independent set, so the problem of finding a minimum vertex cover is 
equivalent to that of finding a maximum independent set. 
Hence, we can use the same mapping as for Max Independent Set
with each $\bar{x}_v$ replaced by $x_v$.  
The resource counts are the same as for Max Independent Set.

\section{Mappings on $k$-Dits}
In this section, we consider constructions for problems with 
configuration space $[k]^n = \{1,\dots,k\}^n$ for $k>2$, i.e., strings of $n$-many $k$-dits.  
We 
show how encoding each $k$-dit in unary (i.e., using $k$-qubits), as opposed to binary, gives implementation advantages for certain problems, 
and we give explicit constructions for three graph coloring optimization problems.  

Graph-$k$-Coloring ($k\geq 2$) is an important NP-complete 
 problem with many applications, such as scheduling \cite{Leighton1979graph,Rieffel2014parametrized} and memory allocation \cite{chaitin1982register}. Given an undirected graph $G=(V,E)$, Graph-$k$-Coloring asks whether there exists an assignment of one of $k$ colors to each vertex such that every edge  is \textit{properly colored} (connects two vertices of different colors). If such an assignment exists, the graph is said to be $k$-\textit{colorable}. 
 Note that every graph is trivially $(D_G+1)$-colorable, where $D_G$ is the maximum 
 degree of any vertex in $G$. 

Several optimization variants of Graph-$k$-Coloring are known. 
We first consider the Max $k$-Colorability problem of 
maximizing the number of properly colored edges 
\cite{alon1992algorithmic,petrank1994hardness}. 
As a coloring specifies a partition, 
this problem naturally generalizes MaxCut, and 
is also known as 
Max $k$-Cut \cite{frieze1997improved,khot2007optimal}, typically for the case of weighted edges. 
We then consider the approximation problems of finding the largest $k$-colorable induced subgraph, and determining a graph's chromatic number. 

The constructions for the three problems are related and 
each extends the previous. 

\subsection{Max $k$-Colorability} 
This optimization version of graph coloring relaxes the hard constraint that every edge be properly colored, and instead we try to maximize the number of such edges. 
This problem is equivalent to MaxCut for~$k=2$. 
\paragraph{Problem:}
Given a graph $G = (V,E)$ with $n$ vertices and $m$ edges, and $\numColors$
colors, 
find a $k$-color assignment 
maximizing the number of properly colored edges.

A random coloring properly colors a fraction $1-1/k$ of edges in expectation. 
For $k>2$,  semidefinite programming gives a 
$(1 - 1/k + \left(2+o(k)\right) \frac{\ln k}{k^2})-$approximation~\cite{frieze1997improved}, which is optimal up to the $o(k)$ factor under the unique games conjecture~\cite{khot2007optimal}. This problem is APX-complete for $k\geq 2$~\cite{papadimitriou1991} with no PTAS unless P=NP~\cite{petrank1994hardness}. 

The domain $\domain$ is the set of colorings $\x$ of $G$, an assignment 
of a color to each vertex.
(Note that here and throughout, the term ``colorings'' includes \emph{improper} colorings.)
The domain $\domain$ can be represented as the set of length $n$ strings
over an alphabet of $\numColors$ characters, 
$\mathbf{x} = x_1x_2\dots x_n$, where $x_i\in [\numColors]$. 
The objective function 
$\objFunc: {[\numColors]}^n \to {\naturals}$ 
counts the number of properly colored edges in a coloring 
\begin{equation}\label{eq:maxColorableSubgraph-objFunc}
\objFunc(\x) = \sum_{(uv) \in E} \textrm{NotEqual}(x_u,x_v).
\end{equation}

There are different ways to represent this problem on a quantum computer, with
various trade-offs. 
We employ a unary \textit{one-hot} encoding for each $k$-dit, consisting of the $k$-many Hamming weight~$1$
bit strings on $k$ bits, $\{100\dots, 010\dots,\dots  \}$,  
in which the position of the single $1$ indicates the assigned color. 
This encoding, which requires  
$k$ qubits per vertex, has been used in quantum annealing
~\cite{LucasIsingNP,Hen2016quantum,Hen2016driver,Rieffel2014parametrized}.  
Thus, overall color assignments are encoded using $kn$ binary variables $x_{v,i}$, 
with $x_{v,i} = 1$ indicating that vertex $v$ has been assigned color $i$. 
For each vertex $v\in V$, a hard constraint is that it 
be assigned exactly one color, 
\begin{equation}
\label{eq:unique}
\sum_{i=1}^k x_{v,i} = 1,
\end{equation}
i.e., the allowed states of the $k$ variables $x_{v,1}, \dots,  x_{v,k}$ encode the $k$-dit $x_v$. 
Feasible strings 
are then 
Hamming weight $n$ strings such that 
these $n$ constraints are satisfied. 
The initial state $\ket{s}$ may be taken to be any feasible state, which requires $n$ many $X$-gates to prepare in depth $1$.

On feasible strings $x$, the cost function may be written 
\begin{equation} \label{eq:costgc1}
f(x) = m - \sum_{(uv) \in E} 
\sum_{i=1}^k x_{u,i}x_{v,i},
\end{equation}
which 
subtracts one for every improperly colored edge. 
Substituting $\frac{1}{2}(I-Z)$ for each binary variable yields a quadratic Hamiltonian of the same form as (\ref{eq:QUBOHam}). 
Furthermore, 
\eqref{eq:unique} implies that each operator $\sum_{i=1}^k Z_{v,i}$ acts as a constant multiple of the identity, 
simplifying the Hamiltonian to 
\begin{equation}
H_f = 
\frac{km}{4} I 
- \frac14 
\sum_{(uv) \in E} 
\sum_{i=1}^k 
Z_{u,i}Z_{v,i}\;.
\end{equation}
Dropping again the identity term from $e^{-i\gamma H_f}$, 
we define  the phase operator 
\begin{equation} \label{eq:phaseOpGC1}
U_P(\gamma)  = \prod_{(uv) \in E}  \prod_{i=1}^k \; e^{i\gamma \frac14 Z_{u,i}Z_{v,i}},
\end{equation}
which 
consists of $km$ many $R_{ZZ}$ operations. 
The $R_{ZZ}$ mutually commute and can be applied in any order. 
Each $R_{ZZ}$ can be implemented with two CNOT gates and one $R_Z$ gate; see Figure \ref{fig:RZZcircuit}.  

Turning to the mixing operator, we seek a mixing Hamiltonian 
that meets the criteria laid out in 
Sec.~\ref{sec:designCrit}, keeping the evolution within the feasible subspace.
Observe that the hard constraints 
\eqref{eq:unique} 
each depend only on a single vertex. Thus it suffices to define mixing Hamiltonians which preserve feasibility locally for each vertex.   
In particular, the constraints 
\eqref{eq:unique} can be written as
$(1-\sum_{i=1}^k x_{v,i})^2 = 0$,  
so that the corresponding 
Hamiltonian  
\begin{equation} \label{eq:A}
H_ A  =  -\frac14 \sum_{v} \left( 2(k-2)\sum_{i=1}^k Z_{v,i} -   \sum_{i\neq j}^k  Z_{v,i}  Z_{v,j} \right)
\end{equation}
admits the feasible subspace exactly as its ground subspace. 

For each 
$v\in V$, we define 
\begin{equation}   \label{eq:mixinggc1}
B_v =  \sum_{i=1}^k X_{v,i}X_{v,i+1} + Y_{v,i}Y_{v,i+1},
\end{equation}
with indices taken modulo $k$. This Hamiltonian is 
known in physics as the \textit{XY Model on a ring} \cite{lieb2004two}.  
The $B_v$ each act on disjoint sets of qubits and hence mutually commute, 
so for the Hamiltonian 
$B=\sum_v B_v$, 
the corresponding simultaneous mixer is 
\begin{equation}
U_M^{(H)} (\beta) =  e^{-i\beta B} = \bigotimes_{v\in V} e^{-i\beta B_v}. 
\end{equation}
It is easy to check that $[B_v,H_A]=0$ for each $v\in V$. 
Note that this requires the presence of both the $XX$ and $YY$ terms in (\ref{eq:mixinggc1}). 
Thus, Propositions  \ref{prop:commutingHams1} and \ref{prop:commutingHams2} 
imply that $U_M^{(H)} (\beta)$ preserves feasibility. 

It is insightful to elaborate on how the operators $e^{- i \beta B_v}$ act on basis states. 
We will use related operators in our subsequent problem constructions. 
For a vertex $v$ assigned a color $j$, indicated by the state $\ket{j}_v$, $1\leq j \leq k$, we have 
$$ B_v \ket{j}_v = \ket{j+1 \; ({\rm mod } \; k)}_v + \ket{j-1 \; ({\rm mod } \; k)}_v .$$
Thus we may identify $B_v = L_v + R_v$, where $L_v$, $R_v$ are left and right circular shift operators. 
From the series expansion, the operator $e^{-i \beta B_v}$ is easily seen to be given by a weighted linear combination of the identity (the zero shift), single left/right shift operators, double shifts, and so on.  

Each operator $\timeEv{\beta B_v}$  can be efficiently compiled to a quantum circuit using the quantum Fourier transform~\cite{verstraete2009quantum}. 
On the other hand, 
we can derive a simple 
mixing operator with low implementation cost. 
It is easy to check that the Hamiltonians
\begin{equation}
  B_{v,i,j} = X_{v,i}X_{v,j} + Y_{v,i}Y_{v,j},
\end{equation}
$v\in V$, $i,j\in [k]$, 
themselves 
preserve feasibility. 
Thus, from (\ref{eq:mixinggc1}) we define the \emph{sequential} 
mixer 
\begin{equation}
\mixUnitary(\beta) =  \prod_{v\in V, j\in [k]} \timeEv{\beta B_{v,j,j+1}} ,
\end{equation}
where 
the order of the product is arbitrary and may be selected a
as desired. 
Each $B_{v,j,j+1}$ commutes with all but $2$ of the other $B_{v',j',j'+1}$, so $\mixUnitary(\beta)$ can be implemented in depth $2$ with respect to the~$\timeEv{\beta B_{v,j}}$. 

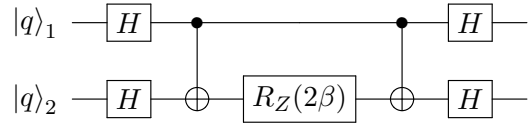
\begin{figure}[H]
\centerline{
\Qcircuit @C = 1.2 em @R = 1.2 em {
\lstick{\ket{q}_1} & \gate{H}    & \ctrl{1} & \qw                              &  \ctrl{1} &  \gate{H}     & \qw \\ 
\lstick{\ket{q}_2} & \gate{H}      & \targ    & \gate{R_Z (2\beta) } & \targ     &  \gate{H}    & \qw \\ 
}
}
\vspace*{8pt}
\caption{Quantum circuit performing 
$R_{X_1X_2}(2\beta)={\rm exp}(-i\beta X_1 X_2)$ on 
two qubits labeled $1,2$. 
} 
\label{fig:RXXcircuit}
\end{figure}

As $[X\otimes X, Y\otimes Y] = 0$, 
each $\timeEv{\beta B_{v,j}}$ can be decomposed into an $XX$-rotation and a $YY$-rotation as 
$\timeEv{\beta B_{v,j}} = e^{-i\beta X_{v,j}X_{v,j+1}}
e^{-i\beta Y_{v,j}Y_{v,j+1}}$. 
The $XX$-rotation can be implemented with~$4$ Hadamard ($\textrm{H}$) gates, $2$ CNOT gates, and an  $R_Z$ gate, 
as shown in Figure \ref{fig:RXXcircuit}.  
A similar construction holds for each $YY$-rotation 
with the Hadamards replaced by $R_X(\pi/2)$ gates; 
 see e.g. \cite{whitfield2011simulation}. 
Thus, $\mixUnitary(\beta)$ can be implemented with $O(nk)$ basic gates. 

\vskip 2pc

{\bf Implementation Cost:} Our construction uses $kn$ qubits.  
Any feasible basis state may be used as the initial state, and prepared using at most $n$ many $X$-gates. 
The operators $U_P(\gamma)$ and  $\mixUnitary(\beta)$ require, respectively, $O(km)$ and $O(kn)$ basic quantum gates. 
 Thus the QAOA$_p$ state can be created with 
 $O(pk(m+n))$ basic quantum gates. 

It is easy to see that for a suitable $U_M(\beta)$, say $\beta = \pi/4$, that  
$r=\lfloor{k/2}\rfloor$ 
repetitions of $U_M(\beta)$ suffice to connect any two states in the sense of our criteria of Section \ref{sec:designCrit}.
Thus, setting $r=O(k)$, the \textit{repeated mixing operator} $U^r_M(\beta)$ 
can be used instead, 
now requiring $O(krn)=O(k^2n)$ basic quantum gates. 
Note that, typically, $k=O(1)$. 
Alternatively, 
the \textit{fully-connected XY model Hamiltonian}
$ B^{(fc)}_v =  \sum_{i<j} X_{v,i}X_{v,j} + Y_{v,i}Y_{v,j}$, 
also satisfies 
$[ B^{(fc)}_v,H_A] =0$ for all $v\in V$, and hence preserves feasibility. 
The corresponding 
sequential mixer 
$U_M^{(fc)}(\beta) = \prod_{v\in V}\prod_{ i< j } \timeEv{\beta B_{v,i,j}}$
similarly  
requires $O(nk^2)$ basic quantum gates. 

With either of these alternative mixers, 
we would hope to obtain better 
performance in exchange for the 
higher implementation costs. Analyzing this trade-off 
between the efficacy of different mixing operators and their implementation costs is an important direction of future work.

The next two graph coloring problems we consider 
 will use similar unary one-hot encodings
for the color state of each vertex. 
It is worthwhile to remark on the trade-offs between this encoding and the binary alternative. 
Firstly, the unary encoding uses $kn$ qubits, whereas the binary encoding requires $\lceil \log_2 k \rceil n$ qubits; as typically $k \ll n$ (e.g., $k=O(1)$), our approach does not add unreasonable overhead. 
Secondly, for the unary encoding, $2$-local 
interactions 
are sufficient to compute the objective function, 
as is evident from (\ref{eq:phaseOpGC1}); in contrast, 
a $2\lceil \log_2 k \rceil$-local interaction 
may be necessary in general 
to compute this in binary, 
as every bit position must be compared to determine if two color labels are the same, 
leading to increased resource requirements. 
Furthermore, if $k$ is not a power of two, then 
some of the one-hot states are redundant or not used and 
must be dealt with somehow, which may be nontrivial.  

\subsection{Max $k$-Colorable Induced Subgraph}
The {\it induced subgraph\/} of a graph $G= (V,E)$ for a subset of vertices
$W \subset V$ is the graph $H = (W, E_W)$, where $E_W$ is the subset of edges in $E$ with both endpoints in $W$.  

\paragraph{Problem:}
Given a graph $G = (V,E)$ with $n$ vertices and $m$ edges, 
find the largest induced subgraph (largest number of vertices) that can be properly $k$-colored.

This problem is as easy and as hard to approximate as Max Independent Set
~\cite{panconesi1990quantifiers,Ausiello2012complexity}. For $k=1$, the two problems are equivalent.  
On bounded degree graphs, Max $k$-Colorable Induced Subgraph 
can be approximated to $(D_G/k +1)/2$, 
but remains APX-complete~\cite{halldorsson1995approximating}.

We represent colorings as in the previous section 
with variables $x_{v,1}, \dots,x_{v,k}$, 
but with one additional variable $x_{v,0}$ per vertex to represent 
an ``uncolored'' vertex, indicating that the vertex $v$ is not included in the 
induced subgraph. 
The state of each vertex thus corresponds to a $(k+1)$-dit.  

In this case, 
feasible strings, in addition to having each vertex uniquely 
colored or assigned as uncolored, correspond to proper 
colorings on the subgraph induced by the colored vertices. 
Thus, the mixing operator 
will be more complicated, 
essentially 
incorporating information that was in the cost function of the previous problem. 

On the feasible subspace, the cost function takes an especially simple form, 
\begin{equation}
\label{eq:costgc2}
f = m - \sum_v x_{v,0},
\end{equation}
which counts the number of included vertices. 
The corresponding objective Hamiltonian, 
after 
dropping the identity terms, 
is 
\begin{equation}
H_f = \frac12 \sum_v Z_{v, 0}
\end{equation}
Hence, the phase separation operator $U_P = e^{-i \gamma H_f}$ can again be implemented with $n$ many $Z$-rotation gates and depth one. 

To design mixing operators, 
consider the allowed transitions between feasible states.
A given vertex can be feasibly colored $i$ only if none of its adjacent
vertices are also colored $i$. Thus, the transition rule at each vertex must
depend on the local graph topology and assigned colors. 
Consider the controlled operation
\begin{equation}
\label{eq:cswap}
(\bar{x}_{v_1,i } \bar{x}_{v_2,i }\dots \bar{x}_{v_{D_u},i}) \cdot \SWAP(x_{u,0}, x_{u,i}),  
\end{equation}
where $v_1, \dots, v_{D_u}$ are the neighbours of vertex $u$. 
This operation swaps the color of vertex $u$ between the uncolored state and color $i$ if and only if none of the neighbours of $u$ are already colored $i$. 
The corresponding Hamiltonian term 
(after dropping the terms for the SWAP that have no effect) is
\begin{equation}  
B_{u,i} =\frac{1}{2^{D_u}} (X_{u,0}X_{u,i}+Y_{u,0}Y_{u,i})\prod_{j=1}^{D_u} (I+Z_{v_j,i}).
\end{equation}
The overall mixing Hamiltonian is $B = \sum_u \sum_i B_{u,i}$.
Since 
$B$ contains the means to color a vertex with color $i$ if none of
its neighbours are colored with color $i$, and a means to uncolor a vertex
(as long as none of its neighbours share its current color, which is always
the case in the feasible subspace), the mixing term enables exploration
of the full feasible subspace starting from any state in that subspace.
A simple 
initial state is 
$\ket{s}=\ket{0}^{\otimes n}$
in which all vertices are uncolored.

 Exponentials of the 
 Hamiltonians $B_{u,i}$ 
 again give controlled unitaries. 
 For the control predicate $f_{u,i} = \prod_{v \in \neighborFunc(u)} \bar{x}_{v,i}$, we define  
\begin{equation}   \label{eq:mixerGCInd}
 U_{M,u,i}(\beta) = e^{-i\beta B_{u,i}} = \Lambda_{f_{u,i}} ( e^{-i\beta (X_{u,0}X_{u,i}+Y_{u,0}Y_{u,i})} ),
\end{equation} 
which is a multiqubit controlled $(XX+YY)$-rotation.

\vskip 2pc
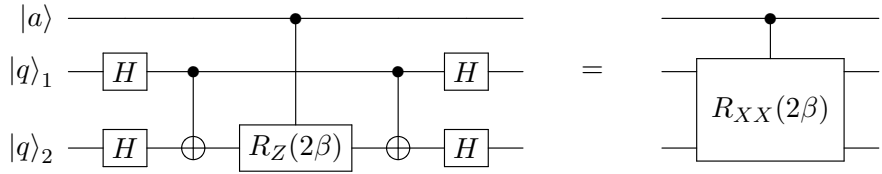
\begin{figure}[H]
\centerline{
\Qcircuit @C = 1.2 em @R = 1.2 em {
\lstick{\ket{a}} & \qw    & \qw & \ctrl{2}                             & \qw & \qw     & \qw    & & & &  & \ctrl{1}   &      \qw    \\ 
\lstick{\ket{q}_1} & \gate{H}    & \ctrl{1} & \qw                              &  \ctrl{1} &  \gate{H}     & \qw   & &= & &  & \multigate{1}{R_{XX}(2\beta)} & \qw  \\ 
\lstick{\ket{q}_2} & \gate{H}      & \targ    & \gate{R_Z (2\beta) } & \targ     &  \gate{H}    & \qw  &  & & &      & \ghost{R_{XX}(2\beta)}    & \qw \\  
}}
\vspace*{8pt}
\caption{Quantum circuit performing the operation controlled $XX$-rotation $U = \Lambda_a (e^{-i\beta X_1X_2})$. Replacing each
$H$ gate with a $R_X(\pi/2)$ gate gives instead $U = \Lambda_a (e^{-i\beta Y_1Y_2})$. } 
\label{fig:ctrlRXXcircuit}
\end{figure}

 From the Hamiltonian $B=\sum_u \sum_i B_{u,i}$, we define 
 the mixers: 
 \begin{itemize}
\item the \emph{simultaneous} (Hamiltonian-based) controlled-$X$ mixer 
$U_M^{(B)} (\beta) = \timeEv{\beta B}$,
\item the \emph{sequential} (partitioned) controlled-$X$ mixer 
$\mixUnitary(\beta) =  \prod_{u\in V}  \prod_{i=1}^k U_{M,u,i}(\beta). $ 
\end{itemize}

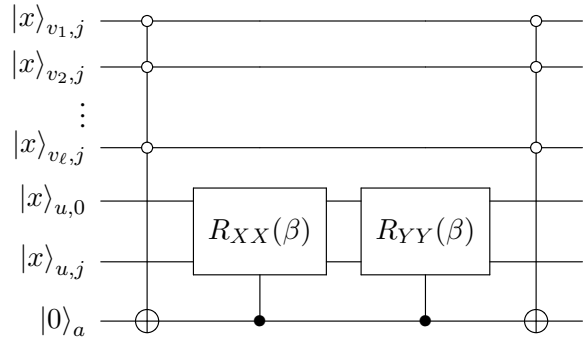
\begin{figure}[H]
\centerline{
\Qcircuit @C = 1.2 em @R = 1.2 em {
\lstick{\ket{x}_{v_1,j}}   & \ctrlo{1} & \qw  &  \qw                              & \ctrlo{1}   & \qw       \\
\lstick{\ket{x}_{v_2,j}}    & \ctrlo{2} & \qw   &  \qw                    & \ctrlo{2}  & \qw   \\
\lstick{\vdots} &  & \\
\lstick{\ket{x}_{v_\ell,j}} & \ctrlo{3} & \qw    & \qw                     & \ctrlo{3}  & \qw   \\
\lstick{\ket{x}_{u,0}}        & \qw    & \multigate{1}{R_{XX}(\beta)}  &  \multigate{1}{R_{YY}(\beta)} & \qw  & \qw \\ 
\lstick{\ket{x}_{u,j}}        & \qw    & \ghost{R_{XX}(\beta)}  &  \ghost{R_{YY}(\beta)} & \qw  & \qw \\ 
\lstick{\ket{0}_a}     &  \targ     & \ctrl{-1}    & \ctrl{-1} & \targ & \qw   \\ 
}
}
\vspace*{8pt}
\caption{Quantum circuit implementing the mixing operator $U_{M,u,j}(\beta)$.  
The ancilla qubit is initialized and returned to $\ket{0}$.
The circuit can be implemented with $O(\ell)$ basic quantum gates.} 
\label{fig:XXYYrotation}
\end{figure}

The sequential mixer requires $nk$ application of  $U_{M,u,i}(\beta)$. Similar to the previous constructions, using an ancilla qubit each $U_{M,u,i}(\beta)$ can be implemented with two  $(D_u+1)$-qubit Toffoli gates, a controlled $XX$-rotation, and a controlled $YY$-rotation.  
The circuits are shown in Figures \ref{fig:ctrlRXXcircuit} and \ref{fig:XXYYrotation}. 
Thus, by a similar argument as for Max Independent Set, we have that 
$U_M(\beta)$ can be implemented with $O(km)$ basic gates.

{\bf Implementation Cost:} 
Our construction uses $(k+1)n+1$ qubits.  
The initial state $\ket{0\dots 0}$ is trivial to prepare. Each application of the QAOA operator
 $Q = \mixUnitary(\beta) U_P(\gamma) $ requires 
 $O(km+n)$ basic gates. 
 Thus the QAOA$_p$ state can be created with 
 $O(p(km+n))$ basic quantum gates.

\vskip 2pc 

\subsection{Min Chromatic Number}   \label{sec:MinChromNum}
A graph $G$ 
that can be $\kappa$-colored but not
$(\kappa-1)$-colored is said to have \textit{chromatic number}
$\kappa$.

\paragraph{Problem:} 
For a graph $G=(V,E)$, 
minimize the number of colors needed to
properly color it. 

The Min Chromatic Number problem has important applications to scheduling  \cite{daniel2004graph} and to physics \cite{vaismanmultilevel}.
The best classical algorithm \cite{halldorsson1993still} achieves an approximation ratio of  
$O\left(n \frac{ (\log \log n)^2}{(\log n)^3} \right)$, and we cannot do better than $n^{1-\e}$ for any $\e >0$ 
in polynomial time 
unless P = NP \cite{zuckerman2006linear}.   

Any graph can be properly $(D_G+1)$-colored, where recall $D_G\leq n-1$ is the maximum vertex degree in $G$.  
We seek a proper coloring (i.e., a certificate) showing 
that fewer colors suffice. 

For the mixing operators we consider below, 
$k = D_G +2$ colors suffice to allow transitions between any two feasible states. 
This follows because any coloring using at most $D_G + 2$ colors can be transformed into any other coloring using at most $D_G + 1$ colors by a series of moves that changes the color of one vertex at a time while maintaining a proper coloring at each step.
Moreover, a $(D_G +2)$-coloring can be trivially constructed 
for any graph. 

Thus, we may use 
$k=D_G+2$ qubits to encode the $k$ 
possible colors of each vertex in the unary one-hot encoding.  We define the feasible domain to be the subset of states encoding proper graph colorings, many of which may use fewer than $k$ colors. 

For a given coloring $x$,
the function 
$y_j (x) = \bigwedge_{u \in V}   \bar{x}_{u,j}$ 
gives $1$ only if no vertex is
colored $j$.  
Thus, we seek to maximize the number of unused colors, encoded by
the $n$-local objective function
\begin{equation}
f(x) = \sum_{j=1}^{D+2} y_j = \sum_{j=1}^{D+2}  \bigwedge_{u \in V}   \bar{x}_{u,j}. 
\end{equation}
The corresponding problem Hamiltonian is
\begin{equation}
H_f = \frac{1}{2^n} \sum_{j=1}^{D+2}  \prod_{u \in V}   (I + Z_{u,j}).
\end{equation}
Expanding the right-hand side gives a sum of $\Omega(2^n)$ terms 
with locality up to $n$, which renders our previous approach 
to implementing $U_P(\gamma)=e^{-i\gamma H_f}$ 
inefficient (the number of $Z$-rotation gates required is exponential in $n$). 

However, we can implement $U_P(\gamma)$ efficiently with the help of an ancilla register.
We append $k$ ancilla qubits $a_1,\dots, a_k$ to our state, 
initialized to $\ket{00\dots 0}$. The ancilla $a_j$ is used to store $y_j(x)$. 
For each color $j$, define the $(n+1)$-local unitary operator to be the multi-controlled CNOT
$$U_j = \Lambda_{y_j} (X_{a_j}).$$ 
Using an ancilla qubit, each $U_j$ can be implemented with $O(n)$ basic gates. 
The $U_j$ act nontrivially on disjoint sets of qubits and mutually commute. 
Thus the operator $$U_a := U_1U_2\dots U_k$$ maps the basis state $\ket{x}\ket{00\dots 0}$ to the state $\ket{x}\ket{y_1 y_2 \dots y_k}$. As $U_i^2 =I$, 
a second application of $U_a$ uncomputes each ancilla. 

After computing the $y_j(x)$, the phase operator may be implemented by applying a $Z$-rotation gate to each ancilla qubit. 
Summing the bits $y_j(x)$ gives 
 the number of unused colors. Dropping the identity term, the corresponding Hamiltonian is given by 
\begin{equation}
 H_g =  - \frac12 \sum_{j=1}^k Z_{a_j}, 
\end{equation} 
 for which $e^{-i\gamma H_g}$ can be implemented with $k$ many $R_Z$ gates. 
 
The overall phase operator is given by 
\begin{equation}
U_P (\gamma) =U_a e^{-i\gamma H_g} U_a=  U_1U_2\dots U_k R_{Z_{a1}} (\gamma) \dots R_{Z_{ak}} (\gamma) U_1U_2\dots U_k,
\end{equation}
 which can be implemented using $k$ ancilla qubits and $O(kn)$ basic quantum gates in constant depth. 

For the mixing operator, we use a controlled operation similar to (\ref{eq:mixerGCInd}). 
In this case, there is no uncolored state. 
A vertex $u$ can be recolored $i$ only if that would produce no conflicts with its neighbours, i.e., the coloring remains proper. Thus, the color bits $x_i$ and $x_j$ for $u$ may be safely swapped if none of the $D_u$ neighbours of $u$ are colored either $i$ or $j$. This gives the Hamiltonian 
%
$$ B_{uij} =   (\bar{x}_{v_1 i } \bar{x}_{v_2 i }\dots \bar{x}_{v_{D_u} i})
 (\bar{x}_{v_1 j } \bar{x}_{v_2 j }\dots \bar{x}_{v_{D_u} j})
  \cdot \SWAP(ui, uj)   $$
for $i,j=1,\dots, \ell$, which after dropping the 
terms having no effect reduces to $$B_{uij} =\frac{1}{2^{2D_u+1}} (X_{ui}X_{uj}+Y_{ui}Y_{uj})\prod_{a=1}^{D_u} (I+Z_{v_a i})\prod_{b=1}^{D_u} (I+Z_{v_b i})  .$$
Exponentials of the 
 Hamiltonians $B_{uij}$ 
 give 
 unitaries controlled by the Boolean predicate 
$f_{uij} = $ $\prod_{v \in \neighborFunc(u)} \bar{x}_{vi}\bar{x}_{vj}$. We define the phase operator   
\begin{equation}
 U_{M,u,i,j}(\beta) = e^{-i\beta B_{uij}} = \Lambda_{f_{uij}} ( e^{-i\beta (X_{ui}X_{uj}+Y_{ui}Y_{uj})} ),
\end{equation} 
which by the previous argument can be implemented with $O(D_u)$ basic gates. 
 
 For $B=\sum_u \sum_{i<j} B_{u,i,j}$ , we define the Hamiltonian-based mixer 
 $U_M^{(B)} (\beta) = \timeEv{\beta B}$, and the \emph{sequential} (partitioned) controlled-$X$ mixer
\begin{equation}
 \mixUnitary(\beta) =  \prod_{u\in V}  \prod_{i<j} U_{M,u,i,j}(\beta).
\end{equation}  
We can implement $U_M (\beta)$ using $O(mk^2)$ basic gates.

{\bf Implementation Cost:} 
Our construction uses 
$nk+k+1$ qubits, 
with $k\geq D_G+2$. 
A proper $(D_G+2)$-coloring can be prepared as an initial state using $n$ many $X$ gates.  Each application of the QAOA operator
 $Q = \mixUnitary(\beta) U_P(\gamma) $ requires at most $O(k^2m+nk)$ basic quantum gates, and $nk+k+1$ qubits. Thus the QAOA$_p$ state can be created with 
 $O(p(k^2m+nk))$ basic quantum gates. 

\section{Mappings on Permutations}
Many important but challenging computational problems have a configuration space that is the set of orderings or schedules of some number of items or events. 
Here, we introduce the machinery for mapping such problems to QAOA, using the Traveling Salesman Problem (TSP) and 
a Single-Machine Scheduling (SMS) problem as illustrative examples. 

The constructions of this sections 
are applicable to many other problems. In particular, 
in \cite{hadfield2017quantum} we show constructions for two other single machine scheduling problem variants.

\subsection{Traveling Salesman Problem}
A {\it vertex tour\/} of a complete graph $G= (V,E)$ is a simple cycle that contains all $|V|=n$ vertices, i.e., 
gives a route for the salesman to visit each \lq city\rq\ exactly once and finish where they started, 
and similarly for directed graphs. 
For both cases, up to symmetries, the set of possible tours is isomorphic to the set of possible 
orderings of the $n$ vertices. 

\paragraph{Problem:} 
Given a complete graph $G = (V,E)$ and distances $d_{u,v} \in \reals$, find the shortest tour. 

The TSP problem is NPO-complete~\cite{orponen1990approximation}.  
MetricTSP, where the distances satisfy the triangle inequality, is APX-complete~\cite{papadimitriou1993traveling} and has a $3/2$-approximation~\cite{christofides1976worst}.
The corresponding MaxTSP problem is approximable within $7/5$ for symmetric distance, and $63/38$ if asymmetric. The TSP has previously been considered for quantum annealing \cite{martovnak2004quantum,LucasIsingNP}.

We represent vertex tours with $n^2$ binary variables 
$\{x_{v j}\}$ indicating whether vertex $v$ is visited at the $j$th stop of the tour, $j=1,\dots,n$,  
which we represent using $n^2$ qubits. 
Feasible 
states are those that encode valid tours, i.e. valid orderings, expressed as the hard constraints that for each $v$ we have $\sum_{j=1}^n x_{v,j}= 1$ (each $v$ visited exactly once), and for each position $j$ we have $\sum_{v \in V} x_{v,j}= 1$ (a single vertex visited at each stop). 

The objective function is the tour length, which may be written 
\begin{equation}
f(x) = \sum_{\{u, v\} \in E} 
d_{u,v} 
\sum_{j=1}^n 
\left(x_{u,j} x_{v,j+1} + x_{v,j} x_{u,j+1}\right). 
\end{equation}
Mapping each term to a Hamiltonian and simplifying using the hard constraints (which alleviates the need for single $Z$ terms) yields the phase operator
\begin{equation}  \label{eq:phaseOpTSP}
U_P(\gamma) = \prod_{\{u, v\} \in E} \prod_{j=1}^{n} e^{-\gamma d_{uv} Z_{u,j}Z_{v,j+1}}  e^{-\gamma d_{uv} Z_{u,j+1}Z_{v,j}},  
\end{equation}
where we have again dropped the terms contributing only global phase. The number of $ZZ$-rotations in $U_P(\gamma)$ is $n^3 - n^2$, or half this amount for directed graphs (where the second rotation in (\ref{eq:phaseOpTSP}) is not needed). Thus $U_P(\gamma)$ can be implemented with 
$n^3 - n^2$ $R_Z$ gates and $2(n^3 - n^2)$ CNOT gates. 

For the mixing operator, it is useful to view feasible states as $n \times n$ matrices with a single~$1$ in every column or row. 
A mixing Hamiltonian may be constructed as a sum of row swaps
$B = \sum_{u=1}^n B_{u,u+1}$ or 
$B = \sum_{u<v} B_{u,v}$, where 
\begin{equation}
B_{u,v} = \prod_{j=1}^n \SWAP((u,j),(v,j)) 
\end{equation}
clearly preserves feasibility. 
The Hamiltonians $B_{u,v}$ contain $2n$-local interactions and are nontrivial to implement.  

Alternatively, we can 
 mix feasible states with the 
$4$-local Hamiltonian 
\begin{eqnarray}
H_{M,u,v,i,j} &=&  
 \ket{0_{u,i}1_{u,j}1_{v,i}0_{v,j}}\bra{1_{u,i}0_{u,j}0_{v,i}1_{v,j}}
 +  \ket{1_{u,i}0_{u,j}0_{v,i}1_{v,j}}\bra{0_{u,i}1_{u,j}1_{v,i}0_{v,j}} \nonumber \\
&=& S^-_{u,i}  S^+_{u,j} S^+_{v,i}S^-_{v,j} + S^+_{u,i}  S^-_{u,j} S^-_{v,i}S^+_{v,j}, 
\end{eqnarray}
where in the second line we have introduced the \textit{spin}%
\footnote{In contrast to the \textit{fermionic} creation and annihilation operators $a^+$, $a^-$ considered in Section \ref{sec:QChem} for the electronic Hamiltonian, the operators $S^+$ and $S^-$ satisfy 
$[S^-,S^+] = Z$ and $\{S^-,S^+\} := S^-S^+ + S^+S^- =I$.}
 creation and annihilation operators $S^+=\frac12(X-iY)=\ketbra{1}{0}$ and $S^-=\frac12(X+iY)=\ketbra{0}{1}$. 
 Similar mixing operations have been considered previously for quantum annealing \cite{martovnak2004quantum}. 
 
\begin{figure}[H]
\centerline{
\Qcircuit @C = 1.0 em @R = 1.0 em {
\lstick{\ket{q}_1} & \gate{H}  & \ctrl{1} & \qw & \qw  & \qw                   &  \qw &  \qw & \ctrl{1} &  \gate{H} & \qw \\
\lstick{\ket{q}_2} & \gate{G}  & \targ   & \ctrl{1} & \qw   & \qw               &  \qw  & \ctrl{1} & \targ &  \gate{G}     & \qw \\
\lstick{\ket{q}_3} & \gate{H}  & \qw    & \targ    & \ctrl{1} & \qw              &  \ctrl{1} & \targ & \qw & \gate{H}  & \qw \\ 
\lstick{\ket{q}_4} & \gate{G}  & \qw & \qw & \targ & \gate{R_Z (2\beta)} & \targ  &  \qw &  \qw & \gate{G} & \qw \\ 
}
}
\vspace*{8pt}
\caption{Quantum circuit performing the operation $R_{X_1Y_2X_3Y_4}(2\beta)={\rm exp}(-i\beta X_1 Y_2 X_3Y_4)$ on 
four qubits labeled $1$ to $4$. 
Generally, the exponential of any tensor product of four $X$ and $Y$ operators 
can be implemented by 
a similar circuit where Hadamard 
 $\textrm{H}$ and $G=R_X(\pi/2)$ gates have been substituted appropriately,  
 corresponding to which of $X$ or $Y$ acts on each qubit. 
} 
\label{fig:spinRotCircuit}
\end{figure}
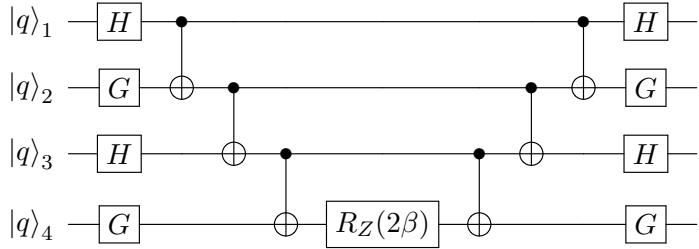

 We define the 
 unitaries 
\begin{equation}
U_{M,u,v,i,j} (\beta) = e^{-i H_{M,u,v,i,j}},
\end{equation}
which can each be implemented using basic gates. 
Indeed, substituting Pauli matrices and expanding yields $H_{M,u,v,i,j}$ as a sum of $8$ terms 
given by (suppressing the qubit indices) 
\begin{equation}  \label{eq:spinOpsPauli}
\frac18(XXXX-YYXX+XYXY+YXYX+XYYX+YXXY-XXYY+YYYY).
\end{equation}
The terms in the sum (\ref{eq:spinOpsPauli}) mutually commute, 
so $U_{M,u,v,i,j} (\beta)$ can be implemented with eight applications of the circuit in Figure \ref{fig:spinRotCircuit}. 
These circuits are similar to, but simpler than, those for the simulation of the electronic Hamiltonian 
\cite{whitfield2011simulation}. 
Thus each $U_{M,u,v,i,j} (\beta)$ can be implemented with $O(1)$ basic gates. 
We define the sequential mixing operator
\begin{equation} 
U_M(\beta) = \prod_{u<v} \prod_{j=1}^{n}    U_{M,u,v,j,j+1} (\beta), 
\end{equation}
which consists of $O(n^3)$ applications of $U_{M,u,v,j,j+1} (\beta)$. 
It is easy to see $U^r_M(\pi/4)$ generates all possible basis state transitions for $r=n-1$, so $U_M(\beta)$ satisfies our design criteria. 

\vskip 2pc

{\bf Implementation Cost:} States are represented with $n^2$ qubits. Any vertex ordering suffices as an initial state which can be prepared with $n$ many $X$ gates. 
Each application of the QAOA operator
 $Q = \mixUnitary(\beta) U_P(\gamma) $ requires at most $O(n^3)$ basic quantum gates. 
 Thus the QAOA$_p$ state can be created with 
$O(pn^3)$ basic quantum gates. 

We remark that it is possible to instead define 
the mixing operator as $\prod_{u<v} \prod_{i<j}^{n}    U_{M,u,v,i,j} (\beta)$, now containing $O(n^4)$ terms and requiring proportionately more resources to implement. The~$O(n^3)$ cost scaling of the proposed $U_M(\beta)$ operator matches that of the phase operator above.

\subsection{Single Machine Scheduling}
We consider 
scheduling jobs on a single machine as to minimize the total (weighted) tardiness. 

\paragraph{Problem:} 
Given $n$ jobs to run on a single machine, 
each with a running time $p_j$, a deadline $d_j\geq p_j$, and a weight $w_j$, 
find a schedule that minimizes the total weighted tardiness 
$T = \sum_{j=1}^n w_j T_j$, where $T_j \geq 0$ gives the amount of time job $j$ is late. 
All times are taken to be integers. 

In scheduling notation \cite{brucker2007scheduling} this problem is denoted $(1|d_j|\sum w_j T_j)$.
There exists an $(n-1)$-approximate algorithm \cite{cheng2005single}. 
The corresponding decision problem is strongly NP-hard \cite{lenstra1977complexity}. 

Assuming a job is always running, job schedules are equivalent to job orderings. 
As in the previous construction, an encoding using $O(n^2)$ qubits may be used to represent the possible schedules. However, 
computing the tardiness for the phase operator is then relatively nontrivial. We propose a different encoding which enables a novel but relatively simple phase operator construction. We will, however, still make use of the equivalence to orderings in the design of the mixing operator.

Clearly, all jobs will finish by time $P= \sum_j p_j$, so the last job will start by time $\tau = P - \min_i p_i $. 
For each schedule, let the binary variables $x_{j,t}$ denote if job $j$ starts at time $t$, where $t=0,1,\dots \tau$. 
We represent these variables with $n\tau$ qubits. 
Feasible strings are those for which each job is assigned a single start time, i.e. Hamming weight $n$ strings satisfying $\sum_t x_{j,t} =1$,
and for which there are no overlapping jobs scheduled. Given an ordering, i.e., a sechule, it is easy to compute the $x_{j,t}$, so an arbitrary ordering can be used as the initial state and prepared using $n$-many $X$ gates. 

Suppose job $j$ starts at time $s_j$; 
its tardiness 
is then defined as $T_j = \max\{0, s_j + p_j - d_j \}$.   
Generally, a maximum function is nontrivial to implement as a Hamiltonian. 
However, 
$T_j $ is simple to implement in the $x_{j,t}$ variables by 
restricting to times $t$ where job $j$ is tardy. 
The cost function, i.e., the weighted total tardiness, then becomes
\begin{eqnarray}
f(x)=\sum_j w_j \sum_{t=d_j-p_j}^\tau x_{j,t} (t+p_j-d_j), 
\end{eqnarray}
which 
maps to a Hamiltonian that is a sum of single-qubit $Z$ operators. Similarly to our previous constructions, we define the 
phase operator 
\begin{equation}
U_P(\gamma) = \prod_{j=1}^n \prod_{t=d_j-p_j}^\tau e^{i w_j (t+p_j-d_j) Z_{j,t}/2},
\end{equation}
which can be implemented with at most $nP$ many $R_Z$ gates.

We construct a mixing operator similar to the previous problem by considering pairwise swaps of jobs in the schedule. It is easy to see sequences of such swaps mix between all possible schedules.  
For consecutive jobs $j$ starting at $t$ and  $j'$ starting at $t+p_j$, swapping their order 
results in job~$j'$ starting at time $t$ and job $j$ starting at $t+p_{j'}$. This exchange is realized by the Hamiltonian 
$
S^-_{j,t} S^-_{j',t+p_j} S^+_{j,t+p_{j'}} S^+_{j',t} + S^+_{j,t} S^+_{j',t+p_j} S^-_{j,t+p_{j'}} S^-_{j',t}  
$. 
Therefore, the mixing Hamiltonian is 
\begin{eqnarray}
B=\sum_{j < j'} \sum_{t=0}^{\tau} S^-_{j,t} S^-_{j',t+p_j} S^+_{j,t+p_{j'}} S^+_{j',t} + 
S^+_{j,t} S^+_{j',t+p_j} S^-_{j,t+p_{j'}} S^-_{j',t} . 
\end{eqnarray}
We thus define the operator
\begin{equation}
U_{M,i,j,t} = e^{-i\beta(S^-_{j,t} S^-_{j',t+p_j} S^+_{j,t+p_{j'}} S^+_{j',t} + S^+_{j,t} S^+_{j',t+p_j} S^-_{j,t+p_{j'}} S^-_{j',t})},
\end{equation}
which can be efficiently implemented as described in the previous section; see Figure \ref{fig:spinRotCircuit}. Hence, we define the mixing operator to be 
\begin{equation}
U_M(\beta) = \prod_{i<j} \prod_{t=0}^\tau U_{M,i,j,t}, 
\end{equation}
which can be implemented using $O(n^2\tau)$ basic quantum gates.

{\bf Implementation Cost:} States are represented with $n\tau$ qubits, where $\tau = P - \min_i p_i \leq P$.  
Any vertex ordering suffices as initial state which can be prepared with $n$ gates. 
Each application of the QAOA operator
 $Q = \mixUnitary(\beta) U_P(\gamma) $ requires at most $O(n^2\tau + n\tau)$ basic quantum gates. 
 Thus the QAOA$_p$ state can be created with 
$O(pn^2P)$ basic quantum gates.

We remark that many important variants of scheduling problems exist, with other parameters (e.g., release dates) and objective functions (e.g., completion time)  \cite{brucker2007scheduling}. Our construction serves as a prototype for these problems. 
We consider 
several other scheduling problems 
 in \cite{hadfield2017quantum}.

\section{Discussion and Concluding Remarks}
We introduced the Quantum Alternating Operator Ansatz (QAOA), a generalization of
the Quantum Approximate Optimization Algorithm, 
and showed how to apply the ansatz to a variety of hard optimization problems.  
The essence of this extension is the consideration of
general parameterized families of unitaries, rather than only those
corresponding to the time evolution of a local Hamiltonian, which allows 
the representation of a larger and potentially more useful set of states than the
original formulation. 
Refocusing on unitaries rather than Hamiltonians in the specification 
leads to a variety of efficiently implementable mixing operators with relatively low resource requirements in terms of the number of qubits and basic gates required for implementation. 
Hence, our constructions provide 
evidence that QAOA may be an especially suitable application for near-term quantum computers. 
Furthermore, the constructions we outline cover a range of problem domains, and may serve as prototypes for mapping other problems of interest. We include a compendium of additional mappings in 
\cite{hadfield2017quantum}. 

For each of the problems we consider, our constructions  
preserve the feasible subspace. 
This requires more sophisticated quantum circuitry than the original QAOA proposal. 
Hence, if we start with a feasible state and create a QAOA$_p$ state, a computational basis measurement is guaranteed to give a feasible solution. 
The feasibility property allows for simplifications to the phase separation operator, typically reducing its implementation cost. The mixing operations, however, are nontrivial to derive and implement. 
We derive mixing Hamiltonian terms controlled by Boolean predicates. 
We further show 
how these terms can be combined in different ways to generate a variety of mixing operators, 
in particular, products of simple controlled unitaries which can be implemented efficiently. 
When the controlled unitaries are local, say for bounded degree problems, they may be implemented with depth much less than their number.  
 
On the other hand, for some optimization problems, it is NP-hard to 
decide if a feasible state exists, let alone find such an initial state; 
see e.g. the problems in \cite{zuckerman1996unapproximable}. 
Designing mixing operators in such cases  
is problematic as there is no obvious way to efficiently ensure that the mixing operations preserve feasibility. Clearly, our approach is not generally 
efficient for such problems. 

The biggest open question is to characterize the performance of QAOA. 
Can QAOA be used to beat classical algorithms for certain problems,  
either in terms of giving a rigorous approximation algorithm that beats all possible classical algorithms, 
or outperforming (say, empirically) all known classical algorithms and heuristics? 
If so, and QAOA does indeed yield practical advantages, then approximate optimization could turn out to be a very important \lq\lq killer application\rq\rq\ for quantum computers, especially near-term ones. 
As evident from the 
results of Chapter \ref{ch:QAOAperformance}, 
obtaining similar results 
for more general problem constructions 
such as those of this chapter 
appears to be a difficult task.
While obtaining further analytic results may
be possible in some cases using our techniques, in general 
this remains an open research direction. 
Improved 
techniques for classically simulating quantum circuits, potentially including  
approaches tailored to QAOA, 
may provide some insight, but the
ultimate test may prove to be experimentation on quantum hardware itself.

While we have successfully shown basic design criteria 
and example constructions for QAOA initial states, mixing
operators, and phase separation operators, we have barely scratched
the surface in terms of which possibilities perform better than others. 
For most of the example problems, we discussed multiple mixers, 
coming from different partitions and orderings of simpler partial mixers. 
Analytic,
numerical, and ultimately experimental work is required to understand 
which of these mixers are most effective, 
in particular with respect to the trade-offs between performance and implementation costs. 
For example, for a near-term quantum device, it may be possible to implement QAOA$_p$ for much larger  $p$ 
with a mixer requiring $n$ gates, 
than when using a more complicated mixer requiring~$n^2$ gates; however, it is 
not at all clear which option would ultimately lead to better performance.  
Furthermore, work is required to investigate 
other trade-offs such as the difficulty of 
finding good algorithm parameters, 
and lower level concerns such as robustness to noise or control error. 
Clearly, similar questions arise with respect to choosing an initial state. 

Effective parameter setting also remains a critical, but mostly open, area
of research. 
While brute-force search
was considered for fixed $p$ in \cite{Farhi2014}, it is practical
only for small p, 
suffering from the curse of dimensionality as $p$ increases; 
see \cite{guerreschi2017practical}. 
In certain simple~\cite{wang2017quantum} or highly symmetric~\cite{Jiang17} cases,
some insights into parameter setting for $p>1$ have been obtained, but even in 
the simplest cases, understanding good choices of parameters seems
nontrivial~\cite{wang2017quantum}. 
Improved parameter setting protocols may come from adapting techniques from existing control theory and parameter optimization methods, and 
by using insights gained from classical simulation of quantum circuits and experimentation on quantum hardware as it becomes available. 
In particular, for problem constructions with local Hamiltonians, early small-scale quantum computers may be used to help characterize the performance of QAOA on much larger problem instances \cite{Farhi2014}. 

To run on near-term quantum hardware, further compilation will be 
required. We have primarily considered compilation to CNOT and general single-qubit gates. 
For other gate sets and architectures, for example where quantum error correction is used, these 
gates will need to be further compiled (and optimized). 
Furthermore, near-term hardware will have additional restrictions,
including which qubits each gate can be applied to, duration and fidelity of the gates, and cross-talk, among others.
This necessitates additional compilation, especially to optimize success probability on pre-fault-tolerance devices.
Other architectures, e.g.\ ones based on higher-dimensional qudits, may prompt other sorts of compilations as well. 
Low-level compilation and optimization of quantum circuits is a rich topic outside the scope of this thesis; see \cite{Venturelli17} for a recent approach.

Going forward, 
we expect some fruitful cross-fertilization 
between research on QAOA and research 
on quantum annealing. 
A promising direction 
is to build on the
results of Yang \ea~\cite{Shabani16}, 
who used Pontryagin's minimization 
principle to show that for quantum annealing, a \lq\lq bang-bang\rq\rq\ schedule  
similar to the form of QAOA is (essentially) always
optimal. 
Unfortunately, their argument does not seem to provide an efficient means
to find such a schedule for a QAOA. 
Similarly, exploiting certain
structural commonalities with variational quantum algorithms such as the variational quantum eigensolver (VQE) \cite{peruzzo2014variational} 
may be fruitful. 
Indeed, as suggested in~\cite{Farhi2017}, it may be possible to take 
advantage of the set of natively available quantum gates 
and to use essentially a VQE approach to optimize the algorithm parameters. 
We are optimistic that tools and results from other aspects of quantum computation may 
prove useful towards a better understanding of the 
power of QAOA circuits. 

Generally, a fundamental question is whether or not quantum computers provide advantages over classical algorithms for approximate optimization. 
Insight into this deep question would have important implications to computational complexity theory.  
We do not generally believe that quantum computers can efficiently solve NP-hard decision problems. However, it remains open whether or not there exists an NP-hard optimization problem such that a polynomial time quantum algorithm can give a better approximation (in the worst case) than any classical algorithms. For problems where we have tight classical algorithms and hardness of approximation results, this appears unlikely. On the other hand, for many problems there is a gap between the best known algorithm and complexity lower bound, and quantum computers may find utility here. Moreover, even if 
it turns out that quantum computers can only provide a quadratic speedup for an optimization problem, meaning finding the same quality solution as a classical algorithm but in reduced time, such a result may 
nevertheless be very important for solving large problem instances. 
Of further interest still is the performance of quantum computers for approximation in the average-case setting, and whether advantages can be realized through quantum heuristics generally. 
While we do not attempt to settle these important but difficult questions here,
the results 
of this thesis 
are 
first steps towards this goal.

%% file: macros2.tex
\renewcommand{\x}{\mathbf{x}}

\newcommand{\ea}{et al.}
\newcommand{\eg}{e.g.\ }
\newcommand{\ie}{i.e.\ }

\newcommand{\B}{{\{0, 1\}}}

\newcommand{\hc}{\mathrm{h.c.}}

\newcommand{\y}{\mathbf{y}}
\newcommand{\w}{\mathbf{w}}

\newcommand{\bqa}{\begin{eqnarray}}
\newcommand{\eqa}{\end{eqnarray}}

\newcommand{\tikzmark}[1]{\tikz[overlay,remember picture] \node (#1) {};}

\newcommand{\colorParity}{\mathrm{CP}}
\newcommand{\timeParity}{\mathrm{TP}}

\newcommand{\OPT}{\mathrm{OPT}}

\newcommand{\QAOA}{\mathrm{QAOA}} 
\newcommand{\QAOAcirc}{Q} 
\newcommand{\domain}{F}
\newcommand{\domainQ}{\mathcal{F}}
\newcommand{\objFunc}{f}
\newcommand{\objHam}{H_{\objFunc}}
\newcommand{\objUnitary}{U_{\objFunc}}
\newcommand{\phaseFunc}{g}
\newcommand{\phaseHam}{H_{\mathrm{P}}}
\newcommand{\mixHam}{H_{\mathrm{M}}}
\newcommand{\mixUnitary}{U_{\mathrm{M}}}
\newcommand{\stdDriver}{H_M}
\newcommand{\initial}{s}
\newcommand{\iniState}{\ket{\initial}}
\newcommand{\hamDist}{\mathrm{Ham}}

\newcommand\phaseUnitary[1][]{
  \ifstrempty{#1}{
    U_{\mathrm{P}}
  }{
    U_{\mathrm{P},#1}
  }
}

\newcommand\Ham[2][]{
  \ifstrempty{#1}{
    H_{#2}
  }{
    H_{#1, #2}
  }
}

\newcommand\unitary[2][]{
  \ifstrempty{#1}{
    U_{#2}
  }{
    U_{#1, #2}
  }
}

\newcommand{\rnv}[1][r]{#1\text{-}\mathrm{NV}}
\newcommand{\ring}{\mathrm{ring}}
\newcommand{\fc}{\mathrm{FC}}

\newcommand{\timeEv}[1]{e^{-i #1}}
\newcommand{\comp}{(\mathrm{comp})}
\newcommand{\enc}{(\mathrm{enc})}
\newcommand{\parity}{\mathrm{parity}}
\newcommand{\even}{\mathrm{even}}
\newcommand{\odd}{\mathrm{odd}}
\newcommand{\last}{\mathrm{last}}
\newcommand{\identity}{I}
\newcommand{\quditSwap}{\mathrm{XY}}
\newcommand{\binary}{\mathrm{binary}}
\newcommand{\controlX}{\mathrm{CX}}

\newcommand{\controlFunc}{\chi}
\newcommand{\proj}{H_{\controlFunc}}

\newcommand{\simMixer}[1]{\mathrm{sim}\text{-}#1}
\newcommand{\seqMixer}[1]{\mathcal{P}\text{-}#1}
\newcommand{\simCX}{\simMixer{\controlX}}
\newcommand{\seqCX}{\seqMixer{\controlX}}

\newcommand{\add}{\mathrm{ADD}}

\newcommand{\controlXY}{\mathrm{CXY}}

\newcommand{\nullSwap}{\mathrm{NS}}
\newcommand{\simNullSwap}{\simMixer{\nullSwap}}
\newcommand{\seqNullSwap}{\seqMixer{\nullSwap}}
\newcommand{\controlSwap}{\mathrm{CS}}
\newcommand{\simControlSwap}{\simMixer{\nullSwap}}
\newcommand{\seqControlSwap}{\seqMixer{\nullSwap}}
\newcommand{\permSwap}{\mathrm{PS}}
\newcommand{\simPermSwap}{\simMixer{\permSwap}}
\newcommand{\seqPermSwap}{\seqMixer{\permSwap}}
\newcommand{\timeSwap}{\mathrm{TS}}
\newcommand{\timeColor}{\mathrm{TC}}

\newcommand{\charSwap}{\mathrm{CS}}
\newcommand{\NONE}{\mathsf{NONE}}

\newcommand{\neighborFunc}{\mathrm{nbhd}}
\newcommand{\quditDim}{d}
\newcommand{\arity}{k}
\newcommand{\numVars}{n}
\newcommand{\numConstraints}{m}
\newcommand{\perm}{\boldsymbol{\sigma}}
\newcommand{\permElmt}{\sigma}
\newcommand{\invPerm}{\perm^{-1}}
\newcommand{\invPermElmt}{\permElmt^{-1}}
\newcommand{\perms}{\Sigma}
\newcommand{\order}{\boldsymbol{\iota}}
\newcommand{\orderElmt}{\iota}

\newcommand{\graphComp}[1]{\overline{#1}}

\newcommand{\create}{S^{+}}
\newcommand{\annihilate}{S^{-}}

\newcommand{\numColors}{k} 
\newcommand{\degree}{D}
\newcommand{\distance}{d}

\newcommand{\strtTime}{s}
\newcommand{\strtTimes}{\mathbf{s}}
\newcommand{\lastStrtTime}{h}
\newcommand{\releaseTime}{r}
\newcommand{\releaseTimes}{\mathbf{r}}
\newcommand{\window}{W}
\newcommand{\deadline}{d}
\newcommand{\deadlines}{\mathbf{\deadline}}
\newcommand{\weight}{w}
\newcommand{\weights}{\mathbf{w}}
\newcommand{\procTime}{p}
\newcommand{\procTimes}{\mathbf{\procTime}}
\newcommand{\horizon}{h}

\newcommand{\NEQ}{\mathsf{NEQ}}
\newcommand{\EQ}{\mathsf{EQ}}
\newcommand{\NOR}{\mathsf{NOR}}
\newcommand{\OR}{\mathsf{OR}}
\newcommand{\AND}{\mathsf{AND}}

\newcommand\allNeighbors[2][]{
  \ifstrempty{#1}{
    \AND(x_{\neighborFunc(#2)})
  }{
    \AND(x_{\neighborFunc(#1), #2})
  }
}

\newcommand\noNeighbors[2][]{
  \ifstrempty{#1}{
    \NOR(x_{\neighborFunc(#2)})
  }{
    \NOR(x_{\neighborFunc(#1), #2})
  }
}

\newcommand{\quditX}{{\breve{X}}}
\newcommand{\quditZ}{{\breve{Z}}}

\newcommand{\SWAP}{\mathrm{SWAP}}
\newcommand{\XYij}{X_iX_j+Y_iY_j}
\newcommand{\SWAPij}{\SWAP_{i,j}}

%% file: _QAOtoolkit.tex
A basic requirement of many 
quantum algorithms 
is the ability to translate between mathematical functions acting on a domain, typically a string of bits, and a quantum Hamiltonian operator acting on qubits. Indeed, 
mapping Boolean and real functions 
to diagonal Hamiltonians has many important 
applications in quantum computing, 
in particular, 
for algorithms solving decision or optimization problems such as 
quantum annealing and adiabatic quantum optimization 
\cite{nishimoriQA,farhi2000quantum,farhi2001quantum}, or QAOA.  
See \cite{LucasIsingNP,hadfield2017quantum} for a variety of problem mappings.  

In this section, we summarize several results which are particularly 
useful for the QAOA constructions of the remainder of the chapter. 
Their proof and details 
are deferred to Appendix \ref{ch:QAOA0}. 
An expanded presentation of these results appears in~\cite{hadfield2018representation}. 
We emphasize that these results have applications to quantum algorithms beyond QAOA. 
This section is self-contained and 
the remainder of the chapter may be read independently.

\subsection*{Boolean Functions}
Boolean 
functions 
can be represented as diagonal Hamiltonians.
We show how every 
such function naturally maps to a Hamiltonian expressed as 
a linear combination of Pauli $Z$ operators, with terms corresponding to the Fourier expansion of the function. For the (faithful) representation on $n$ qubits, this 
mapping is unique. 

\begin{prop}  \label{prop:1booleanFourierRep}
For a Boolean function $f:\{0,1\}^n \rightarrow \{0,1\}$, 
the unique 
Hamiltonian on $n$-qubits satisfying $H_f \ket{x} = f(x)\ket{x}\;$ 
for each computational basis state $\ket{x}$ is 
\begin{equation}    \label{eq:Hfexpan}
H_f \, = \, \sum_{S\subset [n]} \widehat{f} (S) \; \prod_{j\in S} Z_j 
\, = \,  \widehat{f}(\emptyset)I + \sum_{j=1}^{n}\widehat{f}(\{j\})Z_{j} +  \sum_{j<k}\widehat{f}(\{j,k\})Z_jZ_k + \dots 
\end{equation}
where the Fourier coefficients $\widehat{f} (S)  = \frac{1}{2^n} \sum_{x\in \{0,1\}^n} f(x) (-1)^{S\cdot x} \in [-1,1]$ 
satisfy 
\begin{equation}  \label{eq:f0coefficient}
  \sum_{S\subset [n]}  \widehat{f} (S)^2   \, 
=\,  \frac{1}{2^n} \sum_{x\in \{0,1\}^n} f(x) \, = \, \widehat{f}(\emptyset)  .
\end{equation}
\end{prop}
Here we have used the standard notation $S\cdot x := \sum_{j\in S}x_j$. 
The proof of the proposition follows from Theorem \ref{thm:1booleanFourierRep} which is shown in Appendix \ref{ch:QAOA0}.

Thus, computing the Hamiltonian representation (\ref{eq:Hfexpan}) of a Boolean function is equivalent to computing its Fourier expansion. 
By considering functions corresponding to NP-hard decision problems, from (\ref{eq:f0coefficient})  
we have the
following corollary, which is analogous to the 
well-known 
result that deciding equations of Boolean algebra is NP-hard~\cite{cook1971complexity}. 
\begin{cor}  \label{cor:sharpPFourier}
Computing the identity coefficient $\widehat{f}(\emptyset)$ of the Hamiltonian $H_f$ representing an $n$-variable 
Boolean satisfiability (SAT) formula $f$ 
(given in conjunctive normal form and described by ${\rm poly}(n)$ bits) 
is $\#P$-hard. 
Deciding if  $\widehat{f}(\emptyset)=0$ is equivalent to deciding if $f$ is unsatisfiable, in which case $H_f$ is identically (reducible to) the $0$ matrix. 
\end{cor}

Note that the quantity $\widehat{f}(\emptyset)$ is proportional to the trace of $H_f$ and hence is basis independent. 

We emphasize that even if we could compute the value of each Fourier coefficient, 
a Hamiltonian~$H_f$ representing a general Boolean 
function~$f$ may require an exponential (with respect to $n$) number of Pauli $Z$ terms in the sum (\ref{eq:Hfexpan}). 
We define the size of $H_f$,  
$\text{size}(H_f)$, to be the number of (non-zero) terms in the sum (\ref{eq:Hfexpan}), and the degree $\deg(H_f)=\deg(f)$ to be the maximum locality (number of qubits acted on) of any such term. 
We say that a function $f_n$ on $n$-bits is \textit{efficiently representable} as the Hamiltonian $H_{f_n}$ if ${\rm size}(H_{f_n})$ is ${\rm poly}(n)$ and so is the cost for computing the nonzero coefficients.

\subsection*{Pseudo-Boolean functions}
We are particularly interested in 
real functions $f$ 
given as weighted sums of 
Boolean functions~$f_j$, 
$$ f(x) = \sum_{j=1}^m w_j f_j (x) \;\;\;\;\;  w_j \in \reals,$$  
where $f$ acts on $n$ bits and 
$m={\rm poly}(n)$. 
The objective functions for constraint satisfaction problems, considered in QAOA, are typically expressed in this form. 
A different example is the penalty term approach of quantum annealing, 
where the objective function is augmented with a number of high-weight 
penalty terms which perform local checks to see if a state is feasible. 

Note that we do not deal with issues of how the real numbers $w_j$ may be represented and stored. 
The problems considered in the remainder of this chapter will typically have bounded integer weights, in which case 
this issue 
relates to the precision of the QAOA angles.

We have the following useful general result, the proof of which can be found in Appendix \ref{ch:QAOA0}. 

\begin{prop} \label{prop:pseudoBool}
For an $n$-bit real function $f$ given as 
$ f(x) = \sum_{j=1}^m w_j f_j (x) $, $w_j \in \reals$, 
where the $f_j$ are Boolean functions, 
the unique Hamiltonian on $n$-qubits satisfying $H_f \ket{x} = f(x)\ket{x}$ 
is 
\begin{equation} \label{eq:Hfexpan2}
H_f = \sum_{S\subset [n]} \widehat{f} (S) \; \prod_{j\in S} Z_j 
=  \sum_{j=1}^m w_j H_{f_j},
\end{equation}
with Fourier coefficients 
$\widehat{f} (S)  = \frac{1}{2^n} \sum_{x\in \{0,1\}^n} f(x) (-1)^{S\cdot x} 
= \sum_j \widehat{f_j} (S) \in 
 \reals$, 
where the $H_{f_j}$ are defined as in (\ref{eq:Hfexpan}).

In particular, $\deg(H_f) \leq  \max_j \deg(f_j)=:d$ and ${\rm size}(H_f)\leq
\min \{ \sum_j {\rm size}(H_{f_j}), 1+ (e/d)^{d-1}n^d\}$. 
\end{prop}

\subsection*{Constructing Hamiltonians}
The construction of Hamiltonians representing 
standard Boolean functions follows directly from Proposition~\ref{prop:1booleanFourierRep}. 
We summarize mappings of important basic clauses in Table~\ref{tab:basicHams} below. 

We derive formal rules for obtaining Hamiltonians representing more 
complicated expressions such as Boolean formulas or circuits. 
Together with the basic clauses, Hamiltonians for a 
large variety of functions can be easily derived. 

\begin{theorem}[Composition rules]  \label{thm:compositionRules}
Let $f,g$ be Boolean functions represented by Hamiltonians  $H_f,H_g$. 
Then Hamiltonians representing basic operations on $f$ and $g$ are given by 
  \begin{multicols}{2}
    \begin{itemize}
        \item $H_{\neg f} = H_{\overline{f}} = I - H_f$
        \item $H_{f\wedge g} = H_{fg} = H_f H_g$
        \item $H_{f\oplus g} =  H_f + H_g - 2H_fH_g$
                 \item $H_{f\Rightarrow g} = I - H_f + H_fH_g$
        \item $H_{f\vee g} = H_f + H_g - H_fH_g$
        \item $H_{af+bg} = a H_f+b H_g\;\;\;\;  a,b \in \reals.$
    \end{itemize}
  \end{multicols}
\end{theorem}
The proof of this theorem is given in Appendix \ref{ch:QAOA0}. Note that the rules of the theorem hold regardless of whether $f$ and $g$ act on the same or independent variables (which correspond to overlapping or independent sets of qubits). 

Using these results, 
Hamiltonians representing objective functions for 
many important optimization problems can be easily constructed. 
In particular, we use these results to design phase operators for our QAOA constructions in the remainder of the chapter.

\vskip 2pc
\begin{table}[h!]
\begin{center}
\begin{tabular}{| c | c || c | c |}
	\hline
	$f(x)$ &$H_f$ & $f(x)$ & $H_f$  \\ 
	\hline
	\hline
	$x$ & $\frac12 I - \frac12 Z$ & $\overline{x}$ &$\frac12 I + \frac12 Z$   \\
	\hline
	$x_1 \wedge x_2$   &  $\frac14I -\frac14(Z_1+Z_2-Z_1Z_2)$  & $\bigwedge_{j=1}^kx_j$ & $\frac{1}{2^k} \prod_j (1-Z_j)$ \\
	\hline
	$x_1 \vee x_2 $  & $\frac34I -\frac14(Z_1+Z_2+Z_1Z_2)$  & $\bigvee_{j=1}^kx_j$ & $1-\frac{1}{2^k} \prod_j (1+Z_j)$ \\
	\hline
	$x_1 \oplus x_2 $ & $ \frac12 I-\frac12 Z_1Z_2$ & 
	$x_1 \Rightarrow x_2$  & $ \frac34 I + \frac14(Z_1 - Z_2 + Z_1Z_2)$  \\
	\hline
\end{tabular}
\end{center}
\caption{Hamiltonians representing basic Boolean clauses. 
}
\label{tab:basicHams}
\end{table}

\subsection*{Simulating Diagonal Hamiltonians}
It is well known that if a function $f$ can be efficiently computed classically, and if ancilla qubits are available, then the Hamiltonian $H_f$ can be simulated efficiently by computing $f$ in a register and performing controlled rotations; see, e.g., \cite{childs2004quantum}. 
These methods avoid computing $H_f$ explicitly. 
On the other hand, there exist applications where a Hamiltonian-based implementation is desirable, 
such as quantum annealing, or cases where we wish to minimize the need for ancilla qubits, such as, for example, low-resource applications of QAOA.  

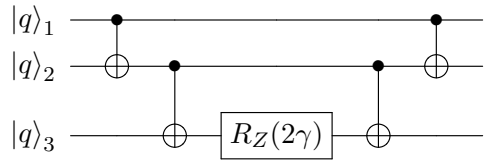
\begin{figure}[H]
\centerline{
\Qcircuit @C = 1.2 em @R = 1.2 em {
\lstick{\ket{q}_1} & \ctrl{1} & \qw      & \qw                             & \qw       & \ctrl{1}  & \qw \\
\lstick{\ket{q}_2} & \targ    & \ctrl{1} & \qw                              &  \ctrl{1} & \targ    & \qw \\ 
\lstick{\ket{q}_3} & \qw      & \targ    & \gate{R_Z (2\gamma) } & \targ     &  \qw    & \qw \\ 
}
}
\vspace*{8pt}
\caption{Quantum circuit performing the operation $U=exp(-i\gamma Z_1 Z_2 Z_3)$ on 
three qubits labeled $1$, $2$, and $3$. The middle operator is a $Z$-rotation gate, and the other gates are controlled-NOT gates with
a black circle indicating the control qubit and cross indicating the target. 
By similar circuits, 
$U=exp(-i\gamma Z_1 Z_2 \dots Z_\ell)$ can be implemented with $2(\ell-1)$ CNOT gates and one $R_Z$ gate. 
Different circuit compilations are possible, including compilation to different gate sets. 
} 
\label{fig:RZZcircuit}
\end{figure}

Efficient circuits simulating products of Pauli $Z$ operators are known as 
shown in Figure \ref{fig:RZZcircuit}. 

\vskip 2pc
Thus, a Hamiltonian $H_f$ representing a general Boolean or pseudo-Boolean function may be simulated efficiently if the number of 
Pauli $Z$-
terms in the sums (\ref{eq:Hfexpan}) or (\ref{eq:Hfexpan2}) are not too many. 
%
Recall 
that we define \textit{basic quantum gates} to be the universal set of CNOT and single-qubit gates. 

\begin{cor}
A Hamiltonian $H_f$ as in (\ref{eq:Hfexpan}) or (\ref{eq:Hfexpan2}) 
can be simulated (i.e., the operation $exp(-i\gamma H_f)$ implemented for $\gamma =O(1)$) with 
$n$ qubits and 
%
$O(\deg(H_f)\cdot \text{size}(H_f))$ 
basic quantum gates. Thus, if $\text{size}(H_f)$ is upper bounded by a polynomial in $n$, then $H_f$ can be simulated efficiently.

In particular, Hamiltonians $H_f$ with bounded maximum degree $d=O(1)$ can be simulated with $O(n^d)$ basic gates. 
\end{cor}
The corollary follows from the above results, the circuits indicated in Figure \ref{fig:RZZcircuit}, and simple counting arguments. 
We remark that the Hamiltonian simulation considered in the 
corollary is exact in the sense that if each of the 
basic gates 
is implemented exactly, then so is $exp(-i\gamma H_f)$. The approximation of quantum gates is an important topic but we do not deal with it here; see, e.g., \cite{NC}. 
Moreover, no ancilla qubits are necessary for the simulation. 

\subsection*{Controlled Hamiltonians and Unitaries}
In many applications we require controlled unitary operations, or, in particular, controlled Hamiltonian simulations. 
Consider two quantum registers of $k+n$ qubits. 
Given a $k$-bit Boolean function~$f(y)$ and a unitary operator $U$ acting on $n$ qubits, 
we define the $(k+n)$-qubit $f$-controlled unitary operator $\Lambda_f (U)$ by its action on basis states 
$$ 
\Lambda_f (U) \, \ket{y}\ket{x} =  \left\{
                \begin{array}{ll}
                  \ket{y}\ket{x}\;\;\;\;\;\;\; \;  f(y)= 0  \\
                   \ket{y}U\ket{x} \;\;\;\;\;  f(y)= 1.
                \end{array}
              \right.
$$
\begin{prop}  \label{prop:controlledHams}
Let $f$ be a 
Boolean function represented by a $k$-qubit Hamiltonian $H_f$, 
and let $H$ be an arbitrary Hamiltonian acting on $n$ qubits. 
Then the $(k+n)$-qubit Hamiltonian 
\begin{equation}  \label{eq:ctrlHam}
\widetilde{H} = H_f \otimes H
\end{equation}
corresponds to 
$f$-controlled evolution under $H$,
i.e., satisfies 
$\; e^{-i \widetilde{H}t} = \Lambda_f ( e^{-iH t}).$  
\end{prop}
The proof follows from exponentiating (\ref{eq:ctrlHam}) directly. 
We will use this result 
many times in the remainder of the chapter 
to construct controlled mixing operators for QAOA 
for constrained optimization problems with feasibility constraints. 
 These operators implement evolution under a local mixing Hamiltonian $H$ only if a Boolean function $f$ is true,  where for each basis state $f$ checks that the action of the mixing Hamiltonian will preserve feasibility.

%% file: _ch_conclusions.tex
\chapter{Conclusions and Future Work}
\label{sec:conclusions}

In this thesis we have studied five problems related to 
quantum algorithms for scientific computing and approximate optimization. 

We first described a modular approach to scientific computing on quantum computers. We showed quantum algorithms and circuits for computing square roots, logarithms, and arbitrary fractional powers, and derived worst-case error and cost bounds. 
In order for future quantum computers to have impact for scientific problems, 
it will be important to develop numerical standards and libraries. Our work represents important first steps in this direction. 
A natural next step is to 
derive efficient quantum algorithms for additional useful numerical functions. Of particular interest are functions where the range of the input may be much smaller than that of the output, or vice versa; two examples are the exponential and arctangent functions. New techniques may be required for such functions to ensure that the 
input and output are represented efficiently while the error remains under control. 
Furthermore, it is important to explore specific applications of our circuits, 
as subroutines of larger quantum algorithms, to problems where quantum computers give advantages over 
their classical counterparts. 

We then considered quantum algorithms for approximating ground and excited state energies, i.e., Hamiltonian eigenvalues. This problem suffers from the curse of dimensionality; 
for an $\ell$-particle system, the cost of the best classical algorithms grows exponentially with $\ell$. 
We showed a general quantum algorithm for approximating a constant number  
of low-order Hamiltonian eigenvalues 
using a perturbation approach. We then applied this algorithm to a 
computationally difficult special case of the Schr\"odinger equation and showed that 
our algorithm succeeds with high probability, 
and with cost polynomial in the number of degrees of freedom and the reciprocal of the desired accuracy. 
Our results significantly extend earlier work showing quantum computers can break the curse of dimensionality for this problem. 
It is important to continue working in this direction, to further weaken the assumptions if possible, and to extend the scope of our algorithm and its applications, in particular, to first-quantized approaches to important problems in physics and chemistry. 

We next considered quantum algorithms for the simulation of quantum mechanical systems. We showed a novel divide and conquer approach for Hamiltonian simulation that takes advantage of the Hamiltonian structure to yield faster simulation algorithms. 
We illustrated our results by applying our approach to the the electronic structure problem of quantum chemistry, and  showed significantly improved cost estimates under very mild assumptions. A next step for chemistry applications is to investigate particular classes of molecules and single-particle basis functions where our approach is particularly advantageous over current methods. Generally, it is important to further investigate the power and limitations of quantum algorithms for Hamiltonian simulation. An 
open problem is whether or not there exist Hamiltonian simulation algorithms for important applications with cost that scales 
polynomially in $\log \|H\|t$, $\log \e^{-1}$ and $n$, where $n$ is the number of qubits the Hamiltonian $H$ acts on. Finally, it remains open whether further improvements can be obtained by using our divide and conquer approach in combination with other Hamilton simulation algorithms as subroutines in place of splitting formulas.

Quantum algorithms for approximate optimization 
are relatively unexplored, with many basic problems open. Indeed, the fundamental question remains far from resolved: do quantum computers provide advantages for the approximation of classically hard combinatorial problems? 
We studied the application of the recently proposed quantum approximate optimization algorithm (QAOA) to the Maximum Cut problem. We showed a general technique, the Pauli Solver algorithm, which we applied to derive analytic performance bounds for the lowest depth realization of the algorithm. The details and proof of our results exemplify the difficulty of obtaining similar performance bounds for other problems or deeper circuits. 
Indeed, characterizing the performance of QAOA for depth $p>1$, 
particularly how the performance improves as a function of $p$, 
remains the most important open problem for QAOA. 
It is important to find ways to further improve our techniques towards obtaining such results. Furthermore, for QAOA of arbitrary fixed depth, it remains open to classify, or give an efficient general procedure for finding, sufficiently good algorithm parameters, which is critical for QAOA to be effective in practice. 

We then showed a generalization of QAOA to wider classes of quantum operators and states, the Quantum Alternating Operator Ansatz. Our approach is especially suitable to optimization problems with feasibility constraints. After specifying design criteria and a design toolkit, we applied our approach to yield efficient constructions for a variety of prototypical 
optimization problems. We derived explicit cost estimates for these constructions, in each case showing appealing resource 
scaling indicative of suitability for early quantum computers. 
Our ansatz allows freedom in the selection of operators and initial states. 
An important future direction is to investigate the trade-off between cost and performance in this selection; 
 e.g., is a more costly mixing operator preferable to a less costly one, in terms of performance, if it means we can only afford to implement fewer rounds of the algorithm? Moreover, can we derive criteria specifying the best possible initial states and mixing operators? Quantum algorithms for approximate optimization remain at an early stage, so many open questions remain. 

Finally, a primary future 
goal is to implement 
the algorithms of this thesis 
on physical quantum computers. As such devices become available, we are optimistic that experimentation, analysis, and testing will empower algorithm designers to discover a variety of new and improved impactful applications of quantum computing.

%% file: _app_QC.tex
\chapter{A Brief Overview of Quantum Computation}
\label{ch:QC}

Quantum computation lives at the intersection of quantum physics, computer science, mathematics, and engineering. In this 
appendix we attempt to give a very brief overview of the key concepts, 
in particular the most relevant for the results of this thesis, and to explain the notation
we will use throughout. 
Many excellent comprehensive introductions to quantum computation exist, notably \cite{NC,KitaevText,preskill1998lecture,rieffel2011quantum}. 
Similarly, see \cite{sakurai1995modern,griffiths2016introduction,Gustafson,shankar2012principles,woitBook} for detailed 
overviews of quantum mechanics and its mathematical structure. 

The idea of using quantum mechanics 
for computation 
is generally attributed to 
Feynman \cite{Feynman} 
who suggested that \textit{universal quantum simulators} 
may provide computational advantages for tasks which appear to require exponential resources classically, such as simulating quantum systems themselves. 
Foundational work by Deutsch \cite{deutsch1985quantum,deutsch1989quantum} and subsequently Yao \cite{yao1993quantum} developed computational models based on quantum counterparts 
to classical Turing machines and Boolean circuits. 
Interest in quantum computation grew dramatically in 1994 when Shor \cite{Shor} showed a 
quantum algorithm for prime factoring (decomposing a number $N$ into its prime factors) requiring only $\textrm{poly}(\log N)$ quantum operations, an exponential speedup over all known classical algorithms. 
The physical realization of such an algorithm could be used to break 
certain public-key cryptography systems such as RSA, whose security is based on the assumption that prime factoring is computationally intractible. 
A second paramount 
result was the algorithm of Grover \cite{grover1997quantum}, which showed that an $N$ element unstructured search problem could be solved a quantum computer with only $O(\sqrt{N})$ queries to the list, whereas any classical algorithm requires $\Omega(N)$ queries. 

Subsequently, quantum algorithms have been developed for a wide variety of discrete and continuous problems. In some cases, significant speedups have been shown, yet it remains open whether or not quantum devices are truly more powerful than classical computers. 
We do not attempt a summary of applications here; see for example  \cite{NC,Qcontinuous,childs2010quantum,lanyon2010towards,montanaro2015quantum} for overviews of quantum algorithms.

\section{Quantum Mechanics of Quantum Computation}
The fundamental equation of quantum mechanics, 
the \textit{Schr\"odinger equation} 
\begin{equation}  \label{eq:SchEqGen}
\frac{d}{dt} \psi (t) = - i H \: \psi (t),
\end{equation}
says that quantum 
systems evolve \textit{unitarily} in time, 
governed by the \textit{Hamiltonian operator} $H$ which encodes the 
system energy levels.\footnote{We use standard \textit{natural units} where the \textit{reduced Planck constant} $\hslash=1$, making (\ref{eq:SchEqGen}) dimensionless; see e.g. \cite{woitBook}. }
States 
$\psi(t)$ are \textit{complex} vectors 
which encode the probability distributions of possible measurement outcomes on the given physical system. 
Thus, as we shall outline, 
 quantum states evolve in a way fundamentally different from classical probabilistic (stochastic) processes. 
Hence, 
the foundational question of quantum computing
is whether or not 
the \lq\lq strange\rq\rq\ behaviour of quantum 
systems 
can be used to give 
computational advantages over classical computers. 

For the purposes of quantum computation, finite dimensional quantum mechanics suffices, which is much simpler and more well-behaved than the fully general theory. Hence, quantum computation reduces to a subset of \textit{matrix mechanics} (i.e., complex linear algebra). We remark that alternative quantum information processing schemes such as continuous-variable quantum computation 
\cite{lloyd1999quantum} have been proposed; 
however, these models are not considered here. 

Analogous to a classical discrete bit $x \in \{0,1\}$, 
a \textit{qubit} 
is defined to be a two-dimensional quantum system, 
which is described in general 
 by a vector in a complex Hilbert space $\ket{\psi} \in \complex^{2}/\{{\bf 0}\}$.  
In the (orthonormal) \textit{computational basis}, which we label 
$\ket{0},\ket{1}$, 
the general state of a qubit may be written
\begin{equation}   \label{eq:defQubit}
\ket{\psi} = a\ket{0} + b\ket{1}.       \;\;\;\;\;\;\; \|\ket{\psi} \|^2 = |a|^2 + |b|^2=1.
%
\end{equation}
All quantum states are physically equivalent under multiplication by a complex scalar  
(i.e., up to \textit{normalization} and \textit{global phase}), so without loss of generality we 
assume states are normalized with 
$\|\ket{\psi} \| = 1$ (we use $\|\cdot\|$ to denote the Euclidean norm). 
%
The standard computational basis is taken to be the eigenvectors of the Pauli $Z$ operator, 
which is typically 
experimentally convenient for facilitating bit readout. 
It acts as
$$ Z\ket{0} = \ket{0}     \;\;\;\;\;\;\;\; Z\ket{1} = -\ket{1} ,$$
so we may write $Z = \ket{0}\bra{0} - \ket{1}\bra{1}$.   
A prototypical example for a systems of electron spins is the identification of 
 $\ket{0}$ with an electron spin-down state, and $\ket{1}$ with the spin-up state.%

If we \textit{measure} $\ket{\psi}$ in the $Z$ 
basis, we obtain outcome \lq$0$\rq\ with probability $|a|^2$ or outcome 
\lq$1$\rq\ with probability $|b|^2 = 1 -|a|^2$, and accordingly the state $\ket{\psi}$
\textit{collapses} to $\ket{0}$ or $\ket{1}$, respectively,
$$
\ket{\psi} 
\xrightarrow[]{\text{Measure Z}}
\left\{
                \begin{array}{ll}
                  \ket{0} \;\;\;\;\;\; \text{ with probability } |a|^2\\
                 \ket{1} \;\;\;\;\;\; \text{ with probability } |b|^2. \\
                \end{array}
              \right.
$$
%
%
%
%
Thus prior to measurement $\ket{\psi}$ gives a probability distribution over the two possible outcomes, whereas immediately after measurement the state is determined, 
given by the basis vector corresponding to the observed measurement outcome. 

Indeed, every possible qubit basis corresponds to a \textit{measurement observable}, which is 
a self-adjoint matrix $M$ constructed so that the two basis vectors are its non-degenerate eigenvectors. Measurement of the observable $M$ on a qubit 
probabilistically returns one of the eigenvalues of $M$, 
which indicates 
a '0' or '1' bit value, 
and collapses the state to 
the corresponding eigenvector. 
 Therefore, 
despite the number of qubit states being uncountable, 
any qubit measurement still only returns one of two possible outcomes $\{0,1\}$, i.e. a classical bit of information. 

It is important to elaborate on our notation. 
We use the standard \textit{bra-ket} notation, where a \textit{ket} $\ket{j}$ is the vector labeled $j$, 
and the \textit{bra} $\bra{k}$ is the adjoint linear functional%
\footnote{For finite dimensional Hilbert spaces, we have a natural isomorphism between $\mathcal{H}$ and its \textit{dual space} $\mathcal{H}^*\simeq\mathcal{H}$, so we can naturally identify a \textit{bra} $\bra{\psi}\in \mathcal{H}^*$ with the map $(\ket{\psi}, \cdot):\mathcal{H}\rightarrow\mathbb{C}$. 
See for example \cite{Gustafson,Folland}.} 
corresponding to the vector labeled $k$, 
 equivalently represented as the complex-conjugated row vector $\bra{k} =  \ket{k}^\dagger$. 
The complex inner product is then compactly represented as the \textit{braket},   $\braket{k}{j} := (\ket{k},\ket{j})=\ket{k}^\dagger \cdot \ket{j} $. 
%
%
For 
matrices $A$ acting on qubits we write 
$\bra{\phi}A\ket{\psi} := (\ket{\phi}, A\ket{\psi}) =  (A^\dagger \ket{\phi}, \ket{\psi}) $. 
The computational basis orthonormality conditions may be written $\braket{0}{0}=1=\braket{1}{1}$ and $\braket{0}{1}=0$, and 
a general (normalized) state $\ket{\psi}$ then satisfies $\braket{\psi}{\psi}=\|\psi\|^2=1$.

The \textit{quantum computational state} of $n$ \textit{qubits} 
is a vector $\ket{\psi}$ in the tensor product Hilbert 
space $(\complex^2)^{\otimes n} \simeq \complex^{2^n} $, with $2^n -1$ complex degrees of freedom (states remain unique up to normalization and overall phase). 
Thus, a general quantum state is described by a number of coordinates exponential in the number of qubits. 
The natural $n$-qubit computational basis is given by the tensor products of the single qubit basis states, which we write as,
e.g., $\ket{0} \otimes \ket{0} \otimes \ket{1} \otimes \ket{0} = \ket{0010}$. 

A primary difference with classical mechanics is that arbitrary linear combinations of quantum states also give states, known 
as the \textit{quantum superposition} principle. 
Furthermore, \textit{quantum entanglement} is the property that 
an arbitrary state $\ket{\psi} \in \complex^{2^n}$ 
cannot be factored as $\ket{\psi} = \ket{\psi_1} \otimes \ket{\psi_2} \otimes \dots \otimes \ket{\psi_n}$, with each $\ket{\psi_i} \in \complex^2$; 
if this was possible generally 
then every quantum state would have an efficient classical description. We remark that both of these nonclassical properties follow directly from the tensor product structure of the underlying state space.

For $n$-qubits, 
it is useful to identify the computational basis vectors 
$\ket{00\dots0},  \ket{00\dots1}, \dots \ket{11\dots 1}$ with the unsigned integers $\ket{0},\ket{1},\dots,\ket{2^n -1}$. (In Chapter \ref{ch:sciComp} we similarly consider more general signed numbers with fractional parts). Thus a general $n$ qubit state may be written
$$ \ket{\psi} = \sum_{j=0}^{2^n -1} c_j \ket{j},$$
where we normalize $\|\psi\|^2 = \sum_j |c_j|^2 = 1$. 
Measurement of such a state in the computational basis 
gives outcome $j$ with probability $|c_j|^2$. 
Thus, quantum states encode the probability of measurement outcomes in the \textit{probability amplitudes} $c_j$. 
We may write $|c_j|^2 = \bra{\psi} P_j \ket{\psi}$, where $P_j = \ket{j}\bra{j}$ is the projector onto a computational basis state $\ket{j}$.

More generally, a measurement observable $M$ is a self-adjoint operator acting on one or more qubits. Measuring $M$ on $\ket{\psi}$ 
returns an single eigenvalue $\lambda$ with probability $\bra{\psi} P_\lambda \ket{\psi}$, where $P_\lambda$ is the projector onto the $\lambda$-eigenspace of $M$. 
For example, measuring the observable $Z_1Z_2$ returns 
$1$ for states where the first two bits are equal and $-1$ for states otherwise; this measurement does not reveal the particular value of the first or second or remaining qubits. 

Turning finally to dynamics, for quantum computation,  
Hamiltonians $H$ are given by $2^n \times 2^n$ 
self-adjoint (Hermitian) matrices. 
When $H$ is time-independent, equation (\ref{eq:SchEqGen}) is solved by
\begin{equation}  \label{eq:SchEqGenSoln}
\ket{ \psi(t)} = e^{-iHt} \ket{\psi(0)}, 
\end{equation}
where the unitary operator $U = {\rm exp}(-iHt)$ preserves state normalization. 
We say such a quantum system $\ket{\psi}$ \textit{evolves} under the Hamiltonian $H$ for time $t$, or equivalently is evolved under the Hamiltonian $Ht$. 
For qubits, any unitary transformation may 
be written as ${\rm exp}(-iH)$ for some Hamiltonian $H$. 
(A measurement, on the other hand, is a non-unitary transformation.)

\subsection*{Pauli Matrices}
A matrix $A$ that is both self-adjoint ($A^\dagger = A$) and unitary ($A^\dagger A = I$) is a square root of the identity, $A^2 = I$. 
The exponential of such a matrix is given by
$e^{-i\gamma A } = \cos (\gamma) I -i \sin (\gamma) A.$ 
A particularly useful set of such matrices are the matrices 
\begin{equation}   \label{eq:PauliMatrices}
I = \left(\begin{array}{cc}1 & 0\\ 0 & 1\end{array}\right), \;\;\;\;
X = \left(\begin{array}{cc}0 & 1\\ 1 & 0\end{array}\right), \;\;\;\; Y = \left(\begin{array}{cc}0 & -i\\ i & 0\end{array}\right), \;\;\;\; Z = \left(\begin{array}{cc}1 & 0\\ 0 & -1\end{array}\right), 
\end{equation}
and their tensor products, which we collectively call the \textit{Pauli matrices}.
The defining relation of the matrices $X,YZ$ is $[X,Y]=2iZ$, and its cyclic permutations, 
where the \textit{matrix commutator} is defined $[A,B]:= AB-BA$. 
We write $X_j$ to indicate the matrix $X$ acting on the $j$th qubit, and do not write identity factors explicitly, e.g. 
$I \otimes Z \otimes Y \otimes I \dots \otimes I$ is written as $Z_2 Y_3$.  

The Pauli matrices give a basis for the vector space of $n$-qubit Hamiltonians. 
Hence, any Hamiltonian $H$ may be expanded as a linear sum 
\begin{equation}  \label{eq:PauliExpansion}
H = a_0 I + \sum_{j=1}^n \sum_{\sigma = X,Y,Z} a_{j\sigma} \sigma_j 
+ \sum_{j \neq k } \sum_{\sigma = X,Y,Z} \sum_{\lambda = X,Y,Z}  a_{jk\sigma\lambda} \sigma_j  \lambda_k + \dots ,
\end{equation}
with 
$a_\alpha \in \reals$. 
Thus, an arbitrary Hamiltonian 
is specified by $4^n$ real coefficients.
Moreover, up to pesky factors of $i$, Hamiltonians give a representation of the 
Lie algebra $i u(2^n)$, i.e., they are closed under ($i$ times) the commutator. 
Thus, exponentials $e^{-iH}$, and in particular 
evolution under sums of Pauli operators, gives the full group $U(2^n)$ of 
unitary transformations on qubits.

We remark that in the computational basis, the operator $X$ acts as $X\ket{0}=\ket{1}$ and $X\ket{1}=\ket{0}$. Hence we identify $X$ as the \textit{bit-flip} operator, i.e. the NOT operation. The eigenstates of $X$ are denoted $\ket{+}$ and $\ket{-}$, with $\ket{\pm}=(\ket{0} \pm \ket{1})/\sqrt{2}$, and related to the computational basis by the \textit{Hadamard gate}
\begin{equation}  \label{eq:Hadamard}
{\textrm H} = \frac{1}{\sqrt{2}} (X+Z)   = \frac{1}{\sqrt{2}}\left(\begin{array}{cc}1& 1\\ 1 & -1 \end{array}\right)
\end{equation}
which transforms between qubit bases ${\textrm H} \ket{0}=\ket{+}$, ${\textrm H} \ket{1}=\ket{-}$, and is self-inverse ${\textrm H}^2=I$.

\section{Quantum Computational Model}
In this thesis we consider the standard \textit{quantum circuit model}. We emphasize that our computational model is abstract; 
we are generally not overly concerned with the underlying physics. 
Many
alternative quantum 
computational models have been proposed, 
such as 
the quantum Turing machine \cite{deutsch1985quantum}, and 
adiabatic \cite{albash2016adiabatic}, measurement-based \cite{briegel2009measurement}, and topological \cite{wang2010topological} quantum computation, 
among others. Typically, these models are equivalent to the quantum circuit model and to each other 
under efficient 
computational reductions \cite{NC}, i.e., they can simulate one another with polynomial overhead. 
Such models are said to be \textit{universal} for quantum computation, an analogy to the many equivalent models of classical computation.  This naturally leads to the \textit{Quantum Church-Turing thesis}, which asserts that \textit{a quantum Turing machine can efficiently simulate any realistic model of computation}.

A \textit{quantum algorithm} is defined to be the product of an ordered sequence of \textit{unitary operations} $U_1,U_2,\dots, U_\ell$, i.e,. algorithmic steps, which 
maps an input state $\ket{s}$ to the output state 
\begin{equation}  \label{eq:quantumAlgorithm}
 \ket{\psi} = U_\ell \dots U_2 U_1 \ket{s}.
\end{equation}
For some algorithms $U=U_\ell \dots U_2 U_1$ 
the goal is to create a state $\e$-close to some target state $\ket{\phi}$ in some norm, 
$\|U\ket{s}-\ket{\phi}\| \leq \e$, or more generally, to $\e$-approximate a target operator $V$, $\|U-V\|\leq \e$. 
Example include Hamiltonian simulation, which we study in Chapter \ref{ch:HamSim}, or distribution sampling problems. 
For other algorithms, such as classical decision or function problems, 
the generation of~$\ket{\psi}$ is followed by a measurement which reveals some classical bit string $x$ with some probability~$p_x$. Without loss of generality, all intermediate measurements may be deferred until the end of a given quantum computation \cite{NC}. 

The qubits for a quantum algorithm are partitioned into  
\textit{quantum registers}. 
Different registers may be used for storing different parts of the computation such as the input, output, or intermediate results. Many computations can be simplified with additional \textit{ancilla} qubits which 
are often used as scratchpad for temporary results. Qubits are an important computational resource, and we say a quantum algorithm is \textit{space efficient} if it requires a number of qubits that is bounded by a polynomial function of its input size.

Like classical algorithms, 
a quantum algorithm must be 
compiled down to a sequence of physically implementable basic operations. 
Each operation 
$U_j$ may be further decomposed 
as $U_j = U_{j \ell_j} \dots U_{j 2} U_{j 1}$, where the unitaries $U_{j k}$
are each from a set of primitive unitaries which act locally on a small number of qubits, called \textit{quantum gates}.
From unitarity, each gate has fan-out equal to fan-in.  
Consecutive gates which act on disjoint sets of qubits 
are commuting operators, and may be 
applied simultaneously, i.e., in parallel. 
We may draw \lq wires\rq, i.e. edge-disjoint paths in a directed acyclic graph, 
which indicate for each qubit the operations applied to it at each step of the computation.   
We call this representation a \textit{quantum circuit}; see Chapter \ref{ch:sciComp} for many example circuits. A gate set is said to be \textit{universal}  
for quantum computation if 
any arbitrary 
$n$-qubit unitary operation can be 
approximated to any desired accuracy using gates from the set \cite{NC}. 
A fundamental result in the quantum gate model is that universal (finite) sets of one-qubit and two-qubit gates exist; these correspond to operations believed experimentally implementable. 

The matrices $X$, $Y$, $Z$ given in (\ref{eq:PauliMatrices}) 
are important examples of single-qubit quantum gates, as are their exponentials the \textit{rotation gates}
 $$ R_Z(\gamma) = e^{-\frac{i\gamma}{2} Z } =\left(\begin{array}{cc}e^{-i\gamma/2} & 0\\ 0 & e^{i\gamma/2}\end{array}\right), \;\;\;\;
 R_X(\beta) = e^{-\frac{i\beta}{2} X } = \left(\begin{array}{cc}\cos (\beta/2) & -i \sin (\beta/2)\\ -i \sin (\beta/2) & \cos (\beta/2) \end{array}\right),
 $$
 and $ R_Y(\gamma) = e^{-i\gamma Y /2}$. 
 These satisfy the useful identity $e^{-i\gamma \Sigma /2} = \cos(\gamma/2) I - i \sin(\gamma /2) \Sigma$, where $\Sigma = I,X,Y,Z$.  
 Furthermore, up to global phase, every single-qubit unitary transformation $U$ can be written 
 $U=R_Z(\alpha) R_Y (\beta) R_Z (\gamma)$ for some $\alpha,\beta,\gamma\in\reals$. 
Other important single-qubit quantum gates include 
the Hadamard gate ${\textrm H} $ given in (\ref{eq:Hadamard}), the
$T$ gate $T=R_Z(\pi/4)$, and the $S$ gate $S=R_Z(\pi/8)$; see \cite[Ch. 4]{NC} for details. 

For a gate set to be universal a multiqubit \textit{entangling gate} is required. 
An important such gate in the controlled-NOT (equivalently controlled-$X$ or \textit{controlled bit-flip}) gate $CNOT=\Lambda_1(X_2) := \frac12 X_2 - \frac12 Z_1 X_2 $, which is drawn as indicated in Fig. \ref{fig:RZZcircuit}. 
On computational basis states $\Lambda_1(X_2)$ flips the second \textit{target} bit only if the first \textit{control} qubit is $1$, i.e.,  
takes $\ket{a}\ket{b}\rightarrow\ket{a}\ket{b\oplus a}$ for $a,b\in\{0,1\}$. 
Two important examples of universal gate sets are 
CNOT with arbitrary single qubit gates, and 
the particular set $\mathcal{G}=\{H,T,CNOT\}$. 
A fundamental result is the \textit{Solovay-Kitaev theorem}, which implies that any quantum circuit consisting of $\ell$ CNOT and  
single qubit gates can be approximated to within accuracy $\e$ using $\ell \cdot \text{poly} \log (\ell/\e)$ gates from 
$\mathcal{G}$ \cite{NC}. Hence, 
without loss of generality 
we will consider circuits drawn from either gate set, 
as gate counts will be the same up to polylogarithmic factors; 
we call such gates \textit{basic}. 

For computation, operators controlled by the values of particular qubits are especially useful. 
Generalizing the CNOT gate, 
the \textit{controlled unitary} $\Lambda_1(U_2)$ applies the operator $U$ to the second qubit on basis states where the first bit is $1$, and otherwise acts as the identity. 
We can implement any $\Lambda_1(U_2)$ using $O(1)$ basic gates.  
A related family of gates, which are particularly useful for intermediate representations of multiqubit gates, are the 
\textit{multi-controlled Toffoli gates} $\Lambda_\ell (X)$, ${\ell \in \naturals}$. These gates act on $(\ell+1)$-qubit basis states as $\Lambda_\ell (X)\ket{x}\ket{a}=\ket{x}\ket{a\oplus \wedge_{j=1}^\ell x_j}$. 
Recall, e.g., the circuits of Figures \ref{fig-InitState1} and \ref{fig:RZZcircuit}. 
The $\ell=1$ gate $\Lambda_1 (X)$ is the CNOT gate.
For $\ell=2$, we have the controlled-controlled not gate $\Lambda_2 (X)$, 
known as the Toffoli gate. 
The Toffoli gate is universal for classical reversible computation, and naturally important for quantum computation. A constant number of basic gates suffices to implement a Toffoli gate \cite{NC}.  
More generally, a $\Lambda_\ell (X)$ gate can be implemented with $O(\ell)$ basic gates, and use of a single temporary ancilla qubit \cite{VBE96}; different compilations are possible. 
We can further combine Toffoli and $\Lambda_1(U)$ gates to create multi-controlled unitaries $\Lambda_\ell(U)$; see e.g. \cite{VBE96} for details. 

For a fixed gate set, the cost of a quantum algorithm $U$ is the minimal number of gates it can be decomposed (compiled) into, or alternatively the depth of such a decomposition, 
in addition to the number of qubits required. 
There typically exist time-space trade-offs in the cost, generalizing those for classical reversible circuits. For example, in some proposed architectures, certain gates may be much more \lq expensive\rq\ than others; this could result from a lower-level quantum error correcting code used to encode the logical qubits, meaning that each quantum gate must typically be further compiled to even lower-level operations. Quantum error correction is a rich topic we do not explore here; see e.g. \cite{preskill1998lecture,NC,lidar2013quantum}. 
We remark that in some cases the cost of a quantum algorithm is taken to be the number of higher-level operations, which provides an indication of cost independent from a specific gate set. For example, in our algorithms of Chapter \ref{ch:sciComp}, the cost 
is taken to be the number of required addition and multiplication operations. A variety of schemes exist in the literature for these basic arithmetic operations, with different trade-offs themselves. 
Furthermore, in some applications certain operators $U_j$ implement \textit{oracle calls}, i.e., access to an unknown black-box operation, in which case the appropriate cost metric is the separate numbers of oracle and non-oracle operations.

We say a quantum algorithm for a given problem is \textit{(time) efficient} if its cost is polynomially bounded with respect to the problem input size $n$. More precisely, 
the quantum circuit for each problem size must be \textit{uniformly} generated;
there must exist a classical deterministic Turing machine which given the input string $1^n=11\dots 1$ 
generates the quantum circuit description 
in polynomial time. 
As is this case with classical circuits, non-uniform quantum circuits appear to be artificially powerful  \cite{AroraBarak}. 

For a decision problem, a quantum algorithm seeks to output a single bit with probability $p > 1/2 + c$ for some constant $c$. If this is achieved, then the problem can be solved by repeating the algorithm and taking the majority vote. For other problems, the output will be a bit string~$x$, which solves the problem with some probability $p$. Using amplitude amplification \cite{brassard2002quantum}, the success probability can be boosted close to unity with $O(1/\sqrt{p})$ repetitions, a quadratic improvement over the classical case. Thus, if $p$ is only polynomially small, i.e., $1/p$ is bounded by a polynomial in $n$, then a polynomial number of repetitions suffice. 

We may define complexity classes for quantum computation analogous to those for classical computation. 
The class of decision problems  efficiently decidable by such a procedure with success probability at least $1/2$ plus a constant is known as BQP, and naturally contains its classical probabilistic analog BPP.  Similarly, the class QMA gives the quantum analog of the classical complexity class MA, which may further be seen as the probabilistic analog of the class NP. 
Quantum complexity classes also have complete problems; as is the case for NP,  
a variety of $QMA$-complete problems have been discovered; see \cite{qmaSurvey,QNPSurvey} for surveys. Moreover, 
complete problems are also know for the complexity class BQP; see \cite{wocjan2006several}. (In contrast, BPP is not believed to have complete problems unless P=BPP.)
Furthermore, natural quantum extensions of \textit{qubit (space), query, circuit} and \textit{communication} complexity can be defined.
Indeed, quantum complexity theory is a rich area which we do not explore further here; 
see e.g. \cite{bernstein1997quantum,Watrous} for details.

Finally, 
we 
emphasize that quantum computation subsumes classical computation. 
As classical computation is no more powerful than classical reversible computation, with polynomial overhead any classical circuit can be converted into a reversible circuit and subsequently efficiently simulated by a quantum circuit \cite{bennett1989time,NC}.


%% file: _app_QAOtoolkit.tex
\chapter{Design Toolkit for Quantum Optimization} 
\label{ch:QAOA0}
In this appendix 
we motivate, derive, and extend the results of Section \ref{sec:QAOtoolkit}, the design toolkit for quantum optimization. 
The goal 
is to provide a suite of basic results which can be used by experts and laymen alike to design quantum algorithms. 
These results can also be found in~\cite{hadfield2018representation}. 
We emphasize that our results are general and have applications beyond QAOA or quantum annealing; see~\cite{hadfield2018representation} for several examples.  

\section{Representing $n$-bit Functions as Diagonal Hamiltonians}  \label{sec:DiagHams}
Many important problems involve Boolean predicates. 
We show how the representation of such functions as Hamiltonians
follows naturally from the Fourier analysis of Boolean functions.

Fourier analysis of Boolean functions has many applications in computer science and related fields such as operations research  \cite{kahn1988influence,linial1993constant,beigel1993polynomial,nisan1994degree,hammer2012boolean},
and is also useful 
a useful tool
for quantum computation \cite{beals2001quantum,montanaro2008quantum}; see \cite{de2008brief,od2014analysis} for comprehensive introductions to the subject. 
Many important combinatoric properties of a given function can be \lq\lq read off\rq\rq\ from its Fourier coefficients~\cite{od2014analysis}. 
However, this presents an obstruction to computing the Fourier representation of a general $n$-bit Boolean functions, or equivalently, 
to computing its Hamiltonian representation explicitly.  
Indeed, Proposition~\ref{prop:1booleanFourierRep} 
shows that computing the first Fourier coefficient $\widehat{f}(\emptyset)$ is as hard as counting the the number of inputs $x \in \{0,1\}^n$ such that $f(x)=1$. For example, if $f$ is an instance of CNF-SAT, then this is $\#P$-hard; see Corollary \ref{cor:sharpPFourier}. 
Hence, for arbitrary Boolean functions, in particular, functions corresponding to instances of 
NP-hard decision problems, 
we cannot hope to efficiently obtain 
their explicit Hamiltonian representations (in the form given in Proposition \ref{prop:1booleanFourierRep}). 

Nevertheless, there is no such difficulty for Boolean functions $f_j$ when $f_j$ acts on a constant number of  bits. Hence we can efficiently construct Hamiltonians representing \textit{pseudo-Boolean} functions of the form $f(x) = \sum_{j=1}^m f_j (x)$, $m={\rm poly}(n)$, which we typically seek to minimize or maximize. 
For such a pseudo-Boolean function, its Fourier coefficients do not 
explicitly reveal its optimal value, so the Hamiltonian representation can often be computed efficiently; see Theorem~\ref{thm:pseudoBool} below.  
For example, solving the optimization problem MaxSAT also solves the decision problem SAT; for $m$ clauses, a string can be found optimally satisfying all $m$ clauses if and only if the conjunction of the clauses is satisfiable. Hence, one approach to solving SAT is to instead try to solve MaxSAT. 
If the clauses each contain at most $k=O(1)$ variables, and there are $\textrm{poly}(n)$ many clauses, then we can efficiently represent the MaxSat instance as a Hamiltonian 
(in the sense of Theorem \ref{thm:1booleanFourierRep} below).  
This avoids the described difficulty, and is a commonly used approach in 
quantum annealing 
to implicitly encode Boolean functions; see, e.g., 
\cite{LucasIsingNP}. 
 
Boolean functions are often encountered as a formula in a \textit{normal form}. 
For example, SAT formulas are given in conjunctive normal form (CNF). 
Many 
normal forms exist such as disjunctive (maxterms), algebraic ($\oplus$), etc. \cite{halmos2009boolean}.
Note that while logically equivalent,
the different forms are often (very much) inequivalent for computational purposes. 
For each form there corresponds a notion of size (which directly relates to 
the number of bits needed to describe a function in such a form). 
We give explicit Hamiltonian representations of basic clauses, and composition rules which can 
be used to construct Hamiltonians for most normal forms. 
This typically allows for easier construction than by working with the Fourier expansion directly.  
Our results may also be applied to other common representations such as Boolean circuits. 

\subsection{Boolean Functions}
The class of Boolean functions on $n$-bits is defined as  $\mathcal{B}_n:=\{ f:\{0,1\}^n\rightarrow \{0,1\} \}$. 
As a vector spaces (over $\reals$), for each $n$ they give a give a basis for the real functions $\mathcal{R}_n=\{f:\{0,1\}^n\rightarrow \reals \}$. 
Moreover, each $\mathcal{R}_n$ is isomorphic to the vector space of diagonal Hamiltonians 
acting on $n$-qubits, or, equivalently, the space of $2^n \times 2^n$ diagonal real matrices. 
Thus, diagonal 
Hamiltonians naturally encode large classes of 
functions.

We say a Hamiltonian \textit{represents} a function $f$ if in the computational basis 
it acts as the corresponding multiplication operator, 
i.e. it satisfies the eigenvalue equations 
\begin{equation}   \label{eq:eigenvalueHf}
    \forall x \in \{0,1\}^n\;\;\;\; H_f\ket{x}=f(x)\ket{x}.
\end{equation}
On $n$ qubits, this condition specifies $H_f$ uniquely. 
Equivalently, we may write $H_f = \sum_x f(x)\ket{x}\bra{x}$, which in the case of Boolean functions becomes
 \begin{equation}
H_f = \sum_{x: f(x) =1} \ket{x}\bra{x}.
 \end{equation}
As Boolean functions are idempotent, both $f^2=f$ and $H_f^2 = H_f$, so $H_f$ is a projector\footnote{In typical constructions \cite{od2014analysis}, Boolean functions $g\rightarrow\{-1,1\}$ are considered, in which case $H_g^2=I$. The analog of 
Proposition \ref{prop:1booleanFourierRep} 
yields $\sum \widehat{g}(S)^2=1$, 
which does not depend on the structure of $g$. In this case, 
multiplication corresponds to the bitwise parity operation, whereas 
for $f\rightarrow\{0,1\}$
it corresponds to bitwise AND.} of 
rank $r=\#f := |\{x:f(x)=1\}| = \sum_x f(x)$.
Hence, 
determining if $f$ is satisfiable is equivalent to determining if $H_f$ is not identically $0$, and determining $H_f$ explicitly in this form is as hard as counting the number of satisfying assignments.

Such a representation is unique (up to change of computational basis). 
Recall that without loss of generality we consider the standard computational basis of eigenstates of Pauli $Z$ operators, 
defined by the relations $Z\ket{0} = \ket{0}$ and $Z\ket{1} = -\ket{1}$. 
Recall $Z_j= I \otimes \dots I \otimes Z \otimes I \dots \otimes I$ denotes
$Z$ acting on the $j$th qubit. 
Products of $Z_j$ over a set of qubits act as 
\begin{equation}
\prod_{j\in S} Z_j\ket{x} = \chi_S(x)\ket{x},
\end{equation}
where 
the \textit{parity function} $\chi_S (x): \{0,1\}^n \rightarrow \{-1,+1\}$ gives the parity of the bits of $x$ in the subset $S\subset[n]$. Identifying each $S$ with its characteristic vector $S\in \{0,1\}^n$ we have $\chi_S (x) = (-1)^{S\cdot x}$. 
Thus $Z_S:=\prod_{j\in S} Z_j$ represents the function $\chi_S(x)$ in the sense of (\ref{eq:eigenvalueHf}).

The set of parity functions on $n$-bits also give a 
basis for the real functions $\mathcal{R}_n$. 
This basis is orthonormal 
with respect to the inner product 
\begin{equation}     \label{eq:innerProduct}
 \langle f,g \rangle := \frac{1}{2^n} \sum_{x \in \{0,1\}^n} f(x) g(x) .
\end{equation}
In particular, every Boolean function $f\in \mathcal{B}_n$ may be written 
\begin{equation}   \label{eq:FourierExpan}
f(x) = \sum_{S\subset [n]} \widehat{f} (S) \chi_S (x),
\end{equation}
called the \textit{Fourier expansion}, 
with 
 \textit{Fourier coefficients} given by the inner products 
\begin{equation}   \label{eq:BooleanFourierCoefficients}
\widehat{f} (S) = \frac{1}{2^n} \sum_{x\in \{0,1\}^n} f(x) \chi_{S}(x) = \langle f, \chi_S \rangle .
\end{equation}
The \textit{degree}  of 
$f$, $\deg(f)$, is defined to be largest $|S|$ such that $\widehat{f}(S)$ is non-zero. Note that if $f$ depends on only $k\leq n$ variables, then $\deg(f) \leq k$.    
We refer to the mapping from $f(x)$ to $\widehat{f}(S)$ as the \textit{Fourier transform} of $f$.

Hence, 
an arbitrary Boolean function $f$ 
is represented 
as a Hamiltonian $H_f$ given by a linear combination of tensor products of $Z_j$ operators 
using the Fourier expansion and the identification $\chi_S = \bigotimes_{j\in S} Z_j$. 
We define the degree of such a Hamiltonian $H_f$, $\deg(H_f)$, to be the largest number of qubits acted on by any term in this sum, 
and the size of $H_f$, ${\rm size}(H_f)$, to be the number of terms.\footnote{
For Boolean functions some authors \cite{boros2002pseudo} define size as the sum of $|S|$ over all subsets such that $\widehat{f}\neq 0$, which is larger than our size quantity by at most a multiplicative factor of $\deg(f)$. Our notion of size is often called \textit{sparsity}.}
By definition,  $\deg(H_f)=\deg(f)$. 
Applying Parseval's identity and some further Fourier analysis 
gives the following theorem, which generalizes Propositon \ref{prop:1booleanFourierRep}. 

\begin{theorem}  \label{thm:1booleanFourierRep}
For an $n$-bit Boolean function $f \in \mathcal{B}_n$ of degree $d=\deg(f)$, the unique $n$-qubit 
Hamiltonian satisfying $H_f \ket{x} = f(x)\ket{x}$ in the computational basis is 
\begin{eqnarray*}
H_f &=& \sum_{S\subset [n], |S| \leq d} \widehat{f} (S) \; \prod_{j\in S} Z_j \\
&=&  \widehat{f}(\emptyset)I + \sum_{j=1}^{n}\widehat{f}(\{ j\})\, Z_{j} + 
\dots + \sum_{j_1< j_2 < \dots < j_d}\widehat{f}(\{ j_1, j_2, \dots, j_d\})\, Z_{j_1}\dots Z_{j_d} 
\end{eqnarray*}
where the Fourier coefficient $\widehat{f} (S)  = \frac{1}{2^n} \sum_{x\in \{0,1\}^n} f(x) (-1)^{S\cdot x}$,  
with $\deg(H_f)=  d$ and 
${\rm size}(H_f) \leq 1+ (e/d)^{d-1} n^d$.
The coefficients satisfy 
\begin{equation}
 0\leq \; \sum_{S\subset [n]}  \widehat{f} (S)^2 =
\widehat{f}(\emptyset) = \frac{1}{2^n} \sum_{x\in \{0,1\}^n} f(x)  = \frac{1}{2^n} {\rm tr}(H_f) \; \leq 1
\end{equation}
and in particular  
$ \sum_{S\subset [n]} \widehat{f} (S) = f(0^n)
,$
where $0^n$ denotes the input string of all $0$s.
\end{theorem}%

Here, 
${\rm tr}(H_f)$ is the \textit{trace} of the matrix $H_f$, i.e., the sum of its diagonal elements, 
which in particular is a basis-independent linear function \cite{horn2012matrix}. 
%
The bound 
${\rm size}(H_f) \leq 1+ (e/d)^{d-1} n^d$
follows from counting arguments and standard bounds for binomial coefficients.  
If $H_f$ acts on a constant number $d=O(1)$ of qubits, then its size is polynomially bounded in the number of qubits, ${\rm size}(H_f)=O(n^d)$. 
We say that a function $f_n$ on $n$-bits is \textit{efficiently representable} as the Hamiltonian $H_{f_n}$ if ${\rm size}(H_{f_n}) = {\rm poly}(n)$ with increasing~$n$. 

\begin{rem}
The Hamiltonian coefficients $\widehat{f}(S)$ depend only on the function values 
$f(x)$, and are independent of how such a function may be represented as input (e.g. formula, circuit, truth table, etc.). 
Many typical representations of Boolean functions as computational input such as Boolean formulas (e.g. CNF, DNF, etc.) or Boolean circuits can be directly transformed to Hamiltonians using composition rules we derive below; see Theorem \ref{thm:compositionRules}.  
\end{rem}

Consider 
a Boolean function $f$ given as a formula in \textit{conjunctive normal form}, the AND of clauses containing ORs of variables and their negations. 
The satisfiability problem (SAT) is to decide 
if there exists a satisfying assignment for $f$. 
It is NP-hard to decide this for an arbitrary such function, even if clauses are restricted to at most $3$ literals ($3$-SAT). 
Theorem~\ref{thm:1booleanFourierRep} implies that computing the single Fourier coefficient $\widehat{f}(\emptyset)$ is equivalent to computing the number of satisfying assignments, which is believed to be a much harder problem. (In fact, this problem $\#$SAT is complete for the counting complexity class $\#$P; see, e.g., \cite{AroraBarak}.) 
Thus, the 
problem of deciding 
if $\widehat{f}(\emptyset)>0$ is NP-hard.  
This result is stated as Corollary  \ref{cor:sharpPFourier} above.
Moreover, arbitrary Boolean functions may have size exponential in $n$, 
in which case, even if we know somehow its Hamiltonian representation, 
we cannot implement or simulate this Hamiltonian efficiently with the usual approaches. 

As remarked, pseudo-Boolean functions, in particular, objective functions for constraint satisfaction problems, often avoid these difficulties. 
We are particularly interested in functions composed of a number $m={\rm poly}(n)$ of clauses $C_j$, where each clause acts on $k=O(1)$ bits, and hence has size~$O(1)$ and degree~$O(1)$ (e.g., Max-$k$-Sat). 
In such cases, we have 
a useful 
lemma, which follows directly from 
 \cite[Thm. 1 \& 2]{nisan1994degree}. 
\begin{lem}
For a function $f\in\mathcal{B}_n$ 
that depends on $k\leq n$ variables,  
represented as a Hamiltonian $H_f$ acting on $n$ qubits 
its degree satisfies  
$$k \geq D(f) \geq \deg(H_f) \geq \log_2 k -O(\log\log k)$$
where $D(f)$ is the decision tree complexity of $f$ and 
$D(f) = {\rm poly}(\deg(H_f))$.
\end{lem}

\subsubsection{Composition Rules}
It is often possible to construct a Hamiltonian 
representing a Boolean function 
much more efficiently than 
by evaluating 
the Fourier coefficients explicitly. 
In general, this 
depends on the \textit{input format} of the given function. 
For example, for a function given as a 
disjunction of clauses, the Hamiltonian can be constructed by
 computing the Hamiltonian for each clause separately and combining them using composition rules.  
These composition rules follow directly from the properties of the Fourier transform.  
 Results for several important basic operations are given in Theorem \ref{thm:compositionRules}. 

\begin{proof}[Proof of Thm. \ref{thm:compositionRules}]
The logical values $1$ and $0$ (i.e., \textit{true} and \textit{false}) are represented as the identity matrix~$I$ and the zero matrix~$0$, respectively. 
Each result follows from the natural embedding 
of $f,g \in \mathcal{B}_n$ into $\mathcal{R}_n(+,\cdot)$, 
the real vector space of real functions on $n$ bits.  
By linearity of the Fourier transform, we immediately have 
$H_{af+bg} = aH_f + bH_g \text{ for } a,b \in \reals.$ 
Using standard logical identities, 
Boolean operations $(\cdot, \vee,\oplus,\dots)$ on $f,g$ 
can be translated into $(\cdot,+)$ formulas, i.e., linear combinations of $f$ and $g$. 
Linearity 
then gives the resulting Hamiltonian in terms of $H_f$ and $H_g$. 

Explicitly, for the complement of a function $\overline{f}$, as  $\overline{f} = 1 -  f$ we have $H_{\overline{f}} = I - H_f$. Similarly, the  identities 
$f\wedge g = fg$, $\; f \vee g = f + g -fg$, $\; f \oplus g = f + g -2fg$, and $f \Rightarrow g = \overline{f} + fg$, respectively, imply the remainder of the theorem. 
\end{proof}

It is straightforward to extend Theorem \ref{thm:compositionRules} to other operations, on any number of Boolean functions, using the same technique of the proof.

\begin{rem}
The Hamiltonians representing basic clauses given in Table \ref{tab:basicHams}
follow directly from the Fourier coefficients as in Theorem \ref{thm:1booleanFourierRep}, or equivalently, from the composition rules in Theorem \ref{thm:compositionRules}. 
The latter approach is generally much easier for constructing Hamiltonians. 
\end{rem}

The rules of Theorem \ref{thm:compositionRules} may be 
applied \textit{recursively}, as desired, 
to construct Hamiltonians representing more complicated Boolean functions, corresponding to, e.g., nested parentheses in logical formulas or wires in Boolean circuits. 
For example, the Hamiltonian representing the Boolean clause $f \vee g \vee h=f \vee (g \vee h)$ is given by
$H_{f \vee g \vee h} = 
H_f + H_{g\vee h}  -H_fH_{g\vee h}$, 
which simplifies to  $H_{f \vee g \vee h} =H_f + H_g +H_h - H_fH_g - H_fH_h - H_gH_h + H_fH_gH_h$. 

Some typical examples of Boolean functions on three variables 
are the 
Majority (MAJ), Not-All-Equal (NAE), and $1$-in-$3$ functions, which are represented as the Hamiltonians
\begin{itemize}
\item $H_{MAJ(x_1,x_2,x_3)} \, = \, \frac12 I - \frac14 (Z_1 + Z_2 + Z_3 - Z_1 Z_2 Z_3) $  
\item $H_{NAE(x_1,x_2,x_3)} \, = \, \frac34 I  -\frac14( Z_1Z_2 + Z_1Z_3 + Z_2Z_3)$ 
\item $H_{1in3(x_1,x_2,x_3)} \, = \, 
\frac{1}{8}(3I +Z_1 +Z_2 +Z_3 -Z_1Z_2 -Z_2Z_3 -Z_1Z_3 -3Z_1Z_2Z_3).$ 
\end{itemize}
%
The higher-order functions $\bigvee_{j=1}^k x_j$ and $\bigwedge_{j=1}^k x_j$ are represented by 
%
Hamiltonians of size $2^k$. 
This 
is analogous to the well-known fact that 
formulas in conjunctive or disjunctive normal form 
that compute the parity function on $k$ bits
have sizes exponential in $k$.  

Finally, observe that for each of the rules of Theorem \ref{thm:compositionRules} we may 
define the right-hand sides as 
binary 
operators on Hamiltonians, e.g., $OR(H_f,H_g):=H_f+H_g-H_fH_g$. 
These rules clearly give homomorphisms (in fact, isomorphisms) between Boolean algebra($\wedge,\vee$) and Boolean ring($\cdot,\oplus$) elements. 
Thus, $2^n \times 2^n$ diagonal Hamiltonians equipped with these operators faithfully represent the $n$-element Boolean algebra and Boolean ring; see, e.g., \cite{halmos2009boolean}.

\subsection{Pseudo-Boolean Functions and Constraint Satisfaction Problems}
%
Real functions on $n$-bits 
 are similarly represented as Hamiltionans via the Fourier transform. 
Every such function $f\in\mathcal{R}_n$ may be expanded (non-uniquely) as a weighted sum of Boolean functions, possibly of exponential size. By linearity of the Fourier transform,  
the Hamiltonian $H_f$ is given precisely by the corresponding weighted sum of the Hamiltonians representing the Boolean functions. Moreover, 
the Hamiltonian $H_f$ is unique, so 
different expansions of $f$ as sums of Boolean functions must all result in the same 
 $H_f$.   

The Fourier coefficients 
are again given by the inner product (\ref{eq:innerProduct}) with the parity functions $ \chi_S$, 
\begin{equation}
\widehat{f} (S) = \langle f, \chi_S \rangle = \frac{1}{2^n} \sum_{x\in \{0,1\}^n} f(x) \chi_{S}(x), 
\end{equation}
and these coefficients again lead directly to 
the Hamiltonian representation.  
We are particularly interested in 
optimization problems with 
 \textit{objective functions}  $f:\{0,1\}^n \rightarrow \reals$ 
of the form 
\begin{equation}
f(x) = \sum_{j=1}^m  w_j f_j (x),
\end{equation}
where $f_j \in \mathcal{B}_n$, $w_j \in \reals$, and $m={\rm poly}(n)$. 
(We call such a function $f\in\mathcal{R}_n$ a \textit{pseudo-Boolean} function, generally.) 
In particular, 
in a \textit{constraint satisfaction problem}, typically 
all $w_j = 1$ and hence $f(x)$ gives the number of satisfied constraints (i.e., clauses). 
We 
will see many examples of such problems in Chapters \ref{ch:QAOAperformance} and \ref{ch:QACOA}.

We have the following theorem which extends the previous 
results for Boolean functions. 

\begin{theorem} \label{thm:pseudoBool}
An $n$-bit 
real function $f:\{0,1\}^n \rightarrow \reals$ is represented as the Hamiltonian  
$$H_f = \sum_{S\subset [n]} \widehat{f} (S) \; \prod_{j\in S} Z_j , \;\;\;\;\;\;\;\; \widehat{f} (S) = \langle f , \chi_s \rangle \in \reals. $$
In particular, a 
pseudo-Boolean function $f=\sum_j w_j f_j$, $w_j\in \reals$, 
$f_j \in \mathbb{B}_n$, 
is represented as 
$$ H_f = \sum_j w_j H_{f_j},$$
with $\deg(H_f) \leq  \max_j \deg(f_j)$ and 
${\rm size}(H_f)\leq
\min \{ \sum_j {\rm size}(H_{f_j}), 1+ (e/d)^{d-1}n^d\}$. 

\end{theorem}

The theorem follows from Theorem \ref{thm:1booleanFourierRep} and the linearity of the Fourier expansion. 
Proposition~\ref{prop:pseudoBool} in Section~\ref{sec:QAOtoolkit} then follows directly from Theorems \ref{thm:1booleanFourierRep} and \ref{thm:pseudoBool}.

\begin{rem}
In contrast to Theorem \ref{thm:1booleanFourierRep}, 
for a constraint satisfaction problem $f=\sum_{j=1}^m  f_j$, $f_j\in \mathbb{B}_f$, 
applying Parseval's identity we have 
$$ \sum_{S\subset [n]}  \widehat{f} (S)^2 
 = \mathbf{E}[f] + 2\sum_{i<j} \langle f_i, f_j \rangle
 = \widehat{f}(\emptyset) + 2 \sum_{i<j} \mathbf{E}[f_i \wedge f_j] \geq \mathbf{E}[f] , $$
where $\mathbf{E}[f]:= \frac{1}{2^n}\sum_{x\in\{0,1\}^n} f(x)$ gives the expected value over the uniform distribution of inputs. 
In particular, $\sum_{S\subset [n]}  \widehat{f} (S)^2 
 = \mathbf{E}[f]$ if and only if $\langle f_i, f_j \rangle = 0$ for all $i,j$.  
 If there does exist an $i,j$ such that  $\langle f_i, f_j \rangle = 0$, then 
 the conjunction of the clauses is unsatisfiable, i.e. $\wedge_j f_j = 0$.
\end{rem}

\section{Controlled Hamiltonian Evolution}  \label{sec:ControlledHams}
In many applications we require controlled Hamiltonian evolutions. 
For example, in quantum phase estimation (QPE) \cite{NC}, we require transformations on $(1+n)$-qubit basis states of the form 
$$ \ket{0}\ket{x} \rightarrow \ket{0}\ket{x}, \;\;\;\;\;\;\;\;\;\; \ket{1}\ket{x} \rightarrow e^{-iHt} \ket{1}\ket{x} ,$$
for various values 
$t=1,2,4,\dots$; see the discussion of QPE in Chapter \ref{ch:ExcStates}. 
Consider such a transformation with fixed $t$. 
Labeling the 
first qubit $a$, 
the overall unitary 
may be written as
\begin{equation}
\Lambda_{x_a} ( e^{-iHt})  = \ket{0}\bra{0} \otimes I + \ket{1}\bra{1} \otimes e^{-iHt}.
\end{equation}
Here the notation $ \Lambda_{x_a} ( e^{-iHt})$ indicates the unitary $e^{-iHt}$ controlled by the classical function $x_a$. 
We obtain the Hamiltonian corresponding to this transformation 
by writing 
$\Lambda_{x_a} ( e^{-iHt})=e^{-i\widetilde{H}t}$, which gives
\begin{equation}
 \widetilde{H} = \ket{1}\bra{1} \otimes H= x_a \otimes H = \frac12 I \otimes H - \frac12 Z_a \otimes H.
\end{equation}
The 
control qubit is assumed precomputed here; its value may or may not depend on~$x$. If $H=H_f$ is diagonal, then so is $\widetilde{H}=x_a H_f=H_{x_a \wedge f}$.

More generally, we can consider Hamiltonian evolution controlled by a Boolean function $g\in \mathbb{B}_k$ acting on a $k$-qubit register. 
In this case we seek to affect the unitary transformation on $(k+n)$-qubit on basis states 
$$ \ket{y}\ket{x} \rightarrow \ket{y}\ket{x} \;\;\;\;\;\;\;\;\;\;\;\;\;\;\; \text{if } g(y) = 0,
$$
$$ \ket{y}\ket{x} \rightarrow e^{-iHt} \ket{y}\ket{x}
\;\;\;\;\;\;\;\;\;\; \text{if } g(y) = 1,$$
which gives the overall unitary
$$ \Lambda_g ( e^{-iH t}) = \sum_{y: g(y)=0} \ket{y}\bra{y} \otimes I + \sum_{y: g(y)=1} \ket{y}\bra{y} \otimes e^{-iH t} = H_{\overline{g}} \otimes I + H_g \otimes e^{-iH t},$$
corresponding to evolution under the Hamiltonian \begin{equation}
 \widetilde{H} = \sum_{y: g(y)=1} \ket{y}\bra{y} \otimes H = H_g \otimes H . 
\end{equation}
These results have been summarized in Proposition \ref{prop:controlledHams}.

As a corollary,  
we show that computing a Boolean function in a register is closely related to computing it as an amplitude. 

\begin{cor}
For an $n$-bit Boolean function $f$, let $G_f$ be 
 the unitary operator on $n+1$ qubits which acts on basis states $\ket{x}\ket{a}$ as
\begin{equation}
G_f \ket{x}\ket{a}=\ket{x}\ket{a \oplus f(x) }.
\end{equation} 
Then $$G_f = H_f \otimes X + H_{\overline{f}} \otimes I,$$ and $G_f=I$ if and only if $f$ is unsatisfiable. 

In particular, if $f$ is given as a 
CNF formula, it is \#P-hard to compute the identity coefficient $\widehat{g}(\emptyset)$ of $G_f$, 
and NP-hard to decide if $\;\widehat{g}(\emptyset)\neq 1$.  
\end{cor}

\begin{proof}
Using Theorem \ref{thm:1booleanFourierRep} for $H_f$, we expand $G_f$ as a sum of Pauli matrices $G_f = (1-\widehat{f}(\emptyset))I + \widehat{f}(\emptyset)X_a + \dots$, where none of the terms to the right are proportional to $I$. Thus computing the $I$ coefficient gives the number of satisfiable assignments for $f$. 
\end{proof}
Observe that with an ancilla qubit $a$ initialized to $\ket{0}$, we can simulate $H_f$ for time $t$ using two applications of $G_f$ and a $Z$ rotation,
\begin{equation}
e^{-iH_ft}\ket{x}\ket{0} =  c\:G_f  \:R_{Z_a}( -t )\:  G_f   \ket{x}\ket{0},
\end{equation}
where $c$ is an unimportant global phase. 
This \textit{phase kickback} is an important and well-known technique in quantum computation. 
Conversely, if we can simulate~$H_f$ in a controlled manner, then we can implement~$G_f$ using a single-bit  
quantum phase estimation, 
which requires two Hadamard gates and a controlled simulation of~$H_f$.  
Indeed, using $e^{-i \pi H_f}=(-1)^{f(x)}$, 
it is easy to check 
\begin{equation}
G_f \ket{x}\ket{0} = {\textrm H}_a  \Lambda_a (e^{-i \pi H_f})    {\textrm H}_a \ket{x}\ket{0} .
\end{equation}
Thus, 
in this sense the 
operators $H_f$ and $G_f$ are closely related. 
We expand on these ideas in \cite{hadfield2018representation}. 

Finally, we remark that if we could simulate arbitrary Hamiltonians $H_f$ efficiently, 
or more precisely, efficiently simulate $H_f$ for each $f$ in some 
class of $n$-bit Boolean functions, 
and if 
we could also somehow 
find and efficiently prepare an initial state $\ket{s}$ with sufficiently large projection $\|H_f \ket{s}\| = 1/{\rm poly}(n)$ for each such $f$, then we could efficiently 
determine if a given $f$ is satisfiable 
using 
QPE. (Smaller state projections proportionally reduce the success probability of QPE, which can be dealt with by an inversely proportional number of algorithm repetitions \cite{NC}.) 
Thus, as 
Hamiltonians representing Boolean formulas given in conjuctive normal form 
can be efficiently simulated 
using ancilla qubits (without knowing the Hamiltonian terms explicitly),  
we conclude that finding such an initial state must be NP-hard. Similar ideas can be applied to real functions.

%% file: _app_sciComp.tex
\chapter{Quantum Algorithms for Scientific Computing}
\label{app:SciComp}
In this appendix we derive theorems showing the worst-case error of the algorithms in 
Chapter \ref{ch:sciComp}. 

The following corollary follows from Theorem B.1 of \cite{Poisson}.

\begin{cor} 
\label{cor0}
For $w>1$, represented by $n$ bits of which the first $m$ correspond to its integer part, and for $b\geq n$, 
 Algorithm \ref{alg:inv} returns a value $\hat{x}_{s}$ approximating $\frac{1}{w}$ with error
\begin{equation}
|\hat{x} - \frac{1}{w}| \leq \frac{2+ \log_2 b}{2^b}.
\end{equation}
This is accomplished by performing Newton's iteration and truncating the results of each iterative step to $b$ bits after the decimal point.
\end{cor}

\begin{proof}
From \cite{Poisson}, 
the Newton iteration of Algorithm \ref{alg:inv} with $b$ bits of accuracy and $s$ iterations produces an approximation $\hat{x}$ of $\frac{1}{w}$ such that
$ |\hat{x} - \frac{1}{w}| \leq \left( \frac{1}{2} \right)^{2^s} + s2^{-b}$.  
For $s=\lceil \log_2 b \rceil$, then 
$$ |\hat{x} - \frac{1}{w}| \leq \frac{1}{2^b} + \lceil \log_2 b \rceil \frac{1}{2^b} \leq  \frac{1}{2^b}  (2  +\log_2 b) .$$
\end{proof}
\begin{theorem} 
\label{SCIthm1}
For $w>1$, represented by $n$ bits of which the first $m$ correspond to its integer part, and for $b\geq \max\{2m,4\}$, Algorithm \ref{alg:sqrt} returns a value $\hat{y}_{s}$ approximating $\sqrt{w}$ with error
\begin{equation}
|\hat{y}_{s}-\sqrt{w}|  
\leq \left( \frac{3}{4} \right)^{b-2m}  \left( 2+ b + \log_2 b  \right).
\end{equation}
This is accomplished by performing Newton's iteration and truncating the results of each iterative step to $b$ bits after the decimal point before passing it on the next iterative step.
\end{theorem}

\begin{proof}
The overall procedure consists of two stages of Newton's iteration, as illustrated in Fig. \ref{fig-SQRToverall} above. We analyze each stage in turn.

Observe that the iteration $x_i = g_1(x_{i-1}):=-w\hat{x}_{i-1}^2 + 2\hat{x}_{i-1}$, $i=1,2,\dots,s_1$, $s_1=\lceil \log_2 b \rceil$, corresponds to Newton's iteration applied to the function $f_1\left(x\right) := \frac{1}{x}-w$  for approximating $\frac{1}{w}$, with initial guess $\hat{x}_0 = 2^{-p}$ where $p\in \nat$ and $2^p > w \geq 2^{p-1}$. 
It has been analyzed in detail in Theorem B.1 of \cite{Poisson}. Here, we briefly review some of the results. 
An efficient circuit for generating the initial state $2^{-p}$ is shown in Fig. \ref{fig-InitState1} above, similar to that in \cite{Poisson}.

The approximations $x_i$ each satisfy $x_i  \leq \frac{1}{w}$, i.e., we underestimate $\frac{1}{w}$. Indeed, $g_1\left(x\right) - \frac{1}{w} = \frac{1}{w}\left(2xw - w^2x^2-1\right) = -\frac{1}{w}\left(wx-1\right)^2 < 0$.  
Accounting for the truncation error, we have 
\begin{equation}
\label{eq:InverseError}
|\xh - \frac{1}{w} | \leq \left(we_0\right)^{2^{s_1}}\frac{1}{w} + 2^{-b}s_1  \leq  \left(\frac{1}{2}\right)^{2^{s_1}} + 2^{-b}s_1 =:E .
\end{equation}
This follows from the facts $w>1$ and $we_0 \leq \frac12$, where $e_0= |2^{-p} - \frac{1}{w}|$, and is shown in \cite{Poisson}.
 The first term in the upper bound corresponds to the error of Newton's iteration, 
while the second term is the truncation error.  
Now we turn to the second stage. 
Iteration $y_j = g_2(y_{j-1}):= \frac12(3y_{j-1} - \hat{x}_{s_1} y^3_{j-1} )$, $j=1,2,\dots,s_2$, $s_2= \lceil \log_2 b \rceil$, is obtained by using Newton's method to approximate the zero of the function $f_2\left(y\right)= \frac{1}{y^2}-\x$, with initial guess $\hat{y}_0 = 2^{\lfloor \frac{q-1}{2} \rfloor}$ where $q\in \nat$ and $2^{1-q} > \hat{x}_{s_1} \geq 2^{-q}$. An efficient circuit for generating the initial state $2^{\lfloor \frac{q-1}{2} \rfloor}$ is shown in Fig. \ref{fig-InitState2}. We have
\begin{eqnarray*}
g_2\left(y\right) - \frac{1}{\sqrt{\x}} &=& y - \frac{1}{\sqrt{\x}} + \frac{1}{2}\left(y-y^3\x\right)\\
&=&  y - \frac{1}{\sqrt{\x}} + \frac{\x y}{2}\left(\frac{1}{\x}-y^2\right)\\
&=&  \left(y- \frac{1}{\sqrt{\x}}\right)   \left(  1 - \frac12 \x y\left(y+ \frac{1}{\sqrt{\x}}\right)   \right)      \\     
&=& - \frac12   \left(y- \frac{1}{\sqrt{\x}}\right)  \left( y^2\x +y\sqrt{\x} -2\right)\\
&=&  - \frac12   \left(y- \frac{1}{\sqrt{\x}}\right)  \left(  \left(y\sqrt{\x} -1 \right)^2  + 3y\sqrt{\x} -3  \right)\\
&=&  - \frac12   \left(y- \frac{1}{\sqrt{\x}}\right)   \left(y- \frac{1}{\sqrt{\x}}\right) \sqrt{\x}   \left(y\sqrt{\x}  -1 + \frac{3y\sqrt{\x} -3}{y\sqrt{\x}  -1}\right) \\
&=&    - \frac12   \left(y- \frac{1}{\sqrt{\x}}\right)^2  \sqrt{\x} \left(y\sqrt{\x} +2\right). \\
\end{eqnarray*}
The last quantity is non-positive assuming $y\geq 0$.
The iteration function $g_2$ is non-decreasing in the interval $[0,\frac{1}{\sqrt{\x}}]$ and $g_2\left(0\right)=0$. 
Since $y_0 = 2^{-\lfloor \frac{q-1}{2} \rfloor} $, we get $y_1 \geq 0$, and inductively we see that all iterations produce positive numbers that are approximations underestimating $\frac{1}{\sqrt{\x}}$, i.e. $y_i\leq \frac{1}{\sqrt{\x}}$ for $i=0,1,2,\dots$.
Then
\begin{equation}
e_{i+1} := |y_{i+1} - \frac{1}{\sqrt{\x}} | = e_i^2 \frac12 \sqrt{\x} | y_i \sqrt{\x}  +2  | \leq \frac32 \sqrt{\x} e_i^2,
\end{equation}
since $| y_i \sqrt{\x}  +2  | \leq 3$ ($y_i$ underestimates $\xh^{-\frac12})$,
where $e_0 = \frac{1}{\sqrt{\x} } - 2^{-\lfloor \frac{q-1}{2} \rfloor}  \leq 2^{-\lfloor \frac{q-1}{2} \rfloor}$.

Let $A=\frac32 \sqrt{\x}$. We unfold the recurrence to obtain
$e_i \leq \frac{1}{A} {\left(Ae_0\right)}^{2^i} $, $i=1,2,\dots$. We have $\sqrt{\x} 2^{\lfloor \frac{q-1}{2} \rfloor} \geq 2^{-q/2} 2^{\lfloor \frac{q-1}{2} \rfloor} $. For $q$ odd, this quantity is bounded from below by $\frac{1}{\sqrt{2}}$, and for $q$ even this is bounded by $\frac12$. 
Thus $\sqrt{\x}e_0 = |\sqrt{\x} 2^{\lfloor \frac{q-1}{2} \rfloor}  - 1 | = 1 - \sqrt{\x} 2^{\lfloor \frac{q-1}{2} \rfloor}   \leq \frac12$ because we have selected the initial approximation $y_0$ to underestimate $\frac{1}{\sqrt{\x}}$. From this, we obtain $e_i \leq \frac{1}{A}\left(\frac{3}{4}\right)^{2^i}$.
Using equation (\ref{eq:InverseError}), we have that $\xh \geq \frac{1}{w}-E$. 
Without loss of generality, $wE \leq \frac12$. This will become apparent in a moment once we select the error parameters. Thus, $\frac1A \leq \frac23 \sqrt{2w}$. Therefore,
\begin{equation}
e_i \leq \frac{2}{3} \sqrt{2w} \left( \frac{3}{4} \right)^{2^i}, \;\; i=1,\dots,s_2.
\end{equation}

We now turn to the roundoff error analysis of the second stage of the algorithm. The iterations of the second stage of the algorithm would produce a sequence of approximations $y_1, \dots , y_{s_2}$ if we did not have truncation error. Since we truncate the result of each iteration to $b$ bits of accuracy before performing the next iteration, we have a sequence of approximations $\hat{y}_1, \dots , \hat{y}_{s_2}$, with
$\hat{y}_{0} = {y}_{0}$, $\hat{y}_{1} = g\left(\hat{y}_{0}\right) + \xi_1$, $\dots$, 
$\hat{y}_{i} = g\left(\hat{y}_{i-1}\right) + \xi_i$,
$\dots$, 
where $|\xi_i | \leq 2^{-b}$, $i\geq 1$.
Using the fact that $\sup_{x\geq 0} |g_2'\left(y\right)| \leq \frac32$, we obtain
\begin{eqnarray*}
|\hat{y}_{s_2}-y_{s_2}| &\leq& |g_2\left(\hat{y}_{s_2 -1}\right) - g_2\left(y_{s_2 -1}\right)| + |\xi_{s_2}|\\
 &\leq& \frac32| \hat{y}_{s_2 -1} - y_{s_2 -1}| + |\xi_{s_2}|\\
 &\vdots& \\
 &\leq& \sum_{j=1}^{s_2} \left(\frac32\right)^{s_2 -j} | \xi_j |
  \leq 2^{-b} \frac{ \left(\frac32\right)^{s_2} -1 }{\frac32 -1}\\
    &\leq& 2^{1-b} \left(\frac32\right)^{s_2}. 
\end{eqnarray*}
Therefore, the total error of the second stage of the algorithm is 
\begin{equation}
|\hat{y}_{s_2} - \frac{1}{\sqrt{\x}}| \leq     \frac{2}{3} \sqrt{2w} \left( \frac{3}{4} \right)^{2^{s_2}}  +  2^{1-b}\left(\frac32\right)^{s_2}.
\end{equation}
For $b \geq \max\{2m,4 \}$ and  $s_2 = \lceil \log_2 b \rceil$, and recall that $w\leq 2^m$. Then we have
\begin{eqnarray}
\label{eq:Stage2Error}
|\hat{y}_{s_2} - \frac{1}{\sqrt{\x}}| &\leq&   \frac{2}{3} \sqrt{2} \; 2^{\frac{m}2} \left( \frac{3}{4} \right)^{b}  +  2^{1-b}\left(\frac32\right)^{\log_2 b +1}   \nonumber \\
&\leq&  \frac{2}{3} \sqrt{2} \; \left(\sqrt{2}\right)^{m} \left( \left( \frac{3}{4} \right)^2\right)^{b/2}  +  2^{1-b}\; 2^{\log_2 b +1} \nonumber \\
&\leq&  \frac{2}{3} \sqrt{2} \; \left(   \frac{\sqrt{2}\;9}{16} \right)^m \left( \frac{9}{16} \right)^{b/2-m}  +  2^{2-b} \; b \nonumber \\
&\leq&  \left( \frac{3}{4} \right)^{b-2m}  +   2^{2-b} \; b,
\end{eqnarray}
Let us now consider the total error of our algorithm,
\begin{equation} \label{eq:TotalError}
|\hat{y}_{s_2}-\sqrt{w}| \leq | \hat{y}_{s_2}- \frac{1}{\sqrt{\x}}| + |  \frac{1}{\sqrt{\x}}-\sqrt{w}|
\end{equation}
We use equation (\ref{eq:Stage2Error}) above to bound the first term. For the second term we have
\begin{equation}  \label{eq:secondTerm}
 |  \frac{1}{\sqrt{\x}}-\sqrt{w}| \leq \frac12|\frac{1}{\x}-w| 
=   \frac12 \frac{w}{\x} |\x - \frac{1}{w}|
 \leq \frac12 \frac{w}{\x} E,
\end{equation}
where the first inequality follows from the mean value theorem ($|\sqrt{a} - \sqrt{b}| \leq \frac12 |a-b|$ for $a,b \geq 1$).
Since $\x \geq \frac{1}{w}-E$,
\begin{equation}
 \x^{-1} \leq \left(1 - \frac{1}{w} - E\right)^{-1} \leq 2w,
\end{equation}
for $wE \leq \frac12$. Then 
(\ref{eq:secondTerm}) becomes 
\begin{eqnarray}
 |  \frac{1}{\sqrt{\x}}-\sqrt{w}| &\leq& w^2 E \leq  w^2 \left( \left(\frac{1}{2}\right)^{2^{s_1}} + 2^{-b}s_1 \right) \nonumber\\
  &\leq&  2^{2m} \left( \left(\frac{1}{2}\right)^{2^{s_1}} + 2^{-b}s_1 \right) \nonumber\\
    &\leq&    2^{2m} \left(  2^{-b} + 2^{-b}  \lceil \log_2 b \rceil \right) \nonumber\\
        &\leq&    2^{2m-b}\left( 1+  \lceil \log_2 b \rceil \right),
\end{eqnarray}
where we set $s_1 = \lceil \log_2 b \rceil$. 
Combining this with equation (\ref{eq:TotalError}), 
\begin{eqnarray}
|\hat{y}_{s_2}-\sqrt{w}|     &\leq &  \left( \frac{3}{4} \right)^{b-2m}  +   2^{2-b} \; b+    2^{2m-b}  \left( 1+  \lceil \log_2 b \rceil \right) \nonumber\\
&\leq&   \left( \frac{3}{4} \right)^{b-2m}  \left( 2+ b + \log_2 b  \right),
\end{eqnarray}
and the error bound in the statement of the theorem follows for $s=s_1=s_2$.

Finally, for completeness we show that for $b\geq \max\{2m,4\}$, $wE\leq \frac12$. Indeed, 
\begin{eqnarray*}
wE &\leq& 2^m \left(  2^{-b} + 2^{-b}  \lceil \log_2 b \rceil \right)\\
 &\leq& 2^{-b+m}   \left( 2+  \log_2 b \right)\\
 &\leq& \left( \frac12\right)^{b/2-m}   \frac{2+ \log_2 b}{2^{b/2}}.
 \end{eqnarray*}
The first factor above is at most $\frac12$ since $b\geq \max\{2m,4\}$, while the second is at most $1$ and this completes the proof.
\end{proof}

\begin{theorem} 
\label{thm2}
For $w>1$, represented by $n$ bits of which the first $m$ correspond to its integer part, Algorithm \ref{alg:2krt} computes approximations $\hat z_1$,$\hat z_2$,\dots,$\hat z_k$ such that
\begin{equation} \label{eq:thm2}
|\hat{z}_{i}-w^{\frac{1}{2^i}}| \leq 2  \left( \frac{3}{4} \right)^{b-2m}  \left( 2+ b +  \log_2 b  \right), \; i=1,2,\dots, k, \rm{\ for\ any\ } k\in \nat.
\end{equation}
This is accomplished by repeatedly calling Algorithm \ref{alg:sqrt}.  Each $z_i$ has $b\ge \max\{ 2m, 4\}$ bits after its decimal point, and by convention its integer part is $m$ bits long.
\end{theorem}

\begin{proof}
We have 
\begin{eqnarray*}
\hat z_1 &=& \sqrt{w} + \xi_1  \\
\hat z_2&=& \sqrt{z_1} + \xi_2  \\
&\vdots& \\
\hat z_k &=& \sqrt{z_k} + \xi_k. 
\end{eqnarray*}
Since $\hat z_i$ is obtained using Algorithm \ref{alg:sqrt} with input $\hat z_{i-1}$, we use Theorem \ref{SCIthm1} to obtain $|\xi_i| \leq \left( \frac{3}{4} \right)^{b-2m}  \left( 2+ b + \log_2 b  \right)$, $i=1,2,\dots,k$. 
We have
\begin{eqnarray*}
|\hat z_k - w^{\frac{1}{2^k}}| &\leq& |\sqrt{\hat z_{k-1}} - w^{\frac{1}{2^k}}| + |\xi_k|\\
&\leq& \frac12 |\hat z_{k-1} - w^{\frac{1}{2^{k-1}}}| + |\xi_k|\\
&\vdots& \\
&\leq& \sum_{j=0}^{k-1}  \frac{1}{2^j} |\xi_{k-j}| \\
&\leq& 2  \left( \frac{3}{4} \right)^{b-2m}  \left( 2+ b + \log_2 b  \right), 
\end{eqnarray*}
where the second inequality above follows again from $|\sqrt{a} - \sqrt{b}| \leq \frac12 |a-b|$ for $a,b \geq 1$.
\end{proof}

\begin{theorem} 
\label{thm3}
For $w>1$, represented by $n$ bits of which the first $m$ bits correspond to its integer part, Algorithm \ref{alg:ln} computes an approximation ${\hat{z}:= \hat{z}_p + (p - 1) r}$ of $\ln w$, where $2^p > w \geq 2^{p-1}$, $p\in \nat$, and $|r-\ln 2|\le 2^{-b}$, such that 
$$ |  \hat{z}  - \ln w | \leq   \left(\frac{3}{4}\right)^{5\ell/2} \left( m+ \frac{32}{9} + 2\left(\frac{32}{9} + \frac{n}{\ln 2} \right)^3 \right),$$
where $\ell\ge \lceil \log_2 8 n\rceil$ is a parameter specified in the algorithm that is used to determine the number $b=\max\{5\ell, 25\}$ of bits after the decimal point in which arithmetic is be performed, and from that the error. 
\end{theorem}

\begin{proof}
An overall illustration of the algorithm is given in Fig. \ref{fig-LogOverall}.
 
Our algorithm utilizes the identity $\ln w = \ln2 \log_2 w$, as well as other common properties of logarithms. For completeness, an example circuit for computing $p$ is given in Fig. \ref{fig-ShiftInteger} above. We proceed in detail. 

If the clause of the second \textit{if} statement evaluates to true, 
in the case $w=2^{p-1}$, 
then the algorithm sets $z_p = 0$ and 
returns $(p-1)r$. This quantity approximates $\ln 2^{p-1}$ with error bounded from above by $(m-1)2^{-b}$, since $p\le m$ in the algorithm.  
We deal with the case 
$w$ not a power of $2$, for which  $z_p\ne 0$.  
Using the same notation as the algorithm, we have
\begin{eqnarray}
\label{eq:lnErrTot}
 | z_p + (p -1)r - \ln(w)| &\le& |z_p + \ln 2^{p-1} - \ln(w)| + |(p-1)r - \ln 2^{p-1}| \nonumber\\
 &=& |z_p - \ln 2^{-(p-1)}w| +(m-1) 2^{-b}   \nonumber \\ 
&\leq&  |z_p - 2^\ell \ln w_p^{\frac{1}{2^\ell}}| + (m-1) 2^{-b}\nonumber \\
 &\leq&  |z_p -   2^{\ell} \ln \hat{t}_p | + | 2^\ell \ln \hat{t}_p- 2^\ell \ln w_p \frac{1}{2^\ell}| + (m-1)2^{-b}\nonumber\\
 &\leq&  | 2^{\ell} \hat{y}_p  - 2^{\ell} y_p | +  | 2^{\ell} y_p - 2^\ell \ln \hat{t}_p| \\
&\qquad& \qquad+   | 2^\ell \ln \hat{t}_p- 2^\ell \ln w_p^{\frac{1}{2^\ell}}| +(m-1)2^{-b}\nonumber,
\end{eqnarray}
where $w_p = 2^{1-p}w$,   
$z_p = 2^\ell y_p$, 
$\hat{z}_p = 2^\ell \hat{y}_p$ and 
by $y_p$ we denote the value $(\hat{t}_p-1) - \frac12 (\hat{t}_p -1)^2$ before it is truncated to obtain $\hat{y}_p$. The first term is due to truncation error and is bounded from above by $2^{\ell-b}$. The last term is bounded from above by $2^\ell |\hat{t}_p - w_p^{\frac{1}{2^\ell}}|;$ this is obtained using the mean value theorem for $\ln$ with argument greater than or equal to one. The second term is bounded from above by $2^{\ell}|2(\hat{t}_p -1)^3|$. Indeed, recall that $\hat{t}_p - 1 = \delta \in (0,1)$ according to line 12 of the algorithm, and we have 
\begin{eqnarray*}
| \ln (1+ \delta)| - (\delta - \frac{1}{2} \delta^2) | 
&=& \left |\sum_{i=3}^\infty (-1)^{i+1}\frac{\delta^{i}}{i} \right| 
= \left| \delta^2 \sum_{i=0}^{\infty} \frac{i+1}{i+3} \int_0^\delta (-x)^i dx \right|\\
&=& \left| \delta^2 \int_0^\delta \sum_{i=0}^{\infty} \frac{i+1}{i+3} (-x)^i dx \right|
\leq  \delta^2 \int_0^\delta \sum_{i=0}^{\infty} |x|^i dx \\
&\leq&  \delta^2 \int_0^\delta \frac{1}{1- |x|} dx
\leq  \delta^2 \int_0^\delta 2 dx = 2\delta^3,
\end{eqnarray*}
assuming $\delta \leq \frac12$.
We now show that in general $\delta$ is much smaller than $\frac12$.
Since we have used Algorithm \ref{alg:2krt} to compute $\hat{t}_p$ which is an approximation of $w_p^{\frac{1}{2^\ell}}$. Algorithm \ref{alg:2krt} uses  Algorithm \ref{alg:sqrt}.
Since the error bounds of Theorem \ref{SCIthm1} and \ref{thm2} depend on the magnitude of $w_p$, the estimates of these theorems hold with $m=1$ because $w_p \in (1,2)$. We have 
$$|\hat{t}_p - w_p^{\frac{1}{2^\ell}}| \leq 2  \left( \frac{3}{4} \right)^{b-2}  \left( 2+ b +  \log_2 b  \right) \leq \frac18,$$
where we have set $b = \max\{5\ell,25\}$. Thus 
\begin{equation}
\label{eq:tp}
\hat{t}_p \leq w_p^{\frac{1}{2^\ell}} + 2  \left( \frac{3}{4} \right)^{b-2}  \left( 2+ b +  \log_2 b  \right) \leq  w_p^{\frac{1}{2^\ell}} + \frac18.
\end{equation}
Now we turn to $w_p^{\frac{1}{2^\ell}}$. Let $w_p = 1+ x_p$, $x_p \in (0,1)$. Consider the function $g(x_p) := (1+x_p)^{\frac{1}{2^\ell}}$. We take its Taylor expansion about $0$, and observing that
 $$g^{(j)}(x_p) = \left(\prod_{i=0}^{j-1} \frac{1-i2^\ell }{2^{\ell}}  \right)\; (1+x_p)^{\frac{1-j2^\ell}{2^\ell}} \leq \frac{(j-1)!}{2^\ell}, \;\; j \geq 1,$$
we have 
\begin{eqnarray*}
w_p^{\frac{1}{2^\ell}} - 1 = (1+x_p)^{\frac{1}{2^\ell}} - 1 &\leq& 
\frac{1}{2^\ell}\sum_{j=1}^\infty  \frac{(j-1)!}{j!} x_p ^j = \frac{1}{2^\ell}\sum_{j=1}^\infty  \frac1j x_p ^j\\
&=& \frac{1}{2^\ell} \sum_{j=1}^\infty \int_0^{x_p} s^{j-1} ds = 
\frac{1}{2^\ell} \int_0^{x_p}  \sum_{j=1}^\infty s^{j-1} ds \\
&\leq&  \frac{1}{2^\ell} \int_0^{x_p} \frac{1}{1-s}ds = \frac{1}{2^\ell} \ln{\frac{1}{1-x_p}}\\
&\leq& \frac{1}{2^\ell}\frac{n-m+p-1}{\log_2 e} \leq \frac{n-m+p-1}{2^\ell} , 
\end{eqnarray*}
because $x_p \leq 1 - 2^{-(n-m+p-1)}$, since $w$ has a fractional part of length $n-m$.
Observing that $n-m+p-1 \leq n$, and by setting $\ell \geq \lceil \log_2 8n \rceil$, we get 
\begin{equation}
\label{eq:wp}
w_p^{\frac{1}{2^\ell}} - 1 \leq \frac{n}{2^\ell} \leq \frac{n}{8n}\leq \frac18.
\end{equation}
Using equations (\ref{eq:tp}) and (\ref{eq:wp}), we get
\begin{equation}
\label{eq:tp2}
\hat{t}_p -1 \leq  \frac{1}{2^\ell} \frac{n}{\ln 2} + 2  \left( \frac{3}{4} \right)^{b-2}  \left( 2+ b +  \log_2 b  \right) \leq \frac14.
\end{equation}
Now we turn to the error of the algorithm. We have
\begin{eqnarray*}
 | z_p + \ln 2^{p -1} - \ln(w)| 
 &\leq&  | 2^{\ell} \hat{y}_p  - 2^{\ell} y_p | +  | 2^{\ell} y_p - 2^\ell \ln \hat{t}_p| +   | 2^\ell \ln \hat{t}_p- 2^\ell \ln w_p^{\frac{1}{2^\ell}}|\\
 &\leq & 2^{\ell - b}  + 2^{\ell}|2(\hat{t}_p -1)^3|   +  2^\ell |\hat{t}_p - w_p^{\frac{1}{2^\ell}}|\\
  &\leq & 2^{\ell - b}  
  + 2^{\ell}2\left( \frac{1}{2^\ell} \frac{n}{\ln 2} + 2  \left( \frac{3}{4} \right)^{b-2}  \left( 2+ b +  \log_2 b  \right) \right)^3\\ 
  &+&  2^\ell  \left(2  \left( \frac{3}{4} \right)^{b-2}  \left( 2+ b +  \log_2 b  \right)\right) 
\end{eqnarray*}
Since $b\geq \max\{5\ell,25\}$, we have
$\left( \frac{3}{4}\right)^{b/2} 2^\ell \leq 1$ and
$ \frac{2+b+\log_2 b}{\left( \frac43 \right)^{b/2}} \leq 1$. This yields
\begin{eqnarray}
\label{eq:lnErr1}
 | z_p + \ln 2^{p -1} - \ln(w)| 
&\leq& 
2^\ell \left( 2^{-b} +  \frac{2}{2^{3\ell}} \left(\frac{32}{9} + \frac{n}{\ln 2}   \right)^3 + \frac{32}{9} \left(\frac{3}{4}\right)^{b/2}    \right)   \nonumber \\
&\leq& 2^{-4\ell} + \frac{2}{2^{2\ell}}\left(\frac{32}{9} + \frac{n}{\ln 2}   \right)^3 + \frac{32}{9} \left(\frac{3}{4}\right)^{5\ell/2  \nonumber }\\
&\leq&  \left(\frac{3}{4}\right)^{5\ell/2} \left( 1+ \frac{32}{9} + 2\left(\frac{32}{9} + \frac{n}{\ln 2} \right)^3 \right).
  \end{eqnarray} 
Finally, from $b\geq 5\ell$ we have 
\begin{equation}
\label{eq:lnErr2}
(m-1) 2^{-b} \leq (m-1) 2^{-5\ell} \leq (m-1)\left( \frac{1}{4}\right)^{5\ell/2} \leq (m-1)\left( \frac{3}{4}\right)^{5\ell/2}
\end{equation}
Combining equations (\ref{eq:lnErrTot}), (\ref{eq:lnErr1}), and (\ref{eq:lnErr2}) then gives
\begin{eqnarray*}
 | z_p + (p -1)r - \ln(w)| &\leq&  \left(\frac{3}{4}\right)^{5\ell/2} \left( 1+ \frac{32}{9} + 2\left(\frac{32}{9} + \frac{n}{\ln 2} \right)^3 \right)+ (m-1)\left( \frac{3}{4}\right)^{5\ell/2}\\
  &\leq& \left(\frac{3}{4}\right)^{5\ell/2} \left( m+ \frac{32}{9} + 2\left(\frac{32}{9} + \frac{n}{\ln 2} \right)^3 \right),
\end{eqnarray*}
which is our desired bound.
\end{proof}

\begin{theorem} 
\label{thm4}
For $w>1$, given by $n$ bits of which the first $m$ correspond to its integer part, and $1\geq f \geq0$ given by $n_f$ bits of accuracy, Algorithm \ref{alg:fracPower} computes an approximation $\hat{z}$ of $w^{f}$ such that
\begin{equation}
|\hat{z} - w^{f}|  \leq \left( \frac{1}{2} \right)^{\ell  - 1  } , 
\end{equation}
where $\ell \in \nat$ is a parameter specified in the algorithm that is used to determine the number 
 $b =\max \{ n,n_f, \lceil 5(\ell + 2m +\ln n_f)\rceil, 40 \}$
of bits after the decimal point in which arithmetic will be performed, and therefore it determines the error.
Algorithm \ref{alg:fracPower} uses Algorithm \ref{alg:2krt} which computes power of 2 roots of a given number. 
\end{theorem}

\begin{proof}
First observe that the algorithm is exact for the cases $f=1$ or $f=0$. Therefore, without loss of generality assume $0<f<1$.

Consider the $n_f$ bit number $f$ and write its binary digits $f_i \in \{0,1\}$ explicitly as $f=0.f_1f_2...f_{n_f} = \sum_{i=1}^{n_f} f_i /2^i$. Denote the set of non-zero digits $\p: =  \{ 1\leq i \leq n_f : f_i \neq 0 \} $, $p:= |\p|$ with $1\leq p \leq n_f$.  
 Observe 
$$w^{f} = w^{0.f_1f_2...f_{n_f}} = w^{ \sum_{i=1}^{n_f} f_i /2^i} = \prod_{i=1}^{n_f} w_i^{f_i}  = \prod_{i\in \p} w_i, $$  
where again $w_i := w^\frac{1}{2^i}$. Let $\hat{w}_i \simeq w^\frac{1}{2^i}$ denote the outputs of Algorithm \ref{alg:2krt}. 
We have 
\begin{equation}    \label{fracPowTotErr}
\left|\hat{z} - w^{f}\right|  \leq \left|\hat{z} - \prod_{i\in \p} \hat{w}_i \right|  + \left|   \prod_{i\in \p} \hat{w}_i -  \prod_{i\in \p} w_i \right| ,
\end{equation}
where these terms give bounds for the repeated multiplication error, and the error from the $2^i$th roots as computed by Algorithm \ref{alg:2krt}, respectively.

Consider the second term. Partition $\p$ disjointly into two sets as $\p_1 : =  \{ i\in \p: \hat{w}_i  \geq w_i \} $ and  $\p_2 : =  \{ i\in \p: \hat{w}_i<w_i \} $. Observe that, for any $i$, from equation (\ref{eq:thm2}) of Theorem \ref{thm2} we have $|\hat{w}_i - w_i| \leq 2  \left( \frac{3}{4} \right)^{b-2m}  \left( 2+ b +  \log_2 b  \right) =: \e$. Observe $w_i, \hat{w}_i \geq 1$. 
First assume $\prod_{i\in \p} \hat{w}_i \geq \prod_{i\in \p} w_i $. Then
\begin{eqnarray*}
\left|   \prod_{i\in \p} \hat{w}_i -  \prod_{i\in \p} w_i \right|  
&=&  \prod_{i\in \p_1} \hat{w}_i \prod_{j\in \p_2} \hat{w}_j  -  \prod_{i\in \p_1} w_i  \prod_{j\in \p_2} w_j \\
&\leq&  \prod_{i\in \p_1} (w_i + \e) \prod_{j\in \p_2} w_j  -  \prod_{i\in \p_1} w_i  \prod_{j\in \p_2} w_j\\
&\leq&  \prod_{i\in \p_1} w_i  \prod_{j\in \p_2} w_j \left(  \prod_{k\in \p_1} \left(    1 + \frac{\e}{w_k}  \right)  - 1    \right)\\
&\leq&  w^f  \left(  \left(1+\e)^p -1 \right)\right) \\
&\leq&  w^f  \left( e^{p\e} -1 \right) 
\end{eqnarray*}
Conversely, assume $\prod_{i\in \p} \hat{w}_i < \prod_{i\in \p} w_i $. Then similarly we have
\begin{eqnarray*}
\left|   \prod_{i\in \p} \hat{w}_i -  \prod_{i\in \p} w_i \right|  
&=&   \prod_{i\in \p_1} w_i  \prod_{j\in \p_2} w_j - \prod_{i\in \p_1} \hat{w}_i \prod_{j\in \p_2} \hat{w}_j  \\
&\leq&   \prod_{i\in \p_1} w_i  \prod_{j\in \p_2} w_j - \prod_{i\in \p_1} w_i \prod_{j\in \p_2} (w_j-\e) \\
&\leq&  \prod_{i\in \p_1} w_i  \prod_{j\in \p_2} w_j \left( 1 - \prod_{k\in \p_2} \left(    1 - \frac{\e}{w_k}  \right)     \right)\\
&\leq&  w^f  \left( 1- \left(1-\e)^p \right)\right) \\
&\leq&  w^f  \left( 1- \left(  2 - e^{p\e}   \right)\right) \\
&\leq&  w^f  \left( e^{p\e} -1 \right) ,
\end{eqnarray*}
where we have used the inequality $\left(1-\e\right)^p -1 \geq 1- e^{pe}$.\footnote{This inequality follows trivially from term by term comparison of the binomial expansion of the left hand side with the Taylor expansion of the right hand side.}
So conclude that $ \left|   \prod_{i\in \p} \hat{w}_i -  \prod_{i\in \p} w_i \right|   \leq  w^f  \left( e^{p\e} -1 \right)$ always.
Furthermore, for $a \geq 0$ we have
$$ e^a -1 = a + a^2/2! + a^3/3!+ \dots = a(1+ a/2! + a^2/3! +\dots ) \leq a(1+a + a^2/2! + \dots)  = ae^a $$
which yields
\begin{equation}
\left|   \prod_{i\in \p} \hat{w}_i -  \prod_{i\in \p} w_i \right| \leq  w^{f} \;p\e \; e^{p\e} 
\end{equation}

Next consider the error resulting from truncation to $b$ bits of accuracy in the products computed in step 13 of the algorithm. 
For each multiplication, we have $\hat{z} = z +\xi$, with error $|\xi| \leq 2^{-b}$. For notational simplicity, reindex the set $\p$ as $\{1,2,\dots,p\}$ so that $\prod_{i\in \p} \hat{w}_i = \prod_{i=1}^p \hat{w}_i$. 
Let $z_i = \hat{w}_1\hat{w}_2\dots\hat{w}_i$, $i=1,2,\dots,p$ be the exact products, and 
let the approximate products be $\hat{z}_i = \hat{z}_{i-1}\hat{w_i}+\xi_i$, $i=2,\dots,p$, $\hat{z}_1 = 1$.
We have 
\begin{eqnarray*}
\hat{z}_1 &=& \hat{w}_1  \\
\hat{z}_2 &=& \hat{z}_1 \hat{w}_2 + \xi_2  \\
&\vdots& \\
\hat{z}_p &=& \hat{z} _{p-1}\hat{w}_p + \xi_p. 
\end{eqnarray*}
 Then we have
\begin{eqnarray*}
 \left|\hat{z}_p - \prod_{i\in \p} \hat{w}_i \right|  &=& | \hat{z}_p - z_p  | 
=  |\hat{z}_{p-1}\hat{w}_p + \xi_p - z_{p-1}\hat{w}_p|
 \leq  |\hat{z}_{p-1}\hat{w}_p - z_{p-1}\hat{w}_p| + |\xi_p|\\
  &\leq&  |\hat{z}_{p-1} - z_{p-1}|\hat{w}_p  + |\xi_p|\\
    &\leq& \left( |\hat{z}_{p-2} - z_{p-2}|\hat{w}_{p-1}  + |\xi_{p-1}| \right)\hat{w}_p  + |\xi_p|\\
        &\leq&  |\hat{z}_{p-2} - z_{p-2}|\hat{w}_{p-1}\hat{w}_p   + |\xi_{p-1}|\hat{w}_p  + |\xi_p|\\
&\vdots& \\
&\leq&  |\hat{z}_{1} - z_{1}|\hat{w}_2 \hat{w}_3 \hat{w}_4\dots \hat{w}_{p-1}\hat{w}_p  +  |\xi_{2}|\hat{w}_3 \hat{w}_4\dots \hat{w}_{p-1}\hat{w}_p    +   \dots + |\xi_{p-1}|\hat{w}_p  + |\xi_p|\\
&\leq&  2^{-b} \left( \hat{w}_3 \hat{w}_{4}\dots \hat{w}_{p-1} \hat{w}_p+ \hat{w}_{4}\dots \hat{w}_{p-1} \hat{w}_p+ \dots + \hat{w}_{p-1} \hat{w}_{p} +\hat{ w}_p +1 \right)\\
&\leq&  2^{-b} \left(p-1\right) w^\frac14
\leq  2^{-b} n_f w^f \leq  2^{-b} n_f w
\end{eqnarray*}
where the last line follows from observing each of the $p-1$ terms in the sum is less than $w_2 = w^\frac14$.

Thus, equation (\ref{fracPowTotErr}) yields total error
\begin{eqnarray*}
\left|\hat{z} - w^{f}\right| &\leq& 
\left|\hat{z} - \prod_{i\in \p} \hat{w}_i \right|  
+ \left|   \prod_{i\in \p} \hat{w}_i -  \prod_{i\in \p} w_i \right|  \\
&\leq& 
  2^{-b} \left(p-1\right) w^\frac14
+   w^{f} \:p\e \: e^{p\e}  \\
&\leq& 
  2^{-b} \; n_f w
+   w \:n_f \e \: e^{n_f\e}.
\end{eqnarray*}
Furthermore, using $\e = 2\left( \frac{3}{4} \right)^{b-2m}  \left( 2+ b +  \log_2 b  \right)$, $1\leq w\leq 2^m$, and as we have chosen $b$ sufficiently large such that $n_f \e \leq 1$ (to be shown later), we have
\begin{eqnarray*}
\left|\hat{z} - w^{f}\right| 
&\leq&  2^{-b} \; n_f w
+   w  \:n_f   \;2 \left( \frac{3}{4} \right)^{b-2m}   \left( 2+ b +   \log_2 b \right) e \\
&\leq&  n_f 2^{m-b}    +  n_f 2^{m} \left( \frac{3}{4} \right)^{b-2m}  2e\left( 2+ b +  \log_2 b  \right).  \\
\end{eqnarray*}
We have selected $b\geq \max\{n, n_f, \lceil 5(\ell + 2m + \log_2 n_f)\rceil , 40\}$ such that several inequalities are satisfied. 
From $b \geq \lceil 5(\ell + 2m + \log_2 n_f)\rceil  \geq  \ell + m + \log_2 n_f$, it follows that $n_f 2^{m-b} \leq 2^{-\ell}$.
Furthermore, for $b \geq 40$ , we have $2e \left( 2+ b +  \log_2 b  \right)  \leq (\frac{4}{3})^{b/2}$. 
Finally, $b \geq 5\left(    \ell + 2m + \log_2 n_f    \right) \geq \left(  \frac{2}{\log_2 4/3} \right)  \left(   \log_2 n_f + \ell + 2m     \right)$ implies that $n_f 2^{2m}  \left( \frac{3}{4} \right)^{b/2} \leq 2^{-\ell}$. Plugging these inequalities into the previous equation yields
\begin{eqnarray*}
\left|\hat{z} - w^{f}\right| 
&\leq&  n_f 2^{m} \frac{1}{2^b}  
  +  n_f 2^{m} \left( \frac{4}{3} \right)^{2m}   \left( \frac{3}{4} \right)^{b/2}    \left(  \left( \frac{3}{4} \right)^{b/2}  2e\left( 2+ b +  \log_2 b  \right)  \right) \\
&\leq&  \left( \frac{1}{2} \right)^{\ell}  
 +  n_f 2^{m} \left( \left(\frac{4}{3} \right)^{2}\right)^m   \left( \frac{3}{4} \right)^{b/2}   \\
&\leq&  \left( \frac{1}{2} \right)^{\ell}  
 +  n_f 2^{2m}  \left( \frac{3}{4} \right)^{b/2}    \\
&\leq& 2  \left( \frac{1}{2} \right)^{\ell}  = \left( \frac{1}{2} \right)^{\ell  -1  }  
\end{eqnarray*}
as was to be shown.

Finally, for completeness, from the above inequalities and 
$$b \geq 5\left(   \log_2 n_f + \ell + 2m     \right) 
\geq   \frac{2}{\log_2 4/3}  \left(  \log_2 n_f  - \log_2 e \right) + 4m, $$
 we have
\begin{eqnarray*}
n_f \e &=& n_f 2\left( \frac{3}{4} \right)^{b-2m}  \left( 2+ b +  \log_2 b  \right) 
\leq  n_f \left( \frac{3}{4} \right)^{b/2-2m}  \left( \frac{1}{e}  \right) \\
&\leq&  n_f \left( \frac{3}{4} \right)^{   \log_{4/3} n_f/e   }  \left( \frac{1}{e}  \right) \leq 1. 
\end{eqnarray*}
\end{proof}

\begin{cor} 
\label{cor1}
Let $w>1$ and $1\geq f \geq0$ as in Theorem \ref{alg:fracPower} above. Suppose $f$ is an approximation of a number $1\geq F \geq 0$ accurate to $n_f$ bits. Then Algorithm \ref{alg:fracPower}, with $f$ as input, computes an approximation $\hat{z}$ of $w^{F}$ such that
\begin{equation}
|\hat{z} - w^{F}|  \leq     \left( \frac{1}{2} \right)^{\ell  - 1  }  + \frac{w\ln w}{2^{n_f}}.
\end{equation}
\end{cor} 

\begin{proof}
Consider the error from the approximation of the exponent $F$ by $f$. We have $|F- f| \leq 2^{-n_f}$. Let $g(f):= w^f$. Then $g'(f) = w^f \ln w$. By the mean value theorem, we have
\begin{eqnarray*}
\left|w^{f} - w^{F} \right|  \leq \sup_{f\in (0,1)} g'(f) |F - f| \leq 2^{-n_f} \: w \ln w,
\end{eqnarray*}
which gives
$$|\hat{z} - w^{F}|  \leq |\hat{z} - w^{f}| + |w^{f} - w^{F} | \leq   \left( \frac{1}{2} \right)^{\ell  - 1  }  + \frac{w\ln w}{2^{n_f}}.$$
\end{proof}

\begin{theorem} 
\label{thm5}
For $0\leq w < 1$, represented by $n$ bits of which the first $m$ correspond to its integer part, and $1\geq f \geq0$ given by $n_f$ bits of accuracy, Algorithm \ref{alg:fracPower2} computes an approximation $\hat{t}$ of $w^{f}$ such that
\begin{equation}
|\hat{t} - w^{f}| \leq   \frac{1}{2^{\ell - 3 }}
\end{equation}
where $\ell \in \nat$ is a parameter specified in the algorithm that is used to determine the number $b = \max \{ n,n_f, \lceil 2\ell + 6m +2\ln n_f \rceil,40\}$ of bits after the decimal point in which arithmetic will be performed, and therefore also will determine the error.
Algorithm \ref{alg:fracPower2} uses Algorithm \ref{alg:fracPower}, which computes $w^f$ for the case $w \geq 1$, and 
also 
Algorithm \ref{alg:inv} which computes the reciprocal of a number  $w \geq 1$.
\end{theorem}

\begin{proof}
First observe that the algorithm is exact for the cases $f=1$, $f=0$, or $w=0$. 
Therefore, without loss of generality assume $0<f<1$ and $0<w<1$. 

Let all variables be defined as in Algorithm \ref{alg:fracPower2}. We shall first consider the error of each variable and use this to bound the overall error of algorithm.

Firstly, the input $0<w<1$ is rescaled  to $x:=2^k w \geq 1 > 2^{k-1}w$ exactly, by $k$-bit left shift,where $k$ is a positive integer. An example circuit for computing $k$ is given in Fig. \ref{fig-ShiftInteger} above. Observe that we have
$ w^f = x^f/2^{kf} = x^f/ ( 2^{\lfloor kf  \rfloor}2^{  \{ kf \}} )$. 
We also have $\log_2 \frac{1}{w} \leq k < \log_2 \frac{1}{w}+1$.

The product $c=kf  < k \leq n-m$ is computed exactly in fixed precision arithmetic because the number of bits after the decimal point in $kf$ is at most
 $n_f \leq b$, where $b$ is the number of bits in which arithmetic is performed.

Next consider $\hat{z}=$ FractionalPower($x$, $f$, $n$, $m$, $n_f$, $\ell$), which approximates $z = x^f$. From Theorem \ref{thm4} we have 
$ e_z: = |\hat{z} - z| \leq \frac{1}{2^{\ell -1}}$.
Similarly, $\hat{y}= $ FractionalPower($2$, ${\{ c \}} $, $n$, $m$, $n_f$, $\ell$) approximates $y=2^{\{ c \}}$, with $e_y:= |\hat{y} - y| \leq \frac{1}{2^{\ell - 1}}$.
Furthermore, for $\hat{s} = $ INV($\hat{y},n,1,2\ell$) which approximates $s =1/\hat{y}$, from Corollary \ref{cor0} we have 
$e_s :=|\hat{s} - s| \leq \frac{2+\log_2\ell}{2^{2\ell}}$, which satisfies $e_s \leq \frac{1}{2^\ell}$ for $\ell \geq 2$.

Finally, observe $2^{- \lfloor c \rfloor}  \hat{z}$ is computed exactly by a right shift of  $\hat{z}$. This is used to compute $t = 2^{- \lfloor c \rfloor}  \hat{z} \hat{s}$, which is again truncated to $b$ decimal bits to give $\hat{t}$ with $e_t:= |\hat{t} - t| \leq 2^{-b}$. 

Now we turn to the total error. 
By our variable definitions, $w^f = 2^{- \lfloor c \rfloor}  z /y $. We have
\begin{eqnarray*}
| \hat{t} - w^f |  &\leq&  |  \hat{t} - t | 
+ | 2^{- \lfloor c \rfloor}  \hat{z}\hat{s} - 2^{- \lfloor c \rfloor}  z\hat{s}  |
+ |2^{- \lfloor c \rfloor}  z\hat{s}   - 2^{- \lfloor c \rfloor}  z s   |
+\left|2^{- \lfloor c \rfloor}  z s -   2^{- \lfloor c \rfloor}  \frac{z}{y}\right|\\
&\leq& 2^{-b}
+    2^{- \lfloor c \rfloor} \hat{s} | \hat{z} -  z  |
+ 2^{- \lfloor c \rfloor}  z  |\hat{s}   -  s   |
+ 2^{- \lfloor c \rfloor}  z \left| s -    \frac{1}{y}\right|\\
&\leq& 2^{-b}
+   \frac{1}{ 2^{\lfloor c \rfloor} } \frac{1}{2^{\ell -1}}
+ w^f 2^{  \{ c \}}  |\hat{s}   -  s   |
+ \frac{x^f}{2^{ \lfloor c \rfloor} } \frac{1}{2^{ \{ c \} } }\left|\frac{2^{ \{ c \} } }{\hat{y}} -  1\right|,
\end{eqnarray*}
where we have used $\hat{s} \leq s = \frac{1}{\hat{y}} \leq 1$ because as remarked in the proof of Theorem \ref{SCIthm1}, the algorithm computing the reciprocal underestimates it value, i.e. $\frac{1}{\hat{y}} \leq \frac{1}{y} = 2^{-\lfloor kf\rfloor } \leq 1$. 
Observe we have $w^f = \frac{x^f}{2^{ \lfloor c \rfloor} } \frac{1}{2^{ \{ c \} } } \leq 1$. Moreover, $2^{\{ kf\}} \leq 2$. Hence, as shown in the proof of Theorem \ref{SCIthm1}, that $\hat{y} \geq 1$.  This, together with the error bounds of Theorem \ref{thm4} and of 
Corollary \ref{cor0} yields
\begin{eqnarray*}
| \hat{t} - w^f |   
&\leq& 
2^{-b} + \frac{1}{2^{\ell -1}}
+ 2 |\hat{s}   -  s   |
+  \frac{1}{\hat{y}}  |  2^{ \{ c \}  } -\hat{y}   |\\
 &\leq& 2^{-b} +  \frac{1}{2^{\ell -1}}
+\frac{2}{2^{\ell}}
+\frac{1}{2^{\ell -1}}\\
 &\leq& 4 \frac{1}{2^{\ell -1 }} = \frac{1}{2^{\ell - 3 }}
\end{eqnarray*}
\end{proof}

\begin{cor} 
\label{cor2}
Let $0 \leq w < 1$ and $1\geq f \geq 0$ as in Theorem \ref{alg:fracPower2} above. Suppose $f$ is an approximation of a number $1\geq F \geq 0$ accurate to $n_f$ bits. Then Algorithm \ref{alg:fracPower2}, with $f$ as input, computes an approximation $\hat{t}$ of $w^{F}$ such that
\begin{equation}
|\hat{t} - w^{F}|  \leq     \left( \frac{1}{2} \right)^{\ell  - 2  }  + \frac{w\ln w}{2^{n_f}}.
\end{equation}
\end{cor} 

\begin{proof}
The proof is similar to that of Corollary \ref{cor1}.
\end{proof}

%% file: _app_HamSim.tex
\chapter{Divide and Conquer Hamiltonian Simulation}  \label{app:HamSim}
In this appendix, 
we review high-order splitting formulas for Hamiltonian simulation, and give the proofs for several results of Chapter \ref{ch:HamSim}. 

\section{Review of High-Order Splitting Formulas}   \label{sec:splittingFormulas} 

Splitting formulas are a family of operator approximations based on the \textit{Lie-Trotter product formula}
\begin{equation}   \label{eqn:LieTrotterFormula}
\lim_{n \rightarrow \infty} (e^{-iH_1t/n} e^{-iH_2t/n}\dots  e^{-iH_mt/n} )^n = e^{-iHt}.
\end{equation}
Using this formula with finite $n$ gives an approximation of $e^{-iHt}$. 
Without loss of generality, and to avoid dealing with absolute values, we will assume $t>0$ here. 
Selecting $n$, often called the \textit{Trotter number}, large enough such that the \textit{time slice} $\Delta t := t/n$ is small $\Delta t \ll 1$, we approximate $e^{-iH\Delta t}$ by 
$e^{-iH_1 \Delta t}e^{-iH_2 \Delta t}\dots e^{-iH_m \Delta t}$
with error $O(\Delta t ^2)$. 
This gives a second-order approximation. 
A third-order approximation is given by the Strang splitting formula
\begin{equation}   \label{eqn:Strang}
S_2(H_1, \dots, H_m, \Delta t) = e^{-iH_1\Delta t/2} 
 \dots e^{-iH_{m-1}\Delta t/2}e^{-iH_m\Delta t}e^{-iH_{m-1}\Delta t/2}\dots e^{-iH_1\Delta t/2},
\end{equation}
with\footnote{For simplicity, when the underlying Hamiltonian decomposition is clear we will use $S_2(\Delta t)$ in place of $S_2(H_1, \dots, H_m, \Delta t)$.} 
$$e^{-iH\Delta t} = S_2(\Delta t)+ O(\Delta t^3),  \;\;\;\;\;\; \textrm{ as }\Delta t \rightarrow 0. $$
Applying $S_2(\Delta t)$ over each time slice $\Delta t$ yields the approximation 
$$ \widetilde{U} =( S_2(\Delta t))^{n} ,$$   
where $ \| U -  \widetilde{U} \| \rightarrow 0$ as $\Delta t \rightarrow 0$. 
Assume for 
the moment that $\Delta t$ is chosen such that the number of time slices $n=t/\Delta t$ is indeed an integer. 
Otherwise, we would have $n=\lceil t/\Delta t \rceil$, and a single different final time slice $\Delta t' := t -\Delta t  \lfloor t/\Delta t \rfloor < \Delta t$.

Suzuki \cite{Suzuki90,Suzuki91} gave 
high-order splitting formulas. These are recursive formulas $S_{2k}$ of order $2k+1$, $k\in\nat$, 
approximating $e^{-iH\Delta t}$ to error $O(\Delta t^{2k+1})$.  
They are defined by 
\begin{equation}   \label{eqn:defS2k}
S_{2k}(\Delta t) =[S_{2(k-1)}(p_k\Delta t)]^2 \; S_{2(k-1)}(q_k\Delta t) \; [S_{2(k-1)}(p_k\Delta t)]^2,
\end{equation}
for $k=2,3,\dots$, with $p_k = (4-4^{1/(2k-1)})^{-1})$ and $q_k = 1 - 4p_k$. 
Applying $S_{2k}(\Delta t)$ 
over each time slice $\Delta t$ 
and unwinding the recurrence relation 
yields a product of $N$ exponentials 
\begin{equation}  \label{eqn:UGen0}
\widetilde{U}= (S_{2k}(H_1,\dots,H_m,t/n))^n
=  \prod_{\ell=1}^{N} e^{-i  H_{j_\ell} t_{\ell} /n },  \;\;\;\;\;\;  j_\ell \in \{1,\dots,m \}, \;\;\;\;  \sum_{{\ell=1}}^{N} t_{\ell}/n =  m t.
\end{equation}

It is important to observe that Suzuki's formulas 
hold asymptotically for sufficiently small $\Delta t$, and do not reveal the dependence of the error on $m$ or the norms $\|H_j\|$, $j=1,\dots,m$. 
Application of these formulas requires explicit calculation of the prefactors in the error bounds.
Typically, cost estimates for splitting methods are expressed as the product of the number of time slices and the number of exponentials required to carry out the simulation within each time slice.  
In particular, the estimates for the simulation error and cost in \cite{PZ12} depend on $m$, $\e$, $k$,
 the largest norm $\|H_1\|$, and the second largest norm $\|H_2\|$. In \cite[Sec. 4]{PZ12} the quantity $M$ is defined as 
\begin{equation}   \label{eqn:def0Mnew}
M=  \left(   \frac{4e m t \|H_2\| }{\e}  \right)^{1/2k}  \frac{4e m}{3}  \left( \frac53 \right)^{k-1},
\end{equation}
and the time slice is given by $\Delta t = (M\|H_1\|)^{-1}$. Hence the number of intervals is  $n=\lceil t/\Delta t\rceil = \lceil M\|H_1\|t\rceil$.  
%
Note that choosing $M$ larger than necessary decreases the simulation error. 
Under the (weak) assumption $4emt\|H_2\|>\e$, 
\cite[Thm. 2]{PZ12} shows an upper bound for the number of exponentials 
\begin{equation}  \label{eqn:PZ12bound}
N  \leq \left( (2m-1)5^{k-1} \right)\cdot  \bigg\lceil  \|H_1\|t  \left( \frac{4emt\|H_2\|}{\e}\right)^\frac{1}{2k} \frac{4em}{3}\left(\frac53 \right)^{k-1}  \bigg\rceil .  
\end{equation}
This bound is derived as the product of two terms. 
The first factor 
is the number of exponentials per 
time slice, which is bounded by $(2m-1)5^{k-1}$. 
The second factor is equal to $n$ 
which bounds  the number of time slices. 
Note that if the argument of the ceiling function is at most one, a single time interval suffices for the simulation. The cost bounds in Section \ref{sec:DCAlgorithms} for Algorithms $1$ and $2$ are generalizations of (\ref{eqn:PZ12bound}). 

Recall  
that the upper bound (\ref{eqn:PZ12bound}) 
does not account for any finer problem structure, such as the possibility that a number of Hamiltonians have norms significantly smaller than $\|H_2\|$. 
Hamiltonians with extremely small norm ($\|H_j\| = O(\e/t)$) can be ignored altogether as indicated in Proposition \ref{prop:discardHamsNew}. 
The remaining Hamiltonians may then be partitioned into groups based on their relative norms, and each group simulated 
independently with our algorithms. This leads us to refined cost estimates which depend not just on $m$, $\|H_1\|$, and $\|H_2\|$, but on  the number of Hamiltonians in each group and largest Hamiltonian norm within each group. 

Furthermore, under weak conditions which guarantee the argument of the ceiling function in (\ref{eqn:PZ12bound}) is at least one (e.g. for sufficiently large $m$, $\|H_1\|$, $t$, or $1/\e$),
 (\ref{eqn:PZ12bound}) may be bounded to obtain 
\begin{eqnarray}  \label{eqn:PZ12boundexpoForm} 
N(k) &\leq& \left( (2m-1)5^{k-1} \right)\cdot 2\|H_1\|t  \left( \frac{4emt\|H_2\|}{\e}\right)^\frac{1}{2k} \frac{4em}{3}\left(\frac53 \right)^{k-1} =:  N(k), 
\end{eqnarray} 
and from this \cite[Sec. 5]{PZ12} shows 
the \lq\lq optimal\rq\rq\ $k$ (in the sense of minimizing the upper bound $N(k)$),
\begin{equation}   \label{eqn:optk}
k^*  := \max \bigg\{ \rm round \left( \sqrt{   \frac12   \log_{25/3}  \frac{4emt\|H_2\| }{ \e }} \right), 1  \bigg\}.
\end{equation}
Setting $k=k^*$ gives the upper bound for the number of matrix exponentials 
\begin{equation}  \label{eqn:PZ12boundOptimalk}
N \leq \frac{8e}{3}  (2m-1) \; m \|H_1\|t  \; e^{2\sqrt{\frac12 \ln \frac{25}{3} \ln \frac{4emt\|H_2\|}{\e}} } \; =: N^{*}. 
\end{equation}
We compare our results against this estimate in Section \ref{sec:speedup}. 

Further observe that $k^*$ is given by an extremely slow growing function of the problem parameters. For example, for the values 
$m=t=\|H_2\|=\e^{-1}=10^{10}$, 
(\ref{eqn:optk}) gives $k^* = 5$. 
Therefore, in most practical cases, one can determine the optimal value of $k$ by inspection, with the need to  carry out a formal analysis.

\section{Divide and Conquer Hamiltonian Simulation}
\subsection*{Discarding Small Hamiltonians}

\begin{proof}[Proof of Proposition $1$]
Using the variation-of-constants formula \cite{jahnkeLubich},  
for any vector $v$ we have 
$$ e^{-iHt}v = e^{-iAt}v - i \int_{0}^{t} \: e^{-iAs}  B e^{-iH(t-s)}v \;ds  \;\;\;\;\;\; (t\geq 0),$$
which implies
$$ \|e^{-iHt} - e^{- i  At  } \| \leq \| B\| t \leq \e/2. $$
Thus, 
$$\| e^{-iHt} - \widetilde{U}\| \leq \| e^{-iHt} - e^{-iAt}\| + \| e^{-iAt} - \widetilde{U}\| \leq \e/2+\e/2 \leq \e.  $$
\end{proof}

\subsection*{Recursive Lie-Trotter Formulas}
We show a generalization of the Lie-Trotter formula. 
For simplicity and to avoid technical details we 
assume that $H$, $A$, and $B$ are complex matrices. 

\begin{lem}[Recursed Lie-Trotter Formula]    \label{lem:LieTrotter}
Let $H_1,\dots,H_m$ be Hamiltonians with 
$H = \sum_{j=1}^m H_j$. 
Consider $A = \sum_{j=1}^{m'} H_j$ and $B = \sum_{j=m'+1}^m H_j$. 
Let 
$$f(n,\alpha,\beta):=\left(
(e^{-iH_1t/\alpha n}  \dots e^{-iH_{m'}t/\alpha n} )^\alpha \;
 (e^{-iH_{m'+1}t/\beta n} \dots e^{-iH_{m}t/\beta n} )^\beta
\right)^n$$
for $n,\alpha,\beta \in \nat$. 
Then for fixed $\alpha,\beta$, we have 
$$ \lim_{n \rightarrow \infty}  f(n,\alpha,\beta) = e^{-iHt}.$$
In particular, for $\alpha=\beta=1$ the usual Lie-Trotter formula (\ref{eqn:LieTrotterFormula}) is reproduced.

Moreover, we may also take limits with respect to $\alpha$ and $\beta$, and in any order, i.e.,
$$ \lim_{n, \alpha, \beta \rightarrow \infty}  f(n,\alpha,\beta) = e^{-iHt}.$$
\end{lem}
\begin{proof}
Fix $\alpha, \beta$. Then we may expand $f$ as 
\begin{eqnarray*}
f(n,\alpha,\beta)&=&\bigg(
\underbrace{(e^{-iH_1t/\alpha n}  \dots e^{-iH_{m'}t/\alpha n} ) 
\dots
(e^{-iH_1t/\alpha n}  \dots e^{-iH_{m'}t/\alpha n} )  }_{\alpha} \dots \\
&\dots&
\underbrace{  (e^{-iH_{m'+1}t/\beta n} \dots e^{-iH_{m}t/\beta n} ) \dots 
 (e^{-iH_{m'+1}t/\beta n} \dots e^{-iH_{m}t/\beta n} ) }_{\beta}
\bigg)^n,
\end{eqnarray*}
to which we may apply the Lie-Trotter formula with respect to $n$ to yield 
\begin{eqnarray*}
\lim_{n \rightarrow \infty}  f(n,\alpha,\beta)&=&
{\rm exp} \bigg( \underbrace{(-iH_1t/\alpha  -  \dots -iH_{m'}t/\alpha)  +
\dots + (-iH_1t/\alpha \dots -iH_{m'}t/\alpha)}_{\alpha} \\
 && + \underbrace{ (-iH_{m'+1}t/\beta  - \dots  -iH_{m}t/ \beta) +  \dots +
(-iH_{m'+1}t/\beta  \dots  -iH_{m}t/ \beta)}_{\beta} \bigg)\\
&=& e^{-\alpha(iH_1t/\alpha ) - \dots -\alpha(iH_{m'}t/\alpha ) -\beta(iH_{m'+1}t/\beta) - \dots  -\beta(iH_{m}t/\beta) } = e^{-iHt}.
\end{eqnarray*}
%
This expression holds for arbitrary but fixed 
$\alpha$ and $\beta$. 
Now suppose we take $\alpha \rightarrow \infty$ while keeping $n$ and $\beta$ fixed. 
Then we have 
\begin{eqnarray*}
\lim_{\alpha \rightarrow \infty}  f(n,\alpha,\beta)&=&
\left( e^{At/ n} 
 (e^{-iH_{m'+1}t/\beta n} \dots e^{-iH_{m}t/\beta n} )^\beta
\right)^n.
\end{eqnarray*}
Taking now $n\rightarrow \infty$ yields 
$$ \lim_{n \rightarrow \infty}   \lim_{\alpha \rightarrow \infty}  f(n,\alpha,\beta) = e^{-iHt}.$$
All possible orderings of the limits with $n,\alpha,\beta \rightarrow \infty$ follow similarly. 
\end{proof}

\subsection*{Error of the Strang Splitting Formula}
Suzuki \cite{suzuki1976generalized} provides error bounds for the Trotter formula, the Strang formula, and other high-order splitting formulas. 
%
%
We will build on the analysis of \cite{wecker2014gate} to derive a useful bound for the error of the Strang splitting formula. 
We note that the original analysis of \cite{wecker2014gate} contains a small error, which has been corrected in 
 \cite[App. A]{wecker2015solving}. 
 We also use results from \cite{suzuki1976generalized}. 

We make frequent use of the inequality (e.g. \cite[Lemma 1]{suzuki1976generalized})
\begin{equation}    \label{eqn:suzukiIdentity}
\|a^n - b^n\| \leq n\|a-b\| (\max(\|a\|,\|b\|))^{n-1},
\end{equation}
 for $a,b$ elements of a Banach operator algebra and $n\in \nat$. 
 Also recall the identity for the commutator operator $\|[X,Y]\| \leq 2\|X\|\|Y\|$, where $[X,Y]:= XY-YX$.

\begin{lem}[Strang Splitting Formula Error]  \label{lem:wecker2}
Let $n\in\nat$, $t>0$, and $\Delta t : = t/n$. 
Let $A,B$ be Hermitian matrices and $H=A+B$. Then 
$$\| e^{-iHt}- (S_2(A,B,\Delta t))^n\|   \leq  \frac{1}{12} \| [[A,B], A+B]  \| t \Delta t^2
\leq \frac{2}{3} \|A\| \|B\| \|C\| t \Delta t^2,$$
where $\|C\|:=\max\{\|A\|\,\|B\|\}.$
\end{lem}
\begin{proof}
Let $H(x):= (1-x)A\Delta t +B\Delta t$, $0\leq x\leq 1$. Then from \cite[Appendix B]{wecker2014gate} 
we have
$$ \| [[A\Delta t,H(x)],H(x)] \| \leq \left( \| [[A,B],A] \| + \|[[A,B],B] \| \right) \Delta t^3.$$
Extending the analysis of \cite[Appendix B]{wecker2014gate}, we get
\begin{eqnarray*}
\| e^{-i(A+B)\Delta t} - e^{-iA\Delta t/2} e^{-iB\Delta t}e^{-iA\Delta t/2}\| 
&\leq& \int_0^1   \bigg  \| \int_0^1  \frac{s-s^2}{2}  [[A\Delta t,H(x)],H(x)] \; ds \bigg \| \; dx\\
&=&   \int_0^1  \frac{s-s^2}{2} \; ds  \;\;  \int_0^1  \;  \bigg \| [[A\Delta t,H(x)],H(x)] \bigg \| \; dx \\
 &\leq& \frac{1}{12}  \left( \| [[A,B],A] \| + \|[[A,B],B] \| \right) \Delta t^3\\
&\leq&  \frac{1}{12} (4 \|A\|^2\|B\|+ 4 \|A\|\|B\|^2 ) \Delta t^3.
\end{eqnarray*}
For $\|C\|=\max\{\|A\|\,\|B\|\}$ this yields
$$ \| e^{-i(A+B)\Delta t} - e^{-iA\Delta t/2} e^{-iB\Delta t}e^{-iA\Delta t/2}\|  \leq  \frac{2}{3}\|A\|\|B\|\|C\|\Delta t^3. $$
Finally, 
$$\|(e^{-i(A+B)\Delta t})^{t/\Delta t} - S_2(A,B,\Delta t)^{t/\Delta t} \| \leq (t/\Delta t)\|e^{-i(A+B)\Delta t} - e^{-iA\Delta t/2} e^{-iB\Delta t}e^{-iA\Delta t/2}\| ,$$
which follows from (\ref{eqn:suzukiIdentity}) with unitary $a$ and $b$. 
\end{proof}

\subsection*{Algorithm $2$}
We give the details of Algorithm $2$, which generalizes Algorithm $1$ by first applying a splitting formula of order $2k+1$; see Figure \ref{fig:flowchart}.  

We apply the results of \cite{PZ12},
which achieves improved bounds to the simulation error and cost by 
rescaling the Hamiltonians to have norm at most $1$. 
Note that such rescalings are equivalent to rescalings of the simulation time. 
Indeed, for Hamiltonians $A$, $B$, $H=A+B$, and $\ell > 0$ we have $U(H/\ell, t) = e^{-i(H/\ell) t} = e^{-iH(t/\ell)}=U(H,t/\ell)$ and $S_{2k}(A/\ell,B/\ell,t) = S_{2k}(A,B,t/\ell)$,
where the definition of $S_{2k}$ is given in (\ref{eqn:defS2k}). 

\begin{proof}[Proof of Proposition $3$]
Recall the preliminary analysis given in Section \ref{sec:GeneralParadigmDetails}. 
Consider the Hamiltonian $H$ as in  (\ref{eq:HamDef}, \ref{eqn:HamOrdering}), partitioned into two groups $H=A+B=(H_1+\dots + H_{m'})+(H_{m'+1}+\dots+H_m)$. 
We have
$$U=e^{-iHt}=e^{-i(A+B)t}=\left(e^{ -i(\frac{A}{\|C\| }+\frac{B}{\|C\| })/M  }\right)^{M\|C\|t} =\left(e^{ -i(A +B ) (M\|C\|)^{-1}  }\right)^{M\|C\|t} ,$$
for $\|C\|:=\max\{\|A\|,\|B\|\}$, and the quantity $M>1$ is sufficiently large and will be defined shortly. 
Also define $\|D\|:=\min\{\|A\|,\|B\|\}$. 
Thus the (algorithm first-step) time slice size is $(M\|C\|)^{-1}$, and the number of (first-step) intervals is $\lceil M\|C\|t  \rceil$. 
Let 
$$n_0:= \lfloor M\|C\| t\rfloor, \;\;\;\;\;\;\;\;\;  \delta := M\|C\|t -\lfloor M\|C\|t \rfloor $$
denote the integer and fractional parts of $M\|C\|t=n_0+\delta$, respectively. 

Recall that our problem is equivalent to simulating 
$H/\|C\|$ for time $t\|C\|$. 
Let 
\begin{eqnarray}  \label{eqn:Uhat2}
\widehat{U} &:=& \left(S_{2k}\left(\frac{A}{\|C\|},\frac{B}{\|C\|},\frac{1}{M}\right) \right)^{n_0} S_{2k} \left(\frac{A}{\|C\|},\frac{B}{\|C\|},\frac{\delta}{M}\right) 
\end{eqnarray}

Unwinding the recurrence (\ref{eqn:defS2k}) defining $S_{2k}$ for two Hamiltonians $X$ and $Y$ and $\tau \in \reals$ yields \cite{wiebe2010higher,PZ12}
$$S_{2k}(X,Y,\tau)  = \prod_{\ell=1}^K  S_{2}(X,Y,z_\ell  \tau) ,$$
where $K=5^{k-1}$ and each $z_\ell$ is defined according to the recursive scheme of (\ref{eqn:defS2k}),  $\ell=1,\dots, K$. 
 In particular, each $z_\ell$ is given as product of $k-1$ factors as $z_\ell = \prod_{r\in I_p} p_r \prod_{r\in I_q} q_r$, where the products are over the index sets $I_p$, $I_q$ defined by traversing the path of the recursion tree corresponding to $\ell$, 
and $\sum_{\ell =1}^K | z_\ell| = 1$; see \cite[Sec. 3]{PZ12} for details. 
Recall that in Section \ref{sec:splittingFormulas} we have defined the quantities $p_k = (4-4^{1/(2k-1)})^{-1})$ and $q_k = 1 - 4p_k$, for $k \in \nat$. 

Let $\widetilde{U}_\mathcal{A}(\tau)$ and $\widetilde{U}_\mathcal{B}(\tau)$ be approximations to $e^{-i(A/\|C\|)\tau}$ and $e^{-i(B/\|C\|)\tau}$, respectively, where 
 $\mathcal{A} = A/\|C\|$ and $\mathcal{B}= B/\|C\|$. 
We approximate $S_2(A/\|C\|,B/\|C\|,\tau)$ by 
$$ \widetilde{S}_2(A/\|C\|,B/\|C\|,\tau) := \widetilde{U}_\mathcal{A}(\tau/2)\widetilde{U}_\mathcal{B}(\tau)\widetilde{U}_\mathcal{A}(\tau/2) ,$$
and this yields 
$$ \widetilde{S}_{2k}(A/\|C\|,B/\|C\|,\tau)  := \prod_{\ell=1}^K \widetilde{S}_2(A/\|C\|,B/\|C\|, z_\ell \tau) = \prod_{\ell=1}^K\widetilde{U}_\mathcal{A}(z_\ell\tau/2)\widetilde{U}_\mathcal{B}(z_\ell\tau) \widetilde{U}_\mathcal{A}(z_\ell\tau/2).  $$
Hence, applying the above to (\ref{eqn:Uhat2}) we get 
\begin{eqnarray}  \label{eqn:Utilde2}
\widetilde{U} &:=& (\widetilde{S}_{2k}(A/\|C\|,B/\|C\|,1/M) )^{n_0} \widetilde{S}_{2k}(A/\|C\|,B/\|C\|,\delta/M) \nonumber\\
&=& \left(\prod_{\ell=1}^K \widetilde{S}_2(A/\|C\|,B/\|C\|, z_\ell/M)\right)^{n_0}  \prod_{\ell=1}^K \widetilde{S}_2(A/\|C\|,B/\|C\|, z_\ell \delta /M ) \nonumber\\
&=&  \prod_{\ell'=1}^{K\lceil M\|C\| t \rceil} \widetilde{U}_\mathcal{A}(z_{\ell'}  /2M)\widetilde{U}_\mathcal{B}(z_{\ell'}  /M)\widetilde{U}_\mathcal{A}(z_{\ell'}  /2M) ,
\end{eqnarray}
where in the last equation 
we have re-indexed the product so that 
$z_{\ell'}=z_{((\ell'\; \rm mod \; K) +1)}$ for $1\leq \ell' \leq n_0 K $, and $z_{\ell'}=z_{((\ell'\; \rm mod \; K) +1)}\delta $ for $n_0K < \ell' \leq (n_0+1)K$. The overall term ordering and time interval sizes are easily computable from (\ref{eqn:defS2k}) and (\ref{eqn:Utilde2}). 
Thus  $\widetilde{U}$ is an ordered product of $(3K\lceil M\|C\|t \rceil )$-many applications of $\widetilde{U}_\mathcal{A}$ and $\widetilde{U}_\mathcal{B}$ (each applied for differing simulation times).

We now turn to the second step splitting formulas, i.e., the ones approximating 
$e^{-i(A/\|C\|)\tau}$ and $e^{-i(B/\|C\|)\tau}$ for $\tau \in \reals$. 
We apply Suzuki's high-order splitting formulas, with different orders in principle. 

Once more simulating 
$\mathcal{A}=A/\|C\|$ for time $\tau$ is equivalent to simulating $A$ for time $\tau/\|C\|$,
and this is further equivalent to simulating $A/\|H_1\|$ for time $\tau\|H_1\|/\|C\|$. 
Thus we define  
$$\mathcal{H}_j :=\begin{cases}
    \frac{H_j/\|C\|}{ \| H_1/ \|C\| \|}  =  \frac{ H_j }{ \| H_1\|}         & (1 \leq j \leq m')\\
    \frac{H_j/\|C\|}{ \| H_{m'+1}/\|C\|\| } = \frac{H_j }{\| H_{m'+1}\|} & (m' < j \leq m) .  \end{cases}
$$ 
Thus we obtain 
\begin{equation}  \label{eqn:UtildeA2}
\widetilde{U}_\mathcal{A}(z_\ell /2M):= S_{2k_A}(\mathcal{H}_{1},\dots, \mathcal{H}_{m'}, 1/M_A)^{ \lfloor  (|z_\ell| /2M) (M_A\|H_{1}\|/\|C\|) \rfloor }  S_{2k_A}(\mathcal{H}_{1},\dots, \mathcal{H}_{m'}, \delta_A/M_A)  ,
\end{equation}
\begin{equation}  \label{eqn:UtildeB2}
\widetilde{U}_\mathcal{B} (z_\ell /M):=S_{2k_B}(\mathcal{H}_{m'+1},\dots, \mathcal{H}_{m}, 1/M_B)^{ \lfloor M_B\|H_{m'+1}\| |z_\ell| / M\|C\| \rfloor } S_{2k_B}(\mathcal{H}_{m'+1},\dots, \mathcal{H}_{m}, \delta_B/M_B),
\end{equation}
where 
$\delta_A:=M_A\|H_{1}\|  |z_\ell| /2M\|C\|  - \lfloor M_A\|H_{1}\|  |z_\ell| /2M\|C\|  \rfloor$ and $\delta_B:=M_B\|H_{m'+1}\| |z_\ell| / M\|C\| - \lfloor M_B\|H_{m'+1}\| |z_\ell| / M\|C\| \rfloor$, and $M_A,M_B>1$. 
%
As before, the quantities $\lceil M_A\|H_1\| |z_\ell| /2M\|C\|  \rceil$ and $\lceil M_B\|H_{m'+1}\| |z_\ell| / M\|C\| \rceil$ give the number of subintervals 
used to further subdivide intervals of length $z_\ell / 2M_A$ and $z_\ell / M_B$, respectively. We define $M_A$, $M_B$ below.  
The reader may wish to recall the text after (\ref{eqn:UtildeA},\ref{eqn:UtildeB}) that deals with the calculation of the number of subintervals and their lengths. 
 
\subsubsection*{Error and Cost} 
Using (\ref{eqn:Uhat2}) and (\ref{eqn:Utilde2}), we bound the overall error by 
\begin{equation*}  
\| U - \widetilde{U} \| \leq \| U- \widehat{U}\| + \| \widehat{U} - \widetilde{U} \|.
\end{equation*}
The first term in the right-hand side corresponds to the error of a splitting formula at the first step of the algorithm, where we pretend that exponentials $e^{-iA\tau}$, $e^{-iB\tau}$, $\tau \in \reals$ are given to us exactly, and the second term corresponds to the error in the second step of the algorithm, i.e. the error introduced by splitting formulas approximating $e^{-iA\tau}$ and $e^{-iB\tau}$. 

As explained in Section \ref{sec:GeneralParadigmDetails}, 
to guarantee $\| U- \widehat{U}\| \leq \e/2$ 
we set the quantity $M$ as in 
(\ref{eqn:defMgeneral}), 
which gives
$$M=M(k) := 
\left(   \frac{16et \|D\| }{\e}  \right)^{1/2k}  \frac{8e}{3} \left( \frac53 \right)^{k-1} .$$
To apply \cite[Thm. 1]{PZ12}
for $H=A+B$ and accuracy $\e/2$, 
the condition of that theorem becomes 
\begin{equation}  \label{eqn:condTopLevThmAlg2}
16et\|D\|\geq \e,
\end{equation} 
which implies $M\geq 1$. 
Hence, the number of ${S}_{2k}$ comprising $\widehat{U}$ in (\ref{eqn:Uhat2}) is at most 
\begin{equation*}  
3 \cdot 5^{k-1} \lceil M\|C\|t \rceil ,
\end{equation*}
 where $\lceil M\|C\|t \rceil$ gives the number of 
time intervals at the first step.
The interesting case is $M\|C\|t \geq 1$, for which it suffices to assume $\|C\|t \geq 1$ (otherwise, as explained in the analysis of Algorithm $1$, we would be dealing with an easy problem). 
Then the above quantity
may be further bounded by 
 $ 3 \cdot 5^{k-1} 2M\|C\|t $.
Let $N_A$ and $N_B$ be upper bounds to the number of exponentials 
comprising  $\widetilde{U}_\mathcal{A}(z_\ell /2M)$ and $\widetilde{U}_\mathcal{B} (z_\ell /M)$, respectively, for any $\ell$. 
Then the resulting total number of exponentials in Algorithm $2$ (in $\widetilde{U}$) is 
\begin{equation} \label{eqn:totalCostGen}
N\leq (2N_A + N_B)5^{k-1} 2 M\|C\|t.
\end{equation}

In order to obtain estimates to $N_A$ and $N_B$ we turn to the second-step error, where we require $\| \widehat{U} - \widetilde{U} \| \leq \e/2$. 
We have 
\begin{eqnarray}  \label{eqn:UhatError2}
\| \widehat{U} - \widetilde{U}\| &=& \| \; S_{2k}(A/\|C\|,B/\|C\|,1/M)^{n_0} S_{2k}(A/\|C\|,B/\|C\|,\delta/M) -  \nonumber \\
& &\;\;\; \widetilde{S}_{2k}(A/\|C\|,B/\|C\|,1/M)^{n_0} \widetilde{S}_{2k}(A/\|C\|,B/\|C\|,\delta/M) \; \| \nonumber\\
&\leq& n_0 \| S_{2k}(A,B,1/M) - \widetilde{S}_{2k}(A,B,1/M) \| \nonumber \\
&+& \| S_{2k}(A,B,\delta /M) - \widetilde{S}_{2k}(A,B,\delta /M) \|.
\end{eqnarray}
Observe the quantity $\| S_{2k}(A,B,1/M) - \widetilde{S}_{2k}(A,B,1/M) \|$ is equal to
$$ \bigg \|  \prod_{\ell=1}^K S_2(A,B, z_\ell /M) - \prod_{\ell=1}^K \widetilde{U}_A(z_\ell /2M)\widetilde{U}_B(z_\ell /2M)\widetilde{U}_A(z_\ell /M)  \bigg \|  $$
which is at most 
$$ 2 \sum_{\ell=1}^K \| e^{-iA z_\ell /2M} - \widetilde{U}_A (z_\ell /2M)  \| + \sum_{\ell=1}^K \| e^{-iBz_\ell /M} - \widetilde{U}_B (z_\ell /M) \|. $$  

The $\widetilde{U}_\mathcal{A} $ and $\widetilde{U}_\mathcal{B}$ are given by splitting formulas over time intervals of size $z_\ell / 2M$, $z_\ell / M$, $z_\ell \delta / 2M$ and  $z_\ell \delta / M$, which vary with the $z_\ell$. 
Since $\delta < 1$ 
we bound the second term in (\ref{eqn:UhatError2}) to get 
\begin{equation}   \label{eq:2ndLevelError}
\| \widehat{U} - \widetilde{U}\| \leq (n_0+\delta) \left(2 \sum_{\ell=1}^K \| e^{-iA z_\ell /2M} - \widetilde{U}_A (z_\ell /2M)  \| + \sum_{\ell=1}^K \| e^{-iBz_\ell /M} - \widetilde{U}_B (z_\ell /M)  \| \right).
\end{equation}
Thus, sufficient conditions for $\| \widehat{U} - \widetilde{U}\| \leq \e/2$ are
  $$\sup_{1\leq \ell \leq K} \| e^{-iA z_\ell /2M} - \widetilde{U}_A (z_\ell/2M)  \|  \leq \frac{\e}{8(n_0+\delta)K} =\frac{\e}{8M\|C\|t 5^{k-1}} \; ,$$    
  $$\sup_{1 \leq \ell \leq K} \| e^{-iBz_\ell /2M} - \widetilde{U}_B (z_\ell/2M)  \|  \leq \frac{\e}{4(n_0+\delta)K} =\frac{\e}{4M\|C\| t 5^{k-1}}\; .$$    

We next explain how to select the 
subintervals for applying 
$\widetilde{U}_\mathcal{A} $ and $\widetilde{U}_\mathcal{B}$, 
keeping in mind that we eventually select the same values of $M_A$ and $M_B$ in all resulting time intervals 
due to the upper bounds (\ref{eqn:MAalg2}, \ref{eqn:MBalg2}) below.  
Note that selecting $M_A$ or $M_B$ to be larger than necessary can only reduce the simulation error.
Thus, for convenience, we select $M_A$ and $M_B$ uniformly and  large enough so that the resulting worst-case errors are sufficiently small. 

In particular, consider $\widetilde{U}_\mathcal{A} (z_\ell /2M)$ which 
approximately simulates $A/\|C\|$ for time $\frac{z_\ell }{2M}$.
This amount of time we further subdivide in $M_A(z_\ell /2M)$ slices. 
From \cite{PZ12}, 
the error will be at most 
$\frac{\e}{8M\|C\|tK}$ 
if using (\ref{eqn:def0Mnew}) 
we set 
\begin{eqnarray*}
M_A\left(\frac{z_\ell}{2M} \right) &:=& \left(   \frac{4em' ( |z_\ell| /2M) \|H_2\|/\|C\| }{(\e/8M\|C\|tK)}  \right)^{1/2k_A}  \frac{4em'}{3} \left( \frac53 \right)^{k_A-1} \\
&=& \left(   \frac{16e K |z_\ell| m' t \|H_2\| }{\e}  \right)^{1/2k_A}  \frac{4em'}{3} \left( \frac53 \right)^{k_A-1}.
\end{eqnarray*}
Importantly, observe that the factors of $M$ and $\|C\|$ have canceled. 
Using that 
\cite[App. A]{wiebe2010higher}
$$ \frac{1}{3^{k-1}}  \leq \; |z_\ell |  \;  \leq \frac{4k}{3^{k}},$$
we have
\begin{equation}    \label{eqn:alg2bound}
\left( \frac53  \right)^{k-1}  \leq   |z_\ell| K \leq \frac{4}{5} k \left(\frac{5}{3}\right)^{k}. 
\end{equation}
Therefore 
\begin{equation}  \label{eqn:MAalg2}
M_A( z_\ell /2M)
\leq  \left( \frac{64e}{5} k \frac{ m' t \|H_2\| }{\e}  \right)^{1/2k_A}  \frac{4em'}{3} \left( \frac53 \right)^{k_A-1+k/2k_A} =: M_A , \end{equation} 
for all $\ell$. 
Hence, we will split every time interval of size  $z_\ell / 2M$, $\ell=1,\dots,K$ into $M_A$ subintervals. 

To bound the cost of each $\widetilde{U}_\mathcal{A} (z_\ell /2M)$, we apply \cite[Thm. 2]{PZ12}. 
The theorem assumes 
$ 4 e m' \left(\frac{|z_\ell|}{2M} \right)\frac{\|H_2\|}{\|C\|}\geq  \frac{\e}{8M\|C\|tK} $, or equivalently 
$ 16e m' \|H_2\| t K  |z_\ell|  \geq \e,$
where again the $M$ and $\|C\|$ factors have canceled. 
Since in the statement of the proposition we have assumed that 
\begin{equation}   \label{eqn:condA2}
16e m' \|H_2\| t \geq \e,
\end{equation}
we can apply 
\cite[Thm. 2]{PZ12}. 
Hence, 
the number of exponentials for each $\widetilde{U}_\mathcal{A} (z_\ell/2M)$, $\ell=1,\dots,K$, is at most
$$(2m'-1)5^{k_A-1} \bigg \lceil M_A \frac{\|H_1\|}{\|C\|} \frac{ |z_\ell |}{2M} \bigg \rceil 
\leq 2m'5^{k_A-1} \bigg \lceil \frac{M_A}{M} \frac{\|H_1\|}{\|C\|}  \frac{2k}{3^k} \bigg \rceil =:N_A. $$ 
%
Note that the argument of this ceiling function may be greater than or less than one, depending on the problem instance and algorithm parameters. 
In the latter case, the time intervals of length $z_\ell/2M$ do not require any subdivision at all. 

We now consider $\widetilde{U}_{\mathcal{B}}(z_\ell / M)$ 
which approximately simulates $B/\|C\|$ for time $z_\ell / M$, 
and proceed similarly. 
to give error at most $\frac{\e}{4M\|C\|tK}$ we select $M_B$ from (\ref{eqn:def0Mnew}) to give
\begin{eqnarray}   \label{eqn:MBalg2}
M_B(z_\ell/M) &=& \left(   \frac{4em' (|z_\ell| /M) \|H_{m'+2}\|/\|C\| }{(\e/4M\|C\|tK)}  \right)^{1/2k_B}  \frac{4e(m-m')}{3} \left( \frac53 \right)^{k_B-1} \nonumber \\
&=& \left(   \frac{16e K |z_\ell| (m-m') t \|H_{m'+2}\| }{\e}  \right)^{1/2k_B}  \frac{4e(m-m')}{3} \left( \frac53 \right)^{k_B-1} \nonumber \\
&\leq &  \left( \frac{64e}{5} k \frac{(m-m') t \|H_{m'+2}\| }{\e}  \right)^{1/2k_B}  \frac{4e(m-m')}{3} \left( \frac53 \right)^{k_B-1+k/2k_B}.
\end{eqnarray}
We define $M_B$ to be the right-hand side of the final equation. 
Observe that the factors of $M$ and $\|C\|$ have again canceled, and $M_B$ is of the same form as $M_A$.  

To apply \cite[Thm. 2]{PZ12} to bound the cost of any $\widetilde{U}_\mathcal{B} (z_\ell /M)$, we require 
$$ 4 e (m-m') \left(\frac{|z_\ell|}{M} \right)\frac{\|H_{m'+2}\|}{\|C\|}\geq  \frac{\e}{4M\|C\|tK} ,$$ 
or equivalently, 
$ 16e (m-m') \|H_{m'+2}\| t K  |z_\ell|  \geq \e,$
which is valid because we have assumed that 
\begin{equation}   \label{eqn:condB2}
16e (m-m') \|H_{m'+2}\|  t \geq \e. 
\end{equation}
The number of exponentials for each $\widetilde{U}_\mathcal{B} (z_\ell/M)$ is at most
$$(2(m-m')-1)5^{k_B-1} \bigg \lceil M_B \frac{\|H_{m'+1}\|}{\|C\|} \frac{|z_\ell|}{M} \bigg \rceil \leq 2m'5^{k_A-1} \bigg \lceil \frac{M_B}{M} \frac{\|H_{m'+1}\|}{\|C\|}  \frac{4k}{3^k} \bigg \rceil =:N_B.$$
Thus, from (\ref{eqn:totalCostGen}) we have that the total cost (total number of exponentials) 
is at most 
\begin{eqnarray*}   \label{eqn:totalCostAlg2inProof}
N &\leq& 2M\|C\|t 5^{k-1} (2N_A + N_B)\\
&\leq& 
2M\|C\|t \; 5^{k-1} \left(  4m'5^{k_A-1} \bigg \lceil \frac{M_A}{M} \frac{\|H_1\|}{\|C\|}  \frac{2k}{3^k} \bigg \rceil + 2(m-m')5^{k_B-1} \bigg \lceil \frac{M_B}{M} \frac{\|H_{m'+1}\|}{\|C\|}  \frac{4k}{3^k} \bigg \rceil    \right).
\end{eqnarray*}
Letting 
$$\widetilde{n}= n_0+\delta=
 M\|C\|t = \|C\|t \left(   \frac{16et \|D\| }{\e}  \right)^{1/2k}  \frac{8e}{5} \left( \frac53 \right)^{k}, $$ 
 which is equal to the lower bound for $n$ as it appears in the statement of the proposition,
and 
$$ \widetilde{n}_A = M_A \|H_1\| t  \frac{2k}{3^k}, \;\;\;\;\;\;\;\;\;\;\;
\widetilde{n}_B = M_B \|H_{m'+1}\| t  \frac{4k}{3^k}, $$
the cost bound becomes (cf. (\ref{eqn:thmTotalCostAlg1}))
$$ N\leq  \;   8m'5^{k+k_A-2}  \widetilde{n} \;  \bigg \lceil \frac{\widetilde{n}_A}{\widetilde{n}}  \bigg \rceil 
+ 4(m-m')5^{k+k_B-2} \; \widetilde{n}  \bigg \lceil \frac{\widetilde{n}_B}{\widetilde{n}}  \bigg \rceil   .  $$
Again applying the inequality  $x \lceil y/x \rceil \leq \max \{ x, 2y \}$ for $x,y > 0$, we have
$$ N\leq  \;   8m'5^{k+k_A-2}  \max \{ \widetilde{n} , 2 \widetilde{n}_A \} 
+ 4(m-m')5^{k+k_B-2} \max \{ \widetilde{n} , 2 \widetilde{n}_B \}    .  $$
Finally, we use the inequality 
\begin{equation}   \label{eq:ineq}
k^{1/2k'} (5/3)^{k/2k'} k/3^k \leq  (35/16)  (3/5)^k \;\;\;\;\;\;\; \text{for } \;  k,k' \in \nat
\end{equation}
to define simpler quantities $n_A$ and $n_B$ as 
\begin{eqnarray*}
2\widetilde{n}_A &=& 2  \|H_1\| t M_A   \frac{2k}{3^k} 
=   \|H_1\| t \left( \frac{64e}{5} k \frac{ m' t \|H_2\| }{\e}  \right)^{1/2k_A}  \frac{16em'}{3} \left( \frac53 \right)^{k_A-1+k/2k_A}  \frac{k}{3^k} \\
&\leq&  m' \|H_1\| t \left(\frac{64e}{5} \frac{ m' t \|H_2\| }{\e}  \right)^{1/2k_A} 7e \left( \frac53 \right)^{k_A-k} =: n_A (k,k_A),
\end{eqnarray*}
and
\begin{eqnarray*}
2\widetilde{n}_B &=& 2  \|H_{m'+1}\| t M_B   \frac{4k}{3^k}  \\
&=&   \|H_{m'+1}\| t \left( \frac{64e}{5} k \frac{ (m-m') t \|H_{m'+2}\| }{\e}  \right)^{1/2k_B}  \frac{32e(m-m')}{3} \left( \frac53 \right)^{k_B-1+k/2k_B}  \frac{k}{3^k} \\
&\leq&  (m-m') \|H_{m'+1}\| t \left(\frac{64e}{5} \frac{ (m-m') t \|H_{m'+2}\| }{\e}  \right)^{1/2k_B} 14e \left( \frac53 \right)^{k_B-k} =: n_B (k,k_B).
\end{eqnarray*}
Applying these estimates, the cost bound becomes
\begin{equation}
N\leq  \;   8m'5^{k+k_A-2}  \max \{ \widetilde{n} , n_A \} 
+ 4(m-m')5^{k+k_B-2} \max \{ \widetilde{n} , n_B \}    .  
\end{equation}
Clearly this inequality remains valid replacing $\widetilde{n}$ by any $n$ such that $n\ge\widetilde{n}$.
\end{proof}

\subsubsection*{Proof of Theorem \ref{thm:alg2generalPartition}}
As the analysis is similar to that of Proposition \ref{thm:alg2}, we give only the important parts. 
Recall the preliminary analysis given in Section \ref{sec:GeneralParadigmDetails}. 

Consider a Hamiltonian as in  (\ref{eq:HamDef}, \ref{eqn:HamOrdering}), partitioned into $\mu$ groups $H=A_1+\dots+A_\mu$ as in Section \ref{sec:GeneralParadigmDetails}, 
labeled such that $\|A_1\|\geq \|A_2\| \geq \dots \geq \|A_\mu\|$.
We approximate $U=e^{-Ht}$ with $\widetilde{U}$ given in (\ref{eqn:UtildeGen}), i.e. 
\begin{equation*}  
\widetilde{U}:= (\widetilde{S}_{2k}(A_1,\dots,A_\mu,t/n))^n = \left( \prod^{N_{k,\mu}}_{\ell=1} \widetilde{U}_{A_{j_\ell}}(t_{\ell}/n)\right)^n,  \;\;\;\;\;\;  j_\ell \in \{1,\dots,\mu \}, \;\;\;\;  \sum^{N_{k,\mu}}_{\ell=1} t_{\ell}= \mu t. 
\end{equation*}
For the first-step error to be at most $\e/2$, we set 
$n=M\|A_1\|t$ with $M$ given as in
(\ref{eqn:defMgeneral}), i.e. 
\begin{equation*}
M=  \left(   \frac{8e \mu t \|A_2\| }{\e}  \right)^{1/2k}  \frac{4e\mu}{3}  \left( \frac53 \right)^{k-1},
\end{equation*} 
where in the statement of the theorem we have assumed $\mu t \|A_2\| \geq \e$.
Note that we do not require $n\in\nat$; however, as evident from 
(\ref{eqn:totalCostGen}) and (\ref{eq:2ndLevelError}), this assumption does not affect our analysis. 

From (\ref{eqn:errorSecondTermGeneralCase}) the second-step error is at most $\e/2$ if the error of each subroutine satisfies 
$$\| e^{-iA_{j_\ell} t_{\ell}/n} - \widetilde{U}_{A_{j_\ell}} (t_{\ell}/n)  \| \leq \frac{\e}{4\mu 5^{k-1}n }   .$$  
Thus for each $j=1,\dots,\mu$ we select
\begin{eqnarray*}
M_{A_j} \left(\frac{z_\ell}{2M} \right)
&=& \left(   \frac{4em_j ( |z_\ell| /2M) \|H_{(j,2)}\|/\|A_1\| }{(\e/4\mu M\|A_1\|tK)}  \right)^{1/2k_j}  \frac{4em_j}{3} \left( \frac53 \right)^{k_j-1} \\
&=& \left(   \frac{8e \mu K |z_\ell| m_j t \|H_{(j,2)}\| }{\e}  \right)^{1/2k_j}  \frac{4em_j}{3} \left( \frac53 \right)^{k_j-1} \\
&\leq&   \left(   \frac{32e}{5} k \frac{\mu  m_j t \|H_{(j,2)}\| }{\e}  \right)^{1/2k_j}  \frac{4em_j}{3} \left( \frac53 \right)^{k_j-1+k/2k_j}
=: M_{A_j},
\end{eqnarray*}
for all $\ell=1,\dots,K$, where in the statement of the theorem we have assumed $\mu  m_j t \|H_{(j,2)}\| \geq \e$, $j=1,\dots,\mu$. 
Hence, 
the number of exponentials in $\widetilde{U}_{\mathcal{A}_j} (z_\ell/2M)$  is at most
$$(2m_j-1)5^{k_j-1} \bigg \lceil M_{A_j} \frac{\|H_{(j,1)}\|}{\|A_1\|} \frac{ |z_\ell |}{2M} \bigg \rceil 
\leq 2m_j 5^{k_j-1} \bigg \lceil \frac{M_{A_j}}{M} \frac{\|H_{(j,1)}\|}{\|A_1\|}  \frac{2k}{3^k} \bigg \rceil =:N_{A_j}. $$ 

Recalling (\ref{eqn:overallCostGen}), from (\ref{eqn:totalCostGen}) we have that the total cost 
is at most 
\begin{eqnarray*}  
N &\leq& 2 n 5^{k-1} \sum_{j=1}^\mu 2 N_{A_j} 
%
\leq 2  n 5^{k-1}   \sum_{j=1}^\mu 4 m_j5^{k_j-1} \bigg \lceil \frac{M_{A_j}}{M} \frac{\|H_{(j,1)}\|}{\|A_1\|}  \frac{2k}{3^k} \bigg \rceil  \\
&\leq&
  8     \sum_{j=1}^\mu  m_j5^{k+k_j-2} n \bigg \lceil \frac{n_{j}}{n}    \bigg \rceil 
%
\leq  8     \sum_{j=1}^\mu  m_j5^{k+k_j-2} \max \{n ,2 n_j \},
\end{eqnarray*}
where $n_j : = M_{A_j} \|H_{(j,1)}\| t \;2k /3^k $. 

We again apply (\ref{eq:ineq}) to give the simpler quantities 
\begin{eqnarray*}
2n_j &=& 2M_{A_j} \|H_{(j,1)}\| t   \frac{2k}{3^k} 
=  \|H_{(j,1)}\| t \left( \frac{32e}{5} k \frac{ \mu m_j t \|H_{(j,2)}\| }{\e}  \right)^{1/2k_j}  \frac{16em'}{3} \left( \frac53 \right)^{k_A-1+k/2k_j}  \frac{k}{3^k} \\
&\leq&  m_j \|H_{(j,1)}\| t \left( \frac{32e}{5}  \frac{ \mu m_j t \|H_{(j,2)}\| }{\e}  \right)^{1/2k_j}  7e \left( \frac53 \right)^{k_j-k} =: n_{A_j} (k,k_j),
\end{eqnarray*}

which gives the cost bound (\ref{eqn:thmTotalCostAlg2gen}), i.e., 
\begin{equation} 
N  \leq 8 \sum_{j=1}^\mu  5^{k+k_j -2 } m_j  \max\{ n(k), n_{A_j} (k,k_j) \} =: \eta (k,k_1,\dots,k_\mu)  . 
\end{equation}

\subsection*{Proof of (\ref{eqn:speedupLimit})}   
\begin{proof} 
Consider all problem parameters to be fixed except $m,m'$. 
Observe that (\ref{eqn:thmTotalCostAlg2}) contains two maximum functions, and hence we have four cases to consider with respect to the relative magnitudes of $n(k)$, $n_A(k,k_A)$, and $n_B(k,k_B)$. 
Recall $n(k)$ 
is given in (\ref{eq:localn}). 
Here we estimate the maximum function by the sum of its arguments to get 
\begin{eqnarray*} 
 \eta(k,k_A,k_B) &\leq&
 8m' 5^{k+k_A-2} \; ( n_A(k,k_A) +  n(k) )  +
4(m-m')5^{k+k_B-2}  \; (n_B(k,k_B) +  n(k) ) \\
&\leq&  8m' 5^{k+k_A-2} \;  n_A(k,k_A) + 4(m-m')5^{k+k_B-2}  \; n_B(k,k_B) \\
&+& (8m' 5^{k+k_A-2}+ 4(m-m')5^{k+k_B-2} ) n(k).
\end{eqnarray*}
Let $\eta^*$ denote the minimum of (\ref{eqn:thmTotalCostAlg2}) with respect to $k,k_A,k_B$. 
Let $k^{(max)},k_A^{(max)},k_B^{(max)}$ be defined as in Proposition~\ref{prop:optks} under the assumptions of Proposition~\ref{thm:alg2}.  
Using $\eta^* \leq \eta(k^{(max)},k^{(max)}_A,k^{(max)}_B)$, 
this gives 
\begin{eqnarray*} 
 N \leq  \eta^* 
&\leq&  8m' 5^{k^{(max)}+k^{(max)}_A-2} \; n_A(k^{(max)},k_A^{(max)}) \\
&+& 4(m-m')5^{k^{(max)}+k^{(max)}_B-2}  \; n_B(k^{(max)}, k_B^{(max)}) \\
&+& (8m'5^{k^{(max)}+k^{(max)}_A -2} +4(m-m')5^{k^{(max)} + k^{(max)}_B -2} ) \; n(k^{(max)})\\
&=& 3^{k^{(max)}} O\left( m'^2 \|H_1\|t \right) \cdot e^{2\sqrt{\frac12 \ln \frac{25}{3} \ln(\frac{64e}{5}m't\|H_2\|/\e)} } \\
&+& 3^{k^{(max)}} O\left( (m-m')^2 \|H_{m'+1}\|t \right) \cdot e^{2\sqrt{\frac12 \ln \frac{25}{3} \ln (\frac{64e}{5}(m-m')t\|H_{m'+2}\| / \e)} } \\
&+&  (5^{k_A^{(max)}} +5^{k_B^{(max)}} ) \: O((m-m')m'\|H_1\|t)\cdot e^{2\sqrt{\frac12 \ln \frac{25}{3} \ln (16e(m-m')t\|H_{m'+1}\|/\e)} }. 
\end{eqnarray*}
The quantities under the square roots are derived as in \cite{PZ12}. 

Next observe that from (\ref{assump:AgreaterThanB}) we have 
$m'\|H_2\| \geq (m-m')\|H_{m'+1}\| \geq (m-m')\|H_{m'+2}\|$ and 
$(m-m')\|H_{m'+1}\| = O(m'\|H_1\|)$. 
Using the bound 
$a^{\sqrt{b}} \leq \sqrt{a^{b+1}}$ for $a,b \geq 1$
we have 
$$\max\{3^{k^{(max)}},5^{k_A^{(max)}},5^{k_B^{(max)}} \} \leq \sqrt{  5^{1+\frac12 \log_{25/3} (16em't\|H_2\|/\e) } } 
\leq \sqrt{5} \; (16em't \|H_2\|/\e)^{0.2},    
$$
which gives 
\begin{eqnarray*} 
 N \leq  \eta^* 
&=& O((m'\|H_2\|t/\e)^{0.2}) \cdot 
O\left( m m' \|H_1\|t \right) \cdot e^{2\sqrt{\frac12 \ln \frac{25}{3} \ln (16em't\|H_2\|/\e)} }.
\end{eqnarray*}
Hence, using $N^*_{prev}$ as 
defined in (\ref{eqn:PZ12boundOptimalk})
we have 
\begin{eqnarray*}
\frac{N}{N^*_{prev}}  \leq \frac{\eta^{*}}{N^*_{prev}} 
 &=&O((m'\|H_2\|t/\e)^{0.2}) \cdot  
  O\left( \frac{m'}{m} \right)  \cdot \frac{e^{2\sqrt{\frac12 \ln \frac{25}{3} \ln (16em't\|H_2\|/\e)} } }{ e^{2\sqrt{\frac12 \ln \frac{25}{3} \ln (4emt\|H_2\|/\e)} } } . 
\end{eqnarray*}

Observe that for $a<b$, the function 
$e^{a \sqrt{ x} - b\sqrt{x}} \rightarrow 0$ as $x\rightarrow \infty$. 
Thus, assuming $m'=O(m^{5/6})$, we have 
\begin{equation}
\frac{N}{N^*_{prev}} \leq \frac{\eta^* }{N^*_{prev}} \xrightarrow[m\rightarrow \infty]{}   0.
\end{equation}
\end{proof}

%% file: _app_QAOA.tex
\chapter{Quantum Approximate Optimization}    \label{app:QAOA}

In this appendix 
we provide proofs for several results from Chapter \ref{ch:QAOAperformance}.

\begin{proof}[Proof of Theorem \ref{thm:maxCutAngles}]
For a $D$-regular triangle free graph, setting partial derivatives of (\ref{eq:triangleFreeExpec}) to zero shows that 
$\tan^2 \gamma = \frac{1}{D-1}$ for $\gamma$ to be stationary, and thus 
every optimal $\gamma^*$ is of the form  $\pm \arctan \frac{1}{\sqrt{D-1}} + \ell \pi$ for some $\ell \in \integers$.
In particular, $\gamma = \ell \pi + \arctan \frac{1}{\sqrt{D-1}}$ is optimal when $\sin(4\beta^*)= 1$,
and $\gamma = \ell \pi - \arctan \frac{1}{\sqrt{D-1}}$ is optimal when $\sin(4\beta^*)=- 1$.
Taking second derivatives shows $(\gamma^*,\beta^*) = ( \arctan \frac{1}{\sqrt{D-1}}, \pi/8)$ indeed gives the smallest maximum. 

To classify the optimal angles for this case, observe that  (\ref{eq:triangleFreeExpec}) may be written as a $\langle C\rangle (\gamma,\beta) = \frac{m}{2} + f(\gamma)g(\beta)$, where $f(\gamma)$ and $g(\beta)$ are odd periodic functions. 
Thus $\langle C\rangle$ is maximized when both $f$,$g$ are maximized, or when both $f$,$g$ are minimized. 
Indeed, the transformations $\beta \to -\beta^*$ and $\gamma \to -\gamma$ flip the signs of $g(\beta)$ and $f(\gamma)$, and hence $(-\gamma^*,-\beta^*)$ is also optimal. 
Moreover, as $\sin (4\beta)$ is $\frac{\pi}{2}$-periodic, the pair $(\gamma^*, \beta^* + \ell \frac{\pi}{2})$ is also optimal for any $\ell \in \integers$.  Combining these facts with the observation that $f(\gamma)$ is $\pi$-periodic when $D$ is even and $2\pi$-periodic when $D$ is odd gives the stated result. 

Finally, observe that $\sum_D D\; n_D = 2m$ for any graph with $m$ edges and $n_D$ vertices of degree $D$. As $D n_D \geq 0$, we may write (\ref{eq:triangleFreeExpec3}) as a convex combination of (\ref{eq:triangleFreeExpec}) for different values of $D=d+1$, from which the third point follows from the second. 
\end{proof}

\begin{proof}[Proof of Theorem \ref{thm:MaxDiCut}]
We compute $\langle C \rangle$ using the Pauli Solver 
algorithm of Section \ref{sec:PauliSolver}, similarly to 
the proof of Theorem \ref{thm:generalMaxCut} for 
undirected MaxCut. 
We consider the general case first. 

Recall $D_u \subset D$ is the subset of directed edges containing $u$, and $U_u \subset U$ are the  \lq undirected\rq\ edges containing $u$, with $d_u=|D_u|$ and $e_u=|U_u|$. 
For convenience, we will write $(uv)\in D$ to indicate that \textit{one} of $|uv)\in D$ or $|vu)\in D$. 
From (\ref{eqn:CMaxDiCut}), 
let $C_u$ denote the 
terms in $C$ that contain vertex $u$, given by 
\begin{equation}  \label{eq:Cu}
C_u :=  \frac14 k_u Z_u -\frac14 \sum_{(uw)\in D_u} Z_uZ_w  -\frac12 \sum_{(ut)\in U_u} Z_tZ_u  .
\end{equation}
Note that with this definition $C \neq \sum_u C_u$. 

Let $c=\cos 2\beta$ and $s=\sin 2\beta$. 
For the single $Z$ terms in (\ref{eqn:CMaxDiCut}), observe that  
$$e^{i \beta B} Z_u e^{-i \beta B} = e^{2i \beta X_u} Z_u
=(I c + i s X_u)Z_u 
 =c Z_u + sY_u .$$
  Let $c' = \cos \gamma $, $s' = \sin \gamma$, 
$c'' = \cos \frac{\gamma}{2}$, and $s'' = \sin \frac{\gamma}{2}$, 
 and for each vertex $u$ let $c'_u = \cos \frac{\gamma k_u}{2}$ and $s'_u = \sin \frac{\gamma k_u}{2}$. 
The $Z_u$ term above commutes with $e^{i \gamma C}$ and thus contributes nothing to the expectation value of $C$ (for $p=1$). 
The $Y_u$ term anti-commutes with each term in $C_u$, and commutes with the remaining terms in $C$, so we have
\begin{eqnarray*}
 \bra{s} e^{i\gamma C} Y_{u} e^{-i\gamma C}  \ket{s} 
 &=&  \bra{s} e^{2i\gamma C_{u}} Y_{u}  \ket{s} \\ 
 &=& \bra{s}  e^{ \frac{i}{2} \gamma k_u Z_u  }  
e^{- \frac{i}{2} \gamma \sum_{(uw) \in D_u} Z_u Z_w } 
e^{- i \gamma \sum_{(uw) \in U_u} Z_u Z_w } 
Y_{u}  \ket{s},  \\
&=& \bra{s} (I c'_u + i s'_u Z_u)   
\prod_{i=1}^{d_u} (I c'' - i s'' Z_uZ_{w_i}) 
 \prod_{j=1}^{e_u} (I c' - i s' Z_uZ_{w_j}) 
  Y_{u} \ket{s}.
\end{eqnarray*}
Expanding the product on the right hand side of the last line again gives a sum of tensor products of Pauli operators. 
Recall that only terms not containing a $Y$ or a $Z$ factor 
contribute to 
$\langle C \rangle$, for initial state $\ket{s}=\ket{+}^{\otimes n}$. 
As the only vertex in common between the edges of $D_u$ and $U_u$ is $u$ itself, the only term that can contribute is proportional to $Z_u * I^{\otimes d_u} * I^{\otimes e_u} * Y_u = -i X_u$,  
so we have
\begin{equation*}
 \bra{s} e^{i\gamma C} Y_{u} e^{-i\gamma C}  \ket{s}  
 = \bra{s} i s'_u  c''^{d_u} c'^{e_u} (-i X_u)  \ket{s} =    s'_u (c'')^{d_u }(c')^{e_u}  ,
\end{equation*}
and hence 
\begin{equation}
 \bra{s} e^{i\gamma C} e^{i \beta B}  Z_u e^{-i\beta B} e^{-i\gamma C}  \ket{s}   =  s  s'_u (c'')^{d_u }(c')^{e_u} .
\end{equation}
Turning to the $ZZ$ terms in $C$ in (\ref{eqn:CMaxDiCut}), we have 
\begin{equation}  \label{eq:MaxDiCutZZterm}
e^{i \beta B} Z_uZ_v e^{-i \beta B} = e^{2i \beta X_u} e^{2i \beta X_v} Z_uZ_v
 =c^2 Z_uZ_v + sc (Y_uZ_v + Z_uY_v) + s^2 Y_uY_v.
\end{equation}
The first term $c^2Z_uZ_v$ on the right  commutes 
with $e^{i \gamma C}$ and 
contributes nothing to $\langle C \rangle$. 
 We conjugate each remaining term in (\ref{eq:MaxDiCutZZterm}) separately by $e^{i \gamma C}$. 
We have
\begin{eqnarray}  \label{eq:YZtermDiMaxCut}
 \bra{s} e^{i\gamma C} Y_{u}Z_v e^{-i\gamma C}  \ket{s} 
 &=&  \bra{s} e^{2i\gamma C_{u}} Y_{u}Z_v  \ket{s}  \\ 
 &=& \bra{s}  e^{ \frac{i}{2} \gamma k_u Z_u  }  
e^{- \frac{i}{2} \gamma \sum_{(uw) \in D_u} Z_u Z_w } 
e^{- i \gamma \sum_{(uw) \in U_u} Z_u Z_w } 
Y_{u}Z_v  \ket{s}  \nonumber\\
&=& \bra{s} (I c'_u + i s'_u Z_u)   
\prod_{i=1}^{d_u} (I c'' - i s'' Z_uZ_{w_i}) 
 \prod_{j=1}^{e_u} (I c' - i s' Z_uZ_{w_j}) 
  Y_{u}Z_v \ket{s},\nonumber
\end{eqnarray}
for which we have two cases depending on if the edge $(uv)$ is directed or undirected.

First suppose 
$(uv) \; \in D$, i.e., one of the edges $|uv)$ or $|vu)$ is in the graph, and $(uv) \; \notin U$. 
Then 
(\ref{eq:YZtermDiMaxCut}) becomes 
$$ 
 \bra{s} (I c'_u + i s'_u Z_u)   
 (I c'' - i s'' Z_uZ_{v}) 
\prod_{i=1}^{d_u-1} (I c'' - i s'' Z_uZ_{w_i}) 
 \prod_{j=1}^{e_u} (I c' - i s' Z_uZ_{w_j}) 
  Y_{u}Z_v \ket{s}. $$
Expanding the product 
again gives a sum of Pauli terms.  
Similar to the undirected case, 
the only term that can contribute is proportional to $I*Z_uZ_v * I^{\otimes d_u -1} * I^{\otimes e_u} * Y_uZ_v = -i X_u$, so 
\begin{equation}
 \bra{s} e^{i\gamma C} Y_{u}Z_v e^{-i\gamma C}  \ket{s}  = \bra{s} c'_u (-i s'') c''^{d_u -1} c'^{e_u} (-i X_u)  \ket{s} =  - s'' c'_u (c'')^{d_u -1}(c')^{e_u}  .
\end{equation}
On the other hand, suppose instead $(uv)\; \in U$. 
Then 
$\bra{s} e^{i\gamma C} Y_{u}Z_v e^{-i\gamma C}  \ket{s}$ is given by 
%
$$ 
 \bra{s} (I c'_u + i s'_u Z_u)   
 (I c' - i s' Z_uZ_{v}) 
\prod_{i=1}^{d_u} (I c'' - i s'' Z_uZ_{w_i}) 
 \prod_{j=1}^{e_u-1} (I c' - i s' Z_uZ_{w_j}) 
  Y_{u}Z_v \ket{s}, $$
where by the previous argument we now have 
\begin{equation}
\bra{s} e^{i\gamma C} Y_{u}Z_v e^{-i\gamma C}  \ket{s}  = \bra{s} c'_u (-i s'') c''^{d_u -1} c'^{e_u} (-i X_u)  \ket{s} =  - s' c'_u (c'')^{d_u }(c')^{e_u-1}  .
\end{equation}

By symmetry, for the $Z_{u}Y_v$ term this gives 
\begin{equation}
\bra{s} e^{i\gamma C} Z_{u}Y_v e^{-i\gamma C}  \ket{s}
=\begin{cases}
              - s'' c'_v (c'')^{d_v -1}(c')^{e_v} \;\;\;\;  (uv)\; \in D\\
              - s' c'_v (c'')^{d_v }(c')^{e_v-1}  \;\;\;\;\;  (uv)\; \in U.
            \end{cases}
\end{equation}

The final term in (\ref{eq:MaxDiCutZZterm}) to consider is  $\bra{s} e^{i\gamma C} Y_{u}Y_v e^{-i\gamma C}  \ket{s} $. 
We 
again have two cases.  
Suppose $(uv) \; \in D$. 
As $[YY,ZZ]=0$, the $Z_uZ_v$ terms in $e^{\pm i \gamma C}$ will commute through $Y_uY_v$ and cancel. 
Hence, 
define $\widetilde{C}_u = C_u + \frac14 Z_uZ_v$, which is $C_u$ as in (\ref{eq:Cu}), but without the $Z_uZ_v$ term. Similarly, define $\widetilde{C}_v = C_v + \frac14 Z_uZ_v$, $\widetilde{D}_u=D_u\backslash \{(uv)\}$, and $\widetilde{D}_v=D_v\backslash \{(uv)\}$.
Then $C_u$ and $C_v$ each anticommute with $Y_uY_v$, and we have 
\begin{eqnarray}   \label{eq:MaxDiCutYYterm}
 \bra{s} e^{i\gamma C} Y_{u}Y_v e^{-i\gamma C}  \ket{s}  
 &=&  \bra{s} e^{2i\gamma \widetilde{C}_{u}}  e^{2i\gamma \widetilde{C}_{v}} Y_{u}Y_v  \ket{s} \nonumber \\ 
 &=& \bra{s}  e^{ \frac{i}{2} \gamma k_u Z_u  }  
e^{- \frac{i}{2} \gamma \sum_{uw) \in \widetilde{D}_u } Z_u Z_w } 
e^{- i \gamma \sum_{(uw) \in U_u} Z_u Z_w } 
 \nonumber\\
 & & \cdot \; e^{ \frac{i}{2} \gamma k_v Z_v  }  
e^{- \frac{i}{2} \gamma \sum_{(vw) \in  \widetilde{D}_v} Z_v Z_w } 
e^{- i \gamma \sum_{(vt) \in U_v} Z_v Z_t } \;
Y_{u}Y_v  \ket{s},   \nonumber\\
&=& \bra{s} (I c'_u + i s'_u Z_u)   
\prod_{i=1}^{d_u-1} (I c'' - i s'' Z_uZ_{w_i}) 
 \prod_{j=1}^{e_u} (I c' - i s' Z_uZ_{w_j})  \nonumber \\
  & & \cdot \; (I c'_v + i s'_v Z_v)   
\prod_{i=1}^{d_v-1} (I c'' - i s'' Z_vZ_{w_i}) 
 \prod_{j=1}^{e_v} (I c' - i s' Z_vZ_{w_j}) 
  Y_{u}Y_v \ket{s}  \nonumber\\
  &=&  \bra{s} \left(I c'_uc'_v + i c'_v s'_u Z_u + i c'_u s'_v Z_v  -  s'_u s'_v Z_u Z_v \right)  \\
  && \cdot 
  \prod_{(ab)\in D_{uv}} (I c'' - i s'' Z_aZ_{b}) 
 \prod_{(ab) \in U_{uv}} (I c' - i s' Z_aZ_{b}) 
   Y_{u}Y_v \ket{s} ,\nonumber
\end{eqnarray}
where in the last line we have defined $D_{uv} = \widetilde{D}_u \cup  \widetilde{D}_v $ and 
$U_{uv}=U_u \cup U_v$, with $d_{uv} := |D_{uv}|= d_u + d_v -2$ and $e_{uv} := |E_{uv}|= e_u + e_v$. 
 
Consider each term in the first parenthesis $\left( c'_uc'_v I + i c'_v s'_u Z_u + i c'_u s'_v Z_v  -   s'_u s'_v Z_u Z_v \right)$ in the last line above. 
It is easy to see that the single $Z_u$ and $Z_v$ terms 
cannot combine with the remaining terms in the product to produce terms composed of $X$ and $I$ factors, so they contribute nothing to the expectation value and can be ignored.  
For the $I$ and $Z_u Z_v$ terms in the parenthesis, 
similar to the argument used in the proof of Theorem \ref{thm:generalMaxCut}, these terms can combine and contribute in many ways, depending on the number of triangles containing $(uv)$ in the graph. The $I$ term contribution depends on the number of ways to pick an odd number of triangles, as was the case in deriving (\ref{eqn:YYtermSumThm1proof}), and the $Z_u Z_v$ term depends on the number of ways to pick an even number of triangles. 
Unfortunately, our present situation is much more complicated, as we now have $4$ different types of triangles (from whether each of $(uw), (vw)$ are directed or undirected), and must consider all possible combinations of different triangle types. This leads to an analysis and result significantly more complicated than that of  (\ref{eqn:YYtermSumThm1proof}). Instead, we consider two special cases leading to simpler results, triangle-free and oriented graphs. 

First, suppose the graph is triangle-free. Then only the $Z_uZ_v$ term from the first parenthesis in (\ref{eq:MaxDiCutYYterm}) can contribute, and we have
$$\bra{s} e^{i\gamma C} Y_{u}Y_v e^{-i\gamma C}  \ket{s} = - s'_u s'_u Z_u Z_v  (c'')^{d_{uv}}(c')^{e_{uv}} Y_uY_v   =     s'_u s'_v (c'')^{d_u + d_v -2}(c')^{e_u + e_v} .$$
%
For the other case $(uv)\in U$, 
repeating the above argument gives 
$$\bra{s} e^{i\gamma C} Y_{u}Y_v e^{-i\gamma C}  \ket{s} =    s'_u s'_v (c'')^{d_u + d_v}(c')^{e_u + e_v -2} $$
Combining this with the above results gives the second part of the Theorem. 

Finally, we consider 
oriented graphs. In this case, every edge is in $D$ and $U$ is empty. 
Thus, the previous results for the edges in $D$ apply. Let $f=f_{uv}$ be the number of triangles in the graph containing edge $(uv)$. Recall that we define a triangle to be any three edges $(uv)$, $(uw)$, $(vw)$, independently of the direction of each edge. 
For this case, the derivation of all quantities up to the last one remain valid by setting $U=\emptyset$. 
For the last quantity $ \bra{s} e^{i\gamma C} Y_{u}Y_v e^{-i\gamma C} \ket{s}$, we have 
$$ 
\bra{s} \left( c'_uc'_v I + i c'_v s'_u Z_u + i c'_u s'_v Z_v  -   s'_u s'_v Z_u Z_v \right)    
\prod_{i=1}^{d_u-1} (I c'' - i s'' Z_uZ_{w_i})  
\prod_{i=1}^{d_v -1} (I c'' - i s'' Z_vZ_{w_i}) 
  Y_{u}Y_v \ket{s} .$$
By the previous argument, only the $I$ and $Z_uZ_v$ terms in the first parenthesis can possibly contribute. 
By inspection, the $I$ term contribution follows identically from the derivation of (\ref{eqn:YYtermThm1proof}) to give (cf. (\ref{eqn:YYtermSumThm1proof}))
$$ \bra{s}    
\prod_{i=1}^{d_u-1} (I c'' - i s'' Z_uZ_{w_i})  
\prod_{i=1}^{d_v-1} (I c'' - i s'' Z_vZ_{w_i}) 
  Y_{u}Y_v \ket{s} 
  =  \frac12 (c'')^{d_u + d_v -2 - 2f}  (1- (c')^f)
  .$$
By similar arguments (now summing even numbers of triangles rather than odd), and using the formula for the sum over even indices  
$\sum_{i=0,2,4,\dots}^{f} a^{f-i} b^i = \frac12 ((a+b)^f + (a-b)^f)$, we have  
$$ \bra{s}    Z_u Z_v
\prod_{i=1}^{d_u-1} (I c'' - i s'' Z_uZ_{w_i})  
\prod_{i=1}^{d_v-1} (I c'' - i s'' Z_vZ_{w_i}) 
  Y_{u}Y_v \ket{s} 
  =  - \frac12 (c'')^{d_u + d_v -2 - 2f}  (1+ (c')^f).
  $$
Hence, 
$$ \bra{s} (e^{i\gamma C} Y_{u}Y_v e^{-i\gamma C} \ket{s} = \frac12 (c'')^{d_u + d_v -2 - 2f} 
\left(  c'_u c'_v (1- c'^f)  + s'_u s'_v  (1+ c'^f)    \right).$$
Thus, 
putting all these quantities together and applying some basic trigonometric identities 
gives the first statement of the theorem. 
\end{proof}